\documentclass[a4paper,12pt]{article}
\pdfoutput=1
\usepackage{geometry}
\usepackage{amsmath,amssymb,amsfonts}
\usepackage{xcolor,graphicx,cite,soul}
\usepackage{array,booktabs,longtable}
\usepackage{caption,microtype}
\usepackage{subcaption}
\usepackage{hyperref}
\usepackage{datetime}
\usepackage{appendix}

\usepackage{multirow}
\usepackage{cancel}

\makeatletter
\providecommand{\tabularnewline}{\\}

\makeatother

\newcommand{\lsim}
{\;\raisebox{-.3em}{$\stackrel{\displaystyle <}{\sim}$}\;}
\newcommand{\gsim}
{\;\raisebox{-.3em}{$\stackrel{\displaystyle >}{\sim}$}\;}

\newcommand\al{\alpha}
\newcommand\be{\beta}
\newcommand\tb{\tan\beta}

\newcommand\CBA{c_{\beta - \alpha}}
\newcommand\SBA{s_{\beta - \alpha}}

\newcommand\ReDiag{\mathop{%
  \raise .5pt\hbox{[}%
  \widetilde{\mathrm{Re}}%
  \raise .5pt\hbox{]}}}
\newcommand\ReOffDiag{\mathop{%
  \raise .5pt\hbox{$\llbracket$}%
  \widetilde{\mathrm{Re}}%
  \raise .5pt\hbox{$\rrbracket$}}}

\newcommand\cL{{\cal L}}

\newcommand\MZ{m_Z}
\newcommand\Mh{m_h}
\newcommand\MH{m_H}
\newcommand\MA{m_A}
\newcommand\MHp{m_{H^\pm}}

\newcommand\msq{m_{12}^{2}}

\newcommand\refeq[1]{Eq.~(\ref{#1})}
\newcommand\refeqs[1]{Eqs.~(\ref{#1})}
\newcommand\refta[1]{Tab.~\ref{#1}}
\newcommand\refse[1]{Sect.~\ref{#1}}

\newcommand\citere[1]{Ref.~\cite{#1}}
\newcommand\citeres[1]{Refs.~\cite{#1}}

\newcommand{\CP}{{\cal CP}}
\newcommand{\cp}{{\CP}}

\newcommand{\tev}{\,\, \mathrm{TeV}}
\newcommand{\gev}{\,\, \mathrm{GeV}}

\newcommand\HB{\texttt{HiggsBounds}}

\newcommand\fb{\ensuremath{\,\mbox{fb}}}
\newcommand\ab{\ensuremath{\,\mbox{ab}}}

\newcommand\iab{\ensuremath{\ab^{-1}}}

\newcommand{\br}{\text{BR}}
\newcommand{\De}{\Delta}

\newcommand{\sig}{\sigma}

\def\order#1{\ensuremath{{\cal O}(#1)}}
\def\reffi#1{\mbox{Fig.~\ref{#1}}}
\def\reffis#1{\mbox{Figs.~\ref{#1}}}

\def\la{\lambda}
\newcommand\kala{\ensuremath{\kappa_{\lambda}}}
\newcommand\laSM{\ensuremath{\lambda_{\mathrm{SM}}}}
\newcommand{\lahhh}{\ensuremath{\la_{hhh}}}
\newcommand{\lahhH}{\ensuremath{\la_{hhH}}}
\newcommand{\lahHH}{\ensuremath{\la_{hHH}}}
\newcommand{\lahAA}{\ensuremath{\la_{hAA}}}
\newcommand{\laHAA}{\ensuremath{\la_{HAA}}}
\newcommand{\lahHpHm}{\ensuremath{\la_{hH^+H^-}}}
\newcommand{\laHHH}{\ensuremath{\la_{HHH}}}

\newcommand{\inter}[2]{\ensuremath{[#1, #2]}}

\definecolor{Orange}{named}{orange}
\definecolor{Purple}{named}{purple}
\definecolor{Lightblue}{cmyk}{0.9,0.1,0.1,0.3}
\definecolor{dgelborange}{cmyk}{0.,0.3,0.5, 0.}
\definecolor{Lila}{rgb}{0.5,0.,1}
\definecolor{Darkgreen}{rgb}{0.,.7,0.2}

\graphicspath{{figs/}}
\allowdisplaybreaks
\begin{document}
\thispagestyle{empty}

\def\thefootnote{\fnsymbol{footnote}}

\begin{flushright}
\mbox{}
IFT--UAM/CSIC-21-020 \\
arXiv:2106.11105 [hep-ph]
\end{flushright}

\vspace{0.5cm}

\begin{center}

{\large\sc 
{\bf Sensitivity to Triple Higgs Couplings via Di-Higgs
  Production\\[.5em]
  in the 2HDM at \boldmath{$e^+e^-$} Colliders}}\\  

\vspace{1cm}

{\sc
F.~Arco$^{1,2}$%
\footnote{email: Francisco.Arco@uam.es}%
, S.~Heinemeyer$^{2,3,4}$%
\footnote{email: Sven.Heinemeyer@cern.ch}%
~and M.J.~Herrero$^{1,2}$%
\footnote{email: Maria.Herrero@uam.es}%
}

\vspace*{.7cm}

{\sl
$^1$Departamento de F\'isica Te\'orica, 
Universidad Aut\'onoma de Madrid, \\ 
Cantoblanco, 28049, Madrid, Spain

\vspace*{0.1cm}

$^2$Instituto de F\'isica Te\'orica (UAM/CSIC), 
Universidad Aut\'onoma de Madrid, \\ 
Cantoblanco, 28049, Madrid, Spain

\vspace*{0.1cm}

$^3$Campus of International Excellence UAM+CSIC, 
Cantoblanco, 28049, Madrid, Spain 

\vspace*{0.1cm}

$^4$Instituto de F\'isica de Cantabria (CSIC-UC), 
39005, Santander, Spain

}

\end{center}

\vspace*{0.1cm}

\begin{abstract}
\noindent
An important task at future colliders is the investigation of the
Higgs-boson sector. Here the measurement of the triple Higgs
coupling(s) plays a special role. 
Based on previous analyses, within the framework of Two Higgs Doublet
Models (2HDM) type~I and~II,  we define and analyze several 
two-dimensional benchmark planes,
that are over large parts in agreement with all theoretical and
experimental constraints. 
For these planes we evaluate di-Higgs production cross sections
at future high-energy $e^+e^-$ colliders, such as ILC or CLIC.
We consider two different channels for the neutral di-Higgs pairs
$h_i h_j=hh,hH,HH,AA$: $e^+e^- \to h_i h_j Z$ and
$e^+e^- \to h_i h_j \nu \bar \nu$.
In both channels the various triple Higgs-boson
couplings contribute substantially. 
We find regions with a strong enhancement of the production channel of 
two SM-like light Higgs bosons and/or with very large production cross
sections involving one light and one heavy or two heavy 2HDM Higgs
bosons, offering interesting prospects for the ILC or CLIC. 
The mechanisms leading to these enhanced production cross sections
are analyzed in detail.  We propose the use of cross section
distributions with the invariant mass 
of the two final Higgs bosons where the contributions from
intermediate resonant and non-resonant BSM Higgs bosons play a
crucial role. 
We outline which process at which center-of-mass energy would be best
suited to probe the corresponding triple Higgs-boson couplings.
\end{abstract}


\def\thefootnote{\arabic{footnote}}
\setcounter{page}{0}
\setcounter{footnote}{0}

\newpage


\section{Introduction}
\label{sec:intro}

The discovery by the ATLAS and CMS collaborations of a new scalar
particle with a mass of $\sim125\gev$
\cite{Aad:2012tfa,Chatrchyan:2012xdj,Khachatryan:2016vau}
 is consistent with the existence
of a Higgs boson in the Standard Model~(SM).
To date, all the measurements performed by the LHC of this
Higgs boson are in agreement with the SM predictions.
However, the uncertainties of the Higgs couplings measurements only
reach a precision of roughly a $\sim 20\%$,
hence there is room to beyond Standard-Model~(BSM)
Physics.
There are plenty of BSM models that lead to scalar sectors
with different features than the SM. 
In consequence, a main goal of the High Energy Physics community is to 
determine the nature of the Higgs mechanism and whether
the discovered Higgs boson belongs to an extended BSM
scalar sector.
One of the main properties of the SM-like Higgs 
that remains yet undetermined is the value of its triple self-coupling,
namely $\lahhh$, that it is only constrained to be inside the
range $-2.3< \lahhh/\laSM < 10.3$ at the 95\%
C.L.~\cite{ATLAS:2019pbo}. Many BSM models can induce important
deviations in $\lahhh$ with respect of the SM, therefore a more
accurate measurement of the triple Higgs coupling will
constitute a strong test of the SM.
For recent reviews on the measurement of the Higgs couplings
at future colliders see for instance \cite{deBlas:2019rxi, DiMicco:2019ngk}.
In the case a BSM Higgs sector manifests itself, it will be
a prime task to measure as well the BSM trilinear Higgs-boson couplings.

One of the most natural extensions of the SM Higgs sector is the 
``Two Higgs Doublet Model"(2HDM) (for reviews see,
e.g.,~\citeres{Gunion:1989we,Aoki:2009ha, Branco:2011iw})
that consists in the addition of a second Higgs doublet to the SM
one, with a ratio of the corresponding two vacuum 
expectation values given by $\tb := v_2/v_1$. This implies the existence
of five physical Higgs bosons: two $\CP$-even bosons $h$ and $H$,
usually with $\Mh<\MH$, one $\CP$-odd boson $A$ and two charged
Higgs bosons $H^\pm$.
In \citere{Arco:2020ucn} we performed an analysis of the possible size
of triple Higgs couplings in the 2HDM being compatible with all the
present theoretical and experimental constraints. 
For that analysis we assumed that the
light $\CP$-even Higgs-boson $h$ is SM-like with a mass of 
$\Mh \sim 125 \gev$. All other Higgs bosons were assumed to be heavier.
To avoid flavor changing neutral currents at
tree-level, a $Z_2$~symmetry is imposed~\cite{Glashow:1976nt},
possibly softly broken by the parameter $\msq$.
Depending on how this symmetry is extended to the fermion sector, four
types of the 2HDM can be realized: 
type~I and~II, lepton specific and flipped~\cite{Aoki:2009ha}.

In \citere{Arco:2020ucn} the 2HDM type~I and~II have been analyzed.
We investigated the allowed ranges for all triple Higgs couplings
involving at least one light, SM-like Higgs boson.  
The analysis was performed in several two-dimensional benchmark
planes (i.e.\ all but two 2HDM parameters were fixed according to
our definitions).  We focused 
on the regions where all relevant theoretical and
experimental constraints were fulfilled. 
For the SM-type triple Higgs coupling w.r.t.\ its SM value,
$\kala = \lahhh/\laSM$, we found allowed intervals of roughly
$\inter{-0.5}{1.5}$ in the 2HDM type~I and $\inter{0}{1}$ in type~II.
For the coupling of two light and one heavy CP-even Higgs bosons we found 
an approximate allowed interval of $\lahhH \in [-1.4,1.5]$ in the
2HDM type~I and $\inter{-1.6}{1.8}$ in type~II.   
Concerning the triple Higgs couplings involving two heavy 2HDM Higgs
bosons, we found large allowed values for both 2HDM type~I and~II.
For $\lahHH$, $\lahAA$ and $\lahHpHm$ 
we found maximum values of up to~15, 16 and~32, respectively.
For further explorations several benchmark points were proposed as
examples that exhibited large deviations from the SM-type triple
Higgs coupling and/or large values of the triple Higgs couplings
involving either two light and one heavy or one light and two
heavy  2HDM Higgs bosons.   Consequently, here our main
motivation is to explore the potential of future $e^+e^-$ colliders for
the measurements of all these BSM triple Higgs couplings.  

Future $e^+e^-$ linear colliders, like the ILC~\cite{Bambade:2019fyw}
and CLIC~\cite{Charles:2018vfv} will play a key role for the
measurement of the Higgs
potential and to detect possible deviations from the SM with high
precision~\cite{Abramowicz:2016zbo,Strube:2016eje,Roloff:2019crr,deBlas:2019rxi,DiMicco:2019ngk}.
In particular,  double Higgs boson production at future  $e^+e^-$
colliders appears to be the best way to explore possible
deviations from the SM Higgs self-couplings (for recent reviews see,
e.g.,  \citeres{Roloff:2019crr,Gonzalez-Lopez:2020lpd},  and
references therein).  Within the 2HDM framework the largest
effects of triple Higgs couplings on $e^+e^-$ cross sections are also
expected to be found in double Higgs production. These and other
processes involving the $125 \gev$ Higgs boson at 
$e^+e^-$~colliders have been explored in
\citeres{Kon:2018vmv,Sonmez:2018smv} in order to find signals from the
2HDM heavy Higgs bosons (for related work before the Higgs-boson
discovery, see  
\citeres{Djouadi:1999gv,Arhrib:2008jp,LopezVal:2009qy,Asakawa:2010xj}).

In this paper we will analyze in detail the production of two neutral
Higgs bosons in the 2HDM for the foreseen center-of-mass  
energies and luminosities of ILC and CLIC in the two main production channels:
\begin{align}
\label{eeZhh}
  e^+e^- &\to Z^* \to Z\,+\, hh, \, hH, \, HH, \,AA ,\\
\label{eenunuhh}
  e^+e^- &\to \nu \bar\nu\,+\, hh, \, hH, \, HH, \, AA.
\end{align}
The first one is similar to the ``Higgs-strahlung'' channel of single
Higgs production. The second one has an important contribution from the
vector-boson fusion mediated subprocess, $W^+W^- \to h_ih_j$, where the
$WW$ pairs (virtual, in general) are radiated from the initial $e^+e^-$
together with the neutrinos: $e^+e^- \to W^*W ^*\nu \bar\nu$. The
processes in \refeq{eenunuhh} also receive a  
contribution from the $Z^{(*)}$ mediated subprocesses,
$e^+e^- \to Z^{(*)} h_ih_j\to \nu {\bar \nu} h_ih_j$,
which are usually smaller than the contribution from $WW$ fusion at the
high energy colliders. 
Our main focus in this paper is addressed to study not just the
cross section as a function of the 2HDM parameters,  but also to
explore the sensitivity to the various
triple Higgs boson couplings via these
two double Higgs production channels,  at both ILC and CLIC.
Our aim is to disentangle the role of these triple Higgs
couplings in the mentioned double Higgs production processes and
to investigate which process at which energy is best suited for an
experimental determination.  

The calculation presented here of the total cross sections for
these double Higgs production channels  is 
performed in several two-dimensional benchmark planes
(i.e.\ with all but two of the 2HDM parameters fixed), based on our
results in \citere{Arco:2020ucn}. The computations done here, as in
\citere{Arco:2020ucn}, are performed 
with the help of the public codes
$\mathtt{MadGraph}$~\cite{Alwall:2014hca},
$\mathtt{FeynRules}$~\cite{Alloul:2013bka}
and $\mathtt{2HDMC}$~\cite{Eriksson:2009ws}.
We demonstrate that it is possible to 
find regions with a strong enhancement of the $hh$ production channels
and/or with  large production cross sections involving one light
and one heavy or two heavy 2HDM Higgs bosons. First,
we analyze in detail the mechanisms leading to these enhanced production
cross sections.  Second,  we illustrate that sizable effects due to
triple Higgs couplings can be seen in the cross section
distributions as a function of the invariant mass of the di-Higgs
final state (where we employed the code
$\mathtt{ROOT}~$\cite{Brun:1997pa}).
Finally,  we discuss which process at which center-of-mass energy
would be best suited to probe the corresponding triple Higgs-boson couplings.

Our paper is organized as follows. In \refse{sec:model} we briefly review
the 2HDM, fix our notation and define the five benchmark planes used
later for our investigation and summarize the constraints, which are
(apart from updates) the same as in \citere{Arco:2020ucn}.
The di-Higgs production cross sections in the
benchmark planes are presented in \refse{sec:xs} and analyzed
w.r.t.\ their dependence on the triple Higgs couplings.
Finally, in
\refse{sec:sensitivity} we analyze how the various triple Higgs couplings
may be accessed experimentally through the invariant mass of
the di-Higgs final state at the high-energy $e^+e^-$ colliders.
Our conclusions are given in \refse{sec:conclusions}.


\section{The Model and the constraints}
\label{sec:model}

In this section we give a brief description of the 2HDM to fix our
notation. We also review the theoretical and experimental constraints,
which are (apart from updates) the same as in
\citere{Arco:2020ucn}. Finally we will define the benchmark planes for
our analysis.


\subsection{The 2HDM}
\label{sec:2hdm}

We assume the $\cp$ conserving 2HDM (see
\citeres{Gunion:1989we,Aoki:2009ha, Branco:2011iw} for reviews) whose potential can be written as:
\begin{eqnarray}
V &=& m_{11}^2 (\Phi_1^\dagger\Phi_1) + m_{22}^2 (\Phi_2^\dagger\Phi_2) - \msq (\Phi_1^\dagger
\Phi_2 + \Phi_2^\dagger\Phi_1) + \frac{\la_1}{2} (\Phi_1^\dagger \Phi_1)^2 +
\frac{\la_2}{2} (\Phi_2^\dagger \Phi_2)^2 \nonumber \\
&& + \la_3
(\Phi_1^\dagger \Phi_1) (\Phi_2^\dagger \Phi_2) + \la_4
(\Phi_1^\dagger \Phi_2) (\Phi_2^\dagger \Phi_1) + \frac{\la_5}{2}
[(\Phi_1^\dagger \Phi_2)^2 +(\Phi_2^\dagger \Phi_1)^2]  \;.
\label{eq:scalarpot}
\end{eqnarray}
\noindent

We denote the two $SU(2)_L$ doublets as $\Phi_1$ and $\Phi_2$, 
\begin{eqnarray}
\Phi_1 = \left( \begin{array}{c} \phi_1^+ \\ \frac{1}{\sqrt{2}} (v_1 +
    \rho_1 + i \eta_1) \end{array} \right) \;, \quad
\Phi_2 = \left( \begin{array}{c} \phi_2^+ \\ \frac{1}{\sqrt{2}} (v_2 +
    \rho_2 + i \eta_2) \end{array} \right) \;,
\label{eq:2hdmvevs}
\end{eqnarray}
where $v_1, v_2$ are the real vevs acquired by the fields
$\Phi_1, \Phi_2$, respectively, with $\tb := v_2/v_1$ and they satisfy the 
relation $v = \sqrt{(v_1^2 +v_2^2)}$ where $v\simeq246\gev$ is the SM vev.
The eight degrees of freedom above, $\phi_{1,2}^\pm$, $\rho_{1,2}$ and
$\eta_{1,2}$, give rise to three Goldstone bosons, $G^\pm$ and $G^0$,
and five massive physical scalar fields: two $\cp$-even scalar fields,
$h$ and $H$, one $\cp$-odd one, $A$, and one charged pair, $H^\pm$.
Here the mixing angles $\al$ and $\be$ diagonalize the $\CP$-even and -odd
Higgs bosons, respectively.

To avoid the occurrence of tree-level flavor
changing neutral currents (FCNC), a $Z_2$ symmetry is imposed, which
is softly broken by the $\msq$ in
the Lagrangian. The extension of the $Z_2$ symmetry to the Yukawa
sector forbids tree-level FCNCs. 
This results in four variants of 2HDM, 
depending on the $Z_2$ parities of the 
fermions. We focus on type~I and~II, where the coupling to fermions are
listed in \refta{tab:types} and \refta{tab:coupling}.

\begin{table}[htb!]
\begin{center}
\begin{tabular}{lccc} 
\hline
  & $u$-type & $d$-type & leptons \\
\hline
type~I & $\Phi_2$ & $\Phi_2$ & $\Phi_2$ \\
type~II & $\Phi_2$ & $\Phi_1$ & $\Phi_1$ \\
\hline
\end{tabular}
\caption{Allowed fermion couplings in 
the 2HDM type~I and~II.}
\label{tab:types}
\end{center}
\end{table}

We will study the 2HDM in the physical basis, where the free parameters
in \refeq{eq:scalarpot} can be expressed in terms of the following set of 
parameters:
\begin{equation}
c_{\be-\al} \; , \quad \tb \;, \quad v \; ,
\quad \Mh\;, \quad \MH \;, \quad \MA \;, \quad \MHp \;, \quad \msq \;.
\label{eq:inputs}
\end{equation}
From now on we use sometimes the short-hand notation $s_x = \sin(x)$,
$c_x = \cos(x)$. 
In our analysis we will identify the lightest $\cp$-even Higgs boson,
$h$, with the one observed at $m_h \sim 125 \gev$.

The couplings of the Higgs bosons to SM particles are modified
w.r.t.\ the SM Higgs-coupling predictions due to the mixing in the Higgs
sector.  In particular,  the couplings of one neutral Higgs boson to
fermions and to gauge bosons are given by:
\begin{eqnarray}
	\mathcal{L} &=&-\sum_{f=u,d,l}\frac{m_f}{v}\left[\xi_h^f\bar{f}fh + \xi_H^f\bar{f}fH +i \xi_A^f\bar{f}\gamma_5fA \right] \nonumber \\
&&+\sum_{h_i=h,H,A}	\left[   g m_W \xi_{h_i}^W  W_\mu W^\mu h_i + \frac{1}{2} g m_Z \xi_{h_i}^Z  Z_\mu Z^\mu h_i\right] ,
\end{eqnarray}
where $m_f$, $m_W$ and $m_Z$ are the fermion mass,  $W$ mass and $Z$ mass, respectively,  and the factors in the couplings  to fermions,  $\xi_{h,H,A}^f$,  and to gauge-bosons, $\xi_{h,H,A}^V$,   
 are summarized in \refta{tab:coupling}, for the 2HDM of type I and of type II\footnote{ 
 Notice that we are using  the same notation for the factors $\xi_{h,H,A}^f$ as in our previous work of \citere{Arco:2020ucn},  and we are correcting here a typo in $\xi_{H}^{u}$ and $\xi_{H}^{d,l}$ of 
Table 2 of that reference.}.

\begin{table}
\begin{center}
\begin{tabular}{c|c|c}
 & type~I  & type~II\tabularnewline
\hline 
$\xi_{h}^{u}$  & $s_{\be-\al}+c_{\be-\al}\cot\be$  & $s_{\be-\al}+c_{\be-\al}\cot\be$\tabularnewline
$\xi_{h}^{d,l}$  & $s_{\be-\al}+c_{\be-\al}\cot\be$  & $s_{\be-\al}-c_{\be-\al}\tan\be$\tabularnewline
$\xi_h^V$ & $s_{\be-\al}$ & $s_{\be-\al}$\tabularnewline\hline
$\xi_{H}^{u}$  & $c_{\be-\al}-s_{\be-\al}\cot\be$  & $c_{\be-\al}-s_{\be-\al}\cot\be$\tabularnewline
$\xi_{H}^{d,l}$  & $c_{\be-\al}-s_{\be-\al}\cot\be$  & $c_{\be-\al}+s_{\be-\al}\tan\be$\tabularnewline
$\xi_H^V$ & $c_{\be-\al}$ & $c_{\be-\al}$\tabularnewline\hline
$\xi_{A}^{u}$  & $-\cot\be$  & $-\cot\be$\tabularnewline
$\xi_{A}^{d,l}$  & $\cot\be$  & $-\tan\be$\tabularnewline
$\xi_A^V$ & 0 & 0 \tabularnewline
\end{tabular}
\caption{Relevant factors appearing in the couplings of the neutral Higgs 
bosons to fermions,  $\xi_{h,H,A}^f$,  and to gauge-bosons,
$\xi_{h,H,A}^V$,  in the 2HDM of type~I and type~II}
\label{tab:coupling}
\end{center}
\end{table}

An important role in this paper is played by the couplings of the
lightest $\cp$-even Higgs boson with the other BSM bosons, concretely \lahhh, 
\lahhH, \lahHH\ and \lahAA.
We define
these $\la_{h h_i h_j}$ couplings such that the Feynman rules are given by:
\begin{equation}
	\begin{gathered}
		\includegraphics{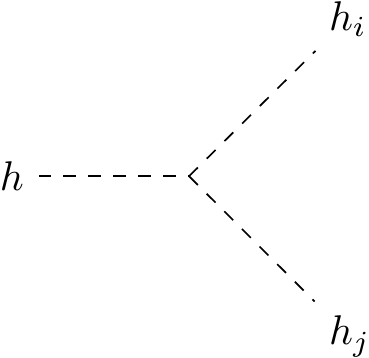}
	\end{gathered}
	=- i\, v\, n!\; \la_{h h_i h_j}
\label{eq:lambda}
\end{equation}
where $n$ is the number of identical particles in the vertex. The
explicit expressions for the couplings $\la_{hh_ih_j}$ can be found in
the Appendix of \citere{Arco:2020ucn}. 
We adopt this convention in \refeq{eq:lambda} so that the light Higgs
triple coupling $\lahhh$ has the same normalization as $\laSM$  in the
SM, i.e. $-6iv\laSM$ with 
$\laSM=\Mh^2/2v^2\simeq0.13$. We furthermore define $\kala := \lahhh/\laSM$.

The couplings of the $\cp$-even Higgs bosons strongly 
depend on $\CBA$. In particular, if $\CBA=0$ one can recover all the
interactions of the SM Higgs boson for the $h$ state, what is known as
the \textit{alignment limit}. However, in the
alignment limit in general one can still have BSM physics related to the
extended Higgs sector, like $hHH$ or $ZHA$ interactions for example.


\subsection{Experimental and theoretical constraints}
\label{sec:constraints}

In this section we will briefly summarize the various theoretical and
experimental constraints considered in our scans.

\begin{itemize}

\item {\bf Constraints from electroweak precision data}\\
Constraints from the electroweak precision observables (EWPO)
can for ``pure'' Higgs-sector extensions of the SM, 
be expressed in terms of the oblique parameters $S$, $T$ and
$U$~\cite{Peskin:1990zt,Peskin:1991sw}.
In the 2HDM the $T$~parameter is most
constraining and requires either $\MHp \approx \MA$ or $\MHp \approx \MH$.
In \citere{Arco:2020ucn} we explored three scenarios:
(A)\;$\MHp = \MA$, (B)\;$\MHp = \MH$ and (C)\;$\MHp = \MA = \MH$.
Here we will focus on scenario~C.
The 2HDM parameter space is explored with the code
\texttt{2HDMC-1.8.0}~\cite{Eriksson:2009ws}.

\item {\bf Theoretical constraints}\\
Here the important constraints come from
tree-level perturbartive unitarity and stability of the vacuum.
They are ensured by an explicit test on the underlying Lagrangian
parameters, see \citere{Arco:2020ucn} for details.
The parameter space allowed by these two constraints can be enlarged, in
particular to higher BSM Higgs-boson mass values by the condition,
\begin{equation}
  \msq = \frac{\MH^2\cos^2\al}{\tb}~.
  \label{eq:m12special}
\end{equation}

\item {\bf Constraints from direct searches at colliders}\\
The $95\%$ confidence level
exclusion limits of all important searches for BSM Higgs bosons
are included in the public code
\HB\,\texttt{v.5.9}~\cite{Bechtle:2008jh,Bechtle:2011sb,Bechtle:2013wla,Bechtle:2015pma,Bechtle:2020pkv},
including Run~2 data from the LHC.
Given a set of theoretical
predictions in a particular model, \HB\ determines which is the most
sensitive channel and determines, based on this most sensitive
channel, whether the point is allowed or not at the $95\%$~CL.
As input the code requires some specific predictions from the model,
like branching ratios or Higgs couplings, that we computed with the
help of \texttt{2HDMC}~\cite{Eriksson:2009ws}.

\item {\bf Constraints from the SM-like Higgs-boson properties}\\
Any model beyond the SM has to accommodate the SM-like Higgs boson,
with mass and signal strengths as they were measured at the LHC.
In our scans the compatibility of the $\cp$-even scalar $h$ with a mass
of $125.09\gev$ with the measurements of signal strengths at Tevatron and 
LHC
is checked with the code
\texttt{HiggsSignals v.2.6}~\cite{Bechtle:2013xfa,Bechtle:2014ewa,Bechtle:2020uwn}. 
\texttt{HiggsSignals} provides a
statistical $\chi^2$ analysis of the SM-like Higgs-boson predictions of
a certain model compared to the measurement of Higgs-boson signal rates
and masses from Tevatron and LHC. Again, the predictions of the 2HDM
have been obtained with {\tt{2HDMC}}~\cite{Eriksson:2009ws}.
Here, as in \citere{Arco:2020ucn}, we will require that for a  parameter
point of the 2HDM to be allowed, the corresponding $\chi^2$ is within
$2\,\sig$ ($\De\chi^2 = 6.18$)
from the SM fit: $\chi_\mathrm{SM}^2=84.73$ with 107 observables. 

Many of the recent LHC Higgs rate measurements are now given in terms of
``STXS observables''. 
Contrary to our previous analysis in \citere{Arco:2020ucn} the
{\tt 2HDMC} output can now allow the application of the (newly in
{\tt HiggsSignals} implemented) STXS observables. This results in
substantially stronger limits on $\CBA$, particularly in the 2HDM type~II.
This will be visible in the fourth benchmark plane, see \refse{sec:eehh}.

\item {\bf Constraints from flavor physics}\\
Constraints from flavor physics have proven to be very significant
in the 2HDM mainly because of the presence of the charged Higgs boson.
Various flavor observables like rare $B$~decays, 
$B$~meson mixing parameters, $\br(B \to X_s \gamma)$, 
LEP constraints on $Z$ decay partial widths
etc., which are sensitive to charged Higgs boson exchange, provide
effective constraints on the available 
parameter space~\cite{Enomoto:2015wbn,Arbey:2017gmh}. 
Here we take into account the decays $B \to X_s \gamma$ and
$B_s \to \mu^+ \mu^-$, which are most constraining. This is done with
the code \texttt{SuperIso}~\cite{Mahmoudi:2008tp,Mahmoudi:2009zz}
where the model input is given by {\tt{2HDMC}}.  We have modified the
code as to include the Higgs-Penguin type corrections in
$B_s \to \mu^+ \mu^-$~\cite{Li:2014fea,Arnan:2017lxi,Cheng:2015yfu},
which were not included in the original version of
\texttt{SuperIso}. These
corrections are indeed relevant for the present work since these
Higgs-Penguin contributions are the ones containing the potential
effects from triple Higgs couplings in $B_s \to \mu^+ \mu^-$.   

\end{itemize}

The effects of the various constraints on the 2HDM parameter space
  is discussed in detail in \citere{Arco:2020ucn}.


\subsection{Benchmark planes}
\label{sec:planes}

Based on the analysis in \citere{Arco:2020ucn} we define five benchmark
planes that exhibit an interesting phenomenology w.r.t.\ the di-Higgs
production cross sections, \refeqs{eeZhh} and (\ref{eenunuhh}).

\begin{enumerate}

\item
  2HDM type I,\\
  $\MHp = \MH = \MA = 1000 \gev$,
  $\msq$ fixed via \refeq{eq:m12special},\\
  free parameters: $\CBA$, $\tb$

\item
  2HDM type I,\\
  $\MHp = \MH = \MA = 650 \gev$,
  $\tb = 7.5$,\\
  free parameters: $\CBA$, $\msq$

\item
  2HDM type I,\\
  $\tb = 10$, 
  $\msq$ fixed via \refeq{eq:m12special},\\
  free parameters: $\CBA$, $\MHp = \MH = \MA$

\item
  2HDM type II,\\
  $\MHp = \MH = \MA=650\gev$, $\CBA=0.02$,\\
  free parameters: $\msq$, $\tan\beta$.
  
\item
  2HDM type II,\\
  $\tan\beta=1.5$, $\CBA=0.02$,\\
  free parameters: $\msq$, $\MHp = \MH = \MA$.

\end{enumerate}



\section{Cross section results}
\label{sec:xs}

\subsection{General considerations}
\label{sec:eehh}

In this section we present our analysis of the various di-Higgs
production cross sections at $e^+e^-$ colliders in the benchmark planes
defined in \refse{sec:planes}.
The analysis will be done for several collider options and
official expected integrated luminosities, as summarized in
\refta{tab:ee}. Concretely, we will take into account the ILC500 and the
ILC1000~\cite{Bambade:2019fyw}, as well as CLIC1500 and
CLIC3000~\cite{Charles:2018vfv}.
The cross section predictions presented in this and the next section  
are calculated at tree-level
precision with the help of the public code
{\tt{MadGraph5-2.7.2}}~\cite{Alwall:2014hca} which generates and
evaluates all contributing diagrams. All the information of the 2HDM
regarding the model Lagrangian required by  
{\tt{MadGraph}} was implemented with the Mathematica package
{\tt{FeynRules-2.3}}~\cite{Alloul:2013bka}. 
In the computation, electrons and positrons were considered massless, 
so the diagrams where a Higgs boson couples directly to a fermionic line 
vanish. Our computation contains all the possible diagrams and does not 
rely at any point on the narrow width approximation. 
The width of all the Higgs bosons were calculated with 
the public code {\tt{2HDMC-1.8.0}}~\cite{Eriksson:2009ws}.  We first
start with a discussion on the cross sections as  functions of the
collider energy and next we continue with the results of the cross
sections in the various benchmark planes introduced in the previous
section for the different projected collider energies.

\begin{table}[h!]
\begin{centering}
\begin{tabular}{c|c|c||c|c|c|c}
  Collider & $\sqrt{s}$ [GeV] & $\cL_{{\rm int}}\;[\iab]$ &
  $\sig(hhZ)$ [fb] & \# events & $\sig(hh\nu\bar{\nu})$ [fb] & \# events
  \tabularnewline
\hline \hline
ILC & 500 & 4 & 0.1576 & 630 & 0.03372 & 135\tabularnewline
\hline 
ILC & 1000 & 8 & 0.1202 & 962 & 0.09734 & 779\tabularnewline
\hline 
CLIC & 1500 & 2.5 & 0.07706 & 192 & 0.2388 & 597\tabularnewline
\hline 
CLIC & 3000 & 5 & 0.03272 & 164 & 0.8194 & 4097\tabularnewline
\end{tabular}
\par\end{centering}
\caption{Anticipated center-of-mass energies, $\sqrt{s}$, corresponding
integrated luminosities, $\cL_{\rm int}$, at ILC and CLIC and 
predicted cross sections and number of events for the di-Higgs production
in the SM.}
\label{tab:ee}
\end{table}

\begin{figure}[htb!]
\begin{center}
	\includegraphics{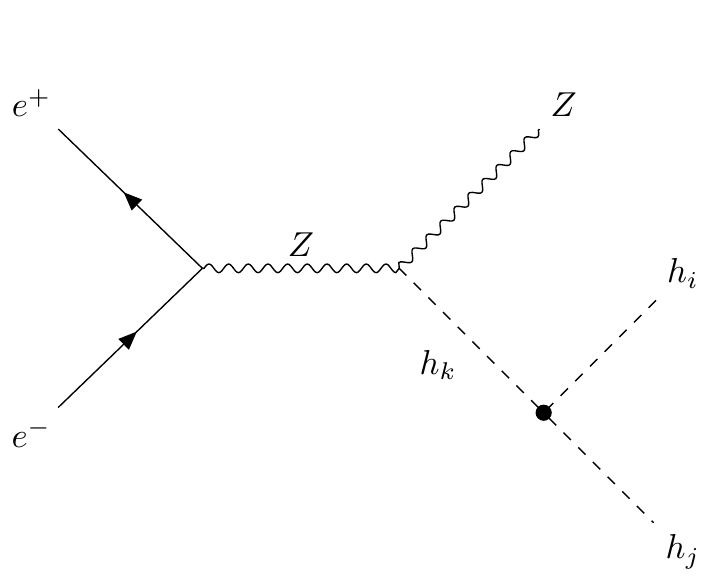}
	\includegraphics{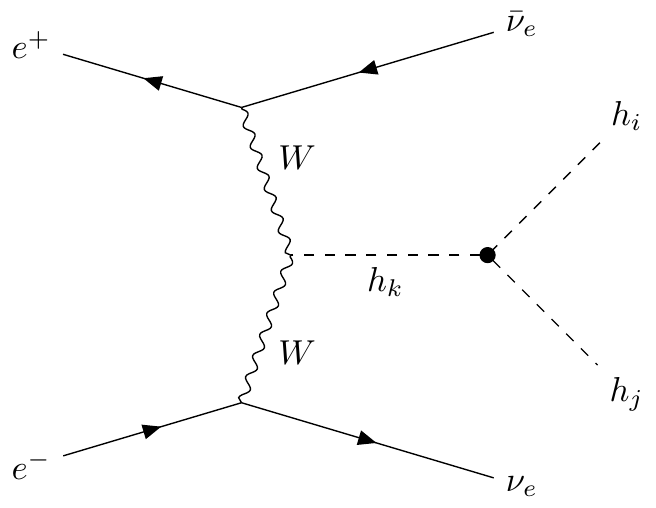}	
\end{center}
\caption{Diagrams that include triple Higgs couplings in the
    di-Higgs production in the studied processes. The diagrams on the
    left contribute to the $e^+e^-\to h_ih_jZ$ channels. The diagrams
    on the left with the final $Z\to\nu\bar{\nu}$ and on the right
    contribute to the $e^+e^-\to h_ih_j\nu\bar{\nu}$ channels.}
\label{fig:diagrams}
\end{figure}

\begin{figure}[htb!]
\begin{center}
	\includegraphics[width=0.48\textwidth]{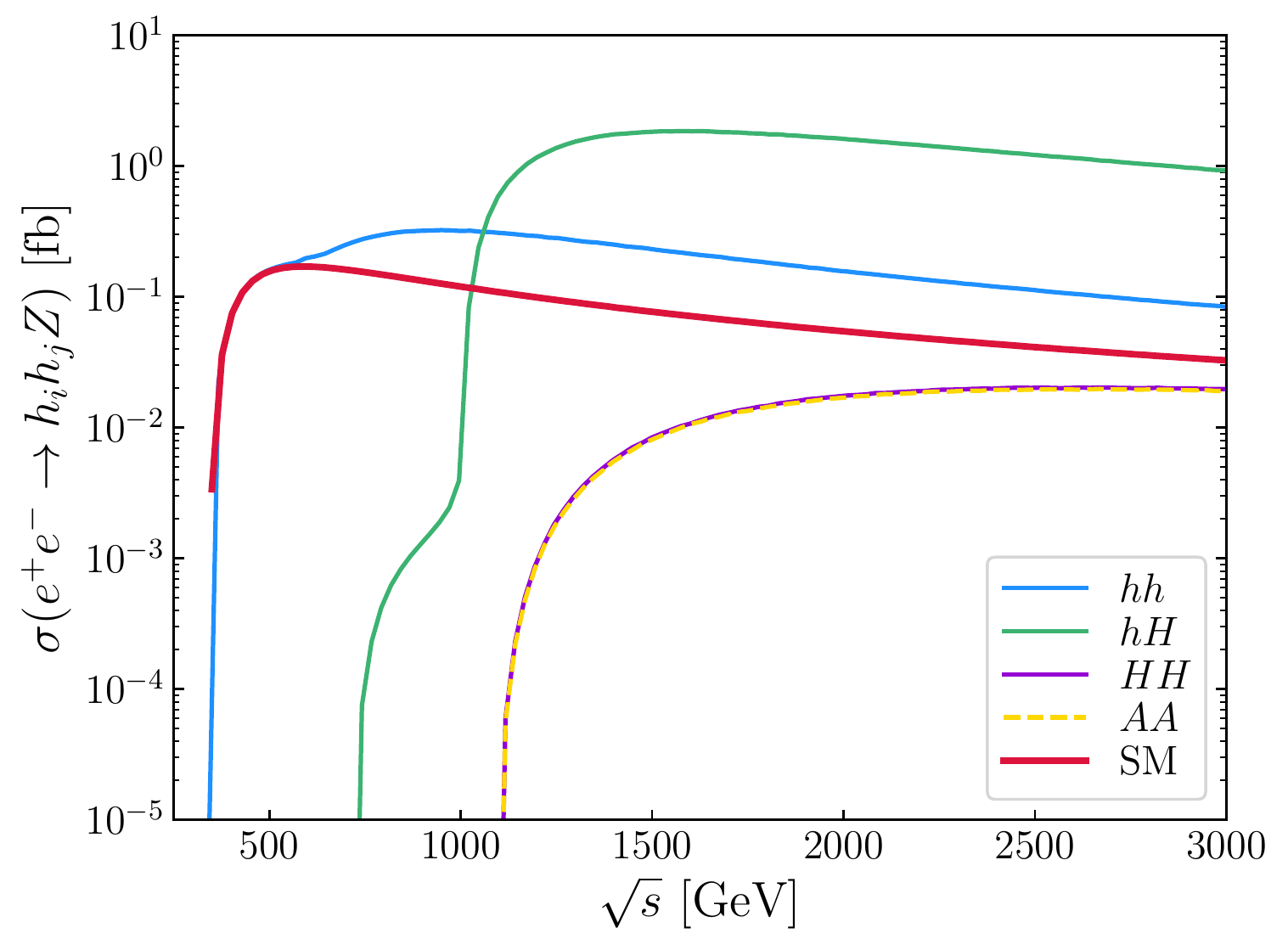}
	\includegraphics[width=0.48\textwidth]{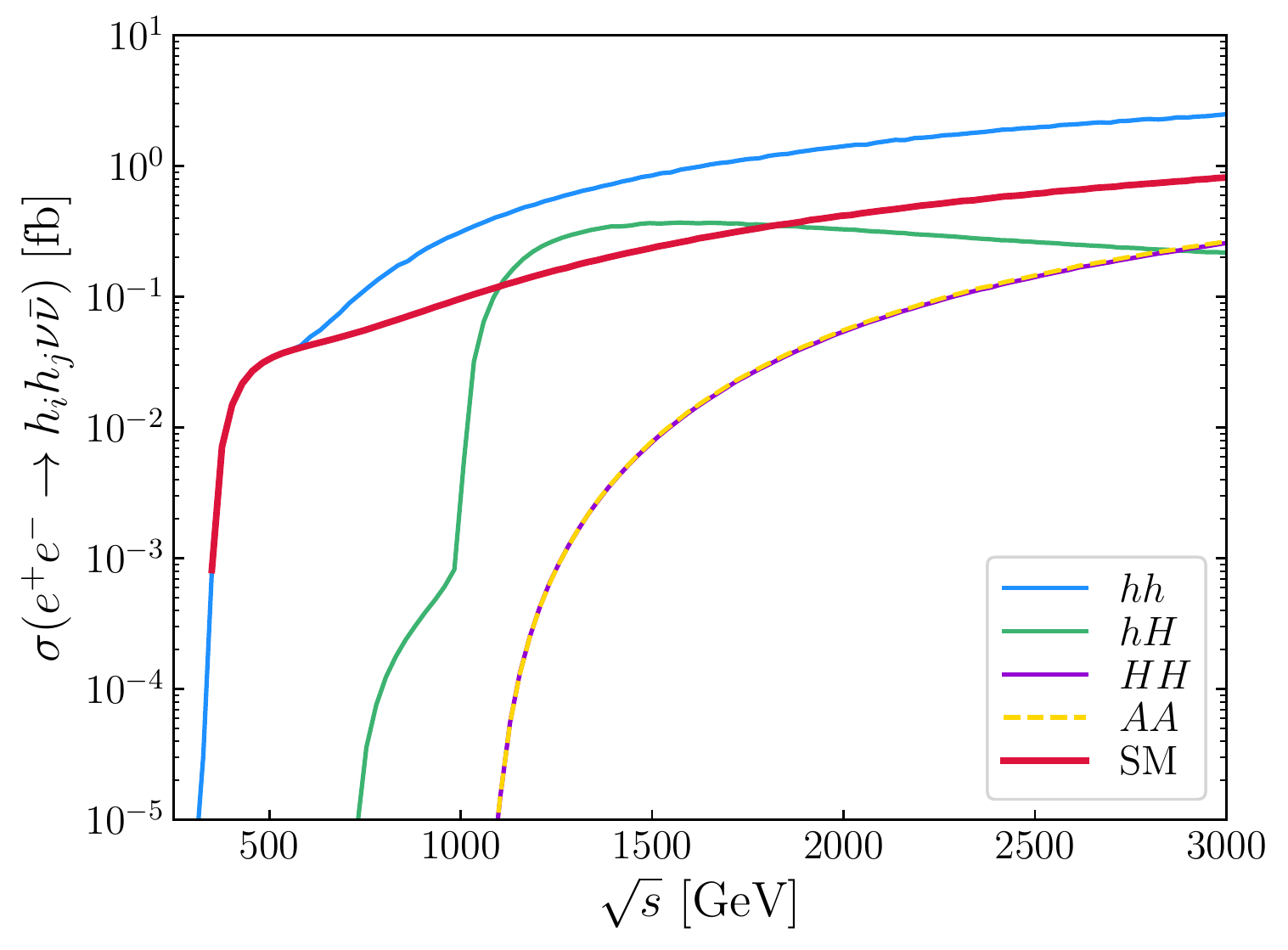}	
\end{center}
\caption{Cross sections as a function of the center-of-mass energy
  $\sqrt{s}$ for the processes $e^+e^-\to h_ih_jZ$ (left) and $e^+e^-\to
  h_ih_j\nu\bar{\nu}$ (right) for the particular point of the 2HDM type
  I defined by the input parameter values: $m_H=m_A=m_{H^\pm}=500 \gev$,
  $\tan \beta= 10$, $c_{\beta-\alpha}= 0.2$,  and $\msq=24000 \gev^2$.
 The numerical values of the triple Higgs couplings corresponding to
 these input parameters  are
 $\kala = 0.99$, $\lahhH = -0.3$, $\lahHH = 4.0$ and $\laHHH = 1.0$.}
\label{fig:XSsqrts}
\end{figure}

The complete set of diagrams that contribute to the two processes of our
interest, $e^+e^- \to h_i h_j Z$ and $e^+e^- \to h_i h_j \nu \bar\nu$ can
be classified according to two different types of configurations: diagrams
mediated by a virtual $Z$~boson and diagrams mediated by virtual $WW$ pairs.
In  $e^+e^- \to h_i h_j Z$ all contributing diagrams are of the first type, with
the virtual $Z$ attached to the initial $e^+e^-$ pair, i.e.,
$e^+e^- \to Z^* \to h_i h_j Z$.
In $e^+e^- \to h_i h_j \nu_e \bar\nu_e$,  in contrast,  the two configurations
contribute. There are diagrams mediated by a
virtual $Z$ attached to the initial $e^+e^-$ , i.e.\
$e^+e^- \to Z^* \to h_i h_j \nu_e \bar\nu_e$,
and diagrams mediated by virtual $WW$ pairs, i.e.\
$e^+e^- \to W^*W^* \nu_e \bar{ \nu_e} \to h_i h_j \nu_e \bar{ \nu_e}$.
In this latter case,  the subprocess involved is the so-called Vector
Boson Fusion (VBF)  given by $WW \to h_i h_j$.  The total number of
diagrams in $e^+e^- \to h_i h_j Z$ are 7, 6, 7, and 7 for $hh$, $hH$,
$HH$ and $AA$,  respectively.  The total number of diagrams in $e^+e^-
\to h_i h_j \nu_e \bar\nu_e$ and their separation in the two mentioned
types ($Z^*$ mediated + VBF mediated) are 14 (7+7), 12 (6+6), 14 (7+7)
and 12 (7+5) for $hh$, $hH$, $HH$ and $AA$,  respectively.   While we do
not show the diagrams here explicitly, they have all been included into
our computations. It is important to note that only one diagram type
among all these diagrams is the one carrying the triple Higgs couplings,
concretely those depicted in \reffi{fig:diagrams}.  Thus,  in order to
get access to these couplings,  a strategy has to be designed to
disentangle the contributions from these particular diagrams among all
the participating contributions in the total cross section.   The
discussion of the accessibility to 
the various triple Higgs-boson couplings will be presented in
\refse{sec:sensitivity}.  In this subsection we focus first on the
total cross sections.   

The importance of the previously commented classification is
 well known in the literature,  within the context of the SM,  since
 their respective contributions to 
the total cross section behave very differently with the center-of-mass
energy of the $e^+e^-$ collisions,  $\sqrt{s}$.
The $Z^*$ mediated configurations decrease
with energy,  whereas  the $W^*W^*$ mediated ones
increase with energy.
This is indeed the reason why for $e^+e^-$ colliders with energies at and
above the TeV scale the diagrams with VBF configuration may dominate the
$h_i h_j \nu \bar{\nu}$ production rates. The VBF dominance is well known
in the case of di-Higgs production $e^+e^-\to hh\nu \bar\nu$ in both the
SM ($h = H_{\rm SM}$), with $\kala =1$,
as well as in BSM models (where $h$ represents the BSM Higgs
  boson at $\sim 125 \gev$) with $\kala \neq 1$
(see, for instance, \citeres{Roloff:2019crr,Gonzalez-Lopez:2020lpd} and
references therein).  Consequently, this is 
expected to also happen in the present case within the 2HDM.  

One relevant
difference with respect to the SM case is that in the 2HDM the extra Higgs
bosons (other than $h$) also participate in these processes: $H$ and $H^\pm$,  participate in the $W^*W^*$ mediated subprocess,  and 
$H$ and $A$ enter in the $Z^*$ mediated subprocess.  One illustrative
example of the different behavior of the cross sections with the collider
energy, for the processes analyzed here, is presented in
\reffi{fig:XSsqrts}.  Here we show $e^+e^- \to Z h_i h_j$ (left plot) and
$e^+e^- \to \nu\bar\nu h_i h_j$ (right plot) for a particular point of 
the 2HDM (type I) parameters given by $m_H=m_A=m_{H^\pm}=500 \gev$,
$\tb = 10$, $c_{\beta-\alpha}= 0.2$,  and $\msq=24000 \gev^2$.
Other examples will be presented in \refse{sec:sensitivity}.  
(See \citere{Arco:2021ecv} for an additional discussion.)
The plots in \reffi{fig:XSsqrts} illustrate the commented different
behavior with energy of the cross sections above about 
$\sqrt{s} \sim 1 \tev$.  In particular,  we see that
$h h\nu \bar\nu$  grows with energy,  whereas $h h Z$ decreases
with energy (compare the two light blue lines in
\reffi{fig:XSsqrts}). Indeed they have a similar behavior with
$\sqrt{s}$ than the corresponding SM cross sections (red lines in
\reffi{fig:XSsqrts}). The 2HDM $hh$ production rates are enhanced
with respect to the SM  
ones by a factor of about 3,  for this particular choice of 2HDM input
parameters.
This enhancement is due to the additional contributions from the
extended 2HDM Higgs sector, including the effects from the new
diagrams with intermediate heavy Higgs bosons ($H$ and $A$ in $hhZ$
and $H$, $A$ and $H^\pm$ in $hh\nu \bar \nu$).  The size of the
enhancement depends strongly on the choice of the
2HDM parameters, most prominently on $\CBA$ away
from the alignment limit.  As will be discussed
below,  this enhancement mainly originates
from the diagrams where the intermediate heavy Higgs bosons, $H$ and $A$,
can be produced on-shell (i.e.\ in the case of $hhZ$ for
$\sqrt{s} \gsim \MH + \MZ$ or $\gsim  \MA + \Mh$, respectively),
but values of $\kala \neq 1$ can also play a role.  Consequently, the
largest contributions to the enhancement w.r.t.\ the relevant invariant
mass distribution occur in the resonant region, 
i.e.\ for $m_{hh} \sim \MH$ or $m_{hZ} \sim \MA$, respectively.
In contrast,  
the contributions from an intermediate $H^\pm$ in $hh\nu\bar\nu$ production,  
which appear only in $t$-channel like diagrams,  
do not exhibit a resonant behavior in this process. 

The case of $hH$ shows a different pattern with  $\sqrt{s}$ for higher
center-of-mass energies above  
$\sim 1200 \gev$ in this particular point.  Both cross sections for
$hHZ$ and $hH \nu \bar\nu$ decrease with energy (green lines in
these plots).  The cross section for $hHZ$ is above the $hhZ$ one by a
factor of about 10,  and the decrease with energy is well understood in
terms of the process being mediated by a virtual $Z^*$ propagating in
the $s$-channel.  The cross section for $hH \nu \bar\nu$ is below the
$hh  \nu \bar\nu$ one, and it shows a qualitatively different behavior
with $\sqrt{s}$ at high center-of-mass energies.  The $hH \nu \bar\nu$
cross section decreases  with $\sqrt{s}$ and follows a similar pattern
as in the $hHZ$ channel, in clear contrast with the  $hh \nu \bar\nu$
cross section that, as discussed above,  increases with  $\sqrt{s}$.
This suggests that at large energies,   $hH \nu \bar\nu$ is dominated by
the $Z^*$ mediated diagrams whereas $hh \nu \bar\nu$  is dominated by
the $W^*W^*$ mediated diagrams. Indeed, we have checked this dominance
explicitly by comparing the cross sections $h_ih_j \nu \bar\nu$ with
electron neutrinos (which include both types of mediated diagrams),
versus those with muon neutrinos  (which include only $Z^*$ mediated
diagrams).  We have found that at high energy,   above around 1 TeV,
$\sigma(hh \nu \bar\nu)\simeq \sigma(hh \nu_e \bar\nu_e)$,  and
$\sigma(hH \nu \bar\nu)\simeq 3 \sigma(hH \nu_\mu \bar\nu_\mu)$,
confirming our reasoning above.  

Besides,  in order to understand better  the VBF mediated contributions
we have explored the cross section of the corresponding subprocess of
the two cases, $hh$ and $hH$. For the case of $hh  \nu \bar\nu$,  we have
explicitly checked that the cross section of the subprocess $WW \to hh$
tends to a constant value at high energies (as in the SM case) due to a
strong cancellation of the contributions from the $t$, $u$ and
contact subprocess diagrams, which are separately growing with energy.
The contribution
from the charged Higgs is very suppressed in this case of large
$m_{H^\pm}$.  For the case of $hH \nu \bar\nu$,  however,  there is no
contact diagram in $WW \to hH$ and the strong cancellations occur
instead between the  $W$ and the charged Higgs boson contributions to
the $t$ and $u$ channels separately.  This leads to a remnant contribution
that decreases with energy, as they do the subprocess diagrams with the
$h$ and the $H$ propagators in the $s$-channel.  The charged Higgs
contribution is important at high energies in this case. 

The production cross sections of $HH$ and $AA$ in both
the $Z$ and the neutrino channels are numerically nearly equal
due to our choice $m_H=m_A$,  
that implies $\lahHH\sim\lahAA$ and $\laHHH\sim\laHAA$. There are
some diagrams present only in the $AA$ production,
involving the interaction vertex $hAZ$, whereas some others
appear only in $HH$ production with $ZZH$ and $WWH$ vertices
(the latter only present in the neutrino channel).
However, all the diagrams mentioned above carry a $\CBA^2$ factor and
consequently, they are suppressed with respect to the rest of the
diagrams, that 
are equivalent for $HH$ and $AA$ productions if $m_H=m_A$.
Furthermore,  the
cases of $HH  \nu \bar\nu$ and $AA  \nu \bar\nu$ production are clearly
dominated by the VBF subprocess and their increase with energy are the
most pronounced ones (purple and yellow lines in \reffi{fig:XSsqrts},
which appear indeed superimposed).  However,  the rates are lower than
for the $hh \nu \bar\nu$ and $hH \nu \bar\nu$ cases due to the large
values of $m_{H,A}$ for this particular point (set in this plot to
$m_{H,A}=500 \gev$. To gain access to them will require the highest
energy option, i.e.\ $3 \tev$ as will be discussed in
\refse{sec:sensitivity}.  Due to the strong similarity of $HH$ and $AA$
production, below we will often discuss only $ZHH$ and $HH\nu\bar\nu$.

Looking at all the production modes,
the important  contribution of VBF in the $h_i h_j \nu\bar\nu$
channels play an important role in the present study. The
high energy collider options, particularly for $3 \tev$, may offer an efficient
window to improve the sensitivity to the triple Higgs couplings, which
are mediated 
via the diagram on the right in \reffi{fig:diagrams}, belonging to
the VBF diagrams subset. This will be analyzed in detail in
\refse{sec:sensitivity}.   In the following,  we present our results for the total cross section. 
We will first discuss the three 2HDM type~I benchmark planes, followed
by a discussion of the two 2HDM type~II planes.


\subsection{2HDM type I}
\label{sec:ee-I}

We present the results obtained in the three benchmark planes for the
various production cross sections as given in \refeqs{eeZhh} and
(\ref{eenunuhh}) and various energies and integrated luminosities
according to \refta{tab:ee}. Some planes are omitted when the results do
not show any relevant variation. In each plane we indicate with (the
interior of) a solid
black line the part of the parameter space that is allowed taking all
theoretical and experimental constraints as given in
\refse{sec:constraints}. Small differences w.r.t.\ \citere{Arco:2020ucn}
are due to updated experimental constraints. 

\begin{figure}[p]
\begin{center}
\includegraphics[width=0.24\textwidth]{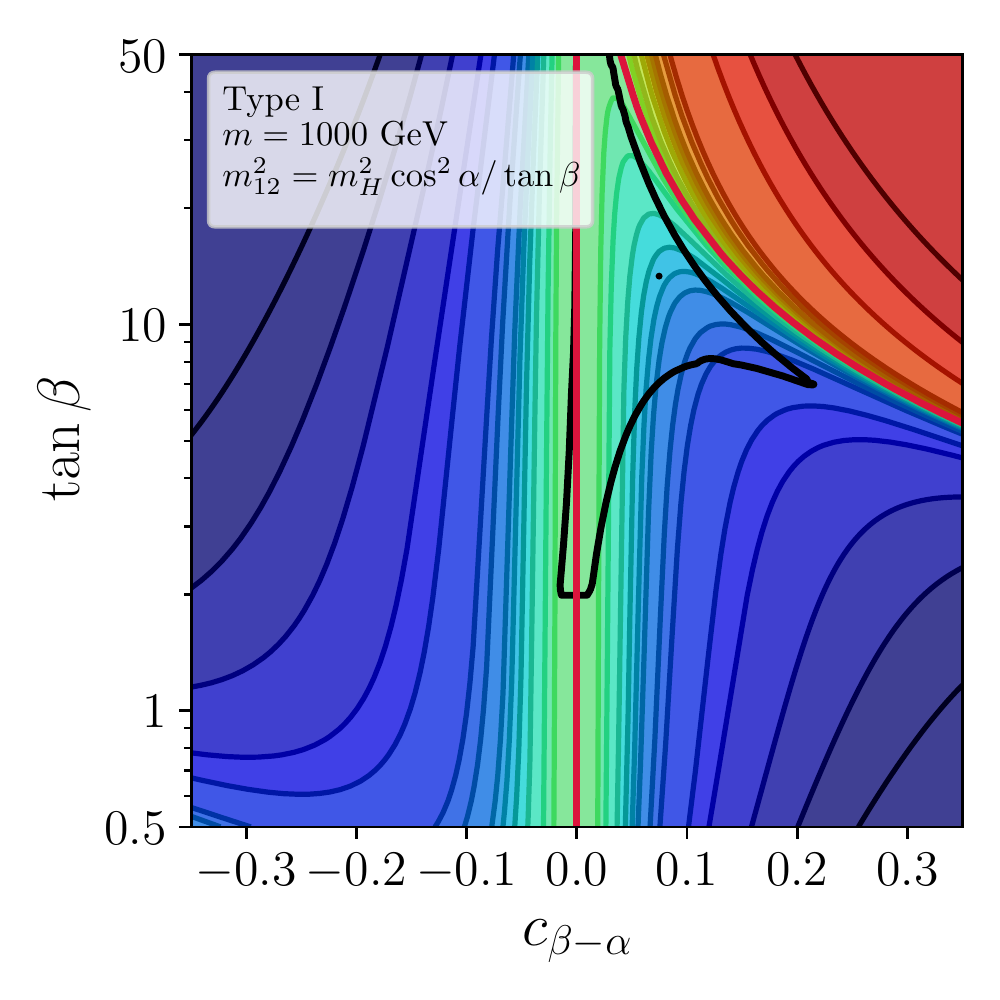}
\includegraphics[width=0.24\textwidth]{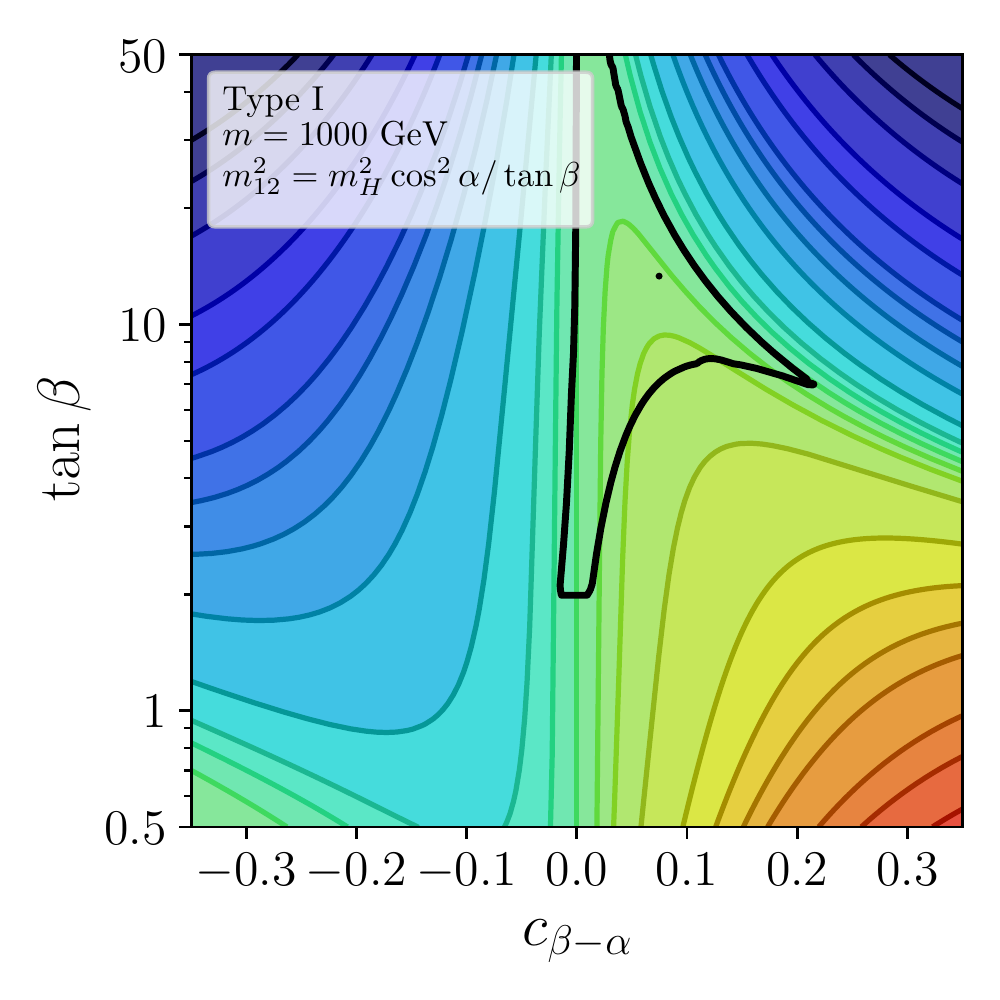}
\includegraphics[width=0.24\textwidth]{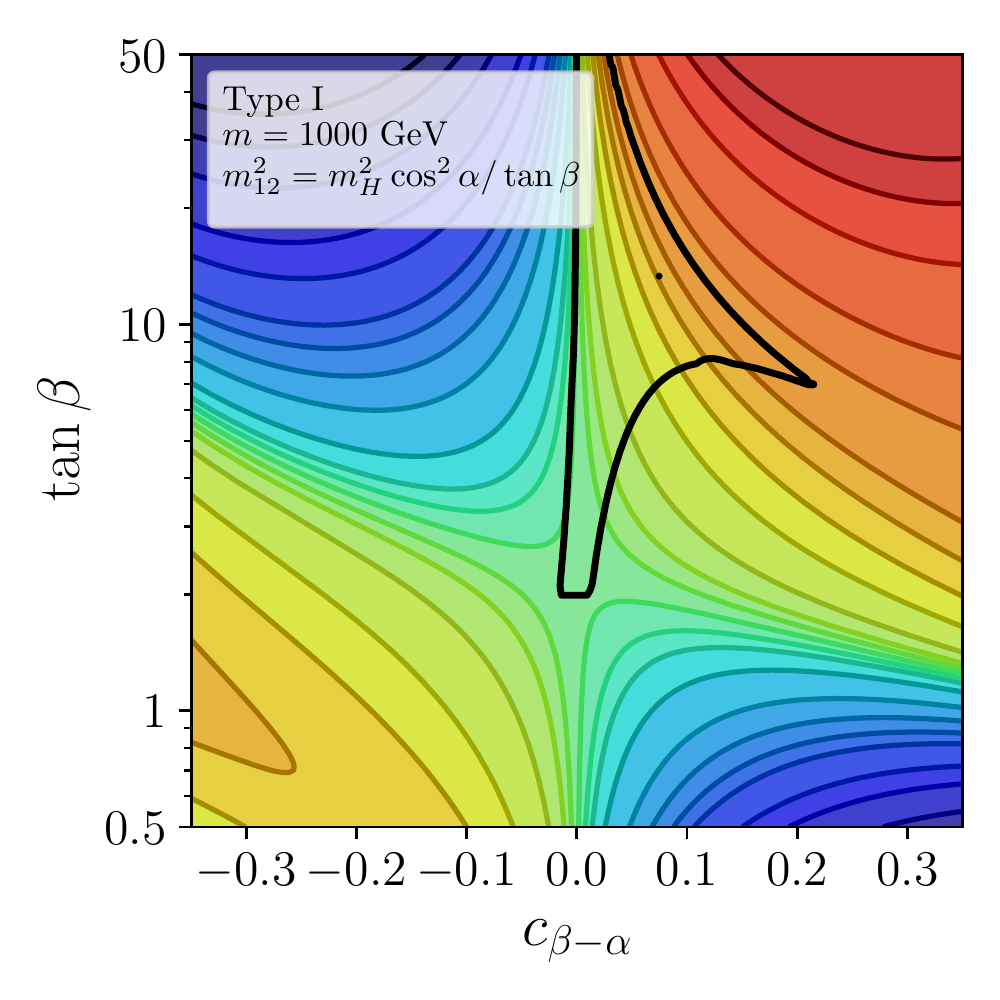}
\includegraphics[width=0.24\textwidth]{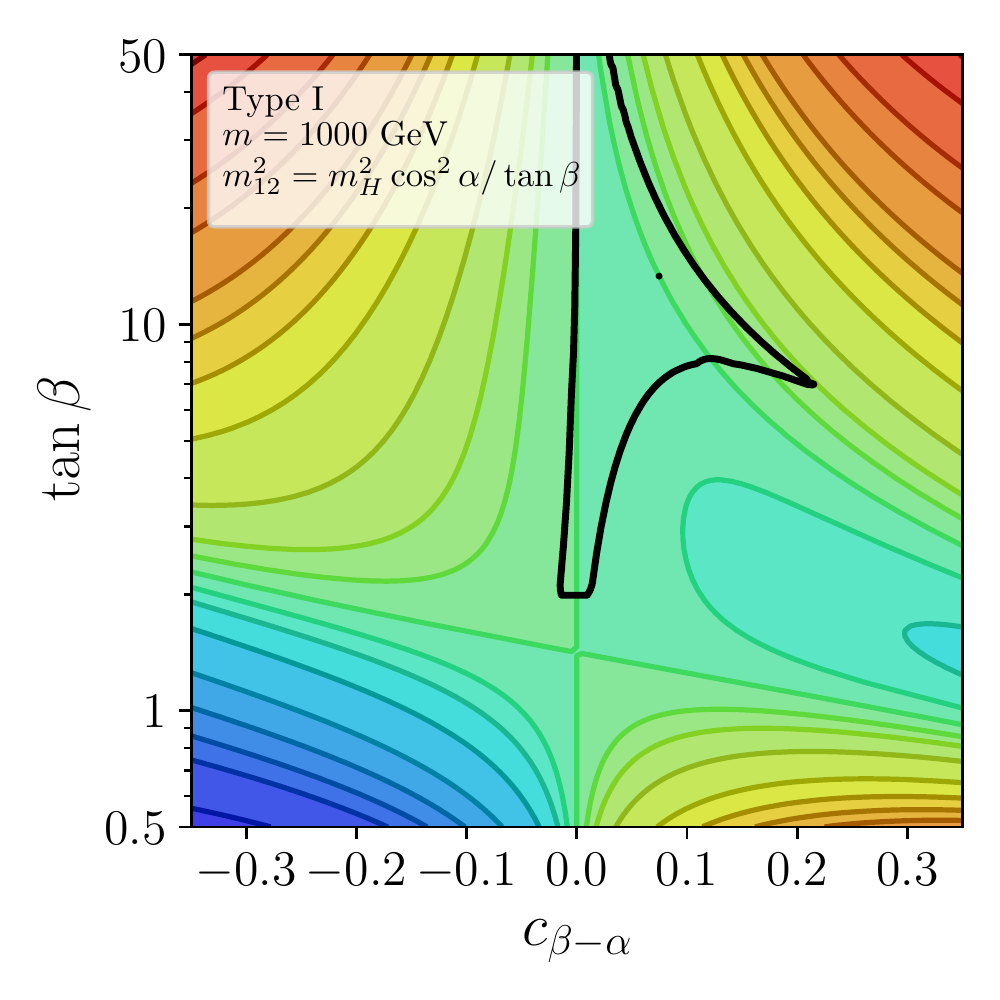}

\includegraphics[width=0.24\textwidth]{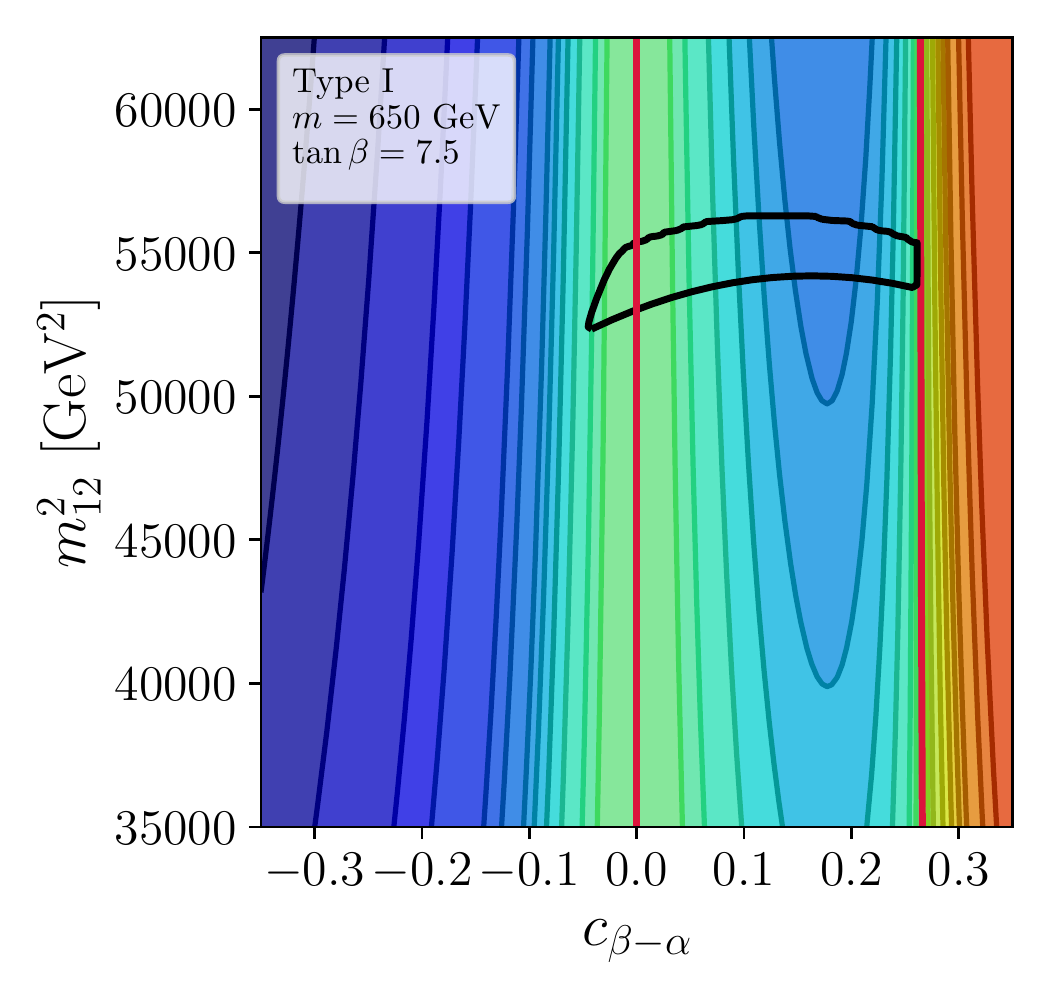}
\includegraphics[width=0.24\textwidth]{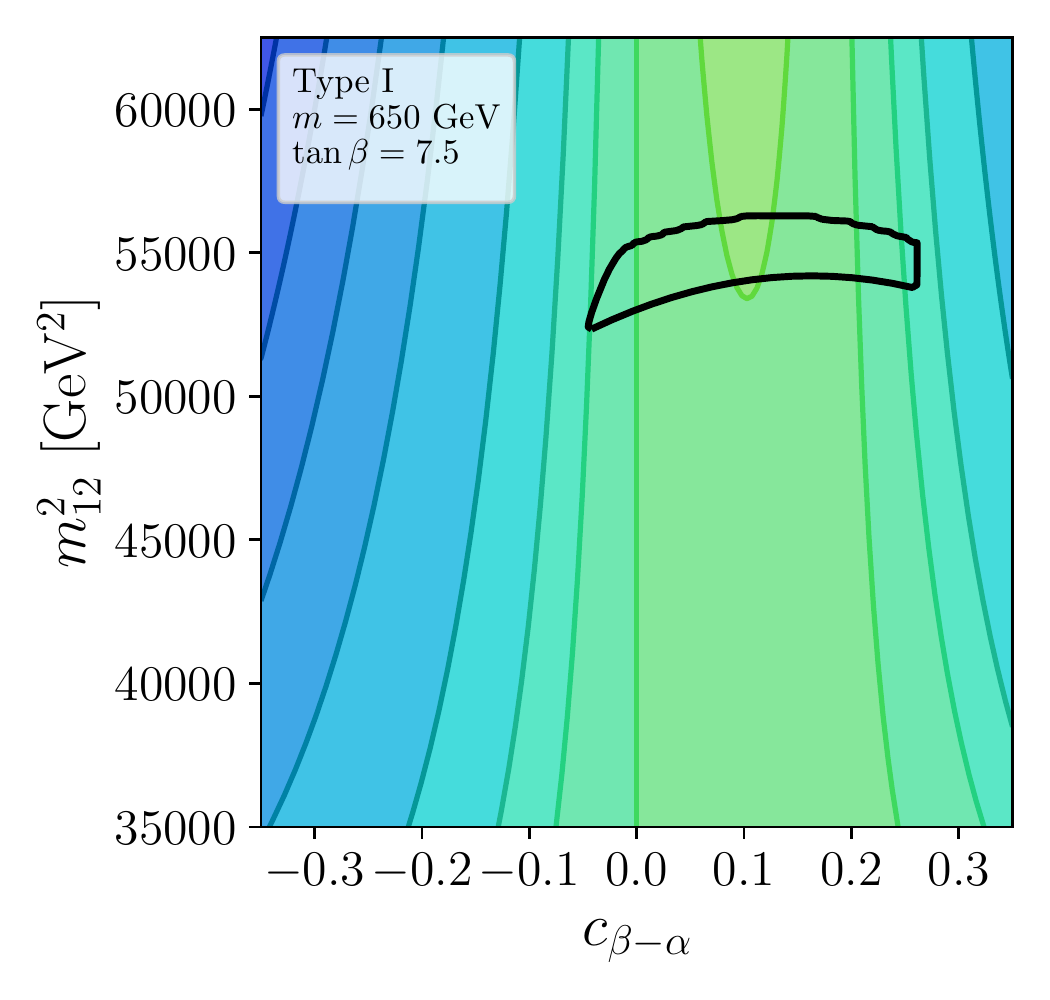}
\includegraphics[width=0.24\textwidth]{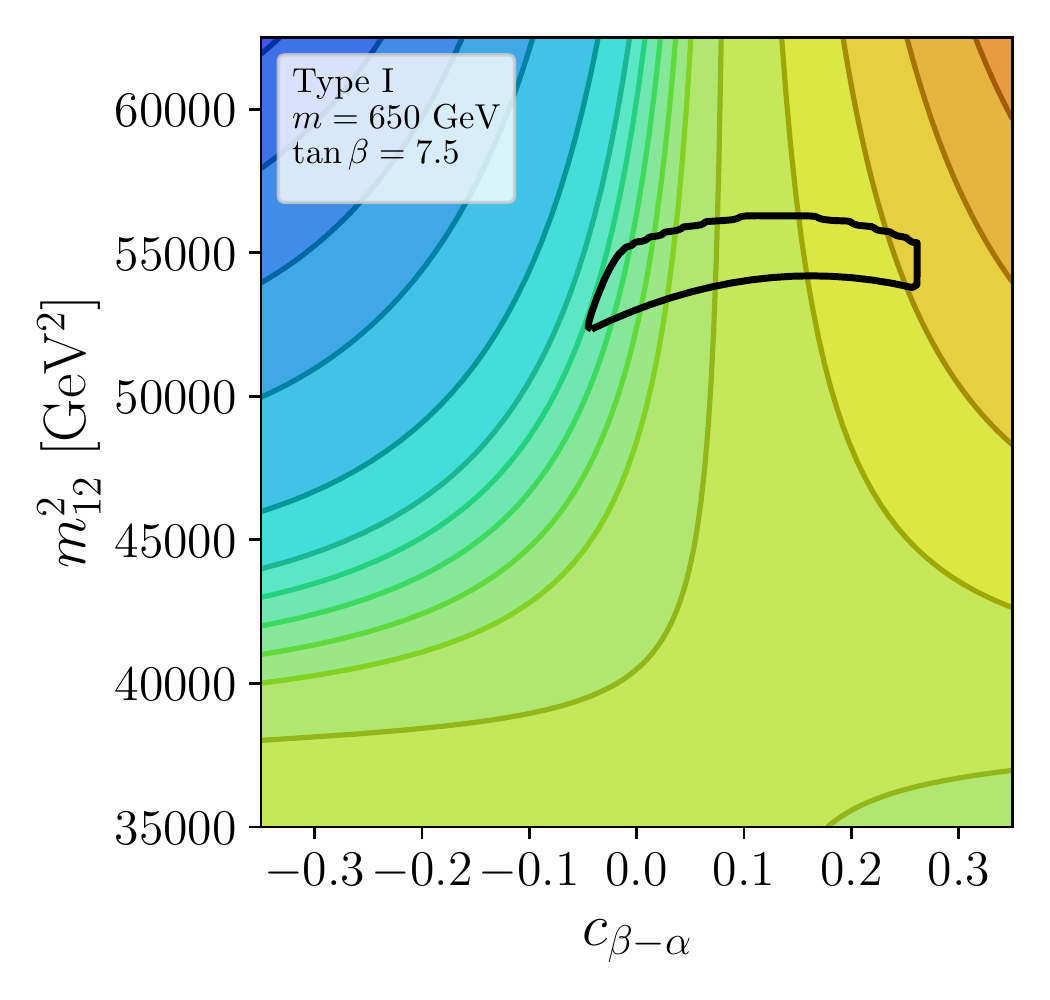}
\includegraphics[width=0.24\textwidth]{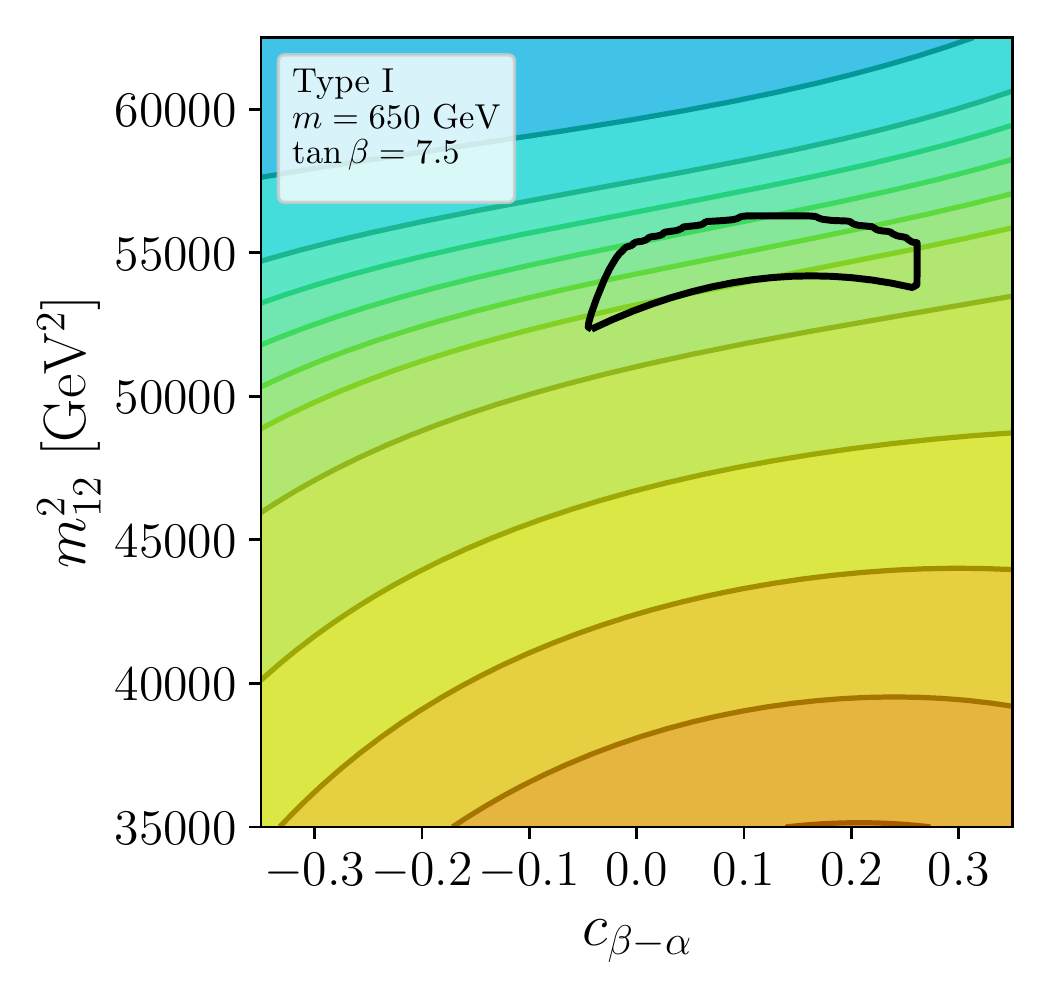}

\includegraphics[width=0.24\textwidth]{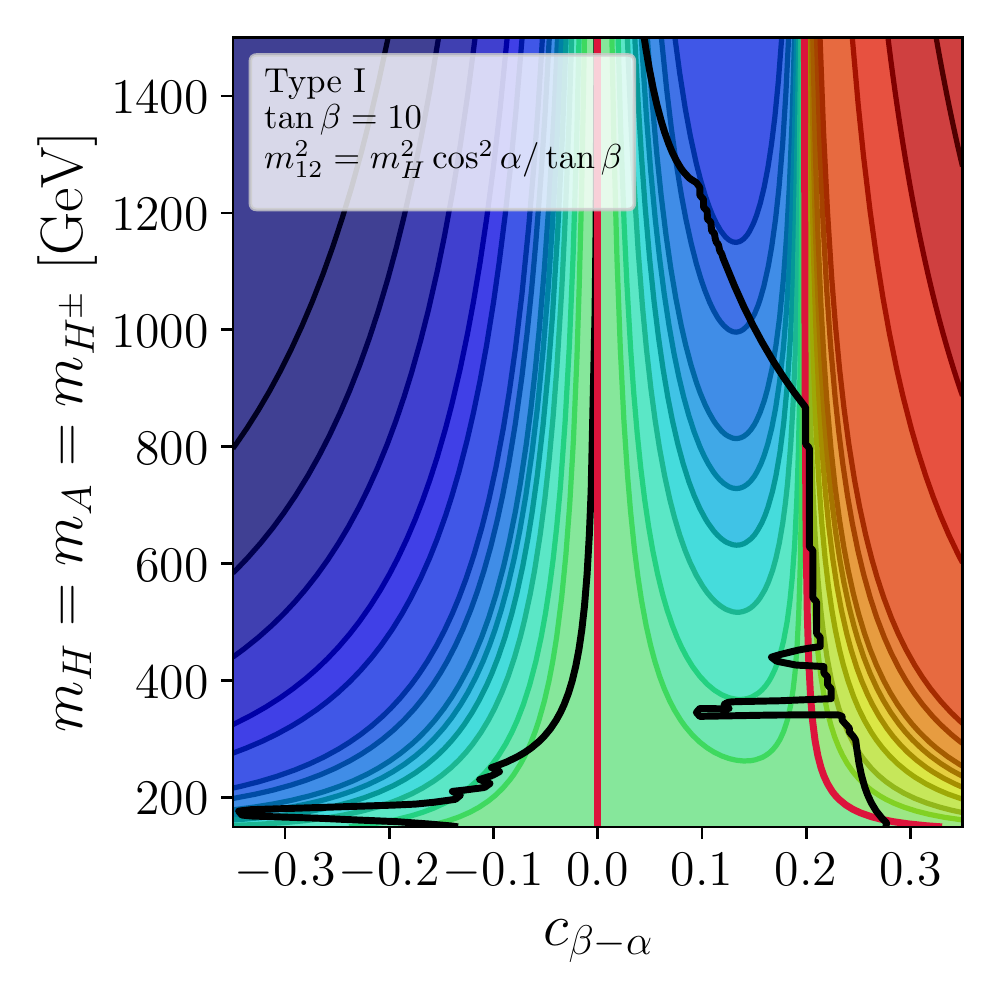}
\includegraphics[width=0.24\textwidth]{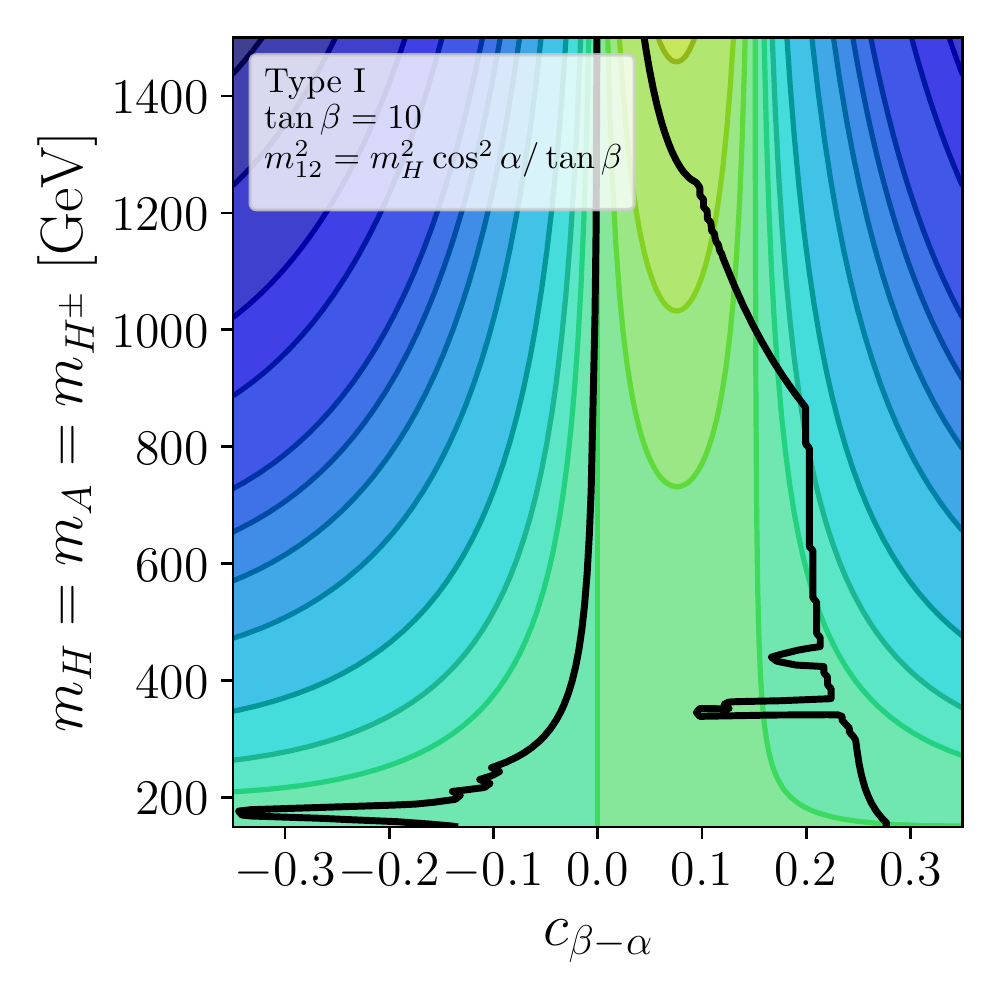}
\includegraphics[width=0.24\textwidth]{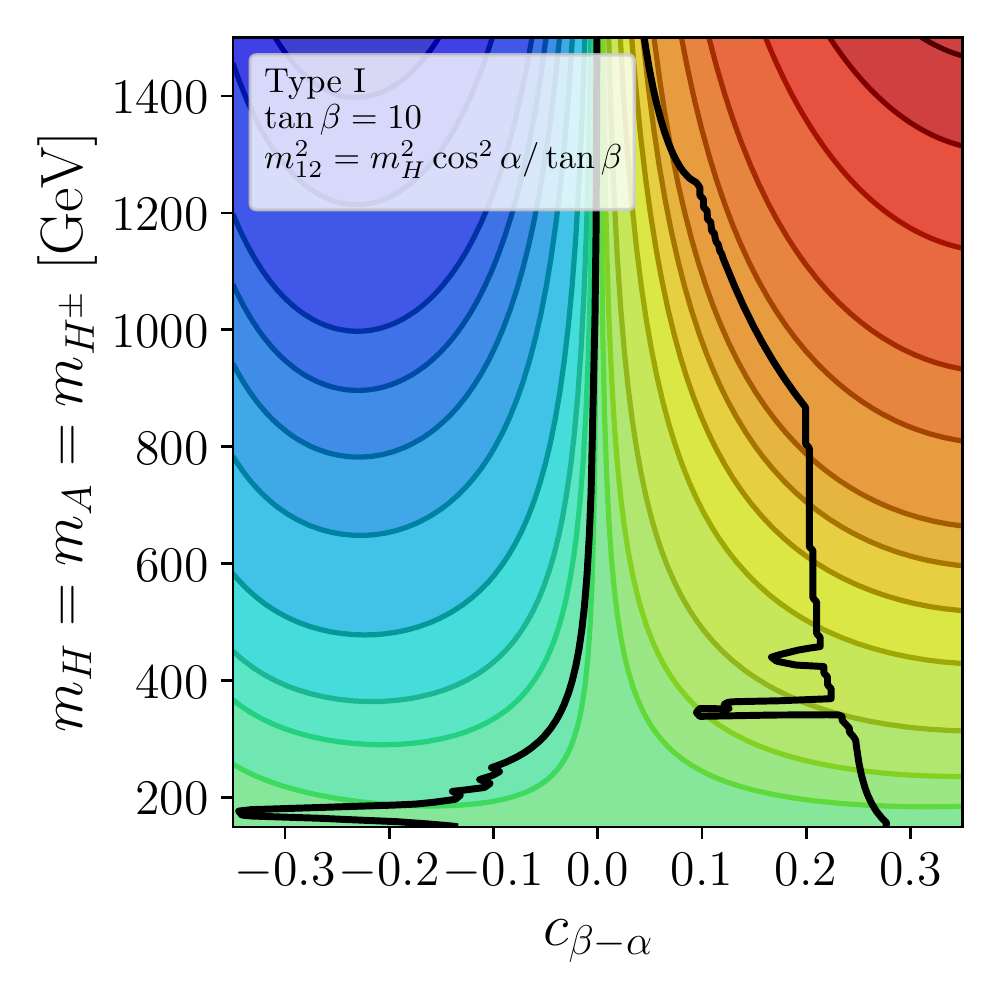}
\includegraphics[width=0.24\textwidth]{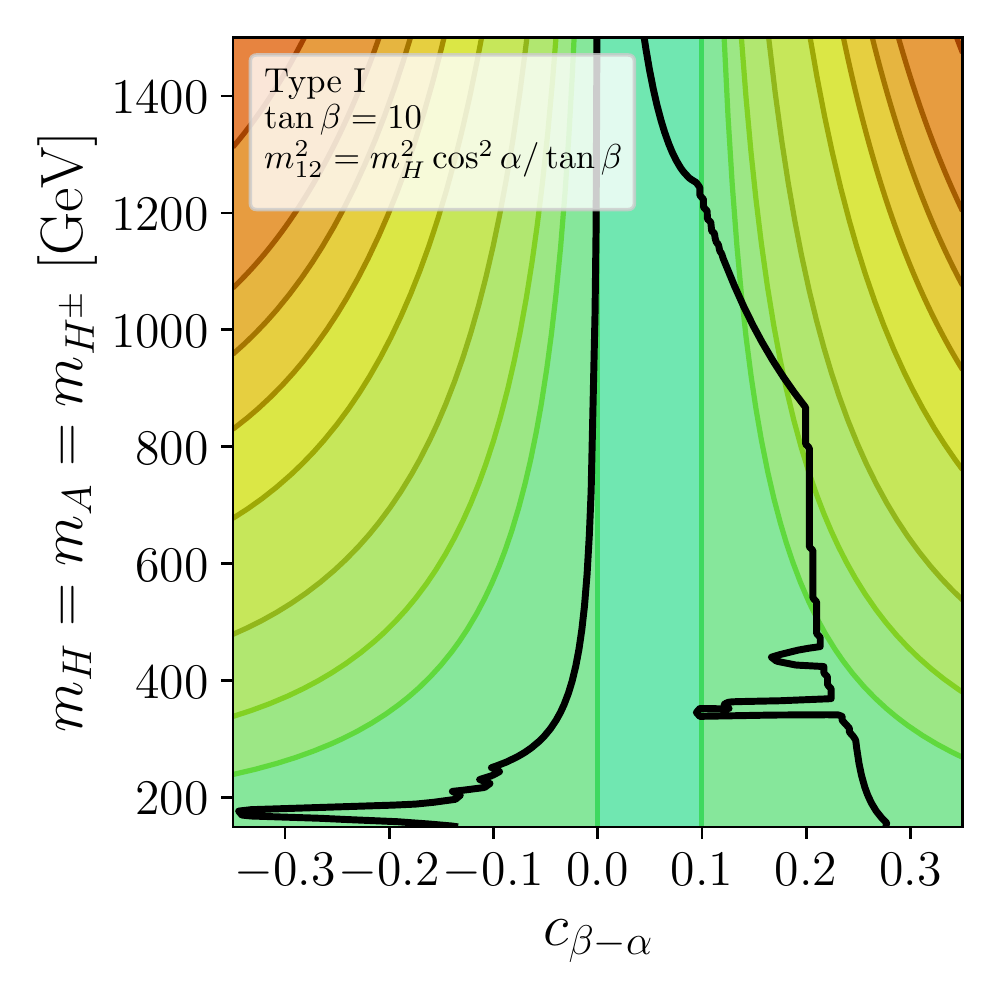}

\includegraphics[width=0.24\textwidth]{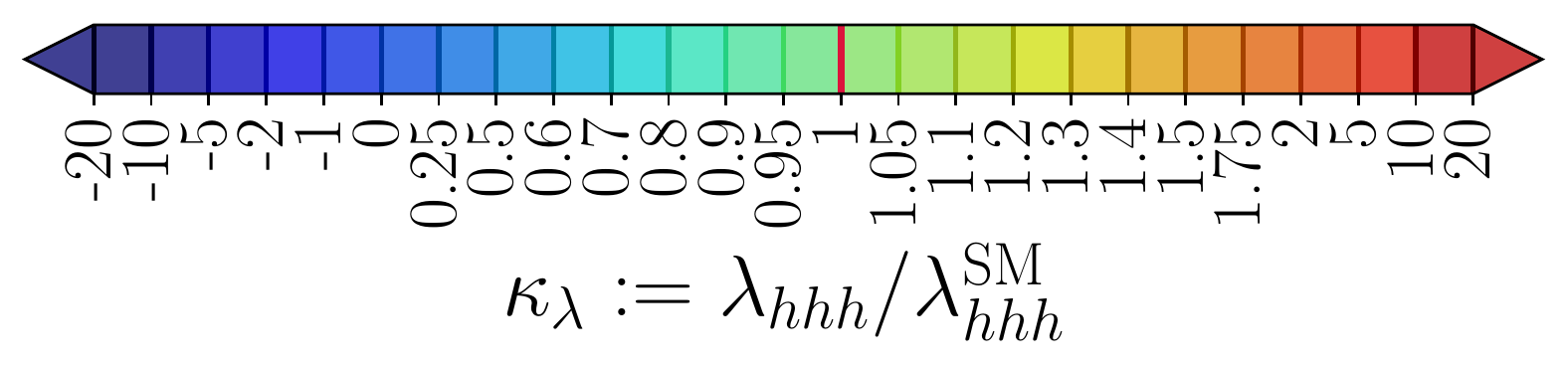}
\includegraphics[width=0.24\textwidth]{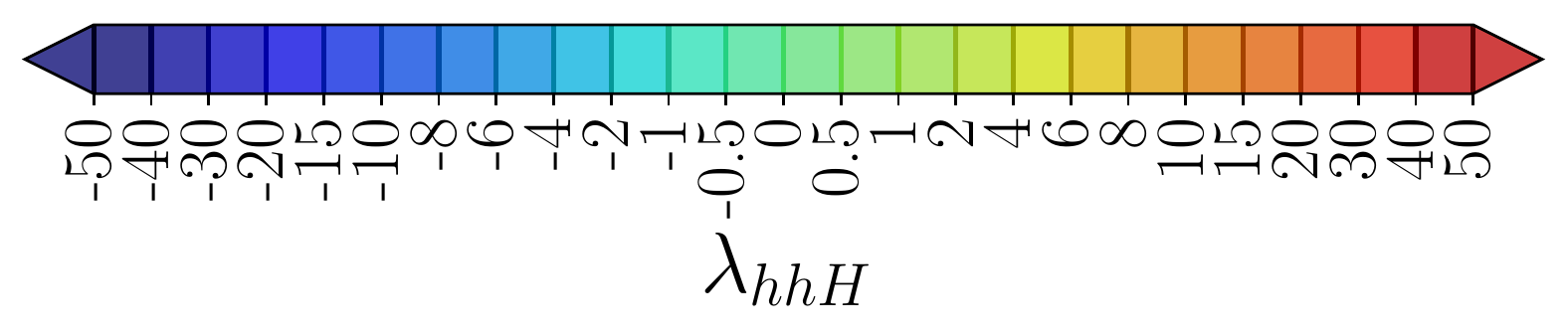}
\includegraphics[width=0.24\textwidth]{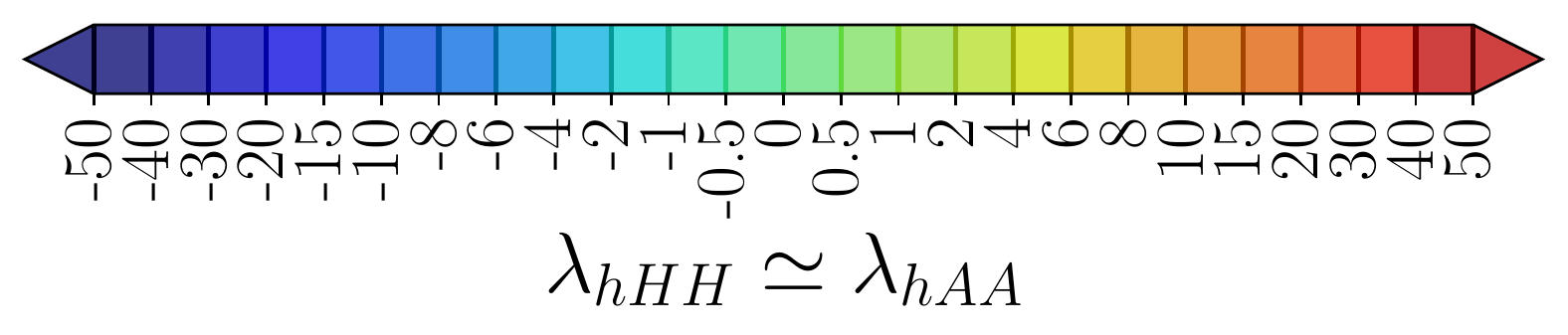}
\includegraphics[width=0.24\textwidth]{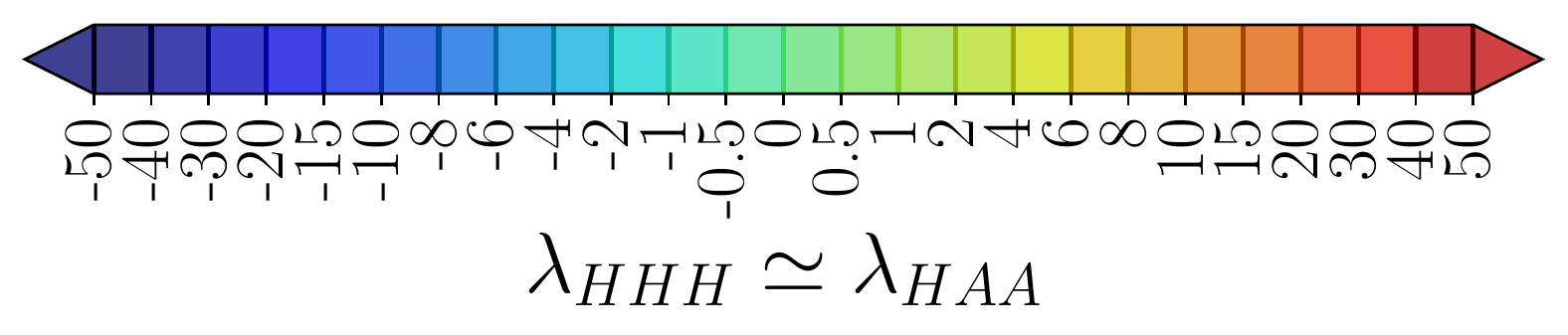}

\end{center}
\caption{Predictions for
  $\kala = \lahhh/\laSM$ (left),
  $\lahhH$ (2nd column), 
  $\lahHH \simeq \lahAA$ (3rd column)
  and $\laHHH \simeq \laHAA$ (right)
  in the 2HDM type~I benchmarks planes~1 (upper row), 2~(middle row),
  3~(lower row).
Shown as solid black lines are the exterior boundaries of the parameter
space regions that are 
allowed taking all theoretical and experimental constraints (see text).
The red lines in the left column indicate $\kala = 1$.}
\label{fig:la-I}
\end{figure}

The predictions for the cross sections depend on the various triple
Higgs couplings. Therefore in \reffi{fig:la-I} we reproduce the results
from \citere{Arco:2020ucn} 
for $\kala = \lahhh/\laSM$ (left), $\lahhH$ (2nd column), 
$\lahHH \simeq \lahAA$ (3rd column) and we add new predictions for 
$\laHHH \simeq \laHAA$ (right) 
in the 2HDM type~I benchmarks planes~1 (upper row), 2~(middle row),
3~(lower row).  The triple Higgs couplings involving $h$ are defined
according to \refeq{eq:lambda}.  The corresponding definitions are set for the other Higgs triple couplings not involving $h$,  which we do not include here explicitely for shortness.   
Shown in black solid lines are the exterior boundaries of the allowed
regions  of the 2HDM parameter space 
(see above). The red line in the left column indicates $\kala = 1$. 

\subsubsection{\boldmath{$hh$} production}
\label{sec:xshh-I}

\begin{figure}[htb!]
\vspace{-2em}
\begin{center}
\begin{subfigure}[b]{0.8\textwidth}
\includegraphics[width=0.5\textwidth]{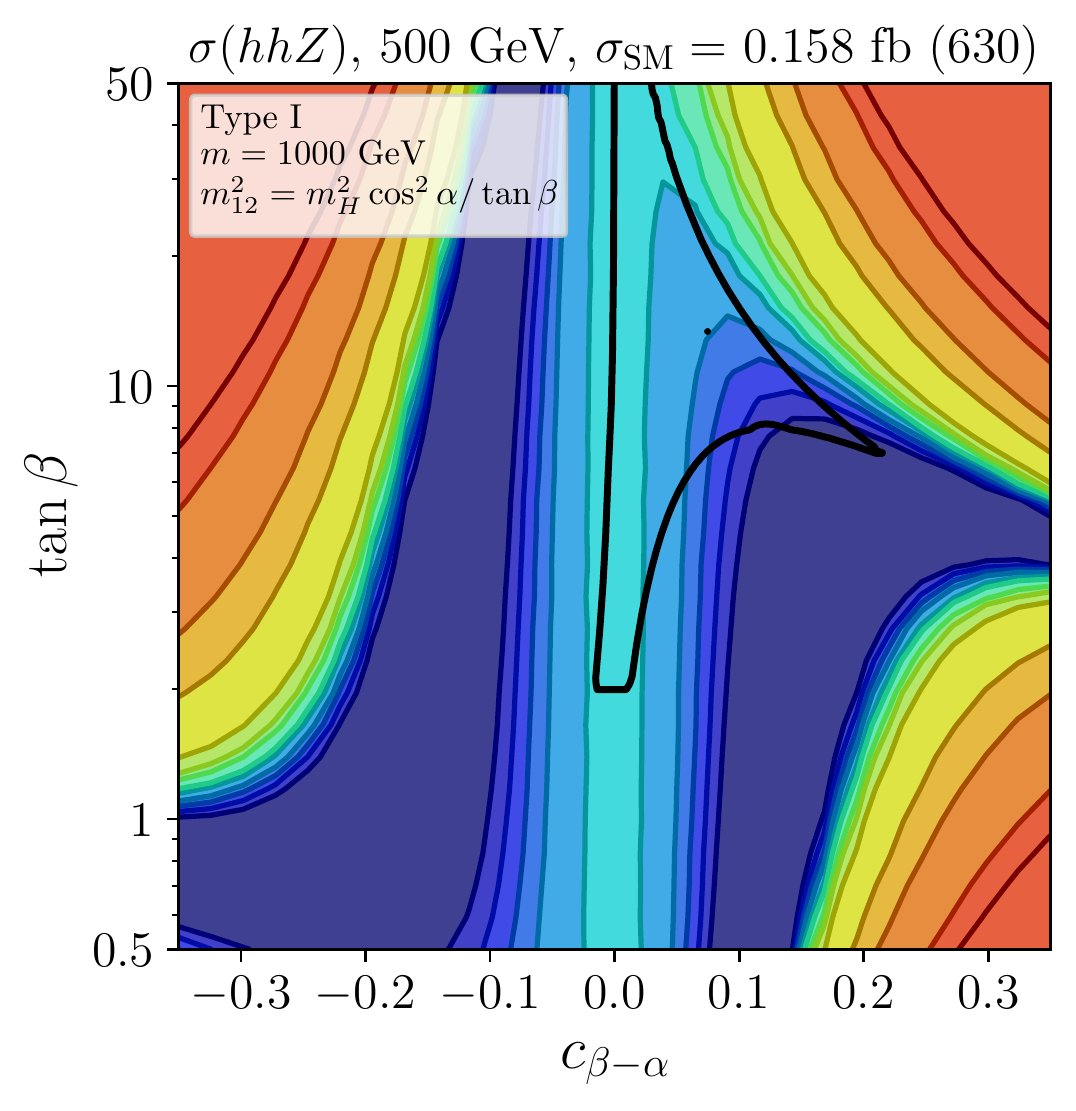}%
\includegraphics[width=0.5\textwidth]{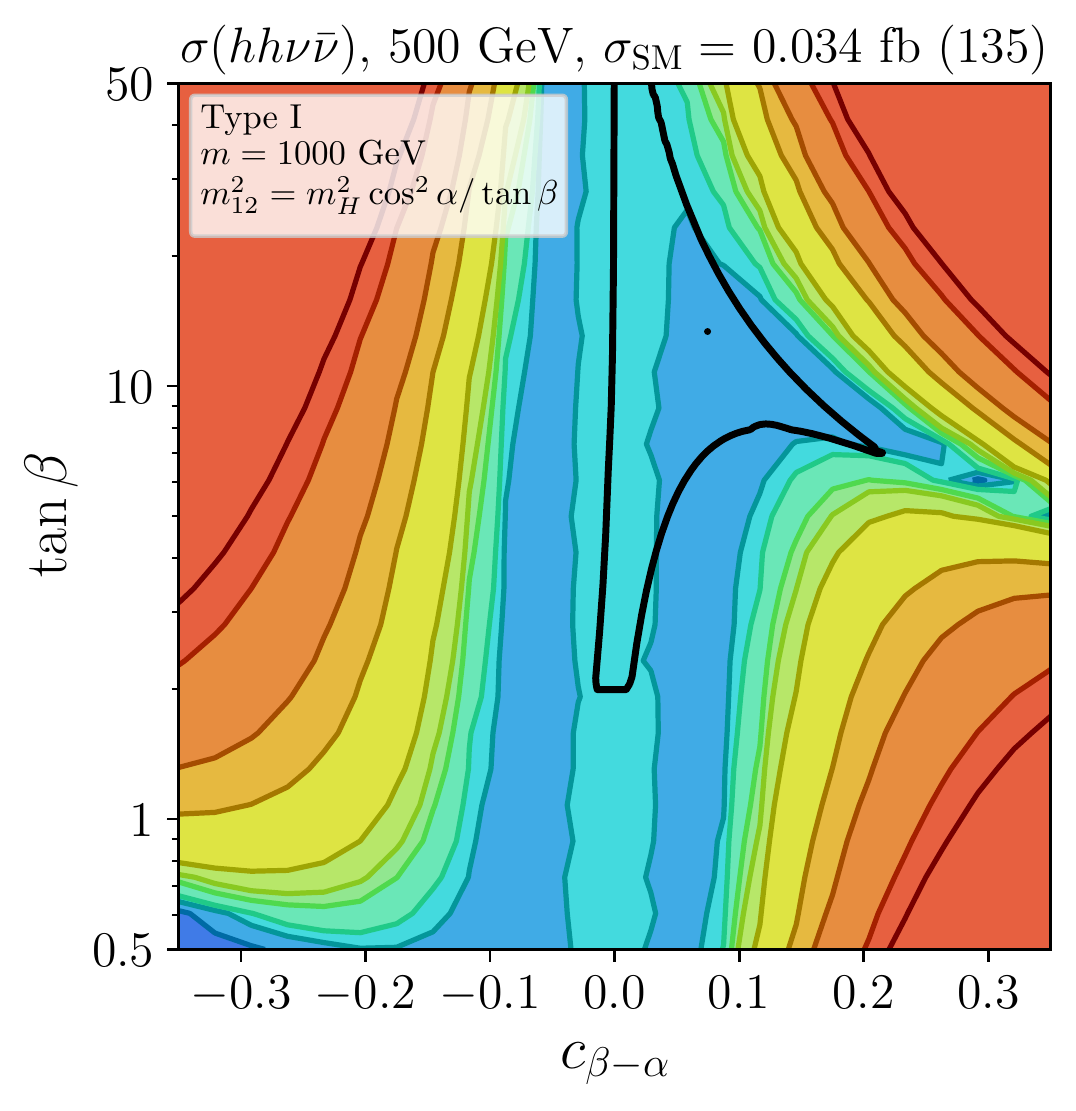}
\includegraphics[width=0.5\textwidth]{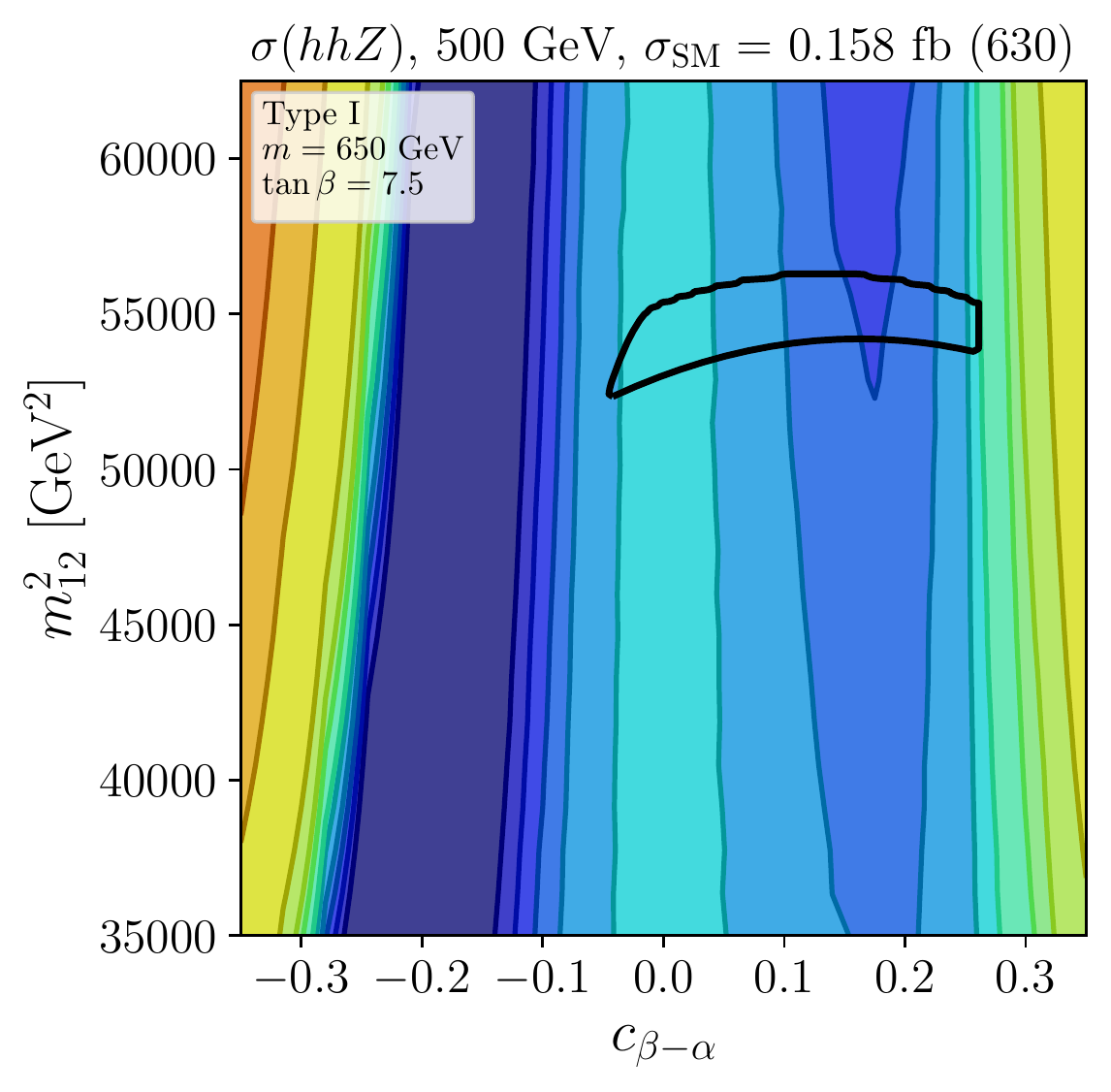}%
\includegraphics[width=0.5\textwidth]{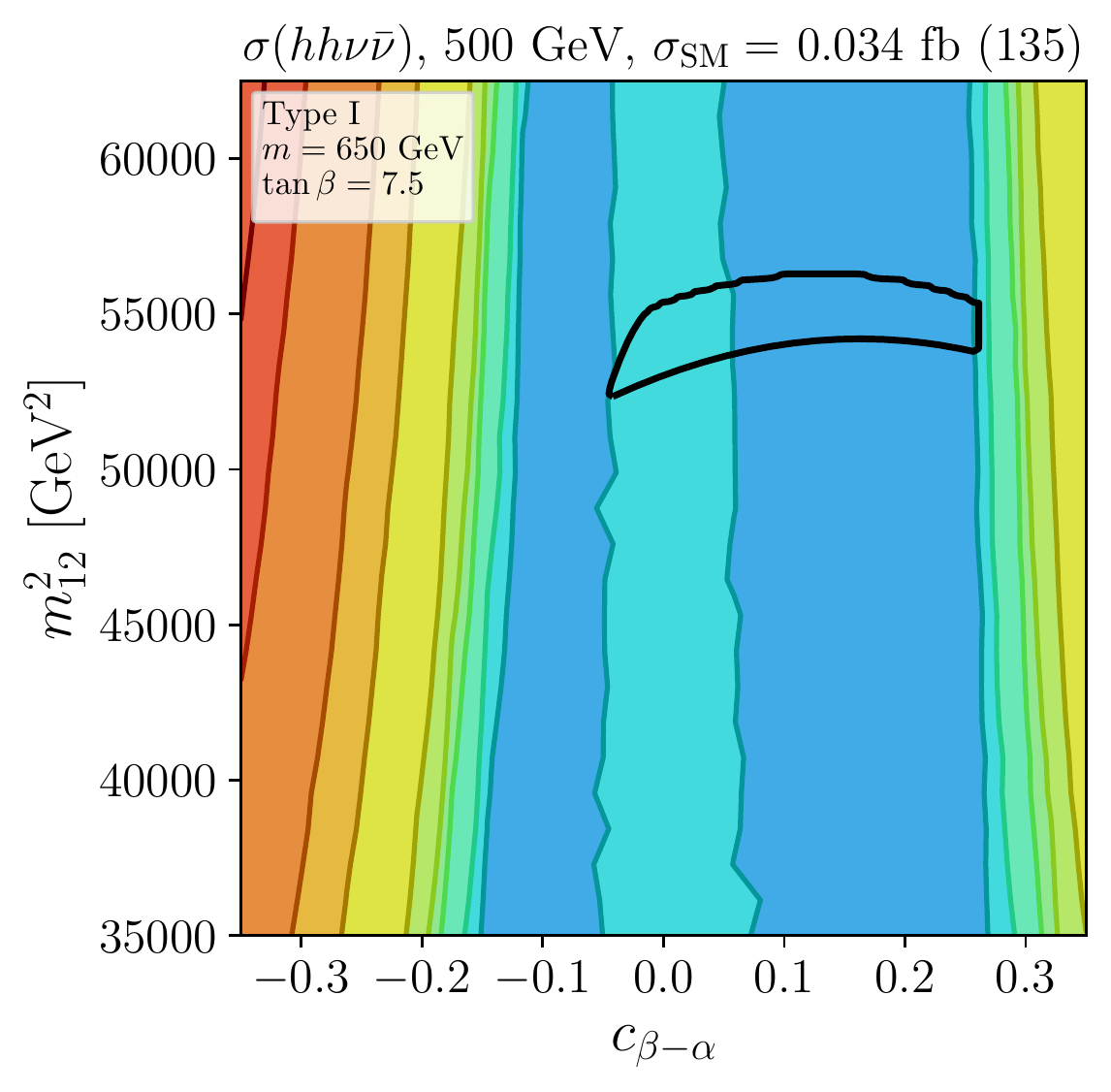}
\includegraphics[width=0.5\textwidth]{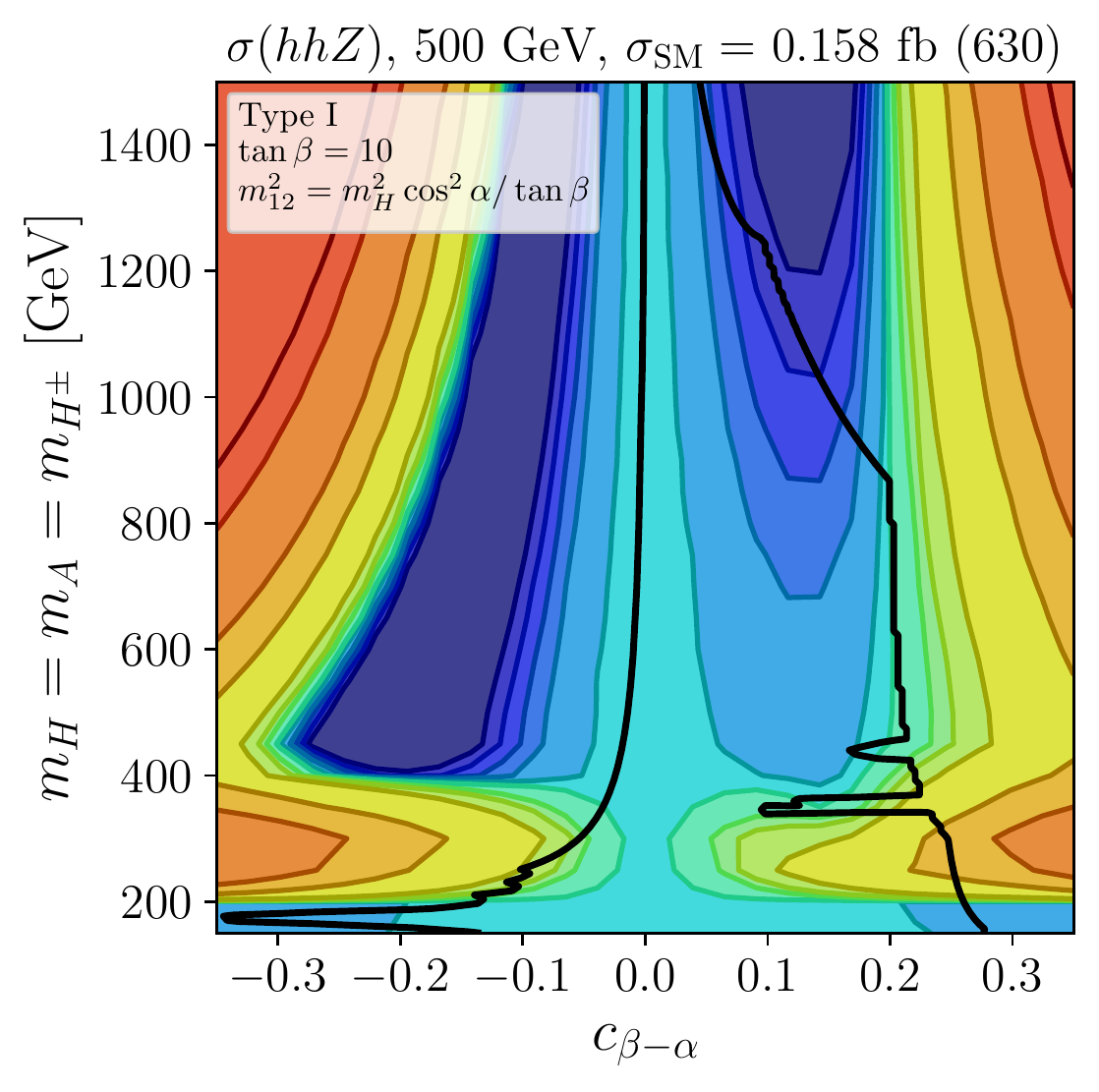}%
\includegraphics[width=0.5\textwidth]{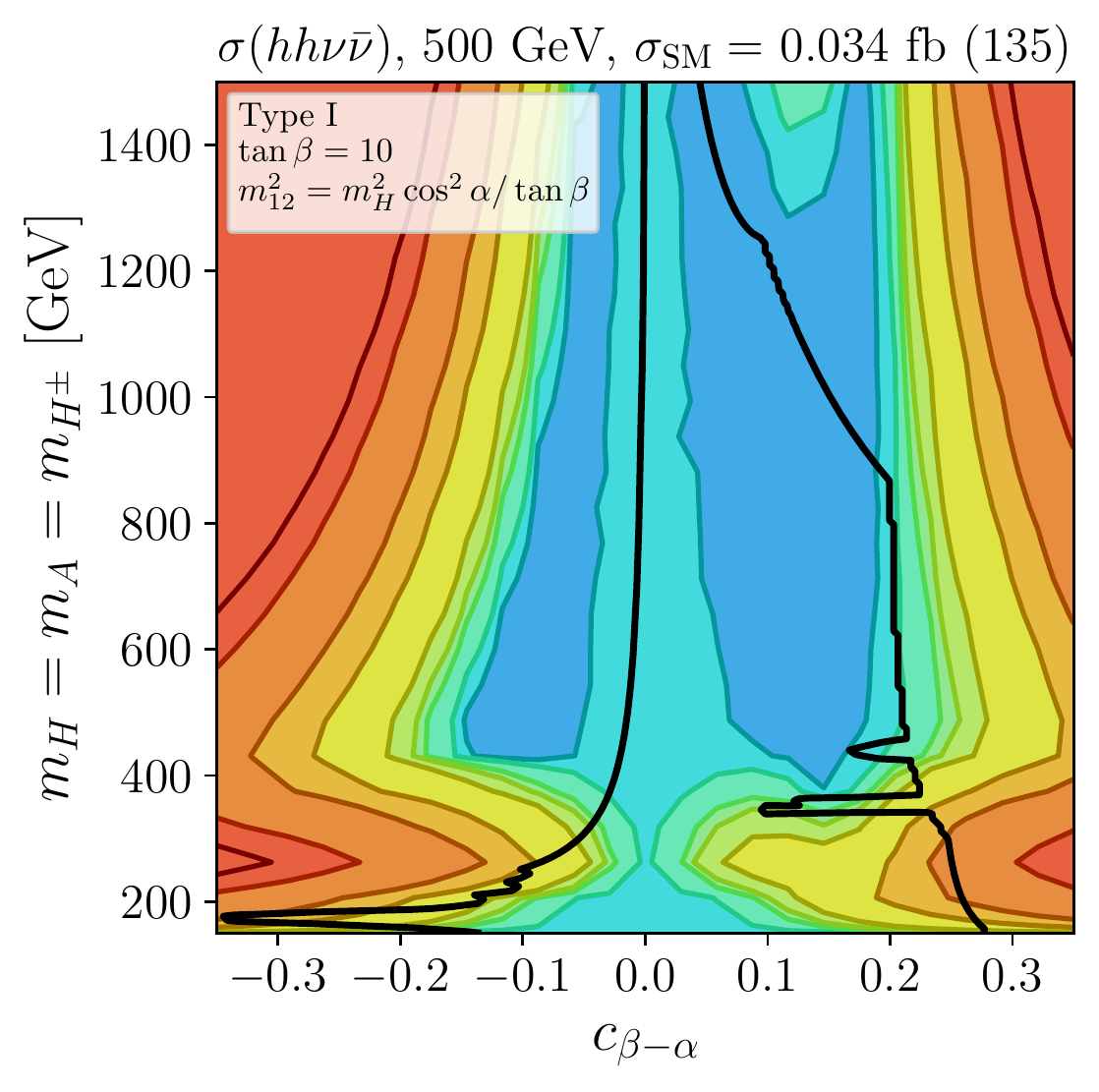}
\end{subfigure}	
\begin{subfigure}[b]{0.18\textwidth}
\includegraphics[height=0.42\textheight]{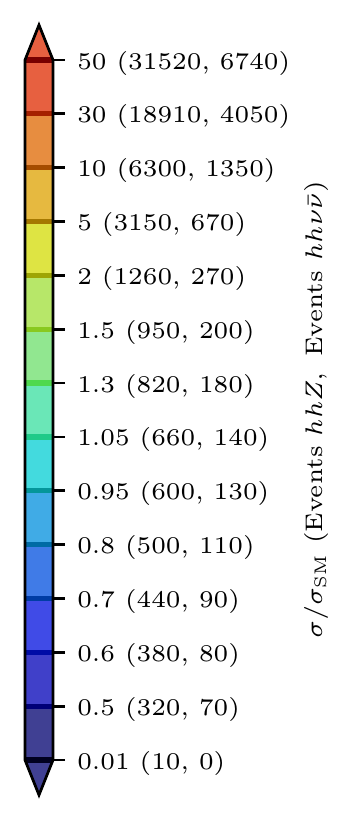}
\vspace{0.25\textheight}
\end{subfigure}
\end{center}
\vspace{-1em}
\caption{Cross sections for $e^+e^-\to hhZ$ (left) and
  $e^+e^-\to hh\nu\bar{\nu}$ (right) for $\sqrt{s}=500\gev$
  for the benchmark planes 1-3 (top to bottom).  The
  total allowed regions are those inside  the solid black lines.
The color code shows the cross section relative to the SM cross
section, with the numbers indicating the number of events for $\cL_{\rm int}$
as given in \refta{tab:ee} for $e^+e^-\to hhZ$ (left) and
  $e^+e^-\to hh\nu\bar{\nu}$ (right).} 
\label{fig:xs_hh_500-I}
\end{figure}

We start with the production of two light (SM-like) Higgs bosons. 
In \reffi{fig:xs_hh_500-I} we show the production cross sections at
$\sqrt{s} = 500 \gev$ for $e^+e^- \to Zhh$ (left) and
$e^+e^- \to hh\nu\bar\nu$ (right) in the benchmark planes 1-3 (top to
bottom). The SM cross sections for the two processes are
$0.158\,\fb$ and $0.034\,\fb$, corresponding to 630 and 135 events,
respectively, for the luminosity given in \refta{tab:ee}.  At this low
energy of $\sqrt{s} = 500 \gev$,  the $e^+e^- \to hh\nu\bar\nu$ channel
is dominated (as in the SM case) by the $Z^*$ mediated configurations,
i.e.\  by $e^+e^- \to Z^* \to hh\nu\bar\nu$, and the VBF configurations
being subdominant do not play a relevant role.   Therefore,  the
potential accessibility to the two involved triple Higgs couplings
$\lahhh$ and $\lahhH$ will mostly come from the diagrams
to the left in \reffi{fig:diagrams}. 

Three main effects can change the 2HDM prediction w.r.t.\ the SM
predictions. The first one are deviations in $\lahhh$ from its SM value,
i.e.\ $\kala \neq 1$.  This will modify the contributions of type
$e^+e^- \to Z^* \to Z h^* \to Z hh$ where the intermediate $h^*$ is
off-shell. The second one are additional diagrams
involving in particular an intermediate $H$,  like
$e^+e^- \to Z^* \to Z H \to Z hh$,  which can be produced on-shell in
contrast to the previous case. 
Correspondingly, large effects are expected for large values of
$\lahhH$, especially when the $H$ can be produced on-shell.
The last contribution
comes from the sub-processes
$e^+e^- \to Z^* \to Ah\to Zhh\ (\to\nu\bar{\nu}hh)$,
that can also give sizable contributions to the cross section
if the virtual $A$ can propagate on-shell,  despite the fact that
  they are $\propto g_{ZhA} \propto \CBA$.
However,  these $A$ mediated diagrams do
not carry any sensitivity to the triple Higgs couplings.

In the alignment limit, reached for $\CBA = 0$, the 2HDM cross sections
reproduce exactly the SM cross sections. Here one has $\kala = 1$ and 
$\lahhH = 0$, as can also be seen in the first two columns of
\reffi{fig:la-I}.  Also,  as mentioned above,
 $g_{ZhA}\propto\CBA$ and therefore the $A^\ast$ mediated diagrams 
also vanish in the alignment limit. However, as
can be seen in the color code of \reffi{fig:xs_hh_500-I}, SM-like cross
sections are also reached away from $\CBA = 0$ in all three planes. In
these regions $\lahhH \sim 0$ (see the 2nd column of \reffi{fig:la-I})
due to cancellations, as has been discussed in \citere{Arco:2020ucn}. 
Furthermore, $\lahhh$ does not
strongly deviate from the SM value, as can be seen in the left column of
\reffi{fig:la-I}, again due to cancellations in the various
contributions to $\lahhh$, see the discussion in \citere{Arco:2020ucn}. 

On the other hand, large enhancements of the cross section predictions
are found for large values of $|\lahhH|$. Furthermore, in particular for
a near resonant di-Higgs production, $\MH \sim 2 \Mh \approx 250 \gev$,
the cross sections are found to be strongly enhanced. Correspondingly,
the largest cross sections inside the allowed regions are found in the
benchmark plane~3 for $\MH \sim 250 \gev$. Here the enhancement goes up
to $\sim 6.5$ (13) times the SM cross section for $hhZ$ ($hh\nu\bar\nu$).
Here it is important to note that in these regions of the
parameter space it is not sufficient to 
calculate the cross section in the narrow width approximation, (NWA)  i.e.\
via $HZ$ or $H\nu\bar\nu$ production with
the subsequent decay $H \to hh$. As will be discussed in detail in
\refse{sec:sensitivity}, the reason for the failure of the NWA to
provide an accurate prediction of the total cross section for the
di-Higgs bosons production in $e^+e^-$ collisions can be
understood as follows.  Outside the resonant region in the invariant
mass of the final Higgs pairs the remaining non resonant diagrams  
are very relevant and account for a sizeable contribution to the total
cross section that cannot be neglected. 

Also cross sections substantially smaller than in the SM can be found.
This can happen in particular for $hhZ$ in the benchmark plane~1 in the
``tip'' of the allowed region where $\sim 0.3$ times the SM cross
section is found. In the planes~2 and~3 the largest suppression are
found for $\CBA \sim 0.15$, where $\sim 0.7$ and $\sim 0.5$ times the SM
prediction is found. These regions correspond to the smallest allowed
values of $\lahhh$, see the left column of \reffi{fig:la-I}.

\bigskip


\begin{figure}[htb!]
\vspace{-2em}
\begin{center}
\begin{subfigure}[b]{0.8\textwidth}
\includegraphics[width=0.5\textwidth]{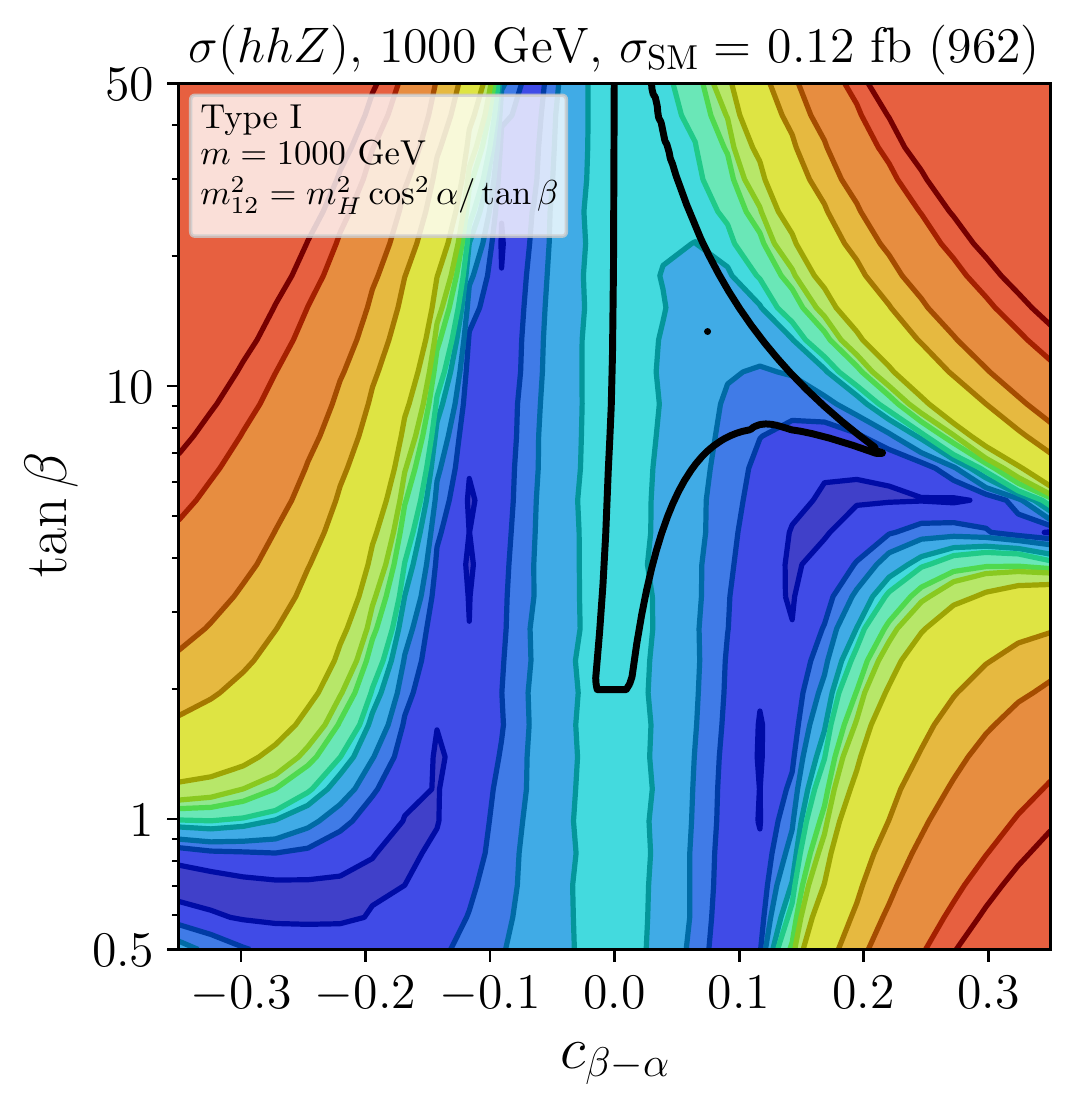}%
\includegraphics[width=0.5\textwidth]{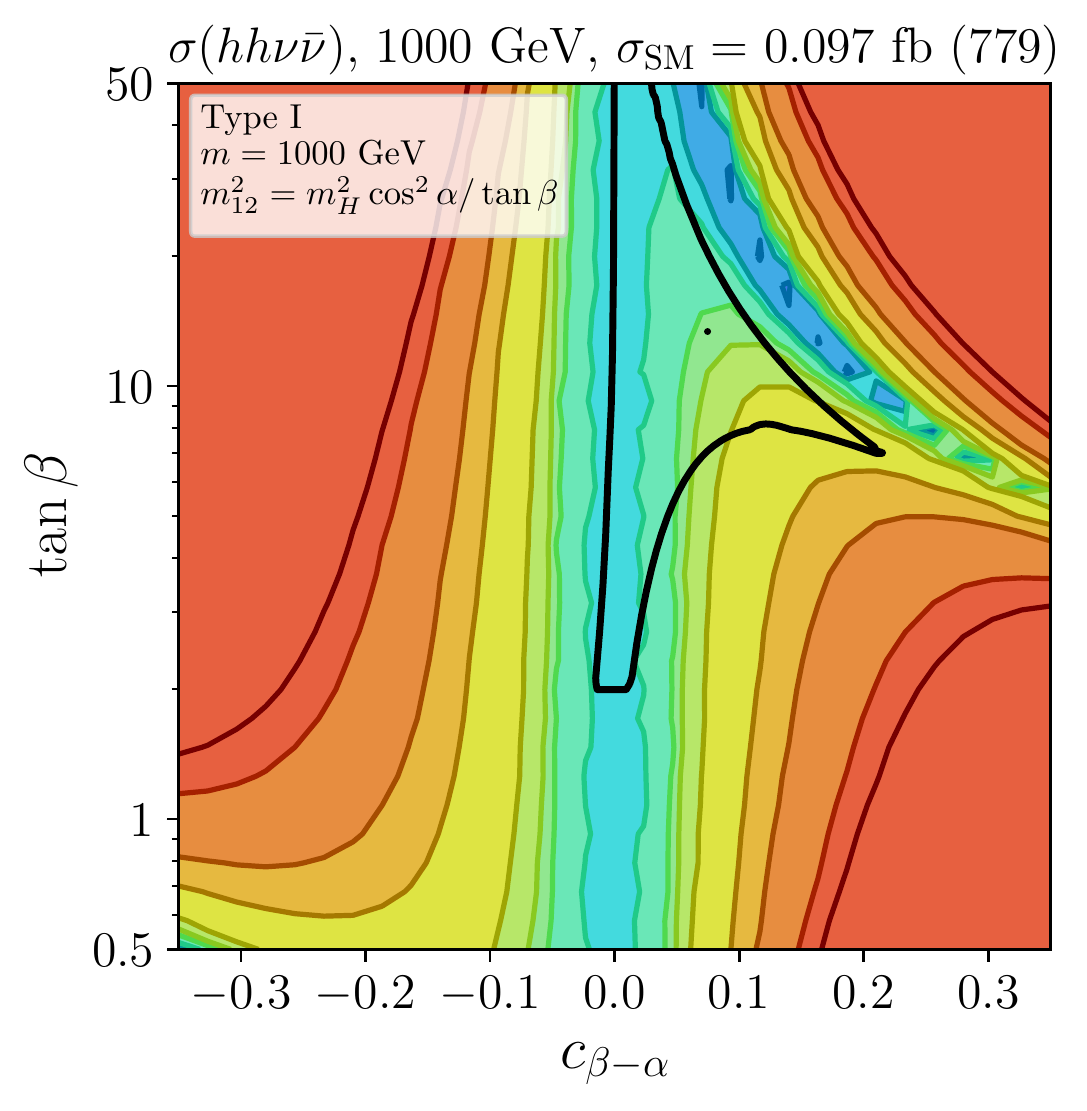}\\[1em]
\includegraphics[width=0.5\textwidth]{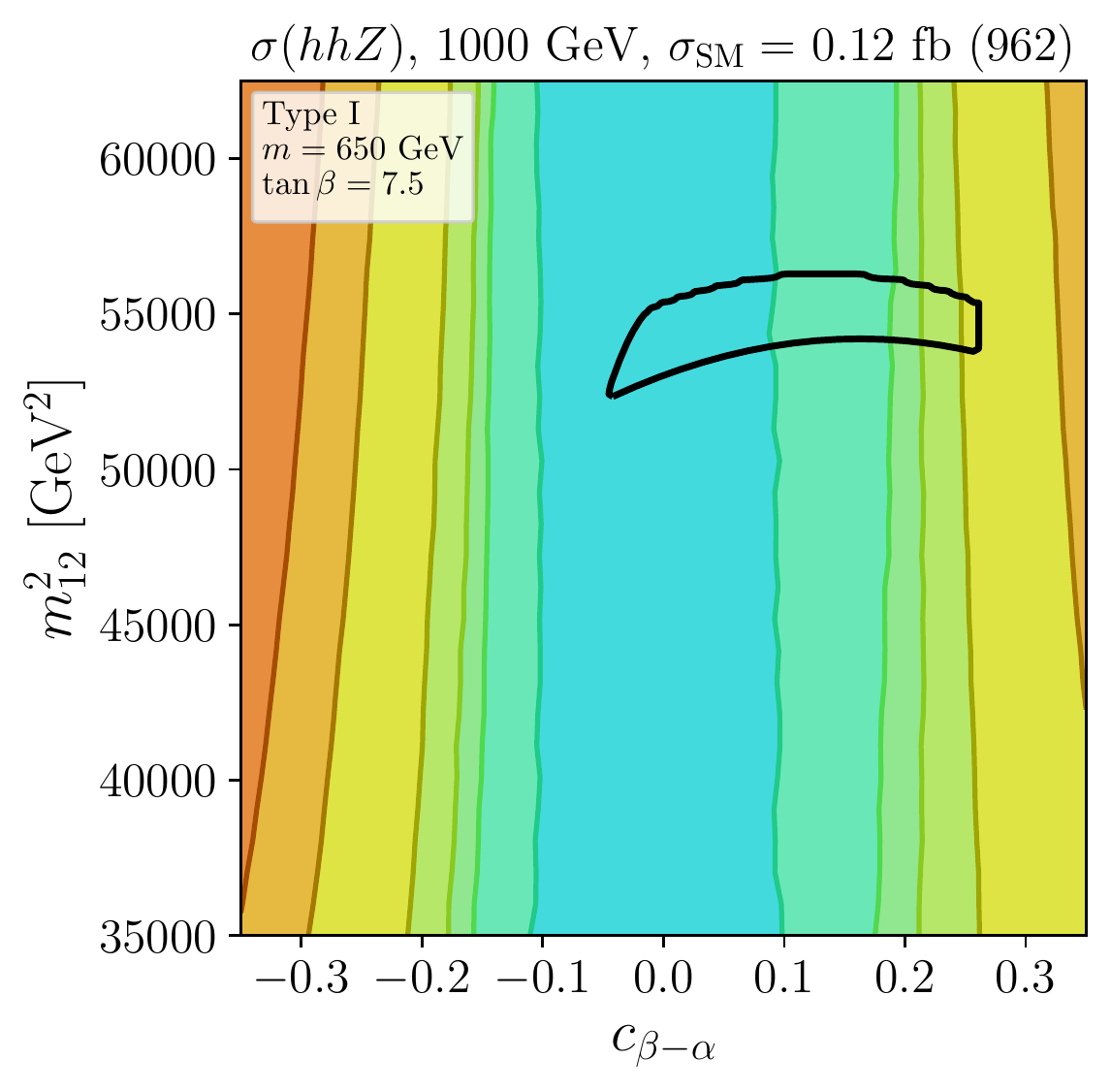}%
\includegraphics[width=0.5\textwidth]{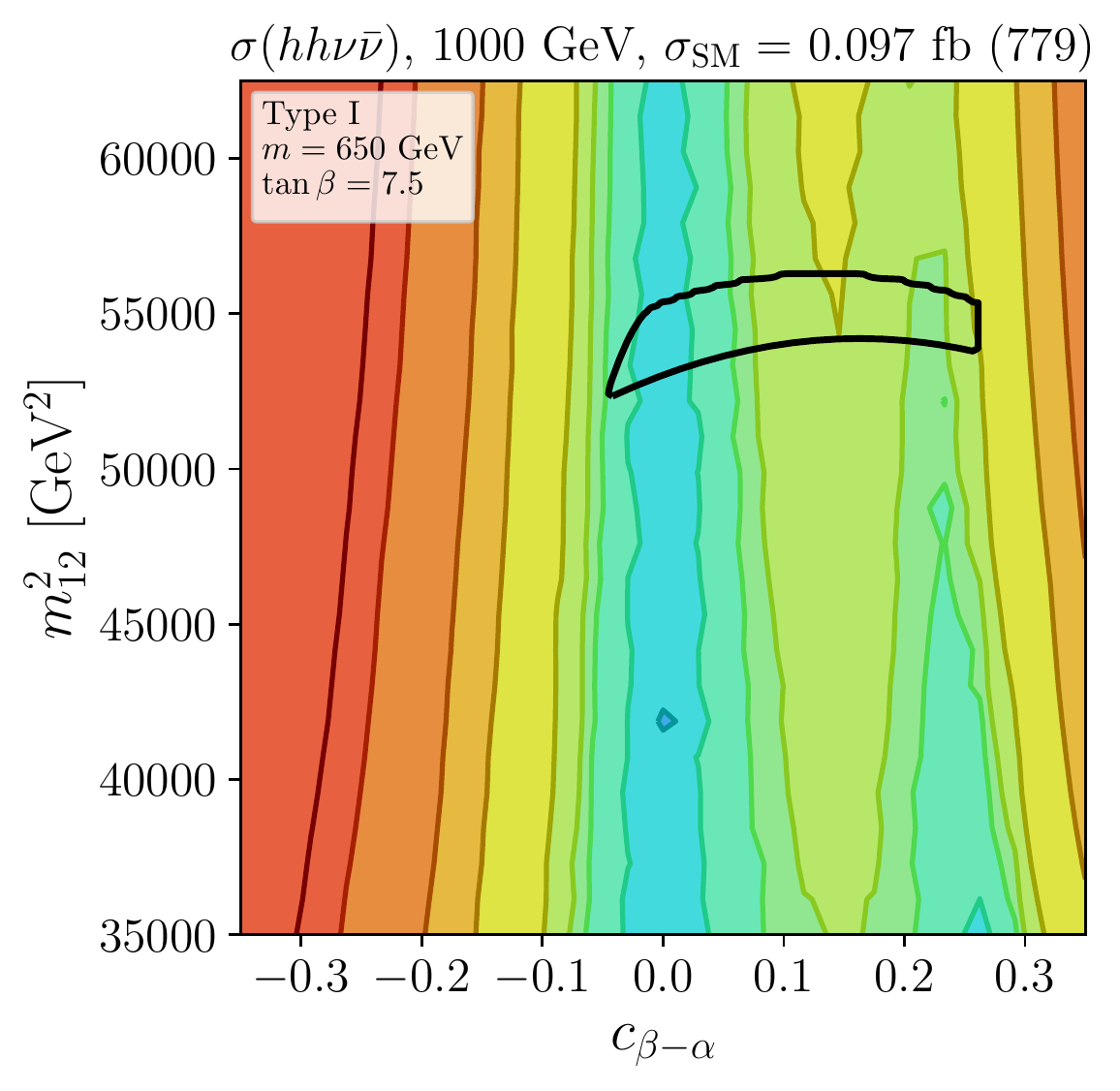}\\[1em]
\includegraphics[width=0.5\textwidth]{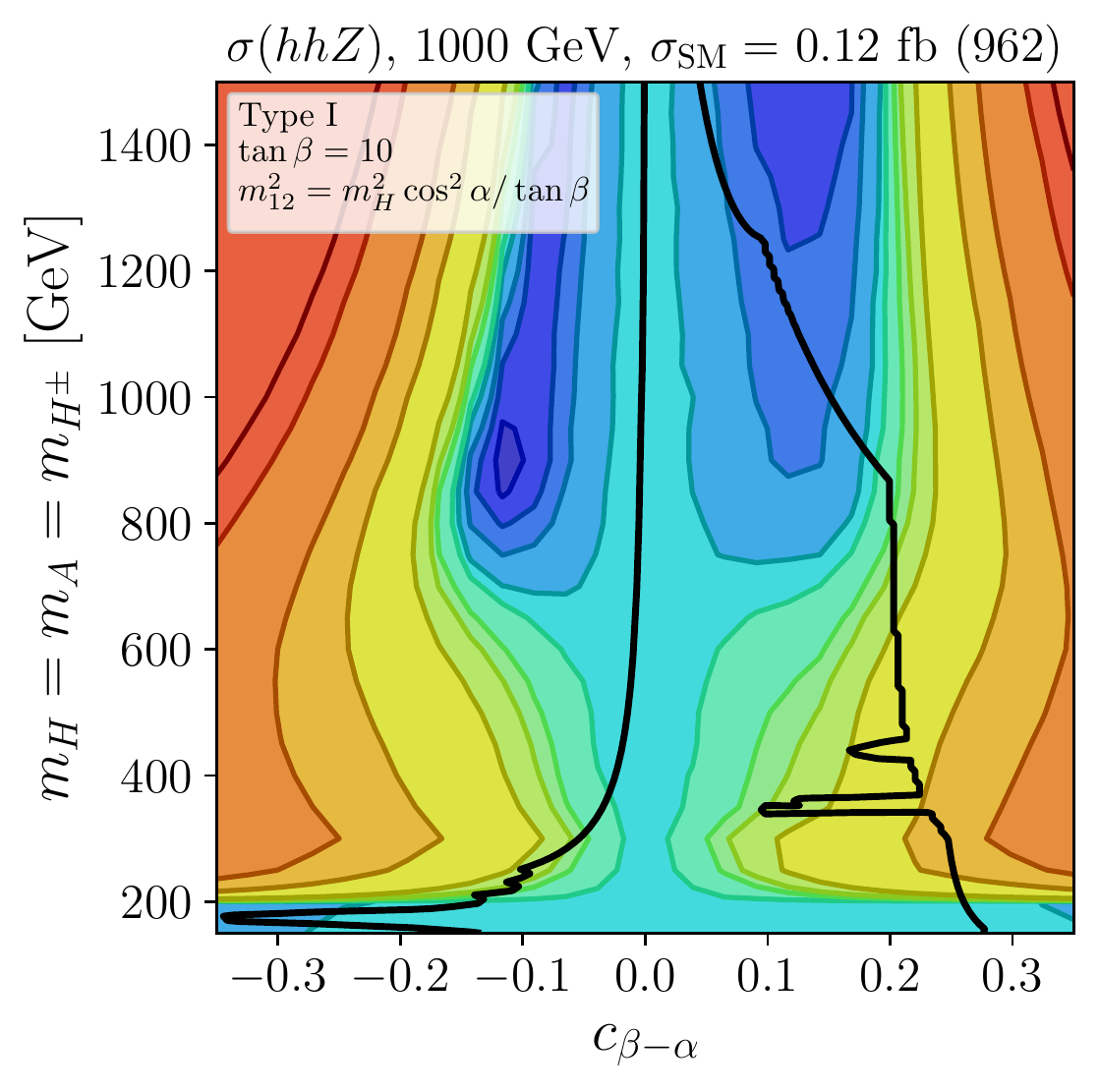}%
\includegraphics[width=0.5\textwidth]{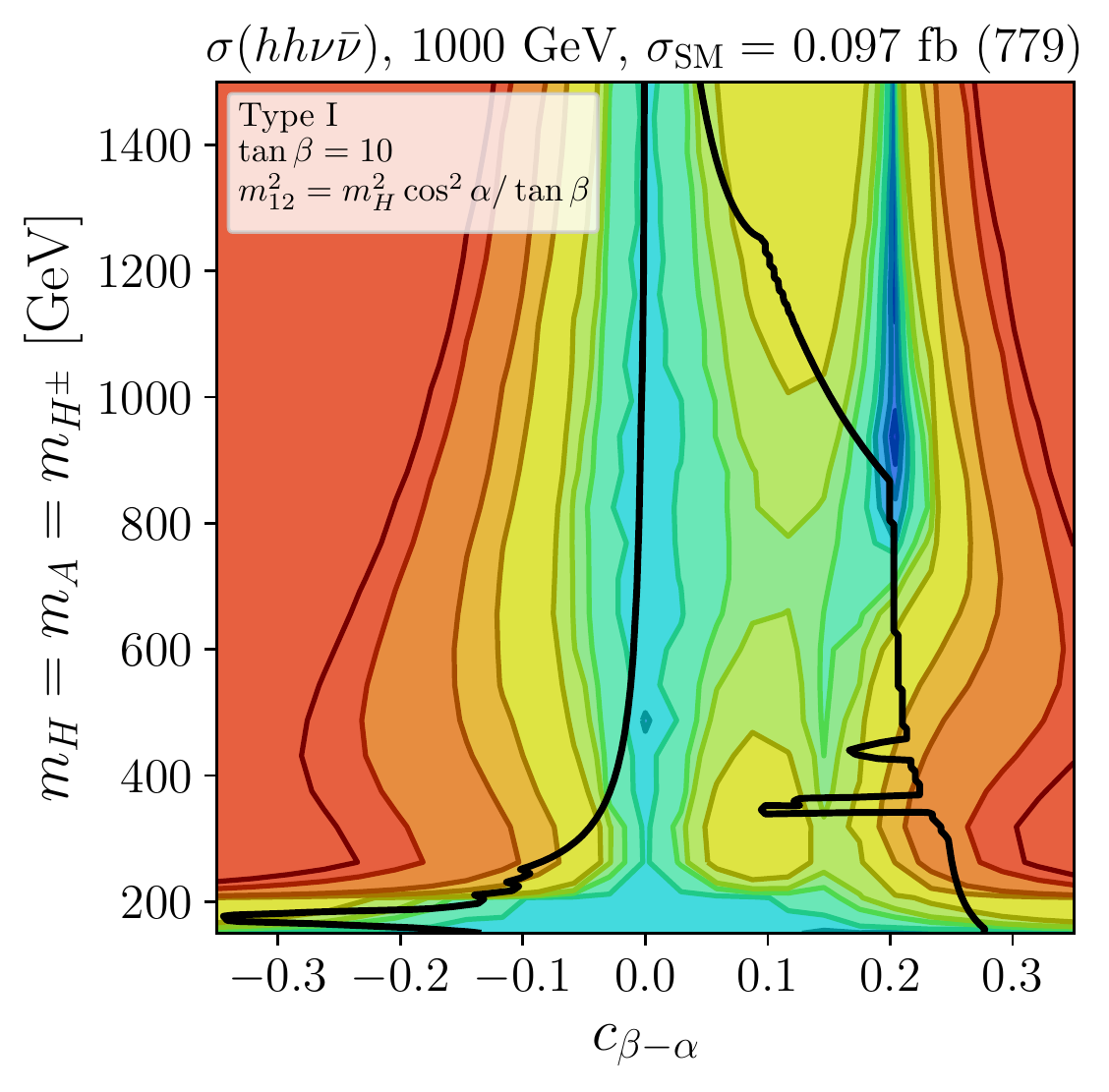}
\end{subfigure}	
\begin{subfigure}[b]{0.18\textwidth}
\includegraphics[height=0.42\textheight]{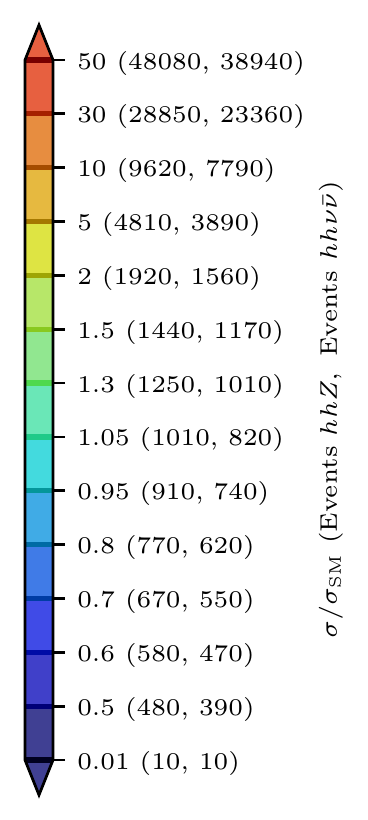}
\vspace{0.25\textheight}
\end{subfigure}
\end{center}
\caption{Cross sections for $e^+e^-\to hhZ$ (left) and
  $e^+e^-\to hh\nu\bar{\nu}$ (right) for $\sqrt{s}=1000\gev$
  for the benchmark planes 1-3 (top to bottom). The
  colors and line styles are as in \protect\reffi{fig:xs_hh_500-I}.}
\label{fig:xs_hh_1000-I}
\end{figure}

We now turn to the results for $\sqrt{s} = 1000 \gev$, which are shown
in \reffi{fig:xs_hh_1000-I} with the same color coding and line styles
as in \reffi{fig:xs_hh_500-I}.
At this center of mass energy, both channels have comparable cross 
sections in the SM. For $hhZ$ and $hh\nu\bar\nu$ production one finds SM
cross sections of  
$0.120\,\fb$ and $0.097\,\fb$, corresponding to 962 and 779 events,
for the luminosity given in \refta{tab:ee}.  One difference respect to
the previous case is that at this higher energy of $\sqrt{s} = 1000
\gev$,  the VBF subprocess enters in  $hh\nu\bar\nu$ more relevantly
than for $\sqrt{s} = 500 \gev$.  Then the sensitivity to $\lambda_{hhh}$
and $\lambda_{hhH}$ enters via the two types of diagrams  (left and
right) in  \reffi{fig:diagrams}. 

The overall dependence of the cross sections on the 2HDM parameters is
similar to the case of $\sqrt{s} = 500 \gev$. Due to the larger
center-of-mass energy here, an on-shell production of the heavy
$\cp$-even scalar is possible for larger values of $\MH$.  This is true
for the intermediate $H$ in both types of diagrams (left and right) in
\reffi{fig:diagrams}. 

As for $\sqrt{s} = 500 \gev$ the largest enhancements within the allowed
regions are found for
$\MH \sim 250 \gev$ and larger values of $\CBA$. In the benchmark
plane~3 we find enhancements of $\sim 6$ and $\sim 13$ w.r.t.\ the SM
predictions for $hhZ$ and $hh\nu\bar\nu$, respectively. 
Within benchmark plane~2 enhancements of $\sim 2$ are found for
$\CBA \sim 0.25$ (0.15) for the two production cross sections,
respectively. 

In benchmark plane~1 with $\MH = 1000 \gev$ the impact of diagrams
involving the~$H$ are negligible, and the cross sections are controlled
by $\kala$. The different signs of interference with the diagrams not
involving the $hhh$ vertex lead to a different numerical behavior of the
two cross sections. In the ``tip'' of the allowed regions, where $\kala$
reaches the smallest possible values (in the allowed region, see the
upper left plot of \reffi{fig:la-I}), we find a
suppression of $\sim 0.6$ w.r.t.\ the SM cross section for $hhZ$, while
for $hh\nu\bar\nu$ we find an enhancement of $\sim 6$. 

\bigskip


\begin{figure}[htb!]
\vspace{-2em}
\begin{center}
\begin{subfigure}[b]{0.8\textwidth}
\includegraphics[width=0.5\textwidth]{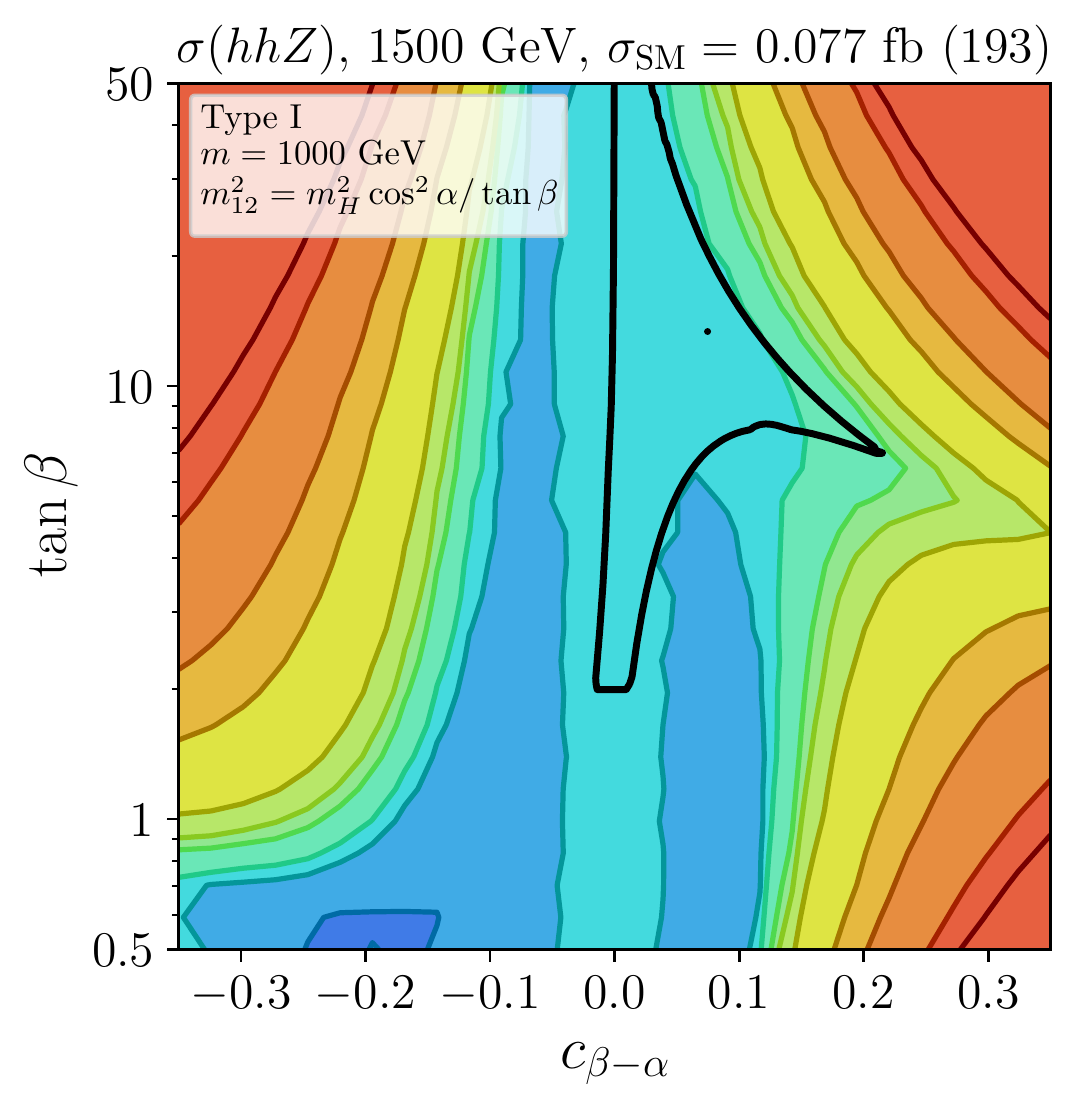}%
\includegraphics[width=0.5\textwidth]{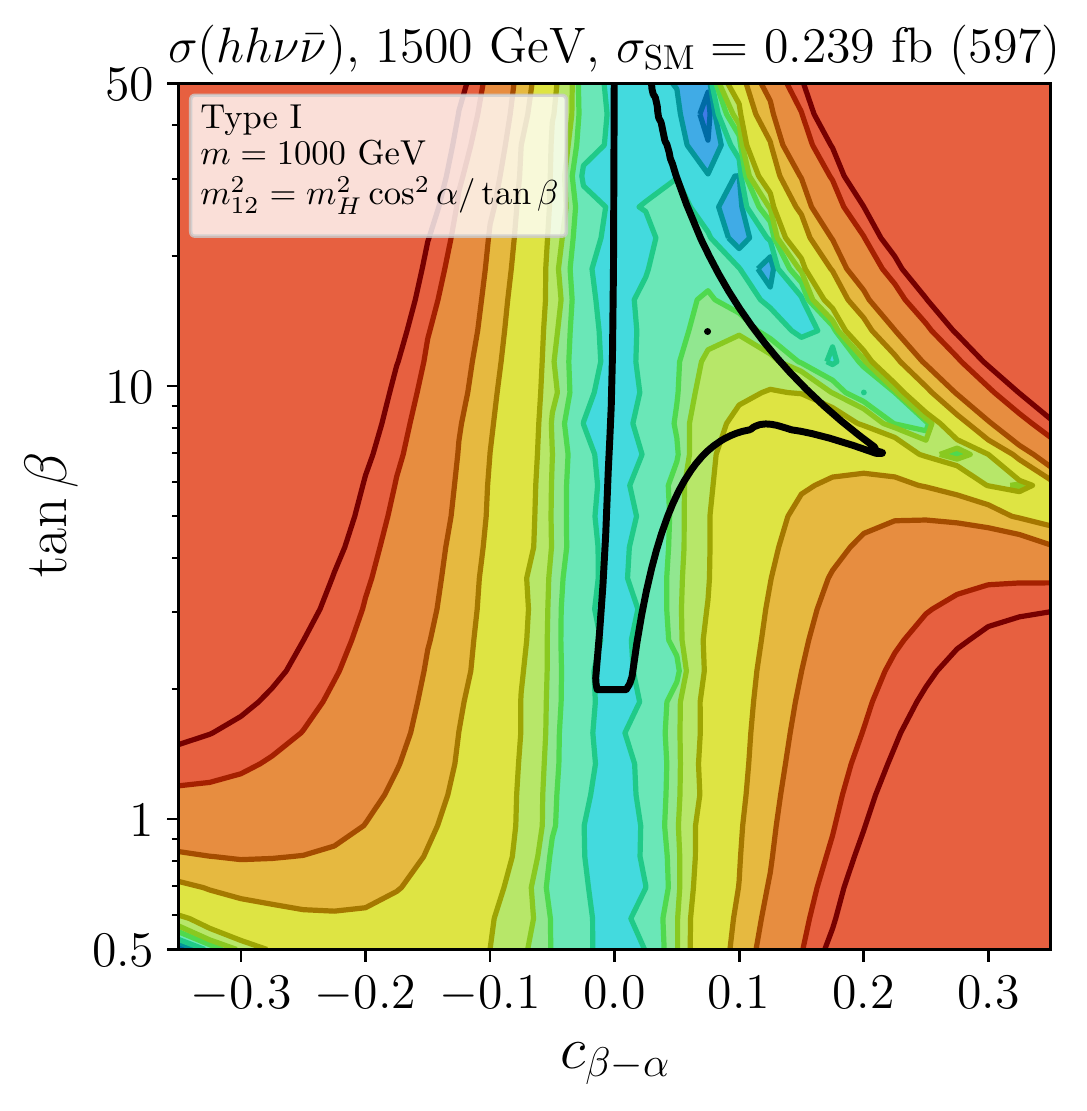}\\[1em]
\includegraphics[width=0.5\textwidth]{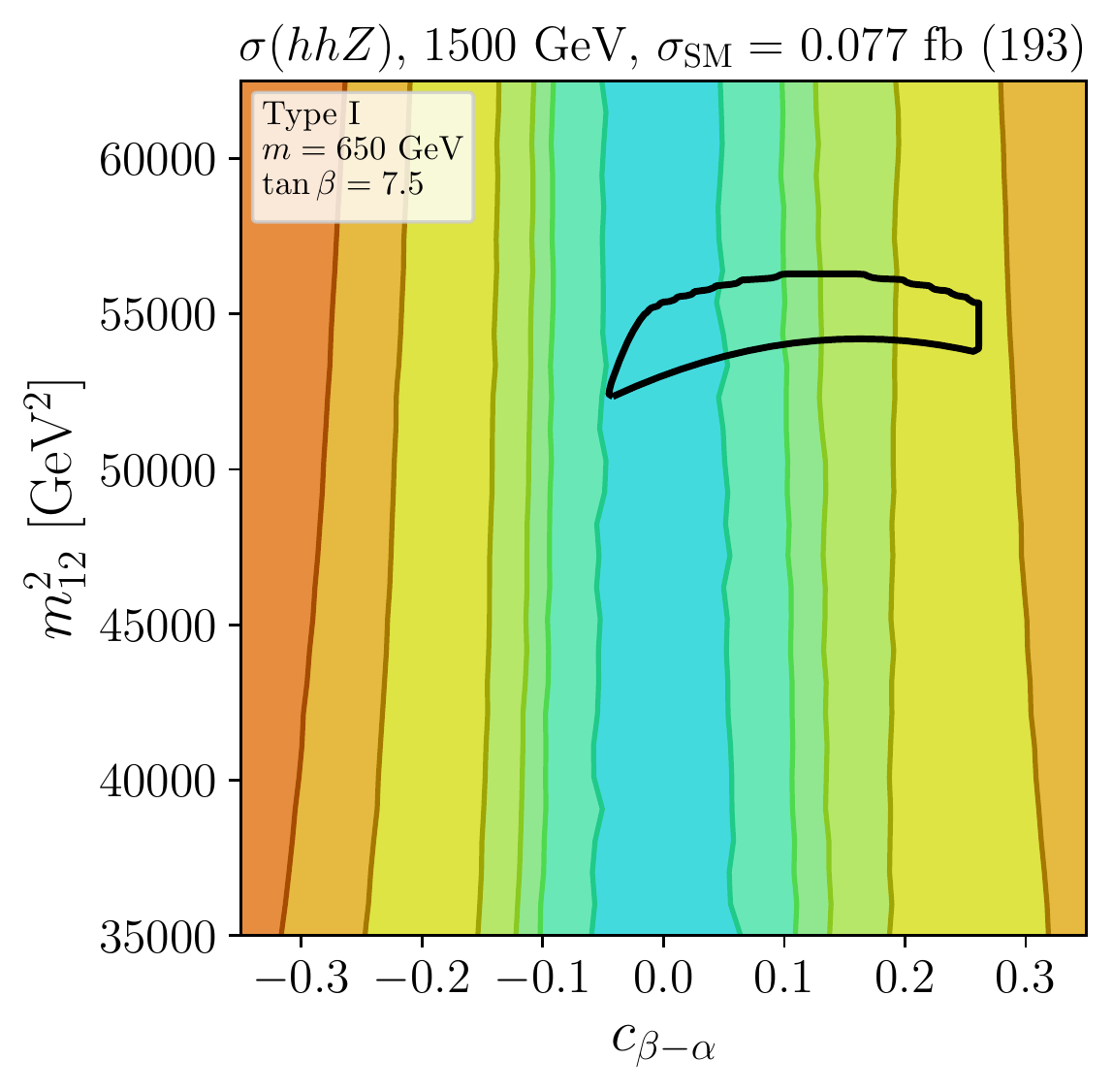}%
\includegraphics[width=0.5\textwidth]{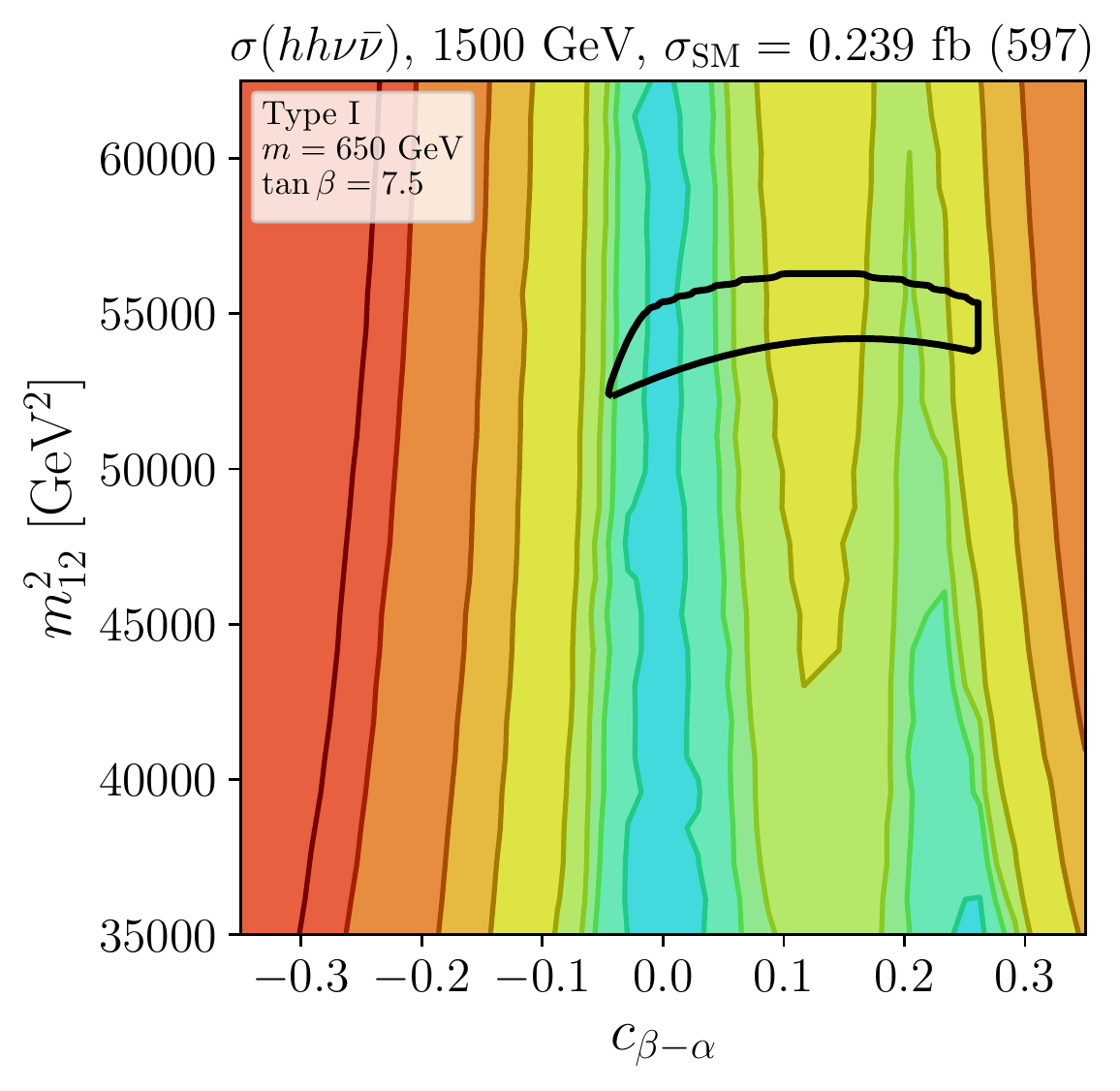}\\[1em]
\includegraphics[width=0.5\textwidth]{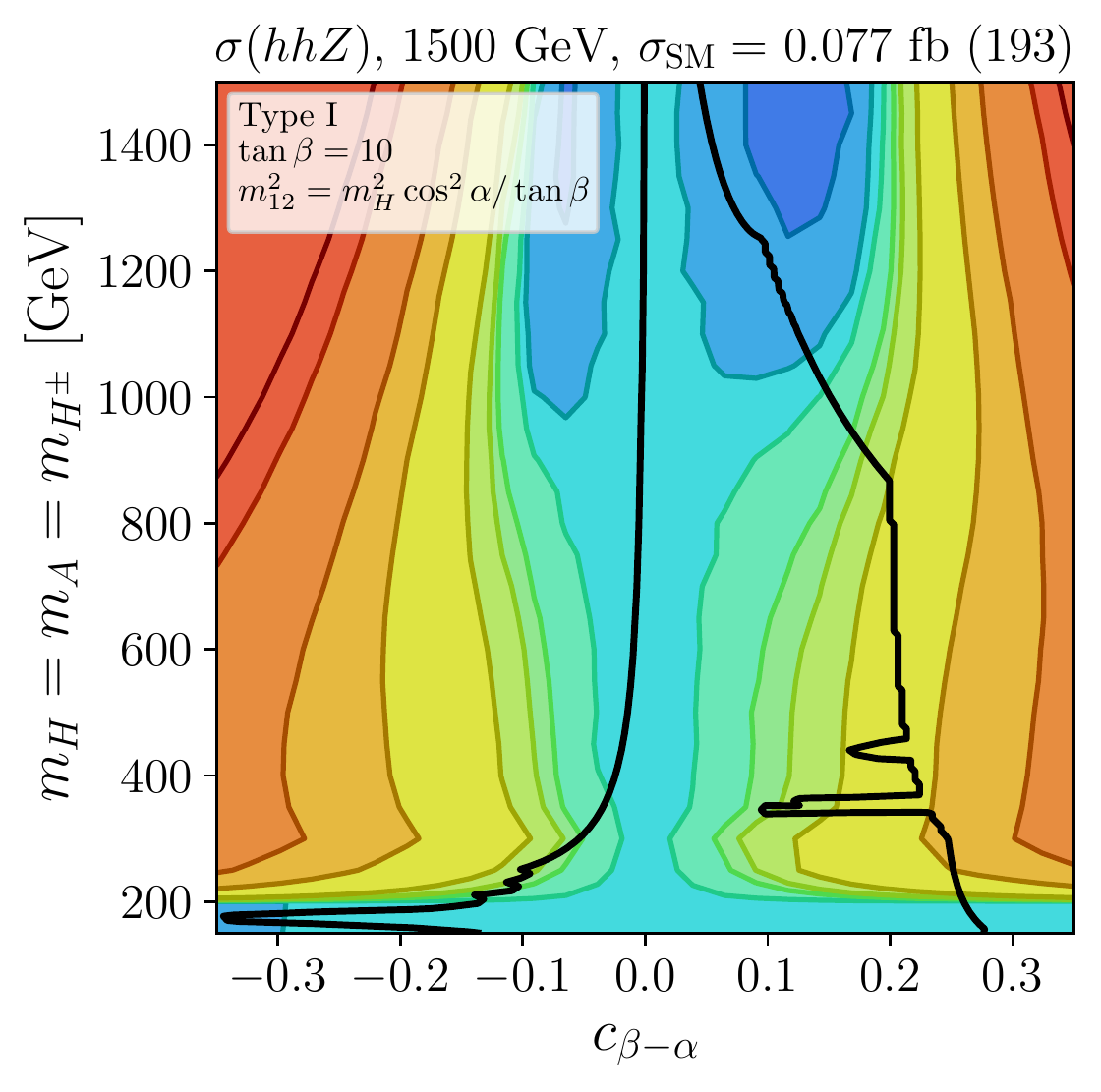}%
\includegraphics[width=0.5\textwidth]{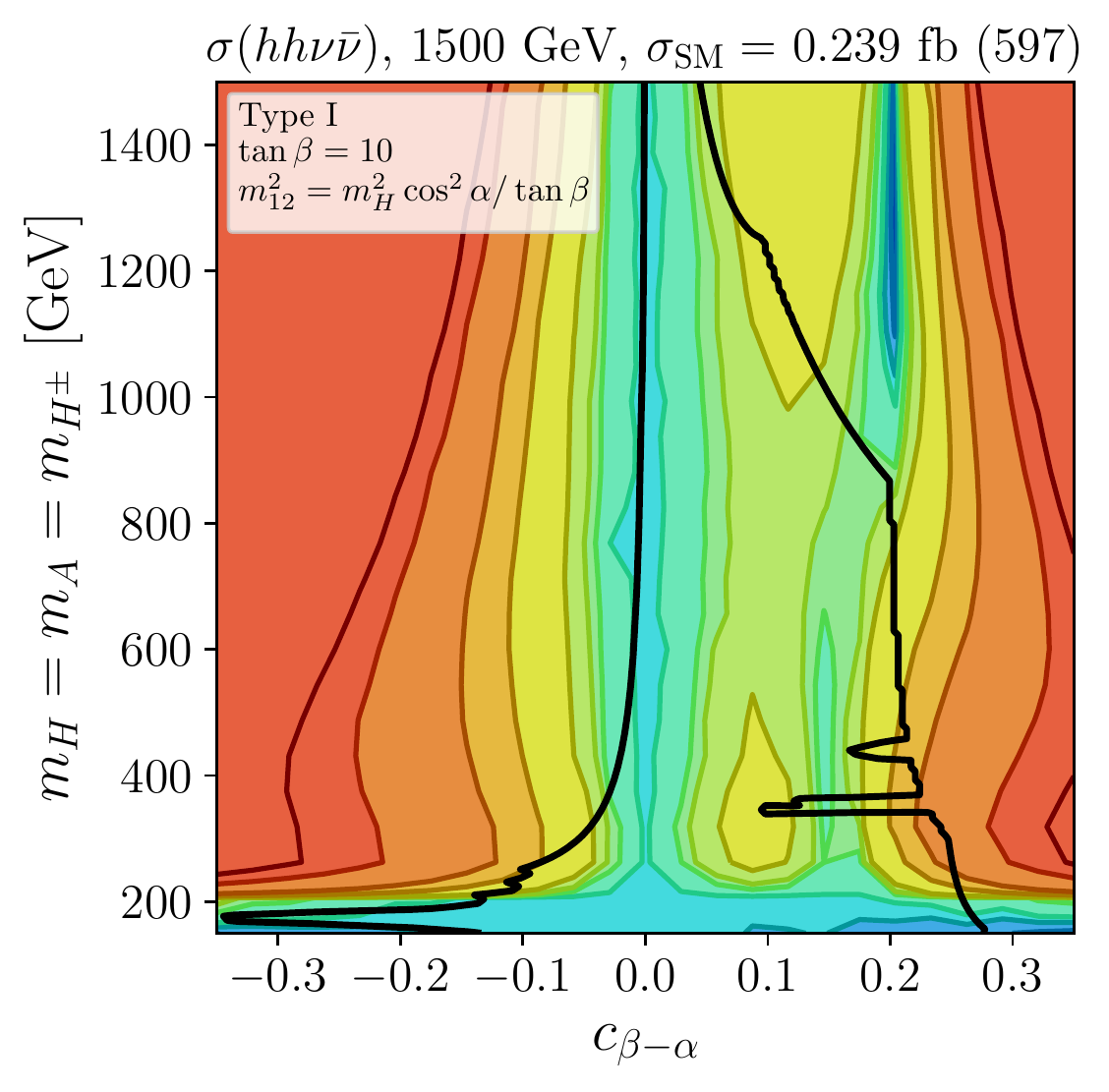}
\end{subfigure}	
\begin{subfigure}[b]{0.18\textwidth}
\includegraphics[height=0.42\textheight]{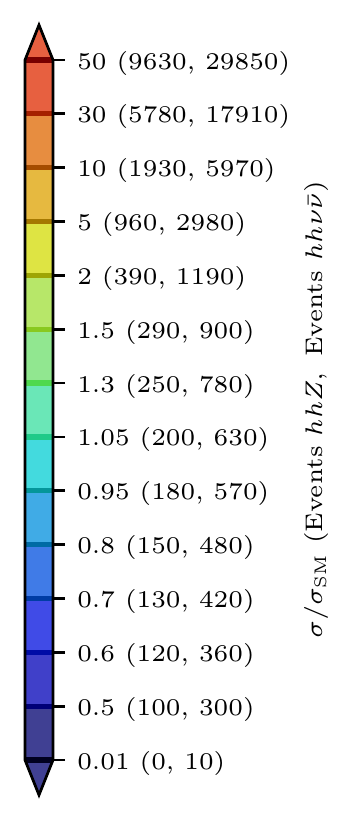}
\vspace{0.25\textheight}
\end{subfigure}
\end{center}
\caption{Cross sections for $e^+e^-\to hhZ$ (left) and
  $e^+e^-\to hh\nu\bar{\nu}$ (right) for $\sqrt{s}=1500\gev$
  for the benchmark planes 1-3 (top to bottom). The
  colors and line styles are as in \protect\reffi{fig:xs_hh_500-I}.}
\label{fig:xs_hh_1500-I}
\end{figure}

In \reffi{fig:xs_hh_1500-I} we show the results for $\sqrt{s} = 1500 \gev$.
At this center of mass energy, the $hh\nu\bar\nu$ channel, due to its
$t$-like-channel nature, is larger than the $s$-like-channel dominated $hhZ$
production. The cross section for $hh\nu\bar\nu$ is clearly dominated by
VBF configurations,  as in the SM case.  For $hhZ$ and $hh\nu\bar\nu$
production one finds SM cross sections of  
$0.077\,\fb$ and $0.239\,\fb$, corresponding to 193 and 597 events,
for the luminosity given in \refta{tab:ee}. However, despite the
substantially larger cross section for $hh\nu\bar\nu$ because of the smaller
anticipated integrated luminosity at $\sqrt{s} = 1500 \gev$ (at CLIC)
w.r.t.\ the high luminosity expected at $\sqrt{s} = 1000 \gev$ (at ILC)
the absolute number of $hh\nu\bar\nu$ events remains smaller.

The overall pattern of the cross sections relative to the SM cross
sections, as given by the color code in \reffi{fig:xs_hh_1500-I} is
qualitatively similar to the one for $\sqrt{s} = 500,\, 1000 \gev$, but
the deviations from the SM tend to be  larger in the $1500 \gev$ case
than in the previous ones of $500 \gev$  and $1000 \gev$. 
In the $\CBA$--$\tb$ plane shown in the first row within the allowed
parameter space we find enhancements up to $1.3$ and $3$ w.r.t.\ the SM
predictions for the $hhZ$ and $hh\nu\bar\nu$ channel,
respectively. These largest values are realized, as before, in the
``tip'' of the allowed region, where $\kala$ deviates most from unity.
While the interference with the SM-type contributions is similar to the
previous center-of-mass energies now also the additional contributions
from an intermediate $H$ play a role, leading, e.g., to an enhancement of
$hhZ$, contrary to the case of $\sqrt{s} = 1000 \gev$.
In the $\CBA$--$\msq$ plane shown in the middle row relative
enhancements of $\sim 5$ can be found. As will be discussed in the next
section, the contribution of an on-shell $H$ with $H \to hh$ plays an
important role here, and thus does the size of $\lahhH$.
We find that the impact of the latter coupling extends up to
$\MH \lsim 1000 \gev$, as can be seen in the $\CBA$--$m$
($m = \MH = \MA = \MHp$) plane in the third row of
\reffi{fig:xs_hh_1500-I}. The maximum effect, as for smaller
center-of-mass energies is found for the resonant di-Higgs
production at $\MH \sim 2\Mh \approx 250 \gev$. Here we find maximum
enhancements of 
$\sim 6$ and $\sim 16$ for the $hhZ$ and $hh\nu\bar\nu$ channel,
respectively. The corresponding values of $\lahhH$ do not vary strongly
in this part of the parameter space and are found to be \order{2}, see
the 2nd column of \reffi{fig:la-I}. 
For $\MH \gsim 1000 \gev$, where $\lahhH$ becomes less relevant, despite
reaching its largest possible values (see the lower plot in the 2nd column of
\reffi{fig:la-I}) we even 
find a decrease of the $hhZ$ channel up to $\sim 0.8$, whereas the
$hh\nu\bar\nu$ channel remains enhanced in the allowed parameter space.

\bigskip


\begin{figure}[htb!]
\vspace{-2em}
\begin{center}
\begin{subfigure}[b]{0.8\textwidth}
\includegraphics[width=0.5\textwidth]{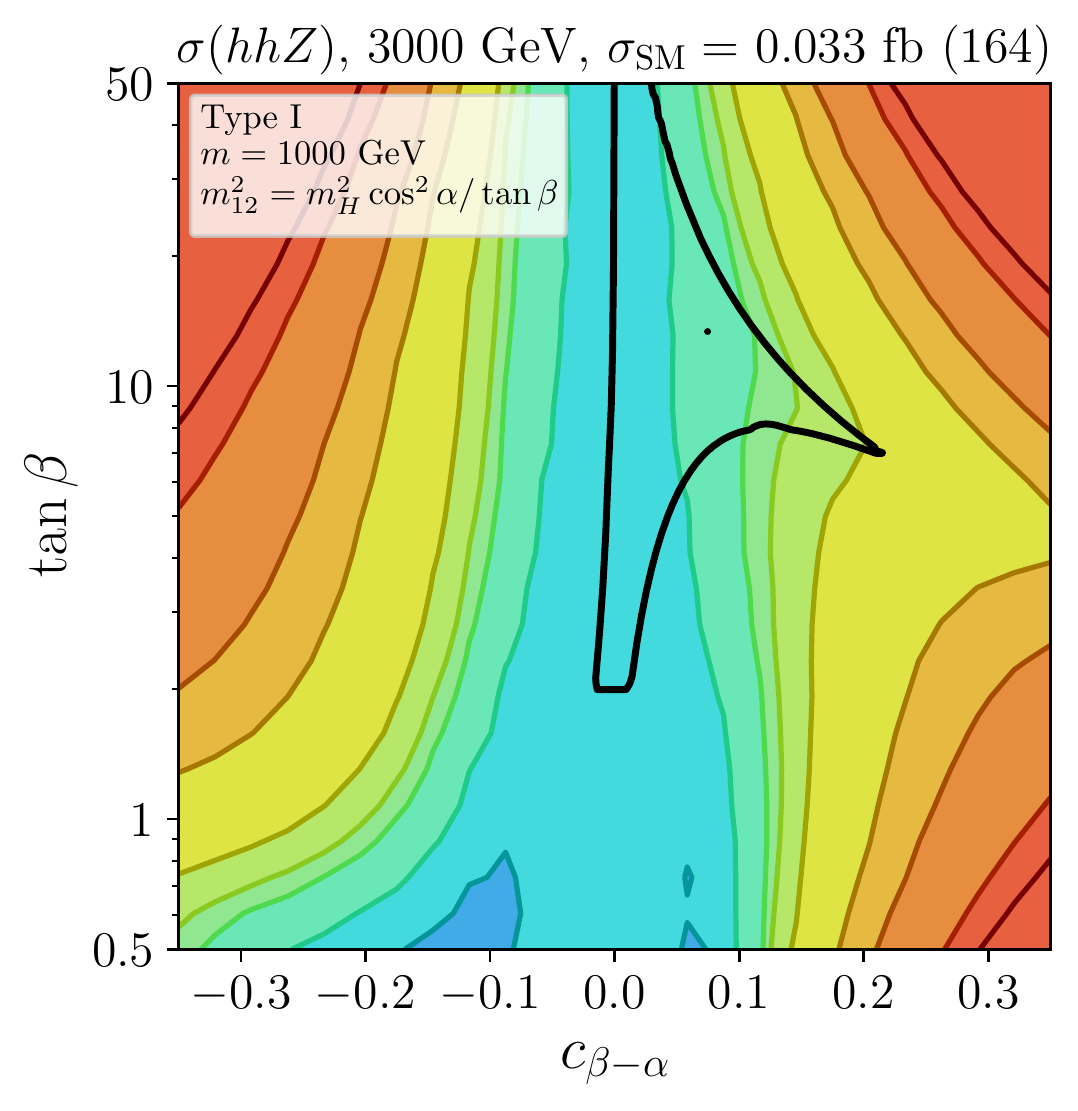}%
\includegraphics[width=0.5\textwidth]{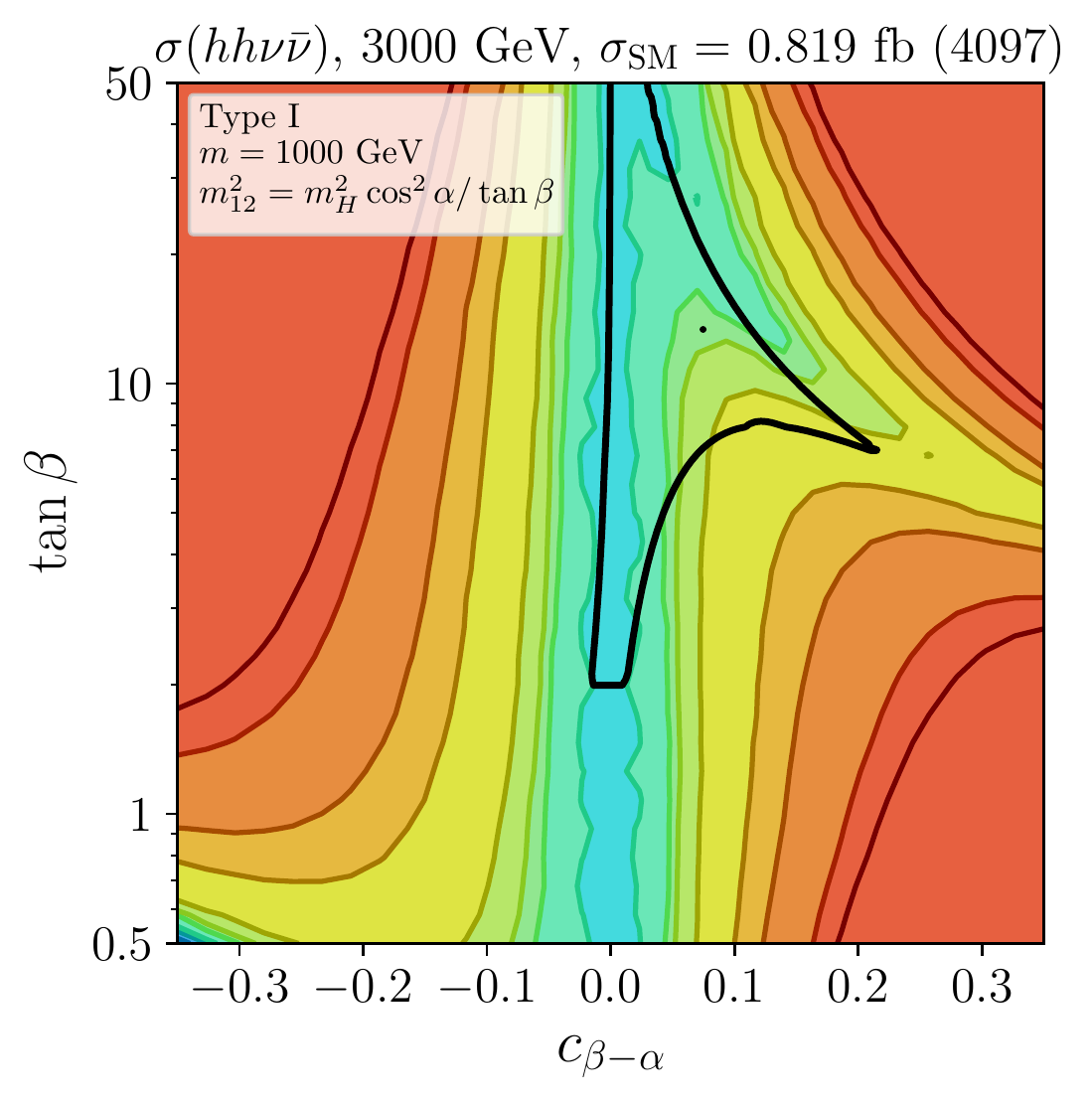}\\[1em]
\includegraphics[width=0.5\textwidth]{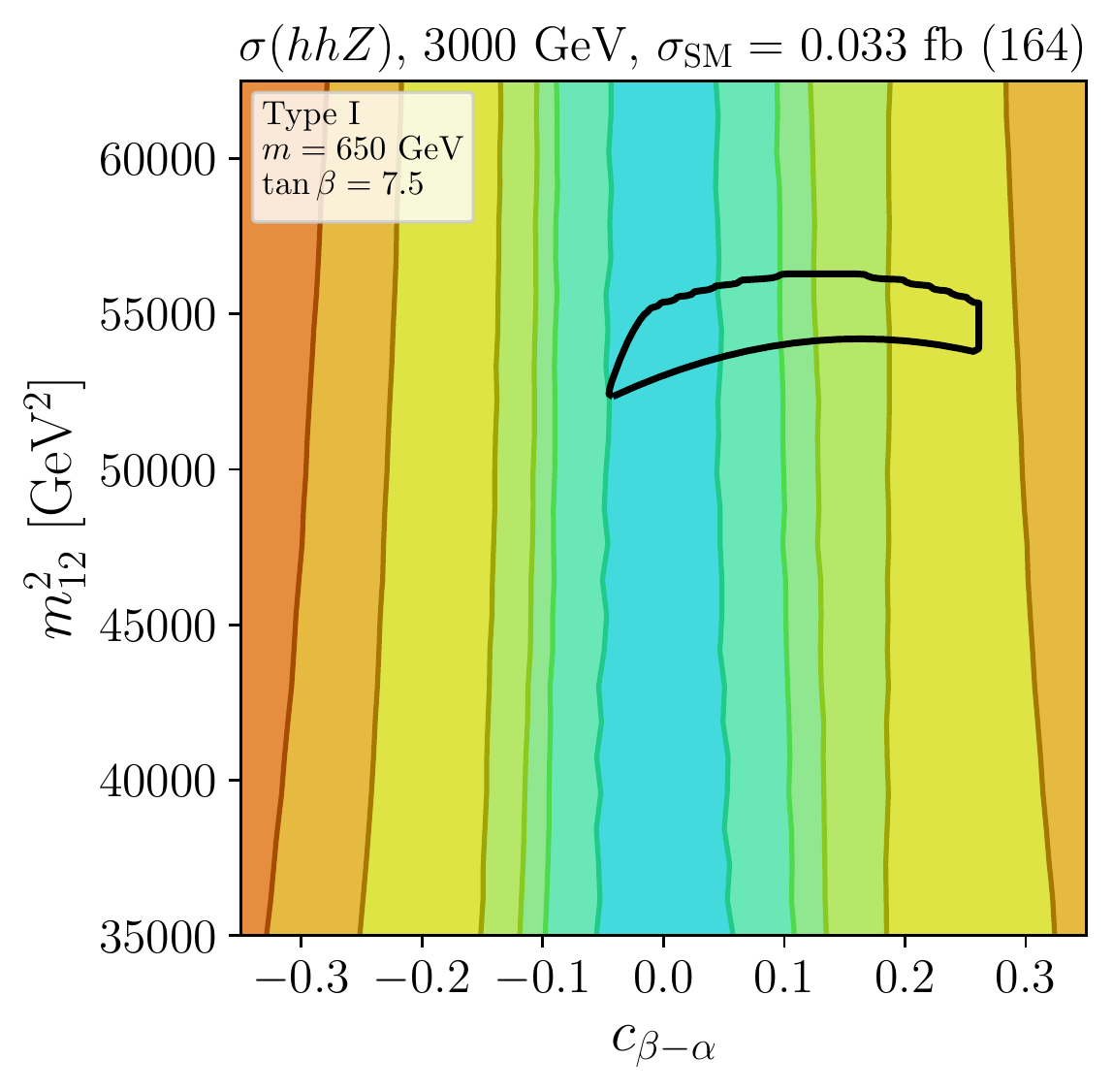}%
\includegraphics[width=0.5\textwidth]{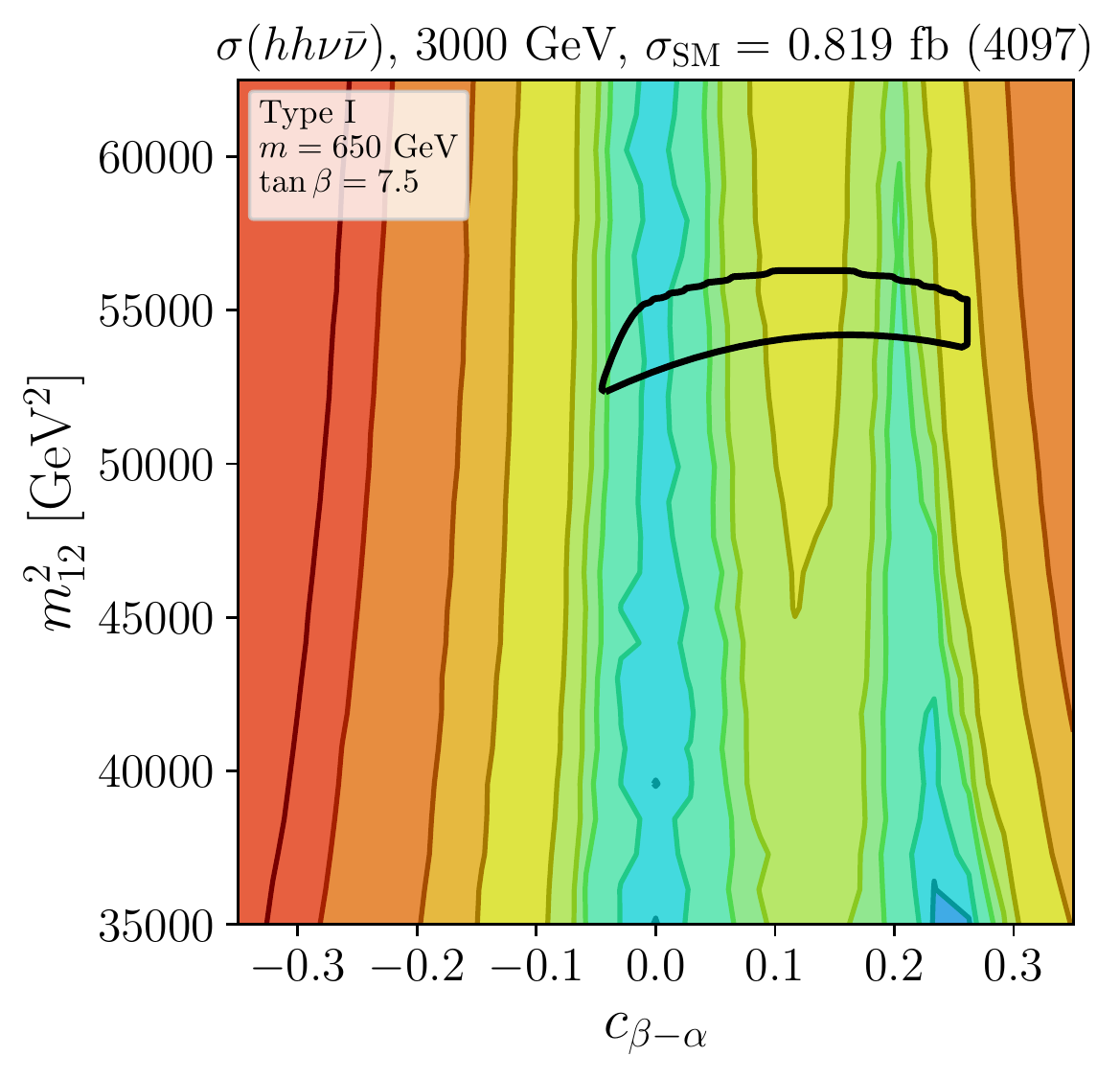}\\[1em]
\includegraphics[width=0.5\textwidth]{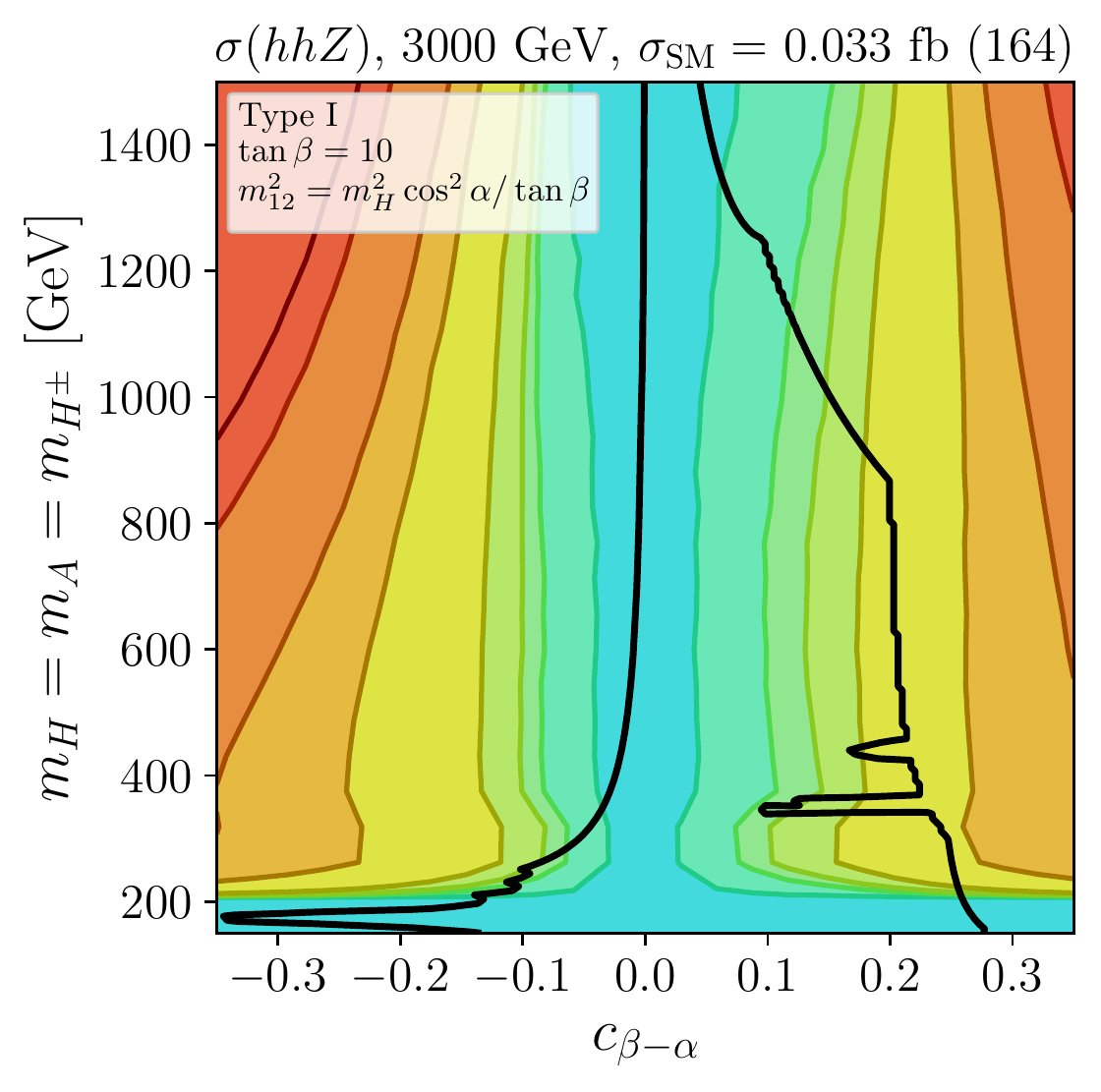}%
\includegraphics[width=0.5\textwidth]{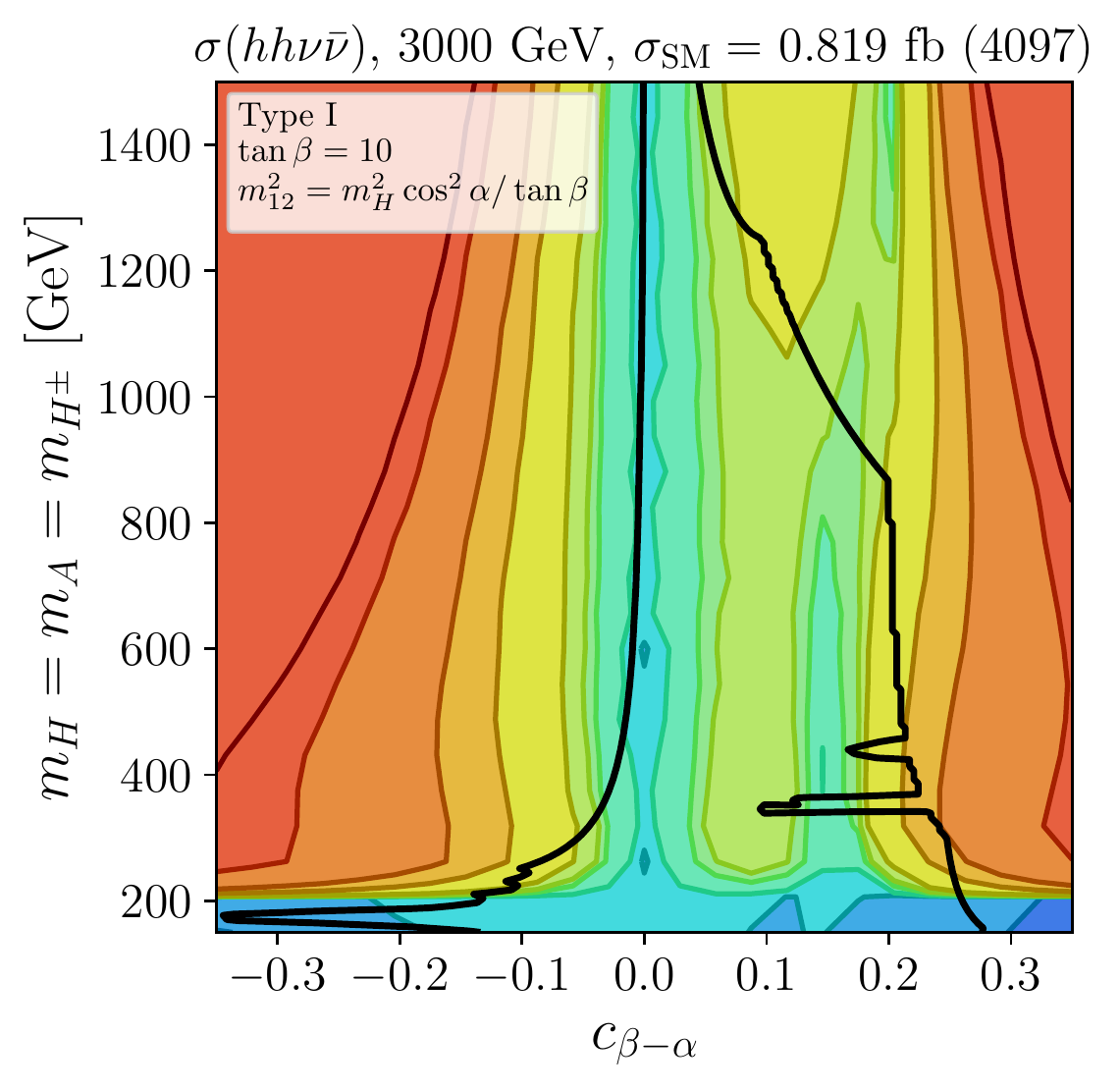}
\end{subfigure}	
\begin{subfigure}[b]{0.18\textwidth}
\includegraphics[height=0.42\textheight]{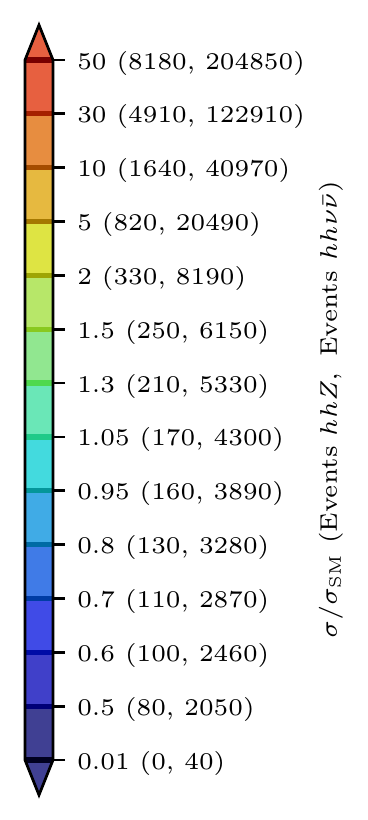}
\vspace{0.25\textheight}
\end{subfigure}
\end{center}
\caption{Cross sections for $e^+e^-\to hhZ$ (left) and
  $e^+e^-\to hh\nu\bar{\nu}$ (right) for $\sqrt{s}=3000\gev$
  for the benchmark planes 1-3 (top to bottom). The
  colors and line styles are as in \protect\reffi{fig:xs_hh_500-I}.}
\label{fig:xs_hh_3000-I}
\end{figure}

We finish the analysis of $hh$ production with $\sqrt{s} = 3000 \gev$,
the highest energy potentially reachable at CLIC, in
\reffi{fig:xs_hh_3000-I}. At this center-of-mass energy $hh\nu\bar\nu$
clearly dominates over $hhZ$, with SM production cross sections of
$0.033\,\fb$ and $0.819\,\fb$, corresponding to 164 and 4097 events,
for the luminosity given in \refta{tab:ee}.  The  $hh\nu\bar\nu$ rates
are now fully dominated by the VBF configurations where the new window
to test $\lahhh$ and $\lahhH$ is clearly open. 

As before, the overall pattern of the cross sections relative to the SM cross
sections, as given by the color code in \reffi{fig:xs_hh_3000-I} is
qualitatively similar to the one for smaller energies,  but now with 
deviations from the SM tending to be larger for $3000 \gev$ than for the $500$, $1000$ and $1500 \gev$ cases.  In the $\CBA$--$\tb$ plane
we find enhancements of up to $\sim 2$ and $\sim 2.5$ for $hhZ$ and
$hh\nu\bar\nu$ production, respectively.  Slightly larger enhancements
are found in the $\CBA$--$\msq$ plane.  As for $\sqrt{s} = 1500 \gev$,  
close to the resonant region,  the
on-shell decay $H \to hh$ contributes substantially to the total cross
section and this will be important to reach sensitivity to  
$\lahhH$,  as will be shown in \refse{sec:sensitivity}.  Due to the very
high $\sqrt{s}$ the dependence of the cross 
sections on $\MH$ is less pronounced than for smaller $\sqrt{s}$. This
holds particularly if we are close to the alignment limit, where $\lahhH$ 
is very tiny.  In the
lowest row of \reffi{fig:xs_hh_3000-I} we find enhancements of up to
$\sim 4.5$ and $\sim 10$, respectively.


\subsubsection{\boldmath{$hH$} production}
\label{sec:xshH-I}

In this subsection we analyze di-Higgs production of one light and one
heavy $\cp$-even Higgs boson. As expected, we find that in
the alignment limit the production cross sections become exactly
(analytically) zero.  Regarding the values of  the two triple Higgs
couplings involved,  $\lahhH$ and  $\lahHH$,  the first
one vanishes in the alignment limit,  as was discussed above,  but the
second one does not vanish in this limit.  The vanishing of the
contribution from the $H^*$ mediated diagram to this  $hH$ cross section,
in the alignment limit,  occurs in this case  because of the vanishing
of the $H$ coupling to the gauge bosons in this limit.
Consequently,  a key
point here is to investigate the accessibility to these $\lahhH$
and  $\lahHH$ away from the alignment limit.

\bigskip

\begin{figure}[t!]
\begin{center}
\begin{subfigure}[b]{0.8\textwidth}
\includegraphics[width=0.48\textwidth]{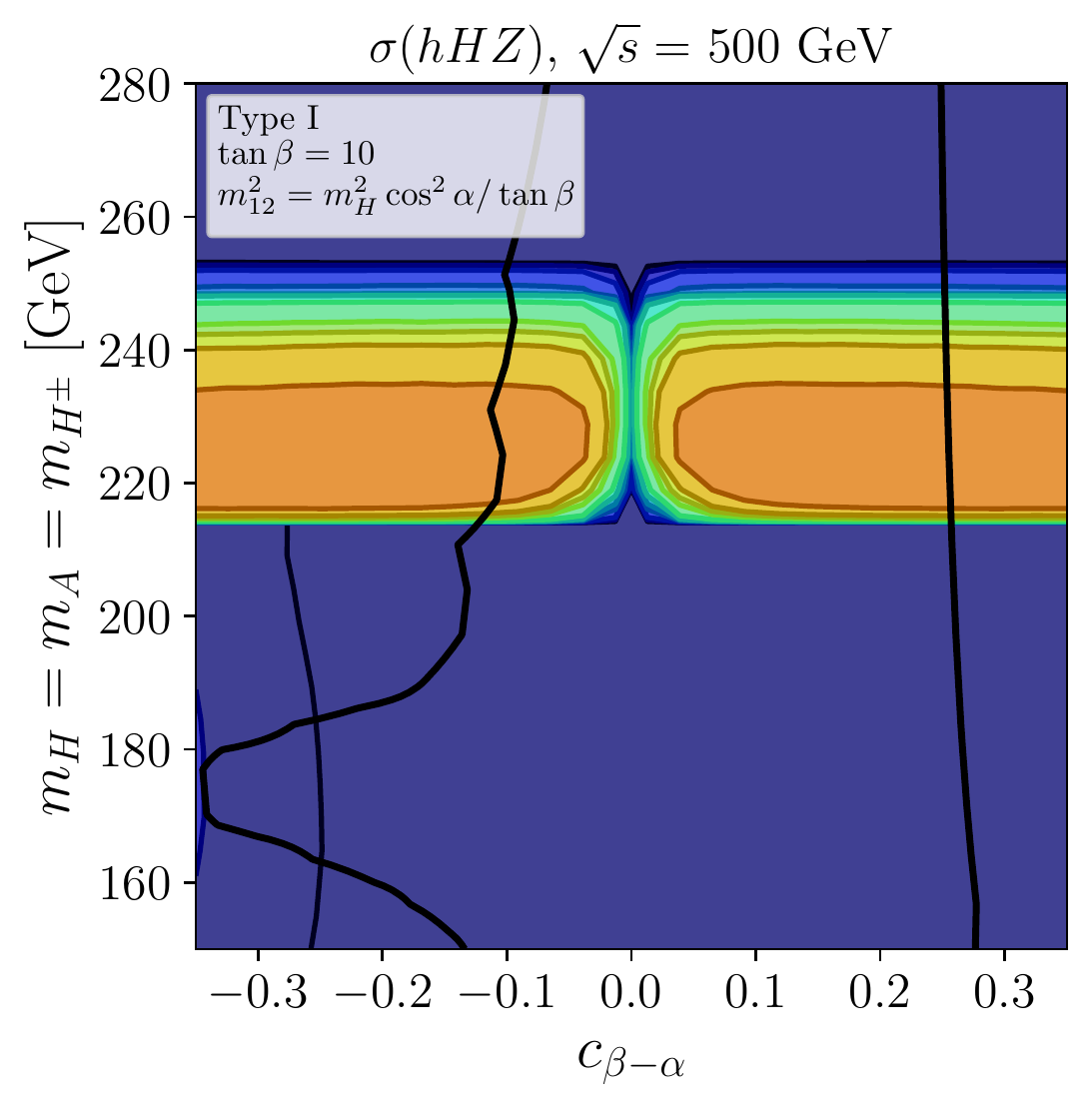}
\includegraphics[width=0.48\textwidth]{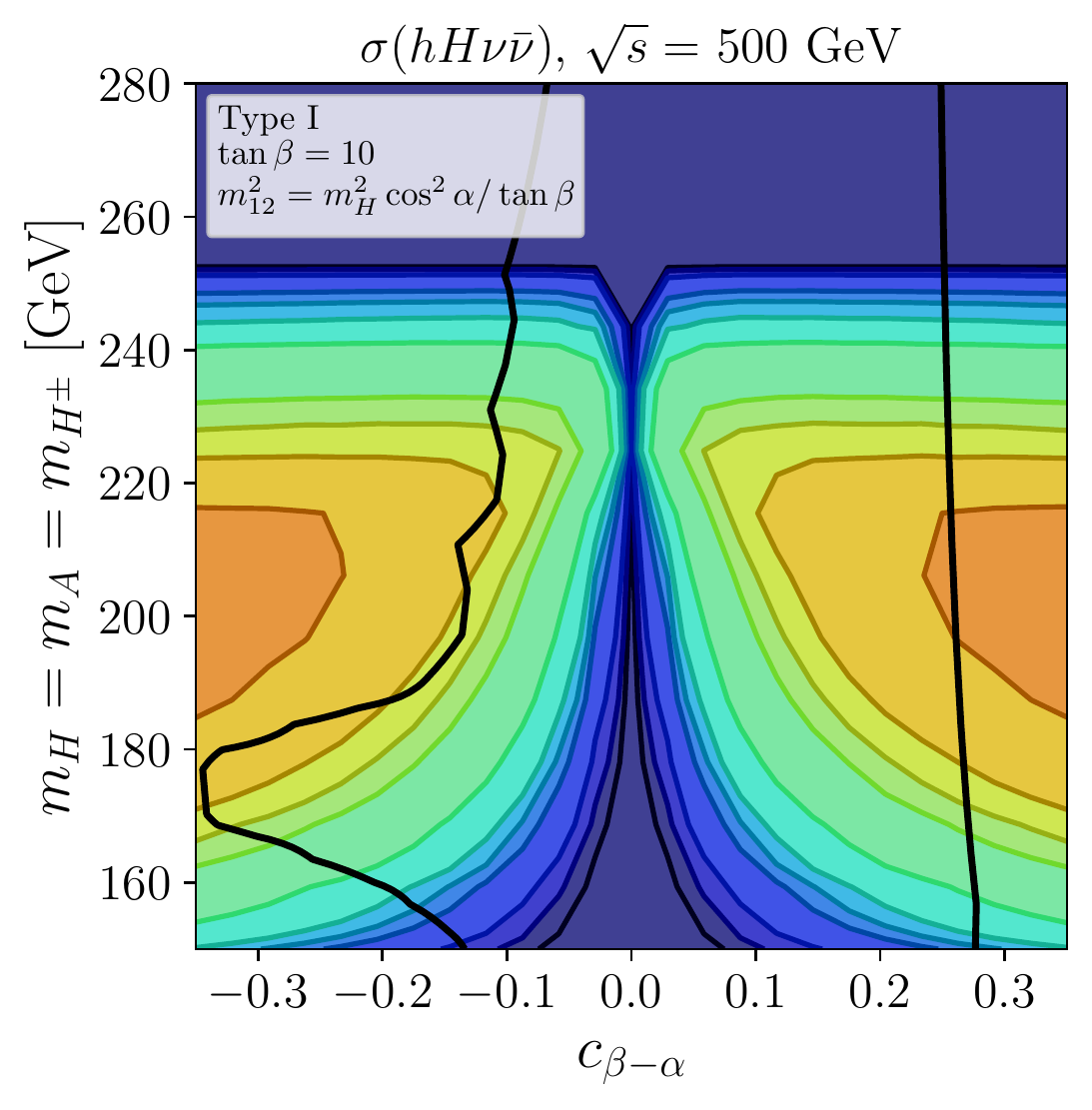}
\end{subfigure}	
\begin{subfigure}[b]{0.18\textwidth}
\includegraphics[height=0.26\textheight]{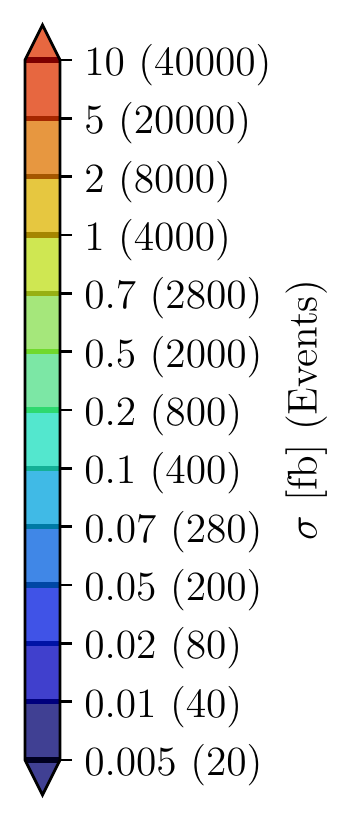}
\end{subfigure}
\end{center}
\caption{Cross sections for $e^+e^-\to hHZ$ (left) and
  $e^+e^-\to hH\nu\bar{\nu}$ (right)
  at $\sqrt{s}=500\gev$ for our benchmark plane~3. The
  total allowed regions is given by the solid black line. The color code
indicates the absolute cross section in fb.} 
\label{fig:xs_hH_500-I}
\end{figure}

In \reffi{fig:xs_hH_500-I} we start with the results for
$\sqrt{s} = 500 \gev$. At this center-of-mass energy, only the benchmark
plane~3, i.e.\ the $\CBA$--$m$ ($m = \MH=\MA=\MHp$) plane exhibits
relevant cross sections. In the other two planes $\MH$ is too large to
be produced on-shell.
The $hHZ$ production is dominated by the subprocess
$e^+e^-  \to Z^* \to H\,A \to H\,hZ$, i.e.\ it becomes sizable above
$\MA \approx \Mh + \MZ \approx 215 \gev$ and below $\MA + \MH \lsim 500 \gev$,
corresponding to $\MA \lsim 250 \gev$. The maximum cross section found
is $\sim 4 \fb$ for $\MH \sim 230 \gev$, corresponding to $\sim 16000$
events for the luminosity given in \refta{tab:ee}. The results are found
largely independent of $\CBA$, with the exception of very small cross
sections for $\CBA \to 0$.

At this low energy,  the $hH\nu\bar\nu$ production is also dominated by
the subprocess 
$e^+e^- \to Z^* \to H\,A \to H\,hZ^{(*)} \to H\,h\nu\bar\nu$.
Because of the possibility of an off-shell decay of the final
$Z^* \to \nu\bar\nu$ 
the cross section can become sizable already for smaller $\MH = \MA$
values. As for the $hHZ$ cross section, the upper limit of
$\MH = \MA \sim 250 \gev$ holds. The maximum cross section inside
the allowed region is found at $\CBA \sim 0.25$ and $\MH = 200 \gev$
with about $\sim 2 \fb$, corresponding to about 8000~events. 
The diagrams involving $\lahhH$ and $\lahHH$ give only a very small
contribution to the cross sections. While the former takes values of
\order{-0.5} for the considered $\MH = \MA$ values, the latter takes
values of \order{1}. 
However, their contributions are suppressed by the off-shell
Higgs boson in the $s$-like-channel.

\bigskip

\begin{figure}[t!]
\begin{center}
\begin{subfigure}[b]{0.7\textwidth}
\includegraphics[width=0.48\textwidth]{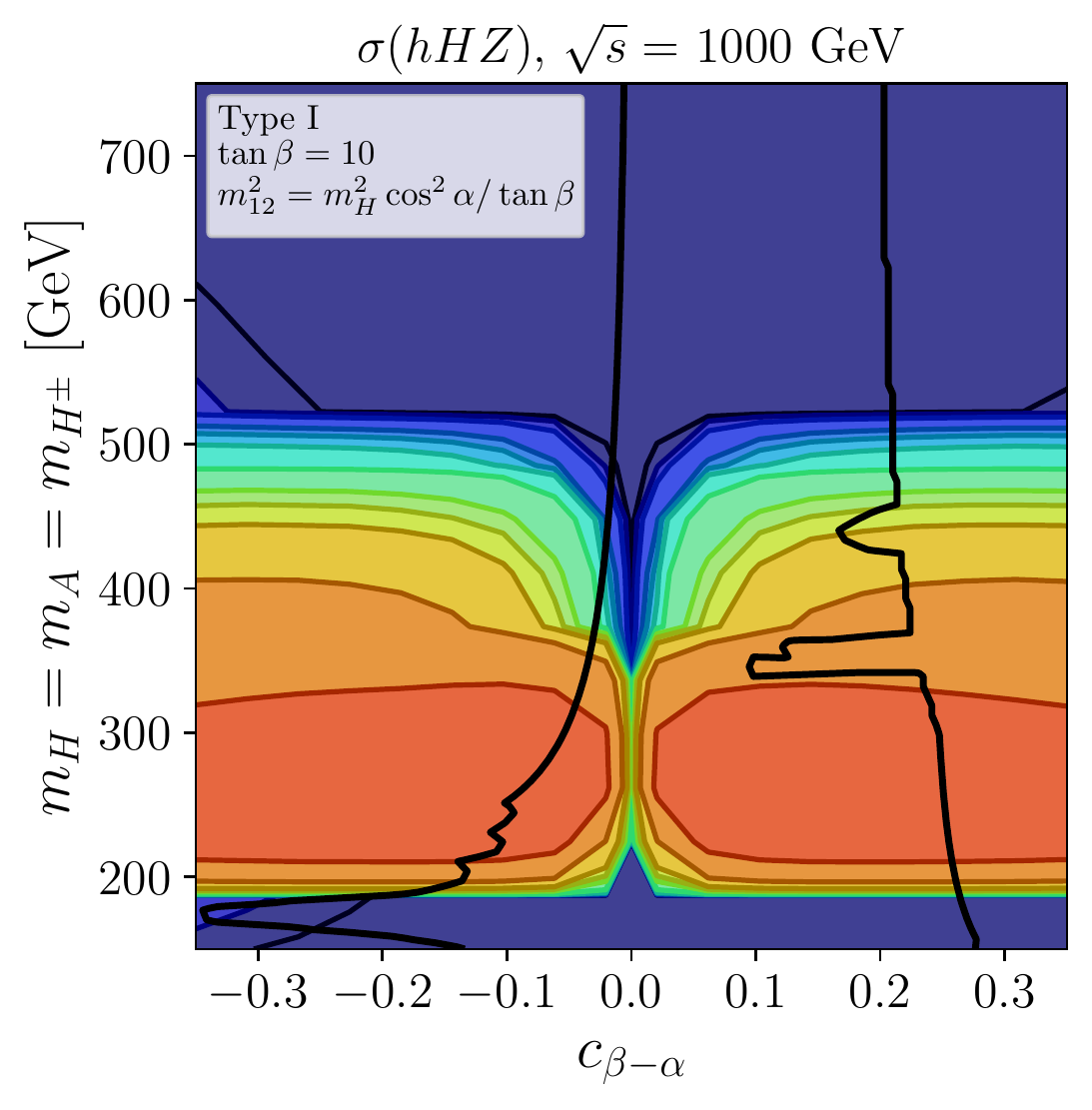}
\includegraphics[width=0.48\textwidth]{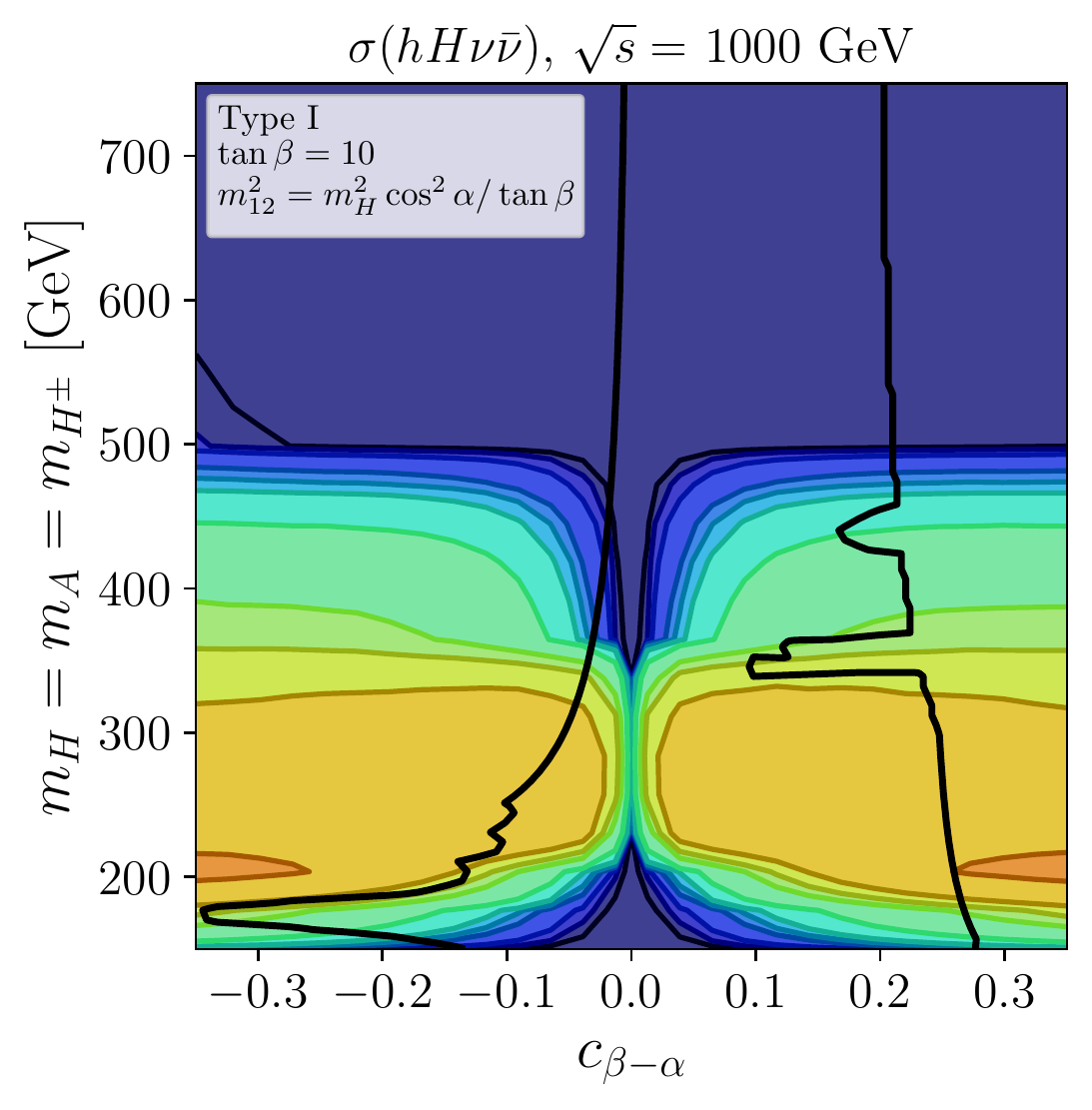}
\end{subfigure}	
\begin{subfigure}[b]{0.18\textwidth}
\includegraphics[height=0.26\textheight]{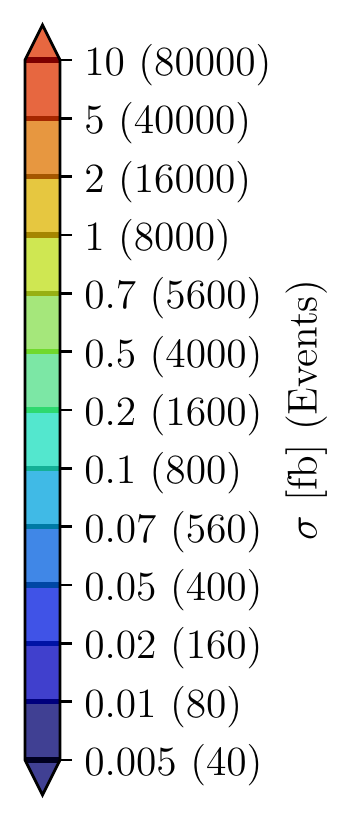}
\end{subfigure}
\end{center}
\caption{Cross sections for $e^+e^-\to hHZ$ (left) and
  $e^+e^-\to hH\nu\bar{\nu}$ (right) at $\sqrt{s} = 1000 \gev$
  for our benchmark plane~3. The
  total allowed regions is given by the solid black line. The color code
indicates the absolute cross section in fb. }
\label{fig:xs_hH_1000-I}
\end{figure}

In \reffi{fig:xs_hH_1000-I} we present the results for the $hHZ$ (left)
and $hH\nu\bar\nu$ (right) production cross sections for
$\sqrt{s} = 1000 \gev$. 
As for $\sqrt{s} = 500 \gev$ only the $\CBA$--$m$
($m = \MH = \MA = \MHp$)
plane shows relevant cross sections. Owing to the higher center-of-mass
energy the regions with relevant cross sections now extends up to
$\MA = \MH \lsim 500 \gev$, i.e.\ half the center-of-mass energy.
The remaining features are similar to the results for $\sqrt{s} = 500 \gev$. 
The largest cross sections that we find in the allowed parameter region
is about $\sim 8 \fb$ and $\sim 2 \fb$ for $hHZ$ and $hH\nu\bar\nu$
production, respectively. As before, $\lahhH$ and $\lahHH$ do not play
an important role here.

\bigskip

\begin{figure}[htb!]
\vspace{-2em}
\begin{center}
\begin{subfigure}[b]{0.8\textwidth}
\includegraphics[width=0.5\textwidth]{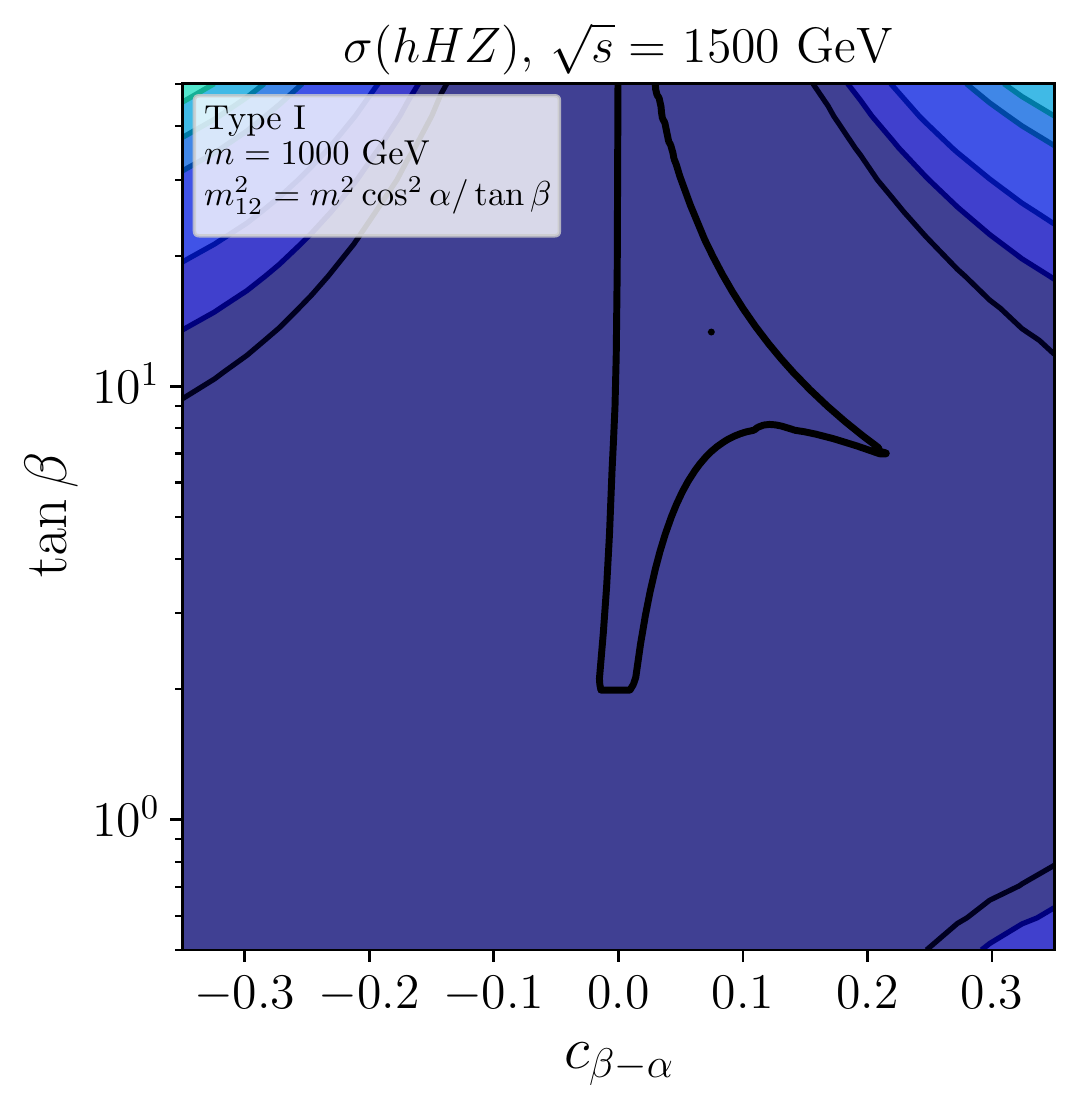}%
\includegraphics[width=0.5\textwidth]{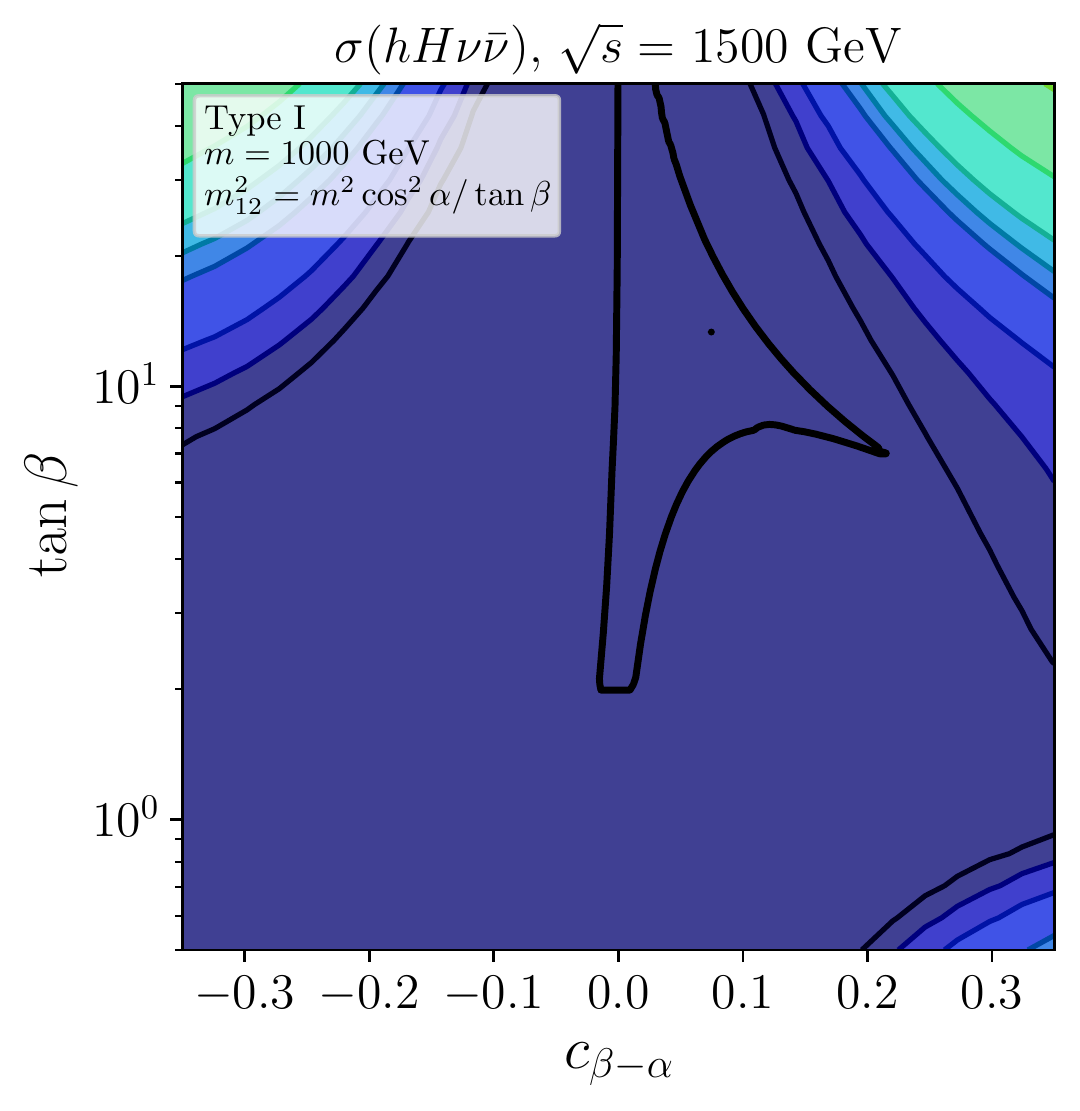}\\[1em]
\includegraphics[width=0.5\textwidth]{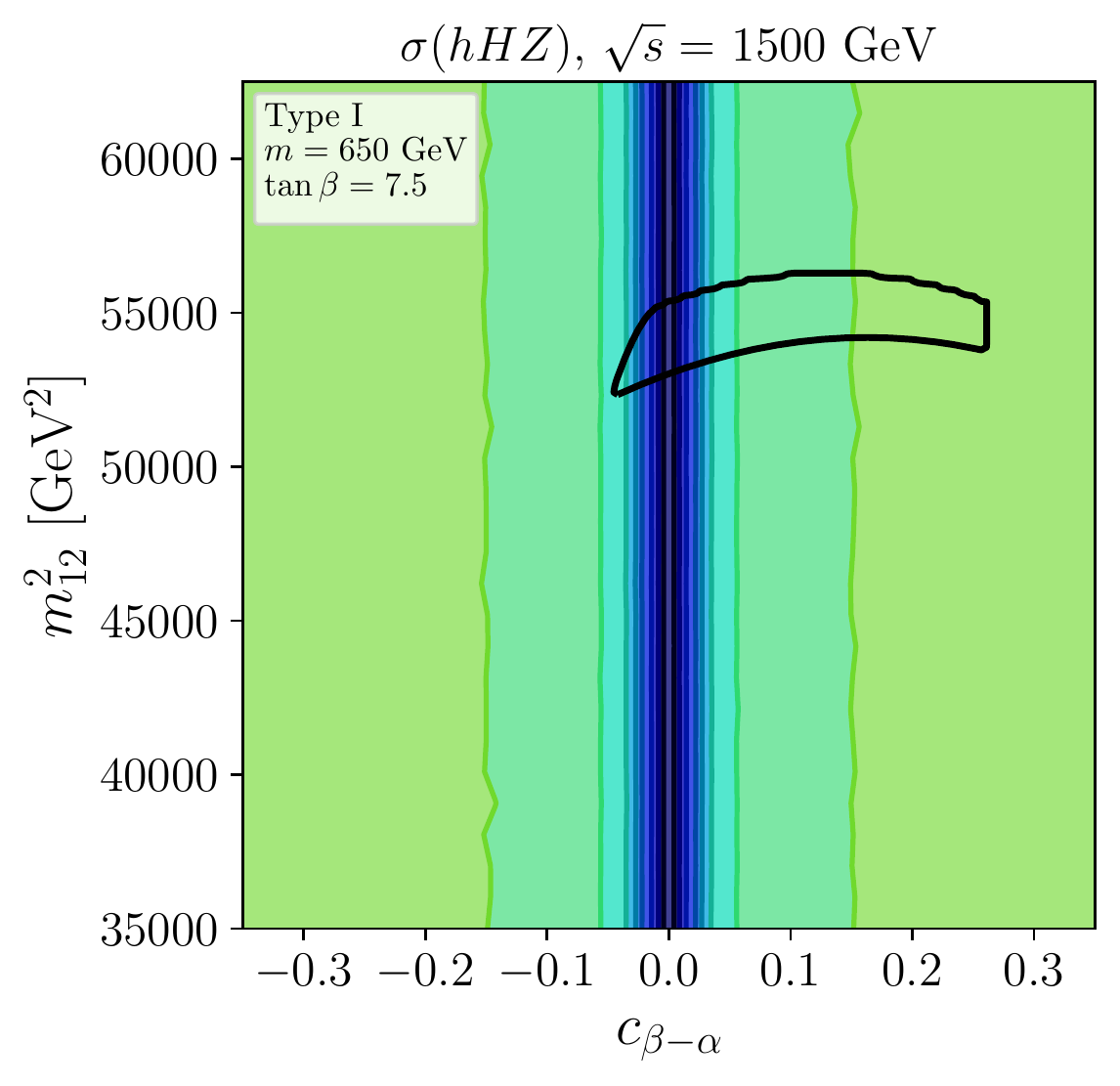}%
\includegraphics[width=0.5\textwidth]{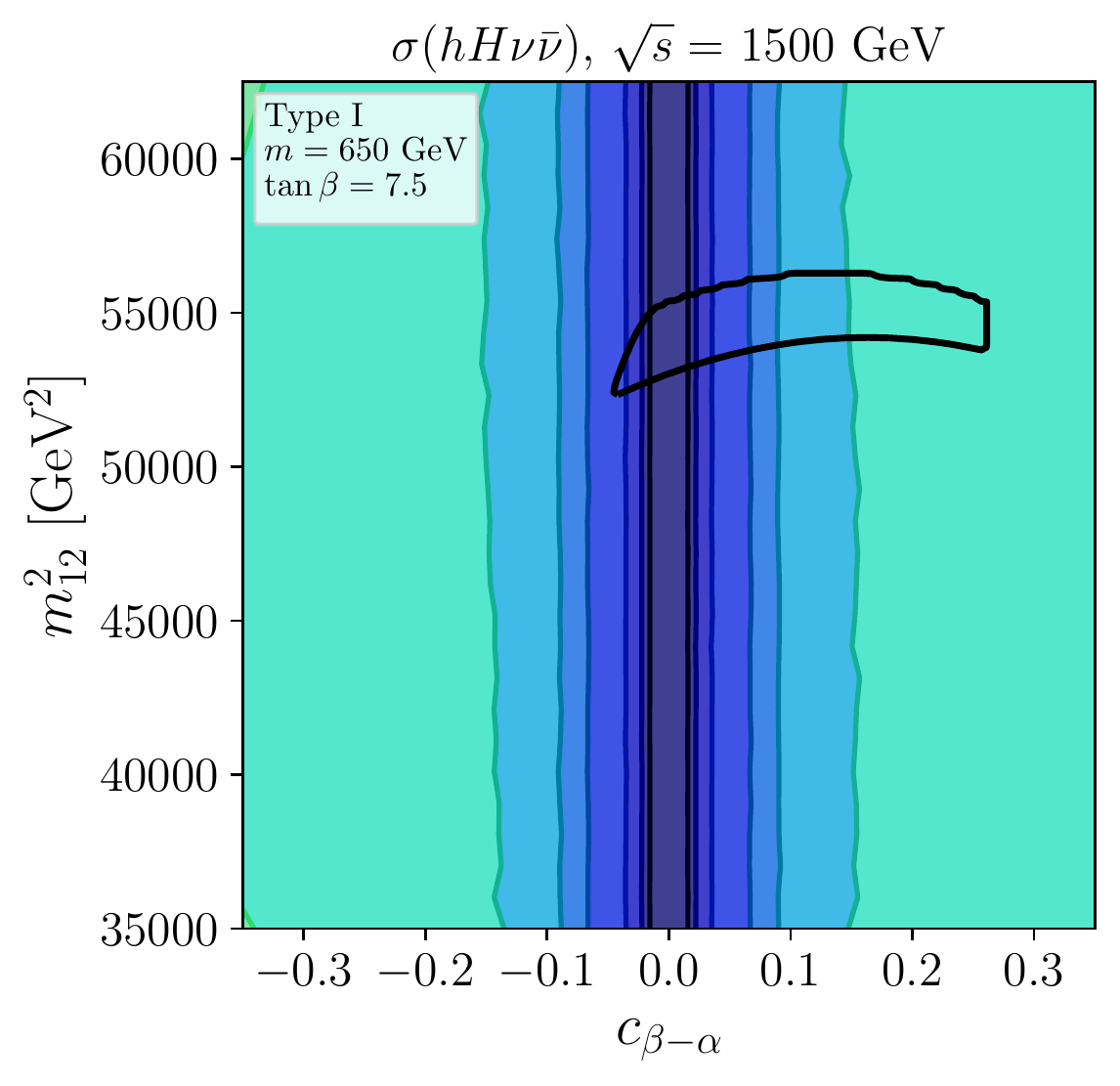}\\[1em]
\includegraphics[width=0.5\textwidth]{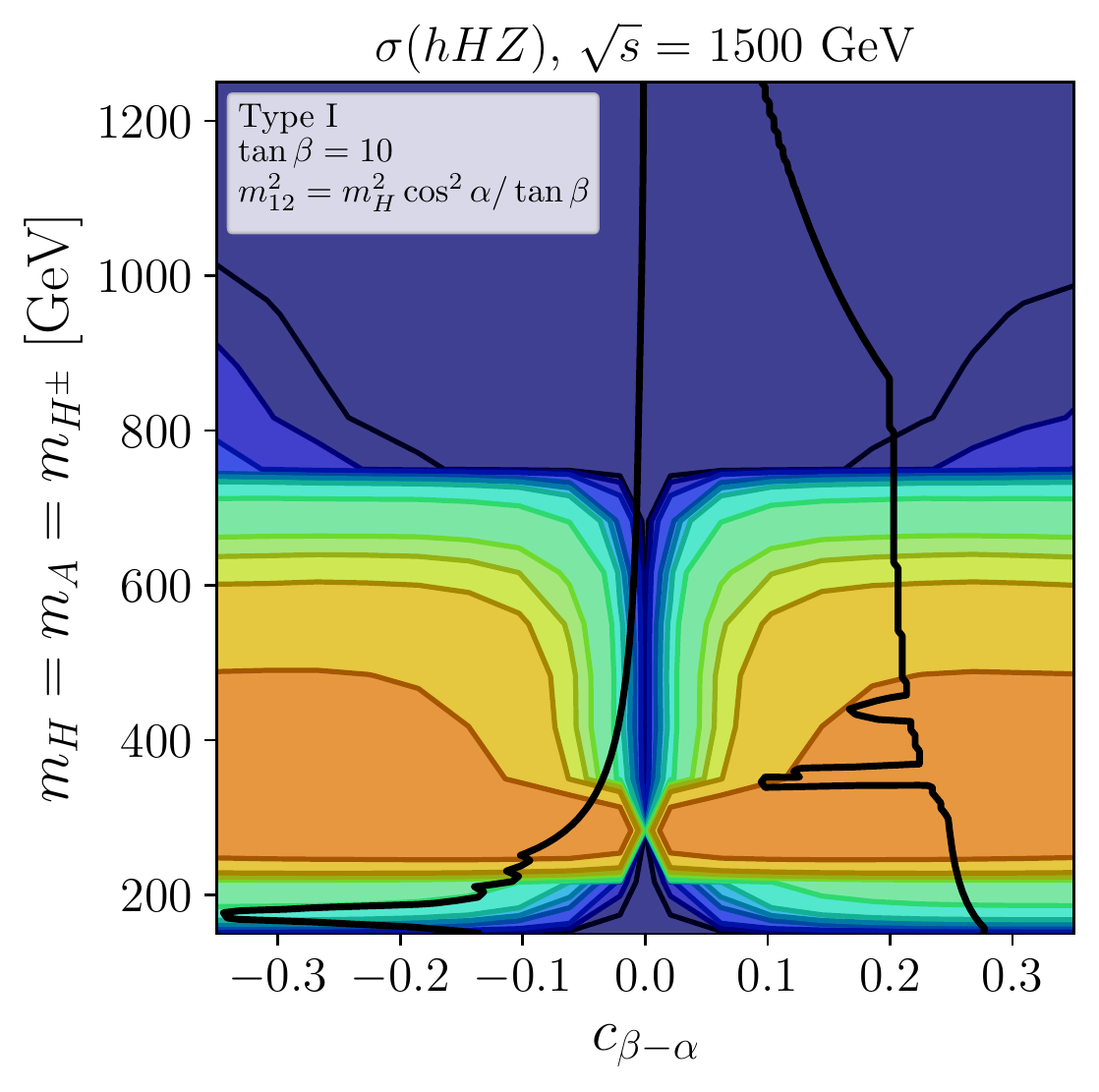}%
\includegraphics[width=0.5\textwidth]{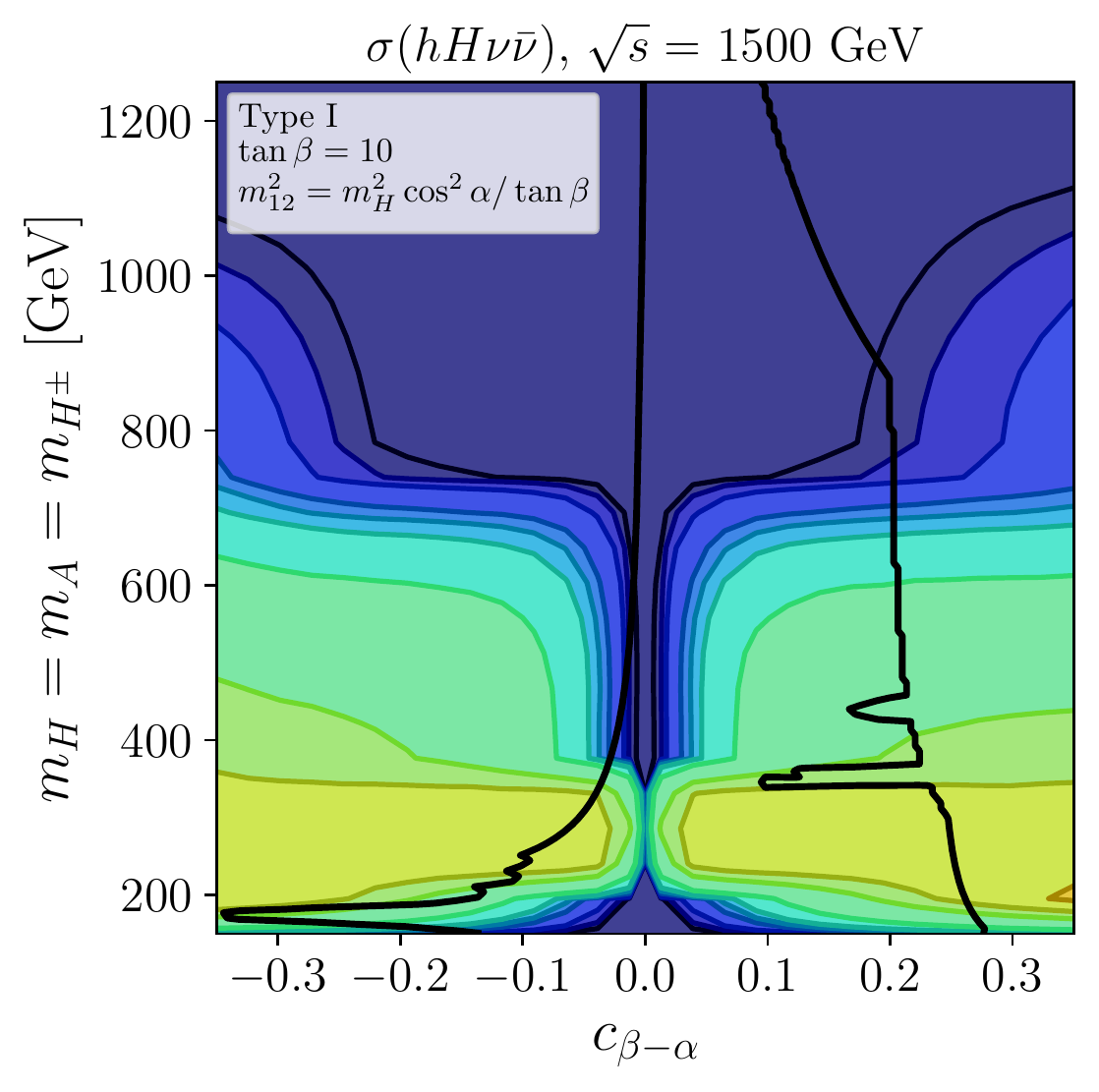}
\end{subfigure}	
\begin{subfigure}[b]{0.18\textwidth}
\includegraphics[height=0.42\textheight]{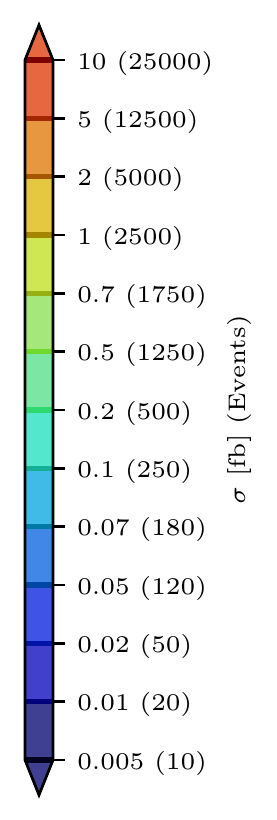}
\vspace{0.25\textheight}
\end{subfigure}
\end{center}
\caption{Cross sections for $e^+e^-\to hHZ$ (left) and
  $e^+e^-\to hH\nu\bar{\nu}$ (right) for $\sqrt{s}=1500\gev$
  for the benchmark planes 1-3 (top to bottom). The
  total allowed regions is given by the solid black line. The color code
indicates the absolute cross section in fb.}
\label{fig:xs_hH_1500-I}
\end{figure}


In \reffi{fig:xs_hH_1500-I} we present the results for
$\sqrt{s} = 1500 \gev$. This center-of-mass energy is sufficiently high
such that all benchmark planes can in principle exhibit relevant
production cross sections.  This large cross sections for both $hHZ$ and
$hH \nu \bar \nu$ processes at 1500 GeV can also be seen in 
\reffi{fig:XSsqrts}.  However, as discussed above,  these large cross
sections found  occur mainly due to the intermediate on-shell $A$
production that decays to $Zh$.  Consequently, we 
find that for the parameters in our benchmark plane~1 with
$\MH = \MA = 1000 \gev$ no large cross section can be observed within
the allowed parameter space. Only for very large values of $\CBA$ and
$\tb$ (outside the allowed region) cross sections reaching the fb region
are reached. This is due to the extreme values that
$\lahHH \simeq \lahAA$ can take.  

In the $\CBA$--$\msq$ plane as shown in the middle row of
\reffi{fig:xs_hH_1500-I} we have $\MA + \MH = 1200 \gev$ and thus
relevant cross sections are indeed found, reaching up to $\sim 0.5 \fb$
for $hHZ$ and $\sim 0.1 \fb$ for $hH\nu\bar\nu$ production. The results
are found to be effectively independent of $\msq$, but depend only on
$\CBA$. This reflects the parametric dependence of the production cross
section on the couplings $g_{ZHA}\,g_{ZhA} \propto \SBA\,\CBA$.
The results in the benchmark plane~3,  presented in the lower row of
\reffi{fig:xs_hH_1500-I} show the same pattern as for the previously
analyzed center-of-mass energies. As expected, relevant production cross
sections are now found up to $\MA = \MH \sim 750 \gev$. The largest cross
sections are found around $\MH = 300 \gev$, reaching $\sim 4 \fb$ and
$\sim 0.8 \fb$ for $hHZ$ and $hH\nu\bar\nu$ production, respectively.
The small but non-zero cross section found above 
$\MA = \MH \sim 750 \gev$ for $|\CBA|\gtrsim0.2$ (i.e.\ outside the
allowed region) originate in regions of the parameter space, where
$\lahHH$ is very large, as can be seen in the lower left plot of
\reffi{fig:la-I}, which can give a relevant contribution to the total
cross section. As it can be seen in \reffi{fig:la-I}, 
this coupling can reach values up to $\sim 15$ in this part of the
parameter space.

\bigskip

\begin{figure}[htb!]
\vspace{-2em}
\begin{center}
\begin{subfigure}[b]{0.8\textwidth}
\includegraphics[width=0.5\textwidth]{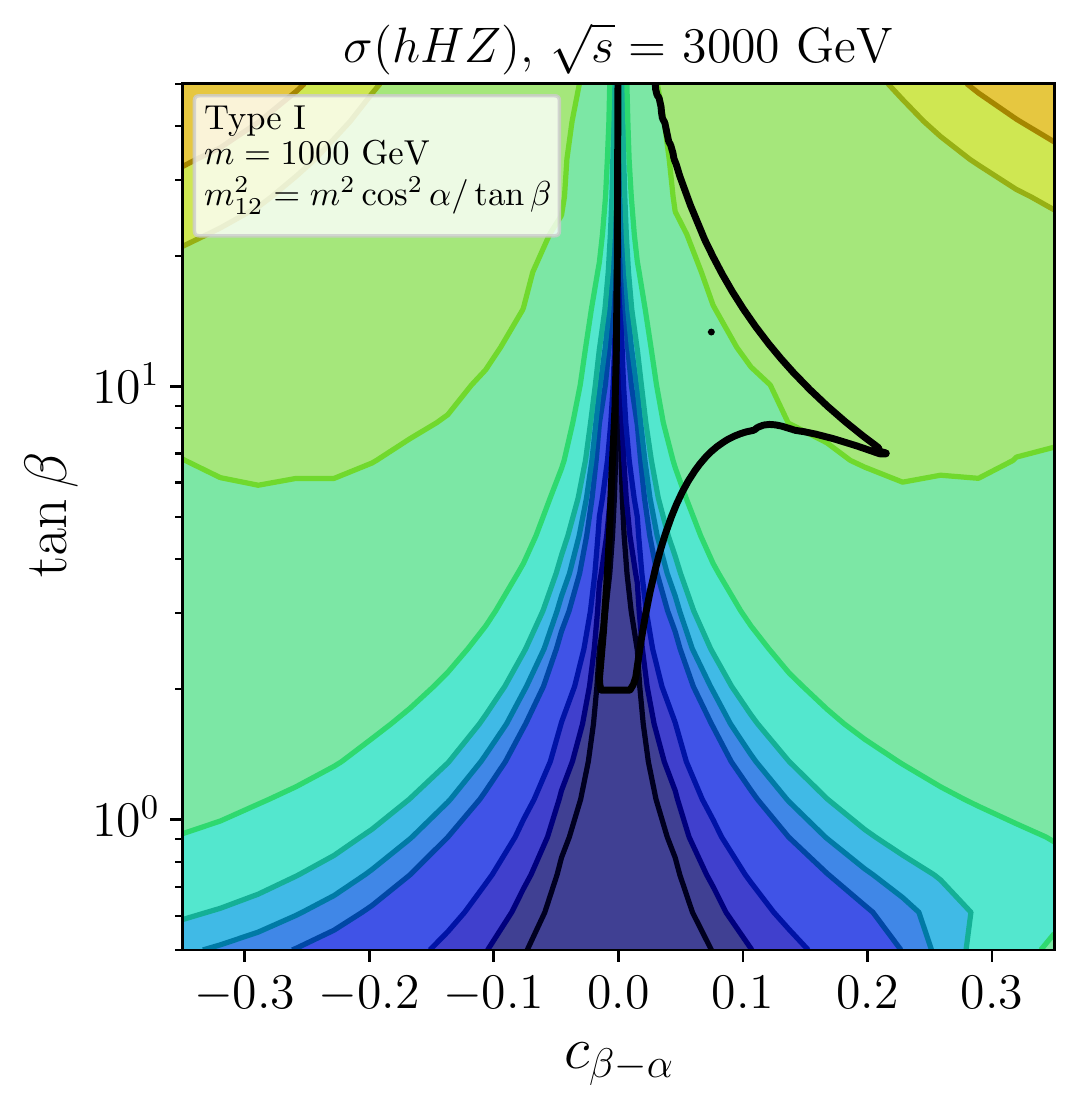}%
\includegraphics[width=0.5\textwidth]{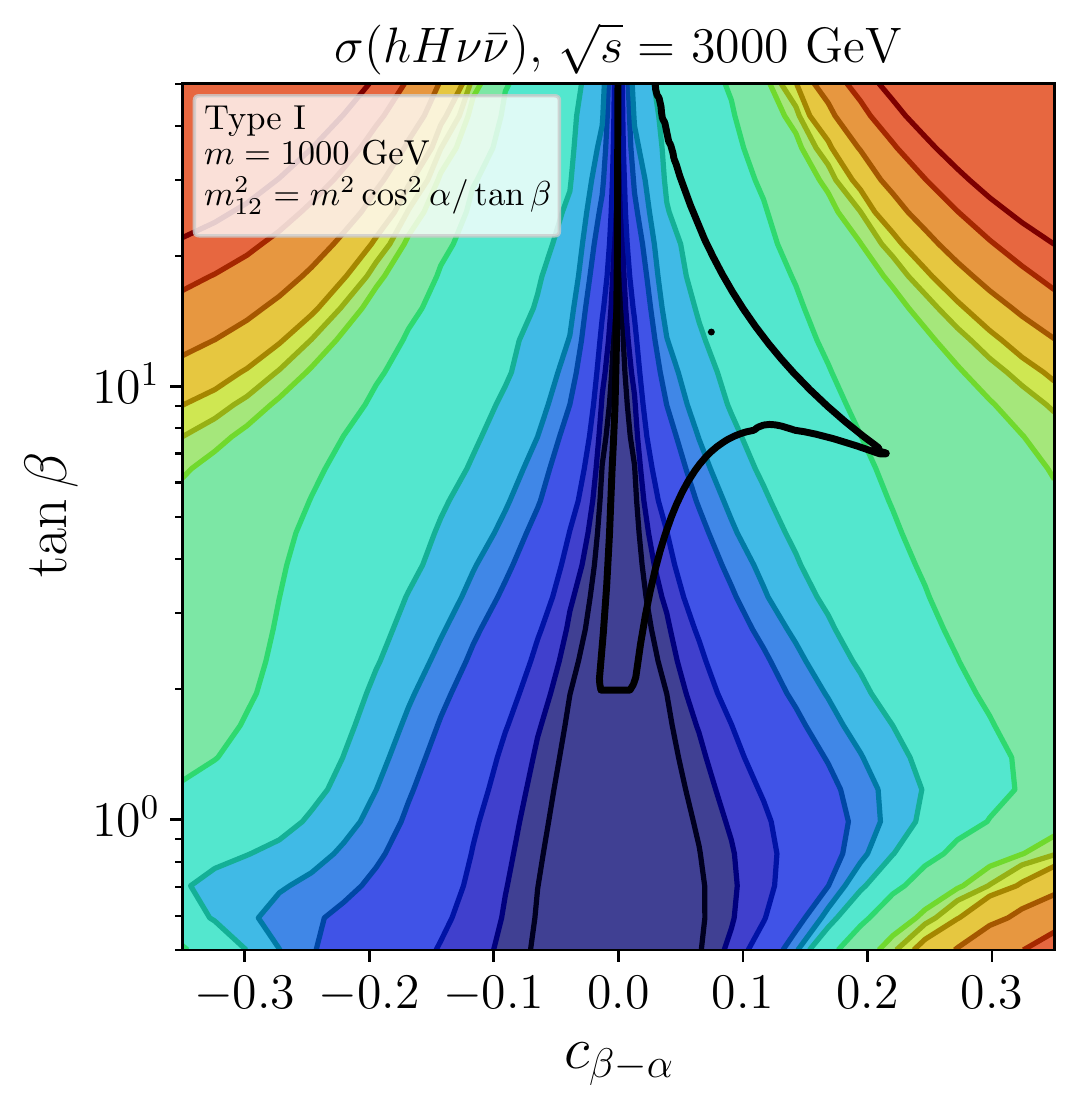}\\[1em]
\includegraphics[width=0.5\textwidth]{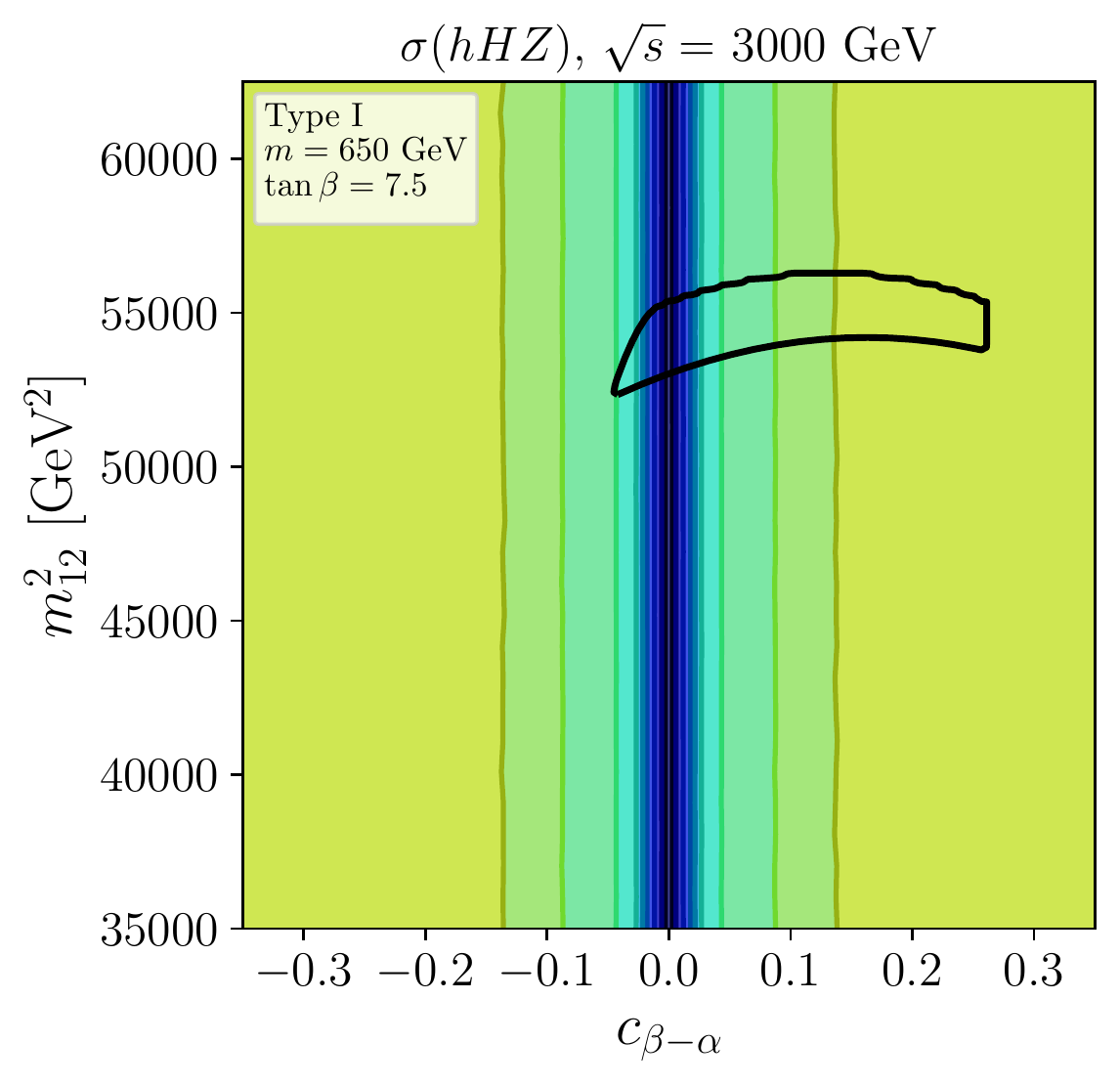}%
\includegraphics[width=0.5\textwidth]{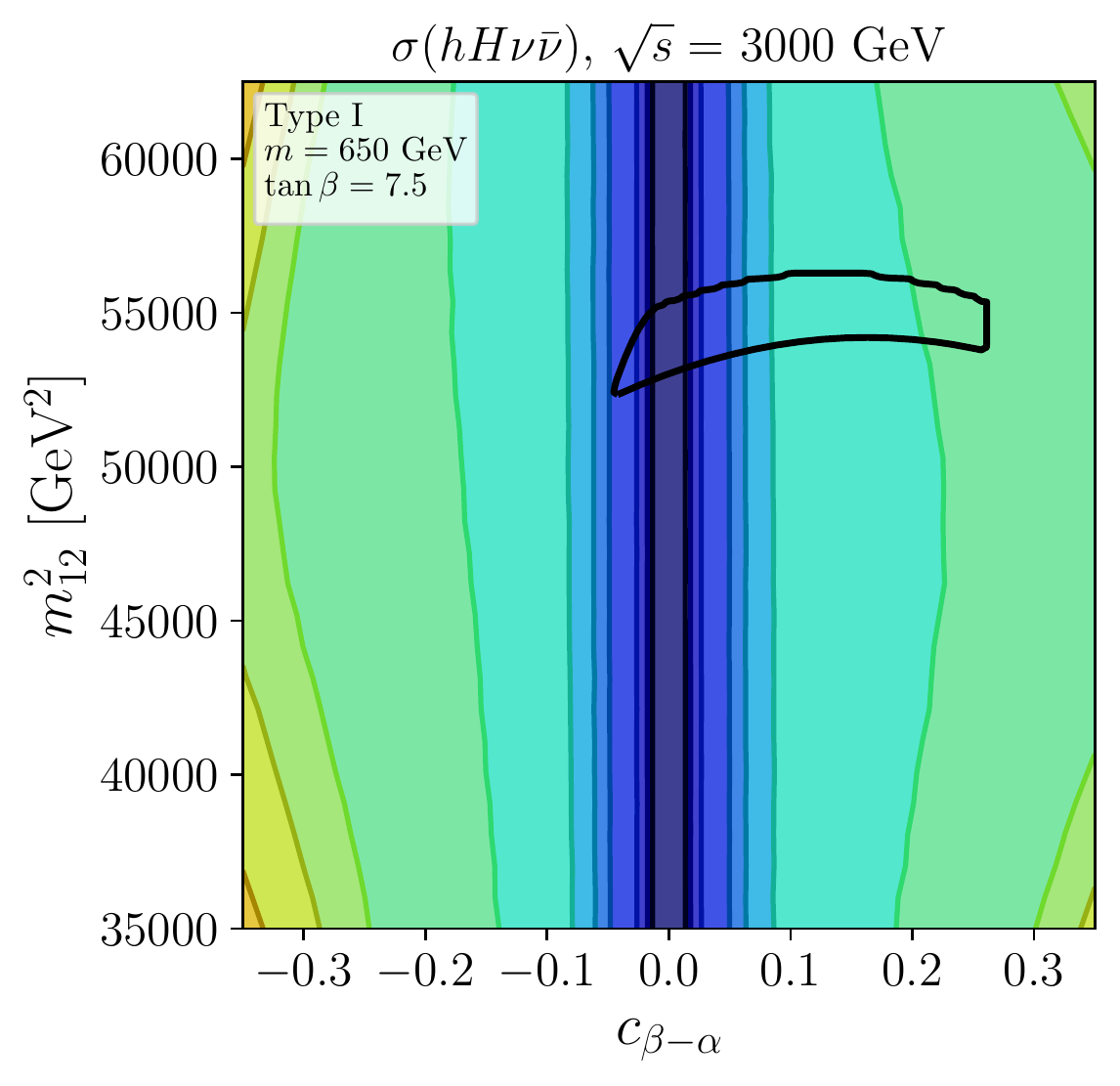}\\[1em]
\includegraphics[width=0.5\textwidth]{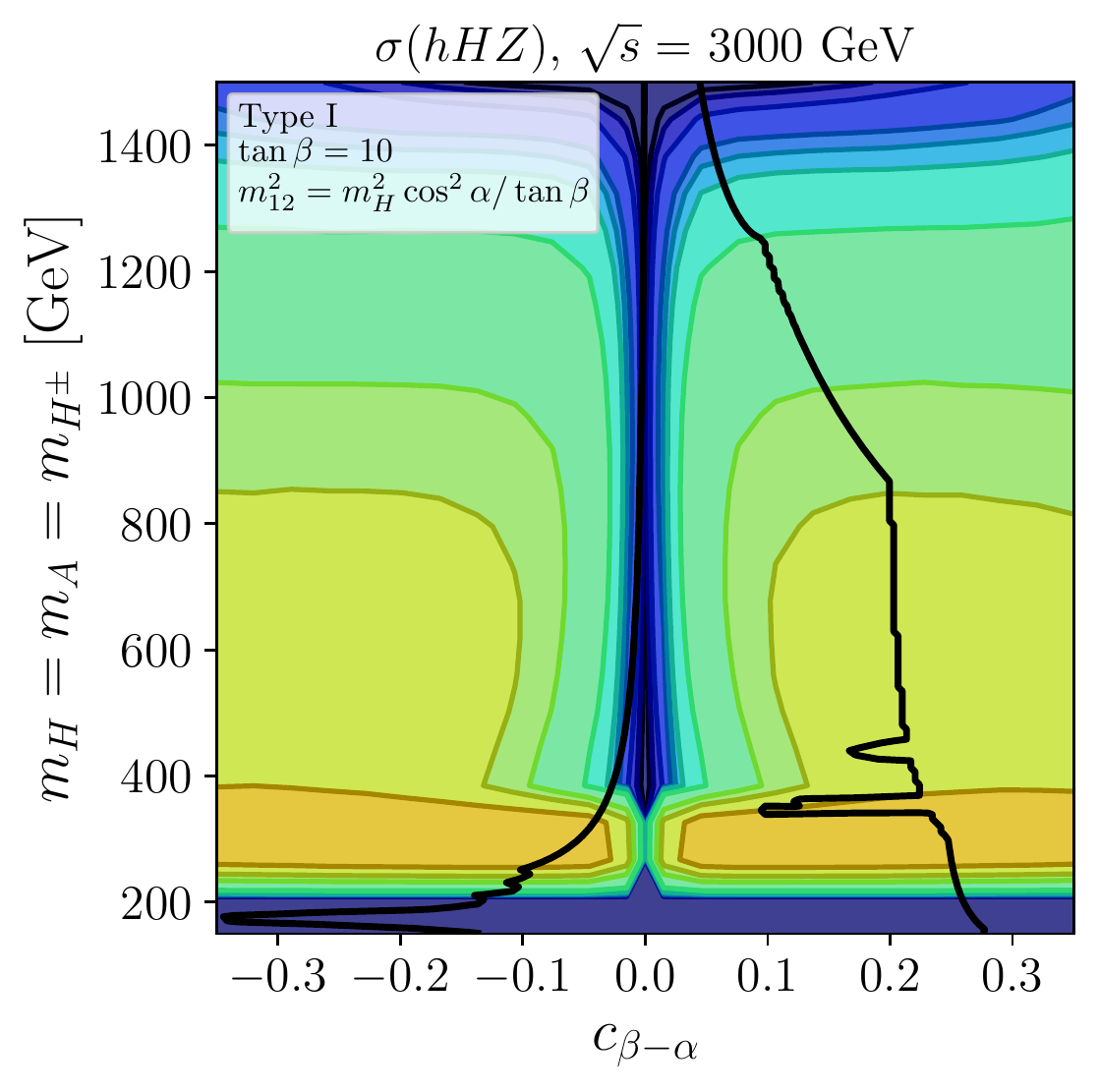}%
\includegraphics[width=0.5\textwidth]{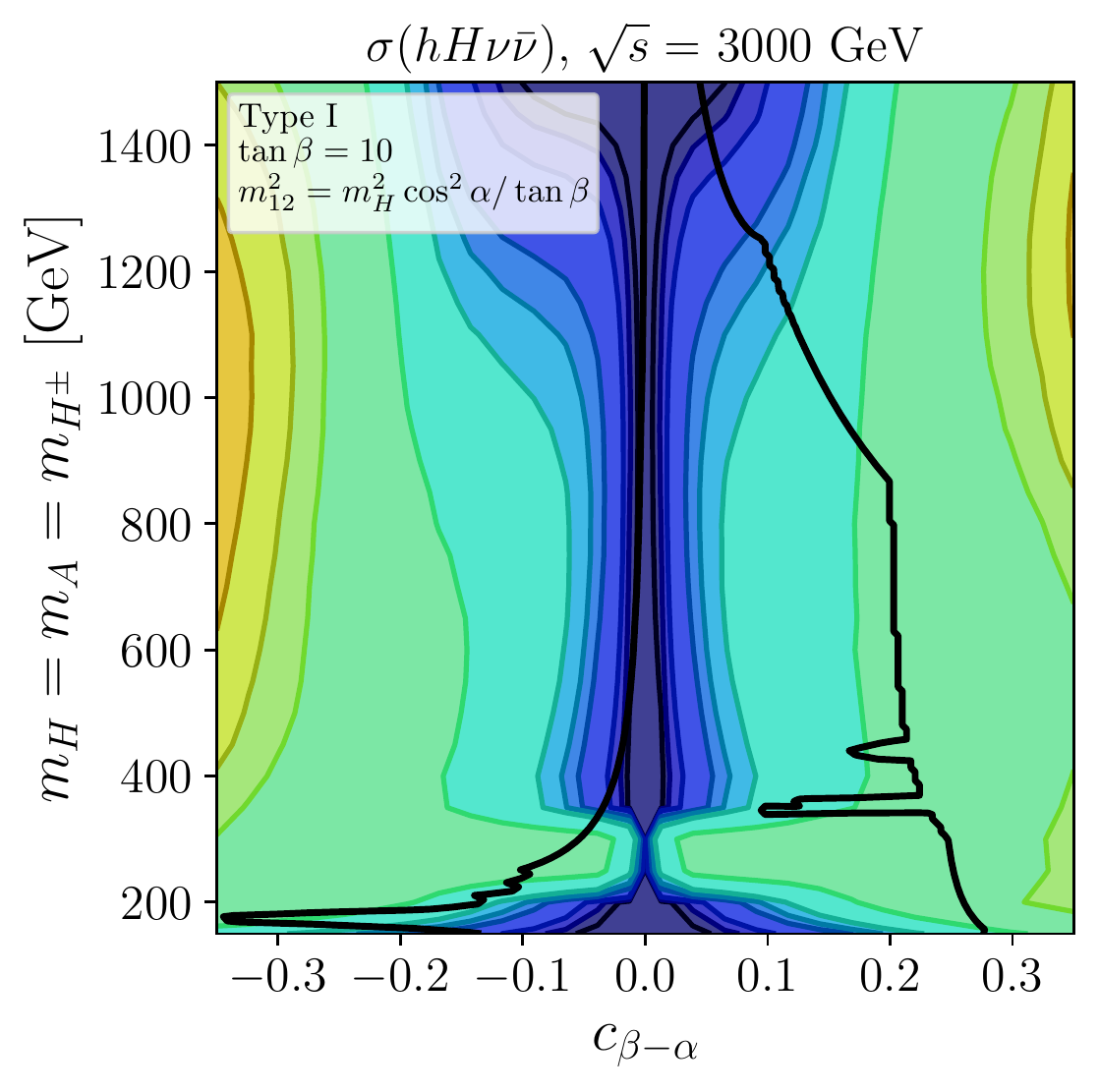}
\end{subfigure}	
\begin{subfigure}[b]{0.18\textwidth}
\includegraphics[height=0.42\textheight]{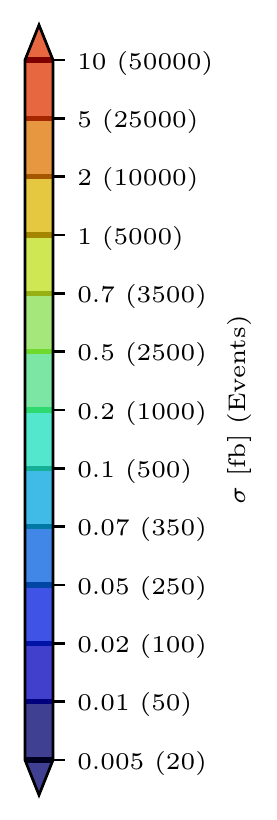}
\vspace{0.25\textheight}
\end{subfigure}
\end{center}
\caption{Cross sections for $e^+e^-\to hHZ$ (left) and
  $e^+e^-\to hH\nu\bar{\nu}$ (right) for $\sqrt{s}=3000\gev$
  for the benchmark planes 1-3 (top to bottom). The
  total allowed regions is given by the solid black line. The color code
indicates the absolute cross section in fb.}
\label{fig:xs_hH_3000-I}
\end{figure}


We finish the $hH$ analysis with \reffi{fig:xs_hH_3000-I}, where we
show the results for $\sqrt{s} = 3000 \gev$. This high center-of-mass
energy allows for on-shell  intermediate $A$ production in all three
benchmark planes 
and consequently leads to relevant cross sections in all plots.
In the $\CBA$--$\tb$ planes shown in the first row we find, as for
$\sqrt{s} = 1500 \gev$ the largest cross sections for large $\CBA$ and
$\tb$ (or very small $\tb$ in the case of $hH\nu\bar\nu$), but outside
the allowed regions. Within these regions the largest cross sections
found are $\sim 0.5 \fb$ and $\sim 0.2 \fb$ for
$hHZ$ and $hH\nu\bar\nu$ production, respectively.
As before, this is correlated with the extreme values that
$\lahHH \simeq \lahAA$ 
take in this part of the parameter space.

The cross sections shown in the $\CBA$--$\msq$ planes in the middle row of
\reffi{fig:xs_hH_3000-I} exhibit the same pattern as for
$\sqrt{s} = 1500 \gev$, i.e.\ dominated by the contributions
originating in $e^+e^- \to HA$ and thus no relevant dependence on $\msq$. 
The largest cross sections are found to reach $\sim 0.9 \fb$ and
$\sim 0.2 \fb$ for $hHZ$ and $hH\nu\bar\nu$, respectively.

In the third benchmark plane we find relevant cross
sections up to $\MH = \MA \sim 1500 \gev$, but for smaller masses,
$\lsim 700 \gev$, now
smaller cross sections are observed w.r.t.\ $\sqrt{s} = 1500 \gev$.
In this part of the parameter space again the contributions
originating in $e^+e^- \to HA$ dominate, which are suppressed with $1/s$.
The largest cross section values are found again for
$\MH = \MA \sim 300 \gev$ and reach $\sim 1.25 \fb$ and $\sim 0.25 \fb$.
However, the overall pattern of the predicted cross sections,  in
particular for $hH\nu\bar\nu$,  
differs from the ones for $\sqrt{s} = 1500 \gev$, which can be
understood as follows.
While the $hHZ$ channel is again dominated 
by the $e^+e^- \to HA \to H(hZ)$ resonant subprocess, in the 
$hH\nu\bar\nu$ channel
it can be seen that there is a similar (or even larger) 
prediction than in the $hHZ$ channel for
$\MH = \MA \gsim  900 \gev$ (albeit outside the allowed area).
The main difference between 
the $hHZ$ and the $hH\nu\bar{\nu}$ channels at high energies
of \order{3 \tev} is the different influence of
the triple Higgs couplings. The $hHZ$ channel
is dominated, as discussed above, by the intermediate $HA$ state and the 
subsequent decay $A \to hZ$,  which are not sensitive to triple Higgs 
couplings.  In contrast, the $hH\nu\bar\nu$
not only receives relevant contributions from the diagrams shown in
the left of \reffi{fig:diagrams} with the subsequent decay
$Z \to \nu\bar\nu$, but as well from VBF
subprocesses such as $WW \to hH$, where the triple Higgs couplings do
enter via the $s$-channel diagrams with either virtual $h$ or virtual
$H$ propagating. This type of topology, mediated by $WW$ fusion, grows
with energy and thus contributes relevantly to
the $e^+e^-\to hH \nu \bar{\nu}$ cross section.
Consequently, at high energies one expects to gain access to 
the triple couplings involved, $\lahhH$ and $\lahHH$.
Since for larger values of $\MH$ one finds
larger values for $\lahHH$ than for $\lahhH$, 
as can be seen in the lower row of \reffi{fig:la-I},
one expects a higher sensitivity to  $\lahHH$ in this part of
the parameter space.
The final reach to these couplings is the combined result of the large
coupling effect vs.\ the suppression of the corresponding virtual
Higgs (either $h$ or $H$) propagating in the $s$-channel of this
VBF subprocess. 
For the parameter region explored in \reffi{fig:xs_hH_3000-I}
$\lahHH$ can reach large values up to $\sim 10$ within the allowed area,
as it can be seen in the lower 3rd column plot of \reffi{fig:la-I}.
The virtual $H$~boson, propagating close to its mass shell, can
thus contribute strongly to the $hH\nu\bar\nu$ production cross section.
Conversely,  diagrams with a virtual $h$ are suppressed by
relatively smaller triple Higgs couplings of \order{1}, as can be seen
in the lower 2nd column plot of \reffi{fig:la-I}, as well by
the propagation of a very off-shell $h$ boson.
Overall, this yields a higher sensitivity to $\lahHH$ than to
$\lahhH$ in the allowed parameter region, as will be discussed in more
detail in \refse{sec:sensitivity}.


\subsubsection{\boldmath{$HH\, (AA)$} production}
\label{sec:xsHH-I}

In this subsection we analyze di-Higgs production of two heavy 
Higgs bosons.  This can be either $HHZ$/$HH\nu\bar\nu$ or
$AAZ$/$AA\nu\bar\nu$. Since for our choice of $\MA = \MH$
the di-$\cp$-odd production is always very
similar to the di-$\cp$-even production we only describe the latter.
In these channels, the alignment limit does not necessarily 
imply a zero cross section, and in general no resonant enhancement of
cross sections can take place in our benchmark planes,  since the
intermediate Higgs bosons are always off-shell.

\begin{figure}[t!]
\begin{center}
\begin{subfigure}[b]{0.7\textwidth}
  \includegraphics[width=0.48\textwidth]{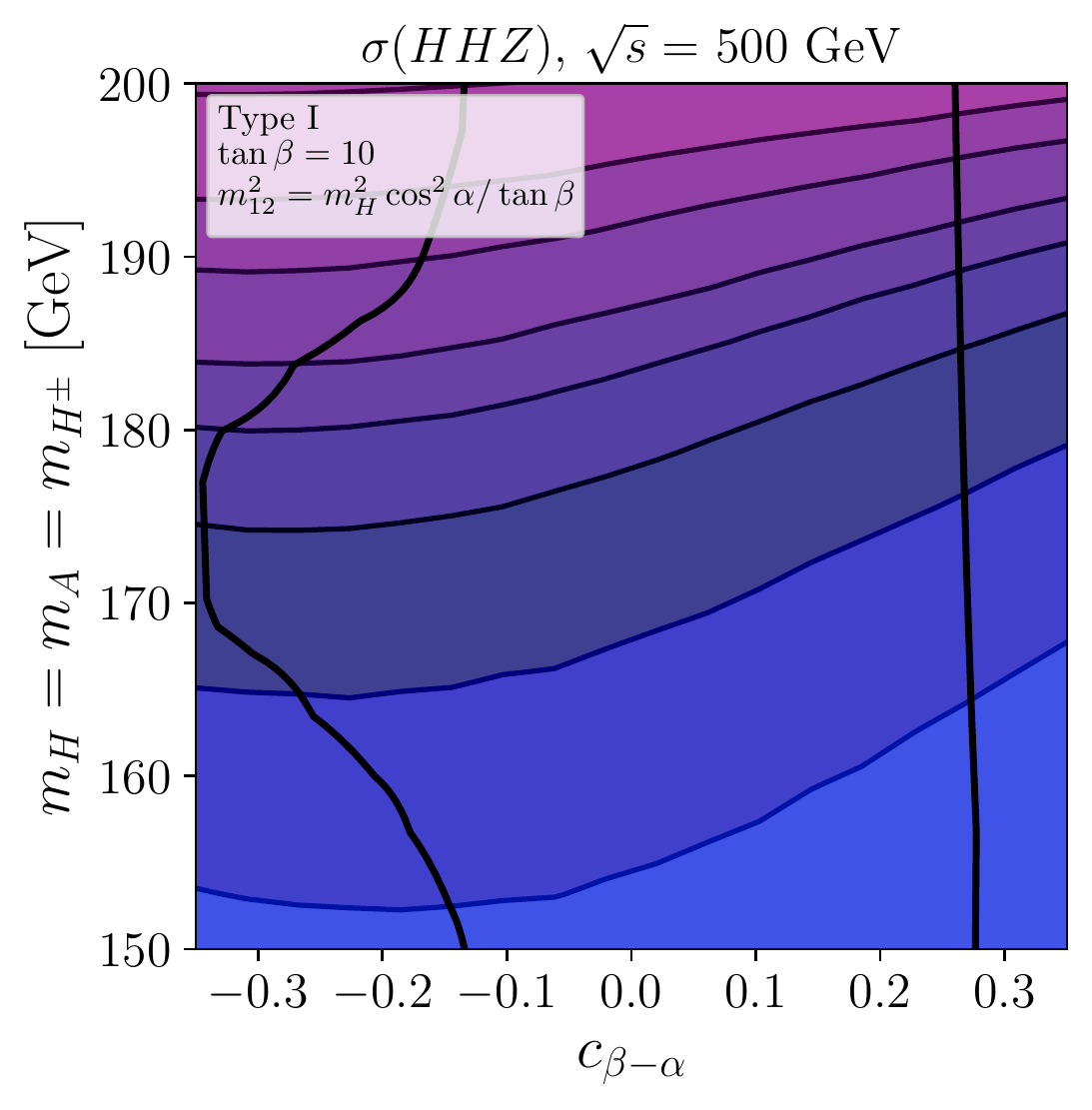}
  \includegraphics[width=0.48\textwidth]{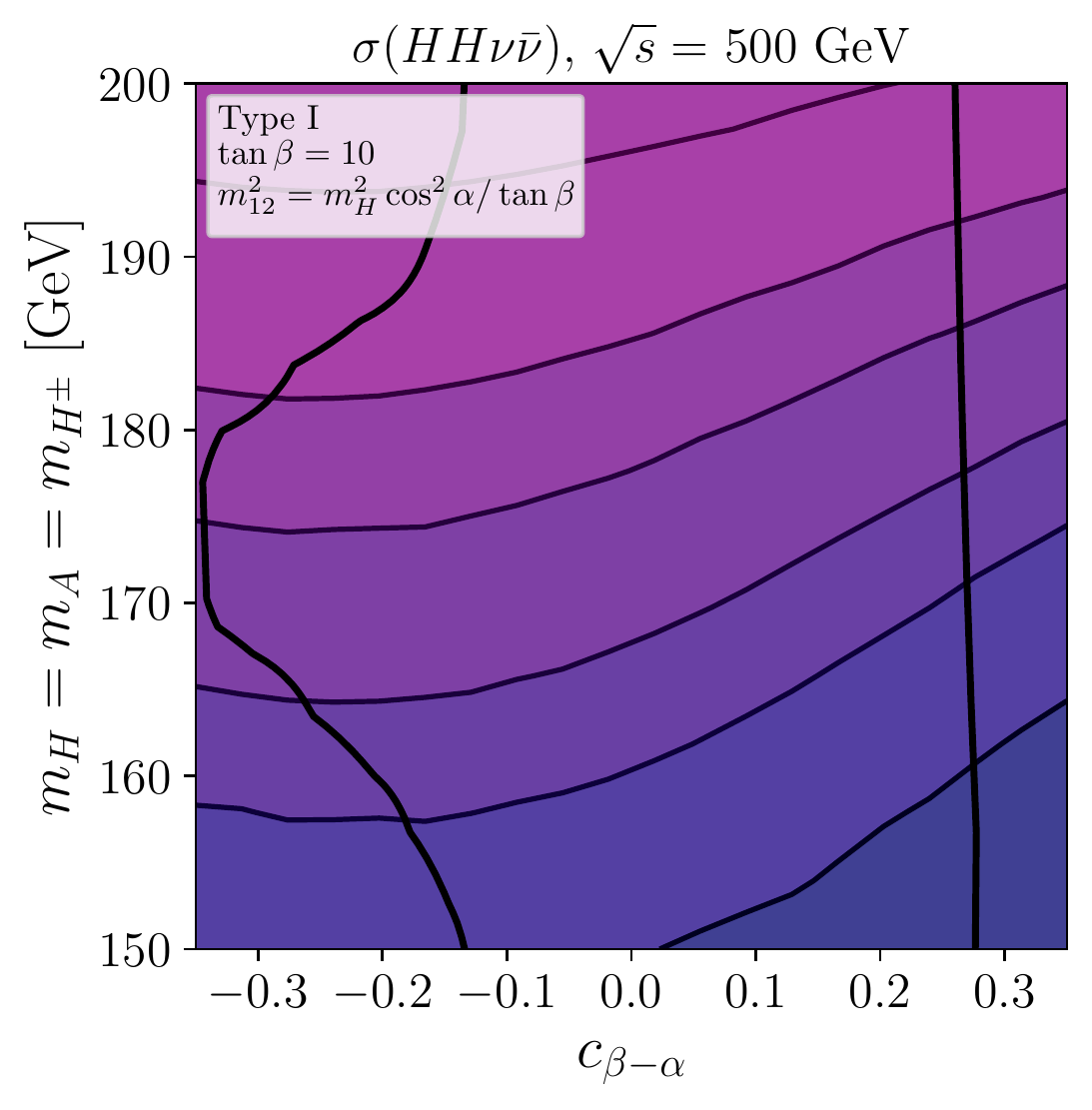}
\end{subfigure}	
\begin{subfigure}[b]{0.18\textwidth}
  \includegraphics[height=0.26\textheight]{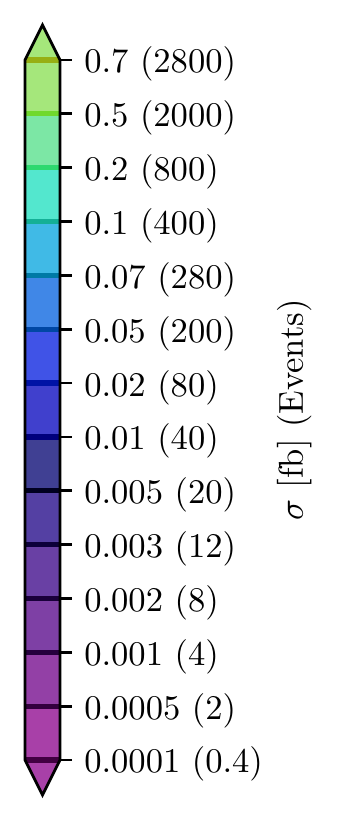}
\end{subfigure}
\end{center}
\caption{Cross sections for $e^+e^-\to HHZ$ and $e^+e^-\to HH\nu\bar\nu$
at $\sqrt{s}=500\gev$ for our benchmark plane~3. The total allowed
regions is given by the solid black line. The color code
indicates the absolute cross section in fb.
}
\label{fig:xs_HH_500-I}
\end{figure}

\bigskip

The results for $\sqrt{s} = 500 \gev$ are shown in
\reffi{fig:xs_HH_500-I}. For this low center-of-mass energy only the low
mass region of benchmark plane~3 yields kinematically allowed production
cross sections. 
For $HHZ$ production we find very small cross sections going up to
$\sim 0.05\,\fb$ in the allowed region for $\MH \sim 160 \gev$.
The $HH\nu\bar\nu$ channel does not exhibit any non-negligible cross
sections. 
In the relevant parameter space $\lahHH$ and $\laHHH$ are of similar
size, but do not exceed $\sim 2$.
The contribution via an $s$-like-channel with an intermediate virtual $H$ involving $\laHHH$, however, is
furthermore suppressed by $\CBA$, with respect to the $h$~boson mediated
contributions being $\propto \SBA\lahHH$.
Diagrams via an off-shell $A$~boson, on the other hand are
$\propto \SBA^2$ (i.e.\ no triple Higgs coupling enters) and can
contribute with similar strength as the $h$~boson mediated one.
Indeed we find that the contribution involving an intermediate
  $A$~boson can be somewhat larger that the $h$ mediated one.

\bigskip

\begin{figure}[t!]
\begin{center}
\begin{subfigure}[b]{0.7\textwidth}
  \includegraphics[width=0.48\textwidth]{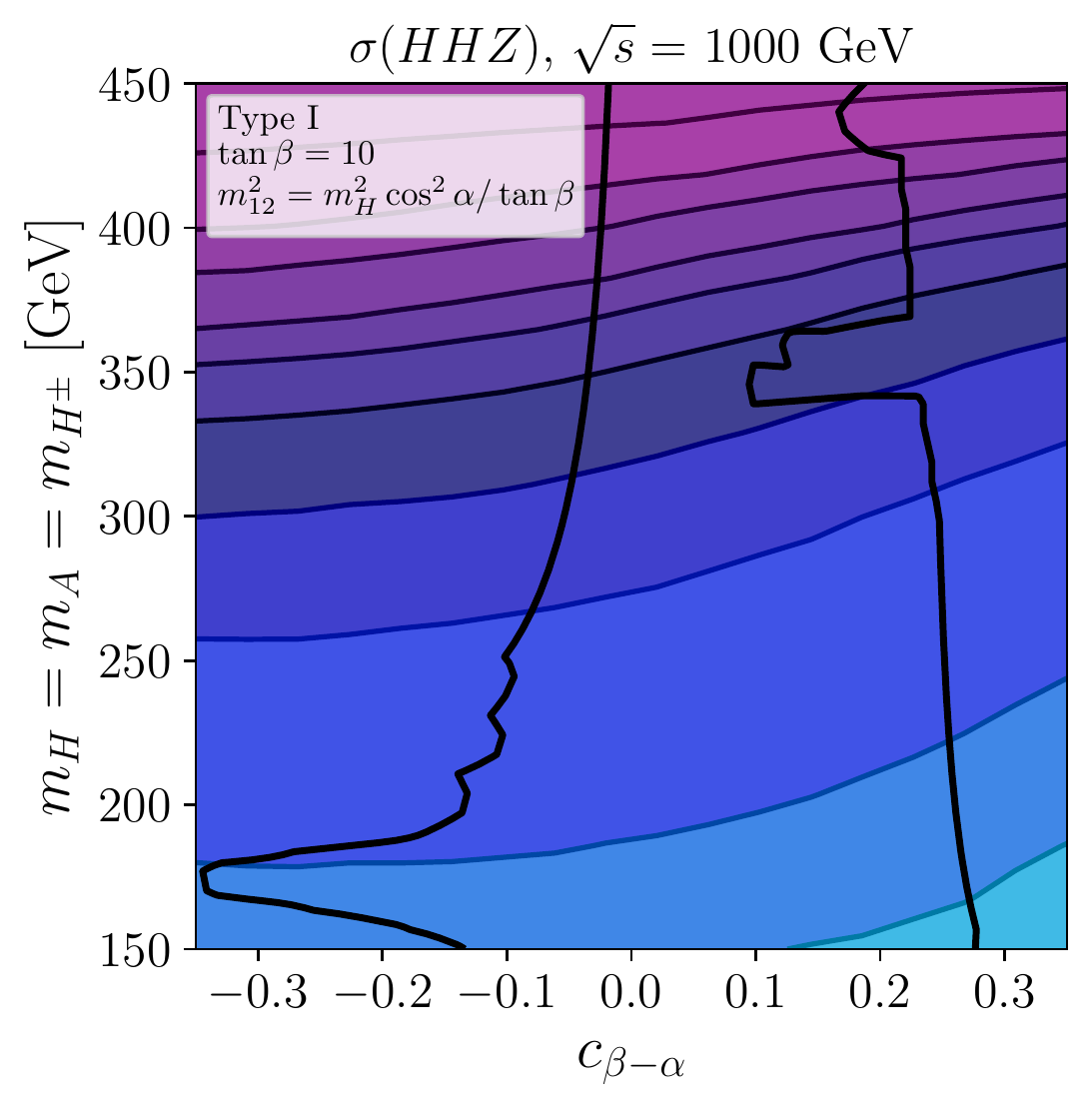}
  \includegraphics[width=0.48\textwidth]{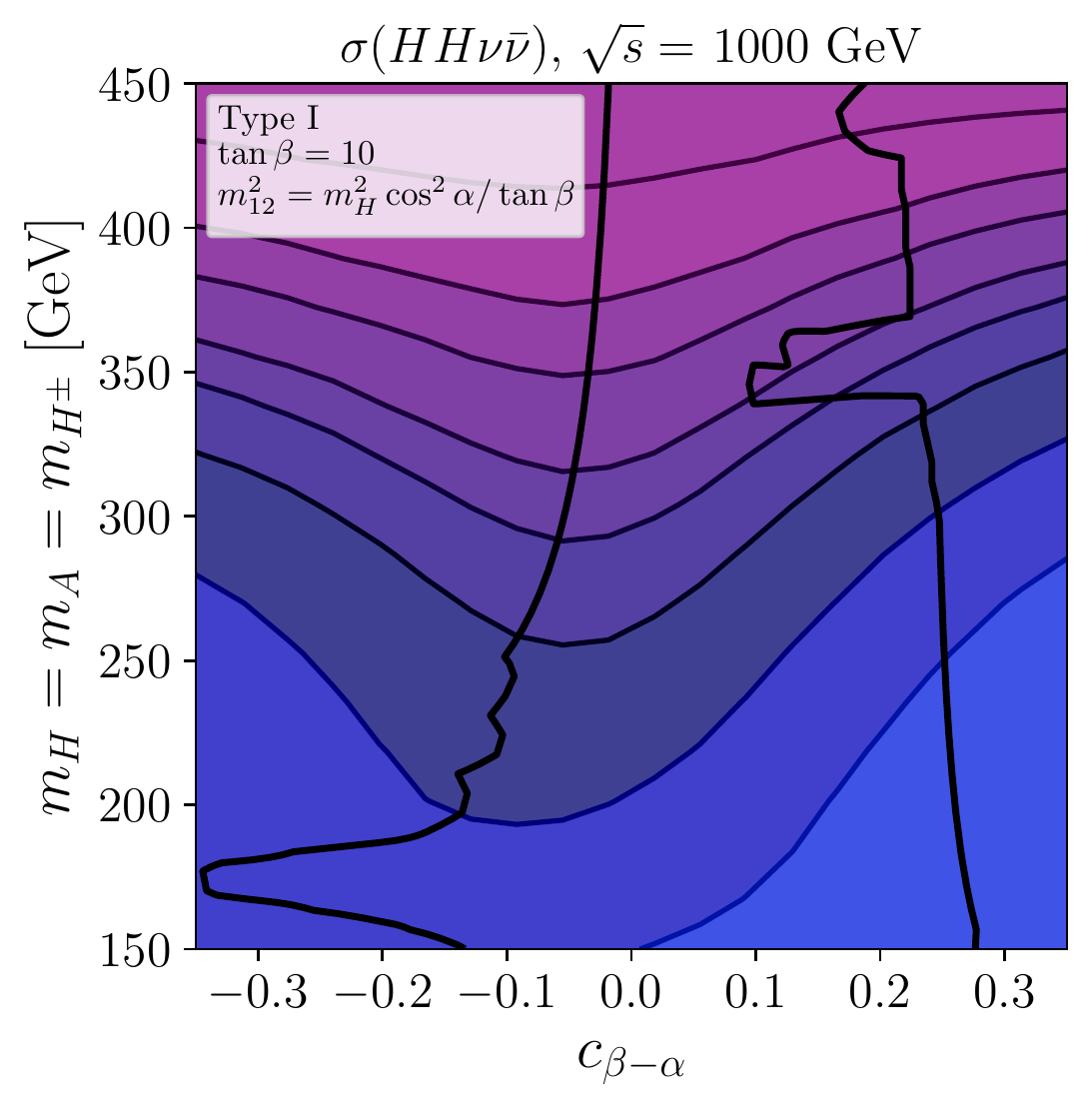}
\end{subfigure}	
\begin{subfigure}[b]{0.18\textwidth}
  \includegraphics[height=0.26\textheight]{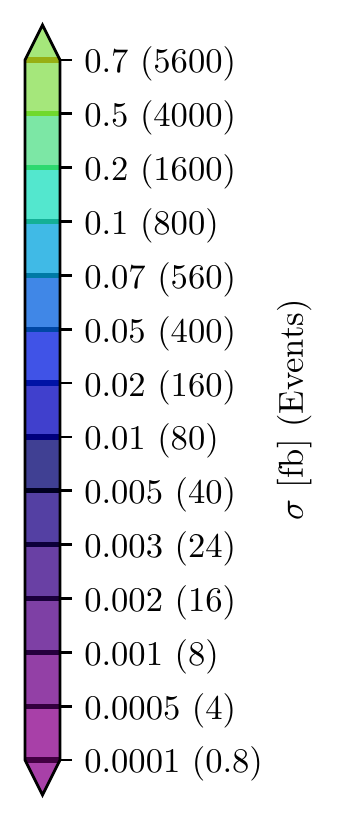}
\end{subfigure}
\end{center}
\caption{Cross sections for $e^+e^-\to HHZ$ and $e^+e^-\to HH\nu\bar\nu$
at $\sqrt{s}=1000\gev$ for our benchmark plane~3. The total allowed
regions is given by the solid black line. The color code
indicates the absolute cross section in fb.
}
\label{fig:xs_HH_1000-I}
\end{figure}

In \reffi{fig:xs_HH_1000-I} the results for $\sqrt{s} = 1000 \gev$ are
shown. As for $\sqrt{s} = 500 \gev$, only the low
mass region of benchmark plane~3 yields kinematically allowed production
cross sections. As before, diagrams involving $\lahHH$ give a larger
contribution than the ones involving $\laHHH$. 
Higher cross sections are found for smaller $\MH$: $HHZ$ production
reaches $\sim 0.1\,\fb$ in the allowed region, whereas 
the $HH\nu\bar\nu$ channel goes only up to $\sim 0.06\,\fb$, both for
the smallest allowed $\MH$ values and the largest allowed $\CBA$ values,
indicating slightly increasing contributions from an $s$-like-channel $H$
coupled to gauge bosons, involving $\laHHH$.

\bigskip

\begin{figure}[t!]
\begin{center}
\begin{subfigure}[b]{0.7\textwidth}
  \includegraphics[width=0.48\textwidth]{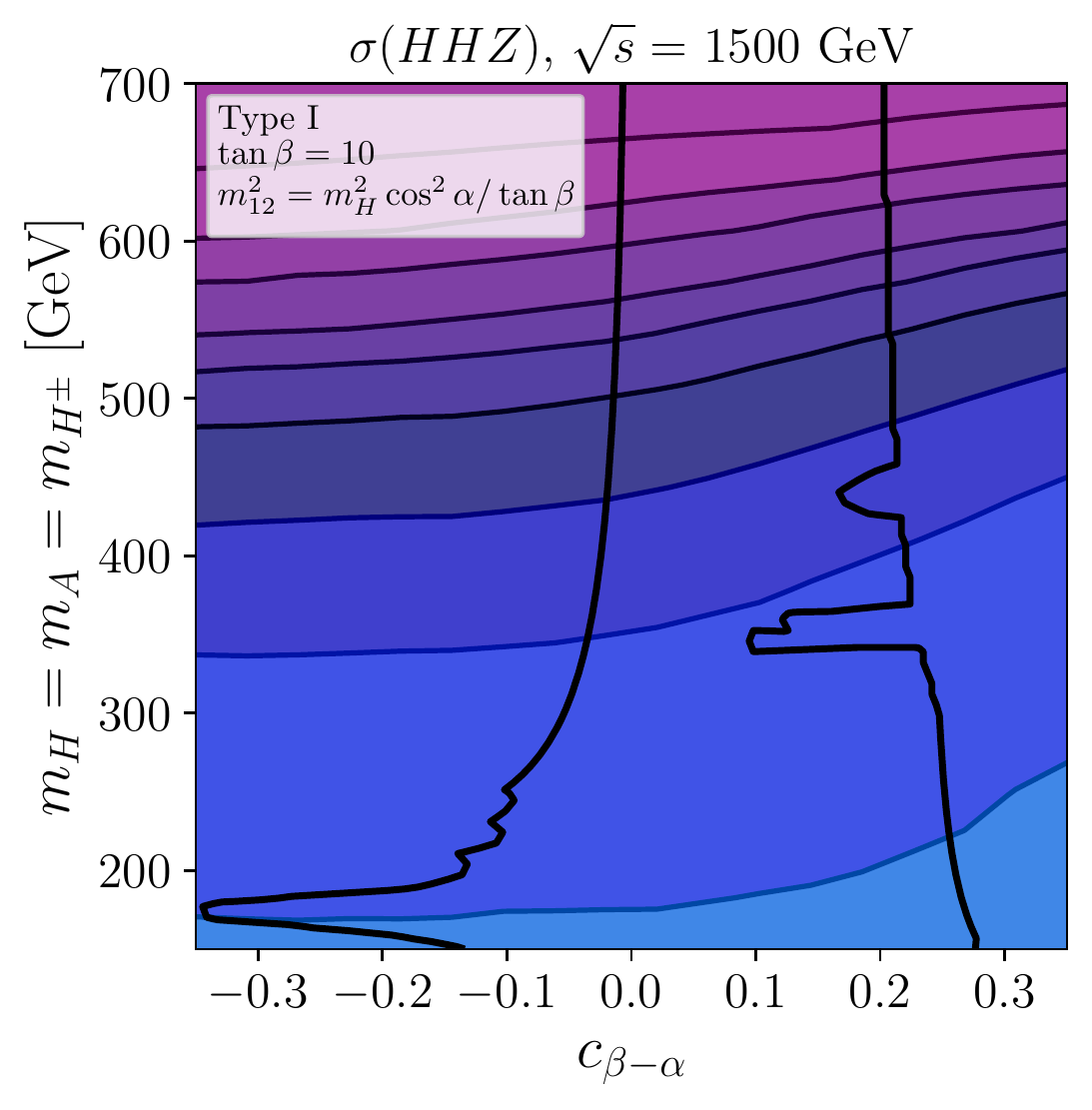}
  \includegraphics[width=0.48\textwidth]{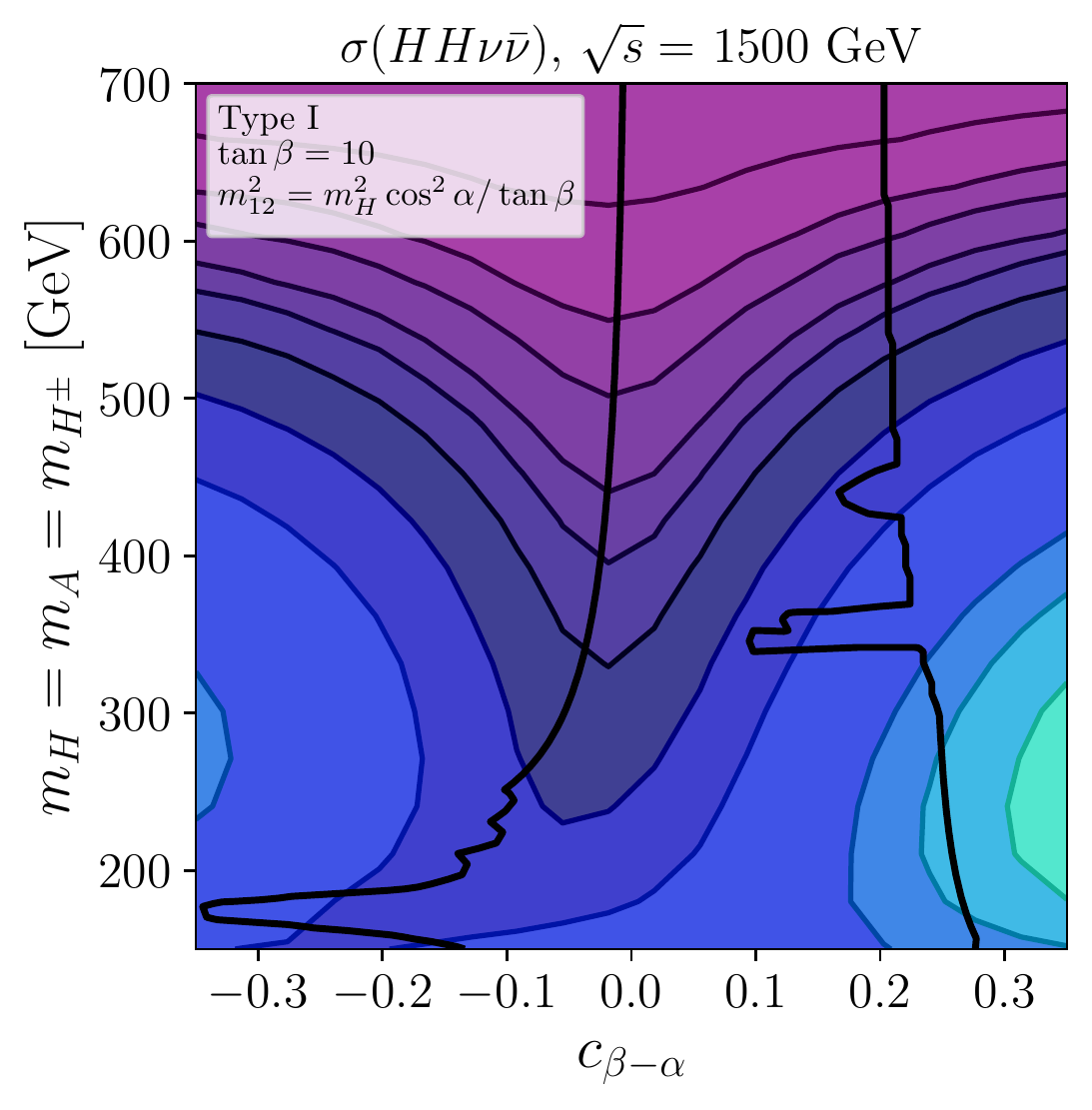}
\end{subfigure}	
\begin{subfigure}[b]{0.18\textwidth}
\includegraphics[height=0.26\textheight]{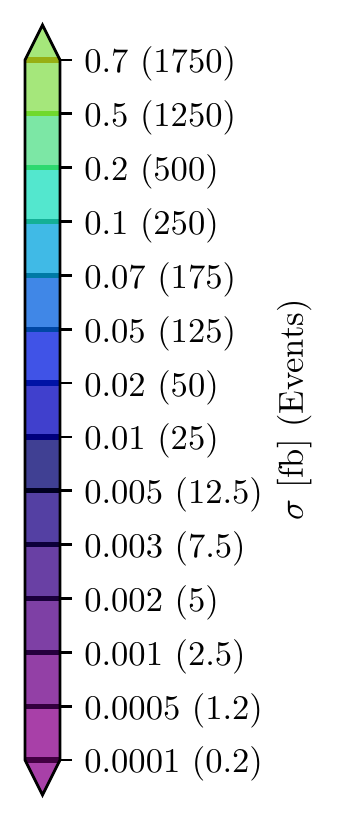}
\end{subfigure}
\end{center}
\caption{Cross sections for $e^+e^-\to HHZ$ and $e^+e^-\to HH\nu\bar\nu$
at $\sqrt{s}=1500\gev$ for our benchmark plane~3. The total allowed
regions is given by the solid black line. The color code
indicates the absolute cross section in fb.
}
\label{fig:xs_HH_1500-I}
\end{figure}

Also for $\sqrt{s} = 1500 \gev$, shown in \reffi{fig:xs_HH_1500-I}, only
the benchmark plane~3 shows non-negligible cross sections in the lower
mass region. In principle also the benchmark plane~2 with $\MH = 650 \gev$
results in kinematically allowed cross sections. However, in agreement
with the results shown for the benchmark plane~3, the calculated cross
sections remain very small, below $0.005\,\fb$. 
While the results for $\sqrt{s} = 1500 \gev$ follow the pattern found
for lower center-of-mass energies, the higher $\sqrt{s}$ here already
tends to yield larger cross sections for the VBF-type mediated
contributions for $HH\nu\bar\nu$ than for $HHZ$ production, mediated by
the Higgs-strahlung type diagrams. 
As before the largest cross sections are reached for small $\MH$
and large $\CBA$. Within the allowed regions the $HH\nu\bar\nu$ cross
section reaches $\sim 0.1\,\fb$.
Several competing effects are relevant here.
The most important contributions to $HH\nu\bar{\nu}$ in this region come
from the diagrams with an intermediate virtual  $h$ or $H$ in the
$s$-like-channel. 
These contributions scale like $\SBA\lahHH$ and $\CBA\laHHH$,
respectively. For growing $\CBA$ also $\laHHH$ itself increases. Thus, 
the contribution from the $s$-channel type $H$ becomes more relevant for
larger $\CBA$, but always stays well below the contribution from the
$h$~boson,  even for large values of $\CBA$ beyond the allowed region.

\bigskip

\begin{figure}[p!]
\begin{center}
\begin{subfigure}[b]{0.7\textwidth}
  \includegraphics[width=0.48\textwidth]{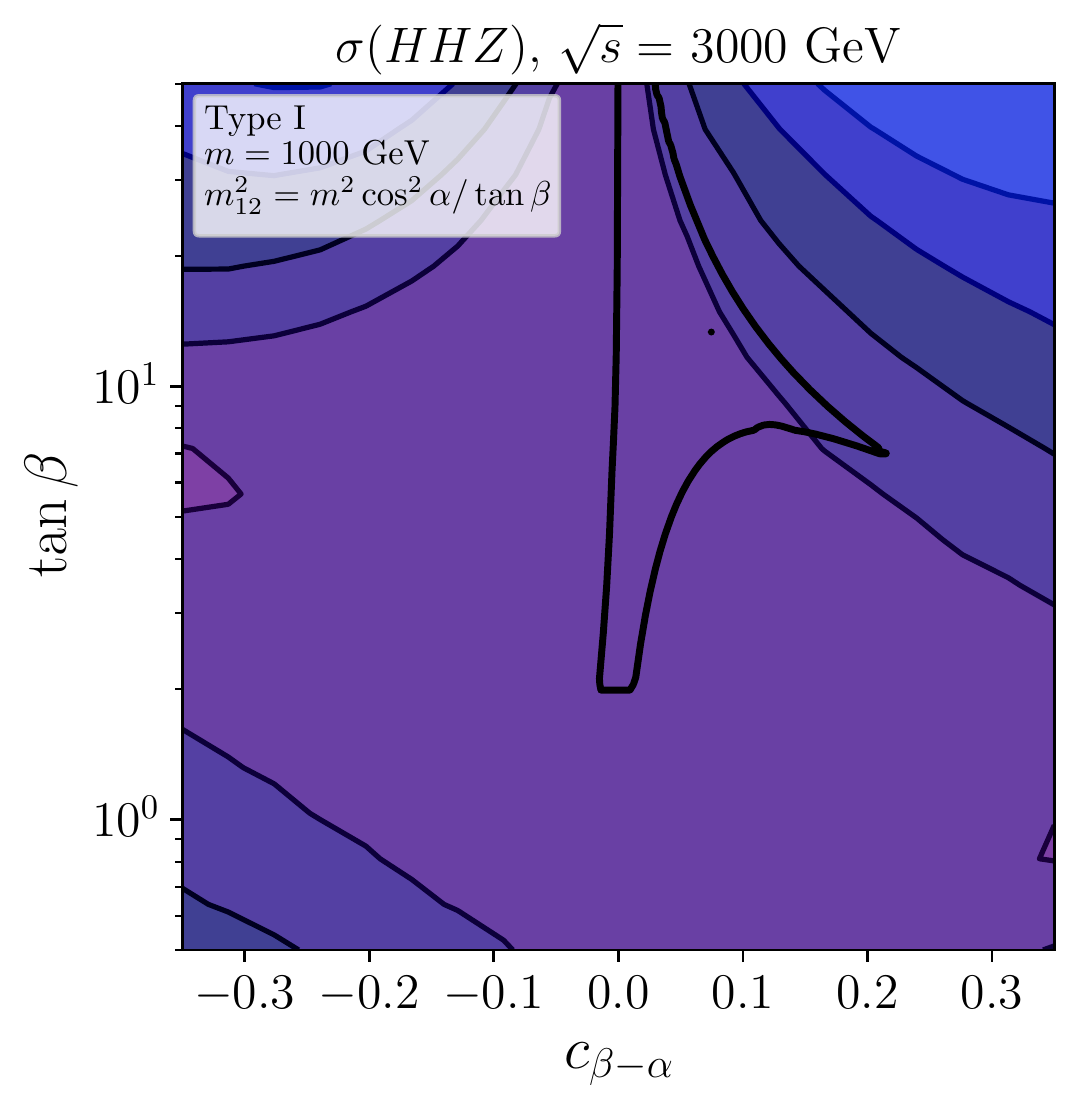}
  \includegraphics[width=0.48\textwidth]{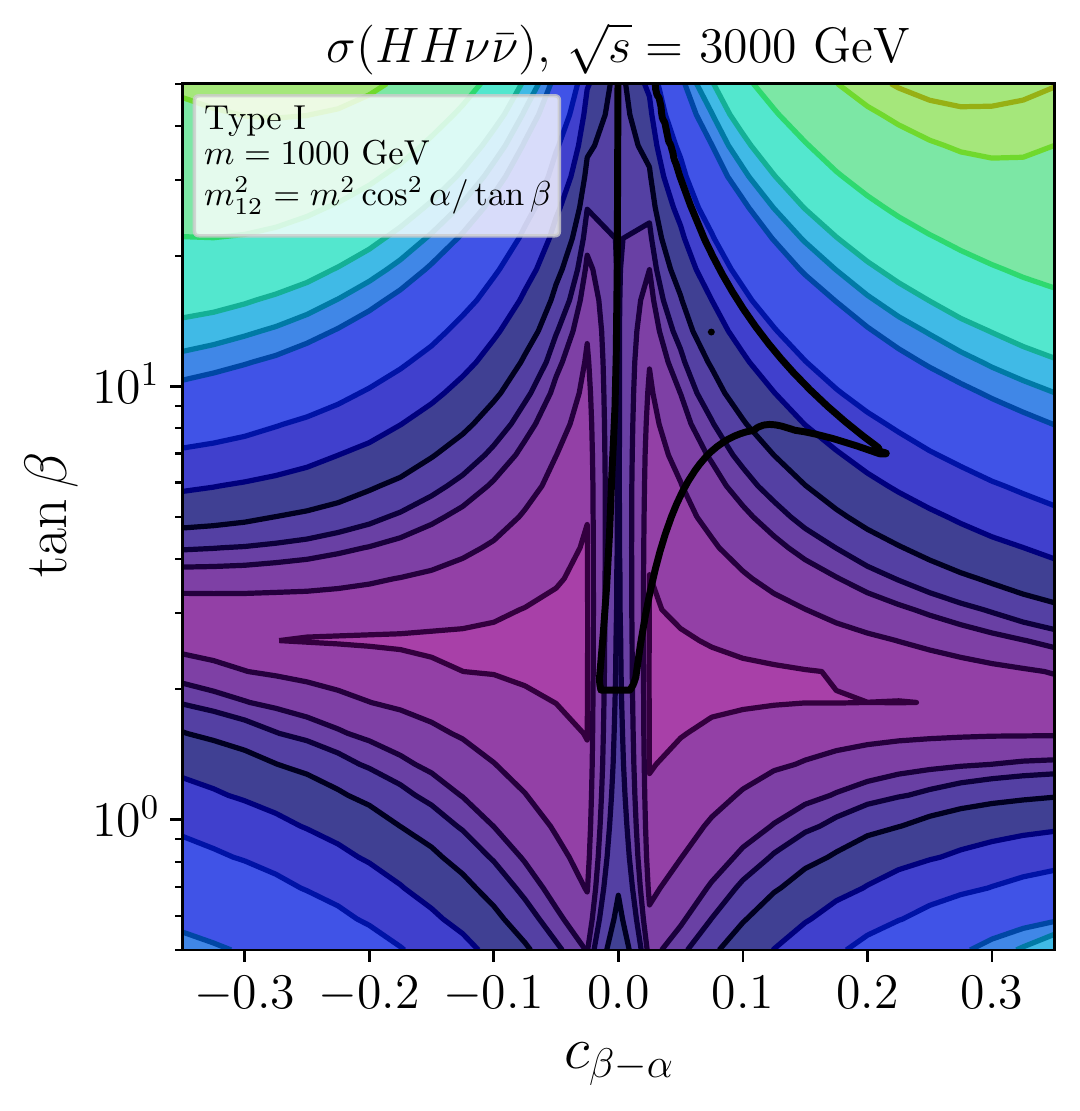}\\
  \includegraphics[width=0.48\textwidth]{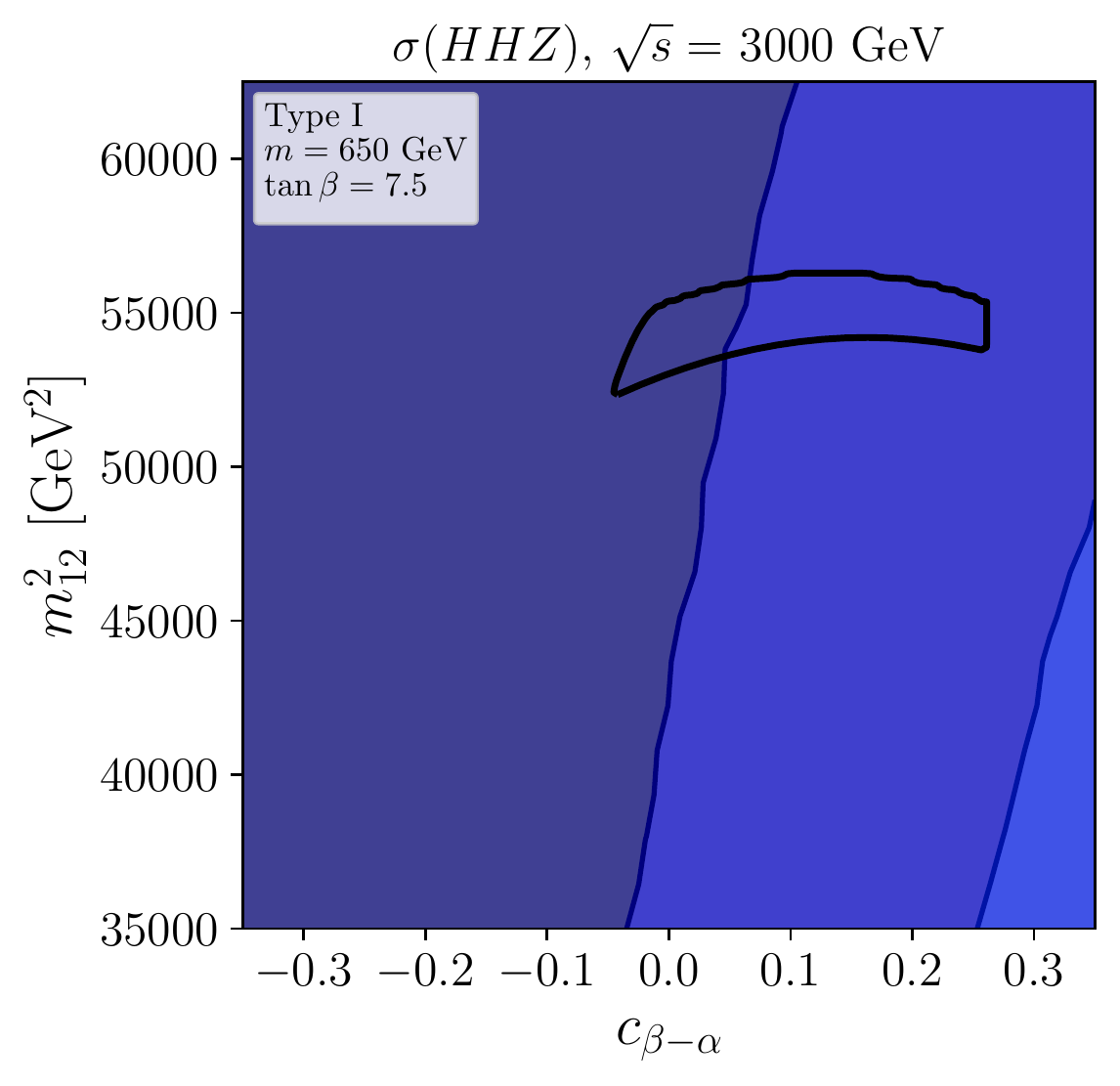}
  \includegraphics[width=0.48\textwidth]{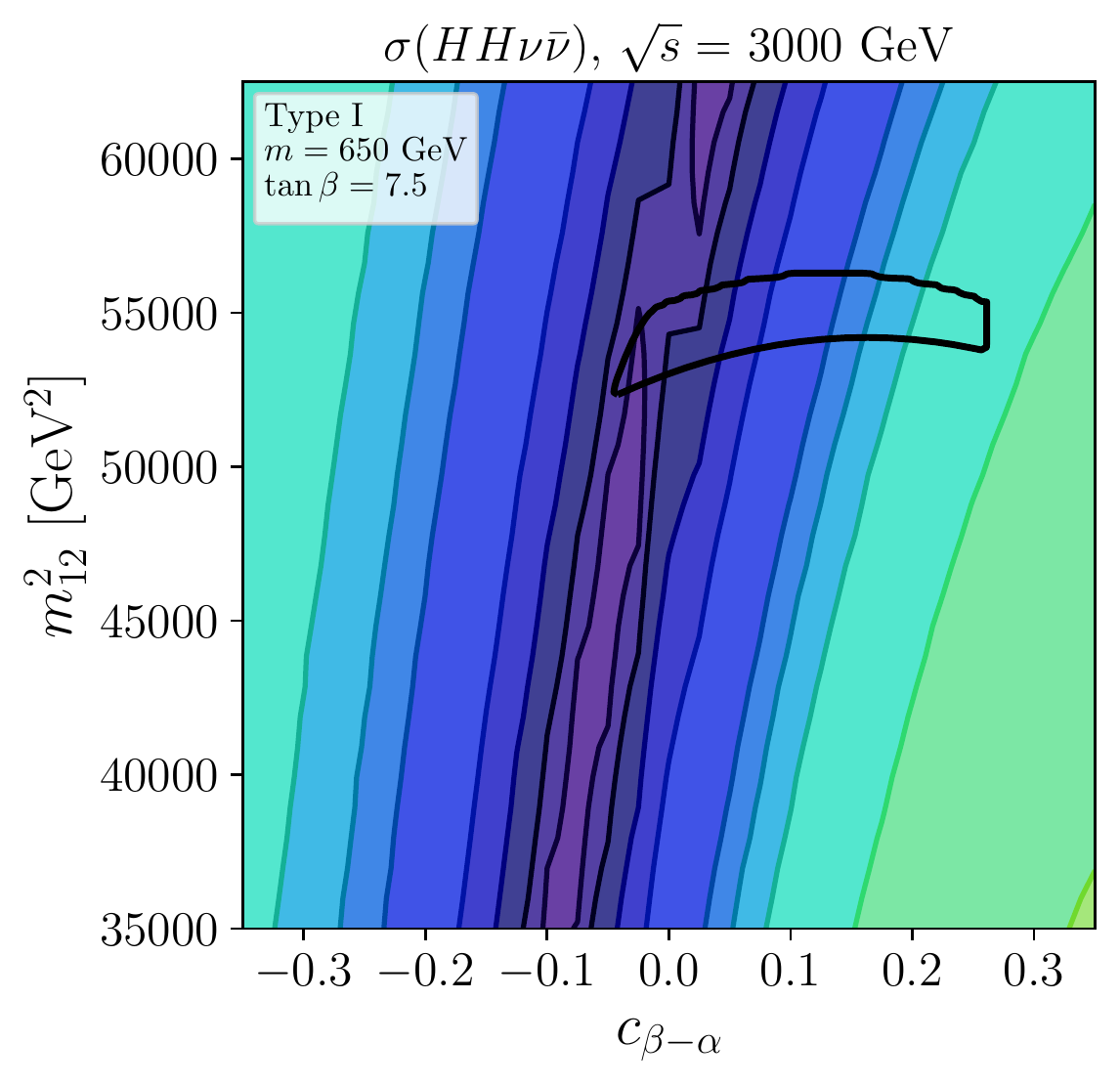}\\
  \includegraphics[width=0.48\textwidth]{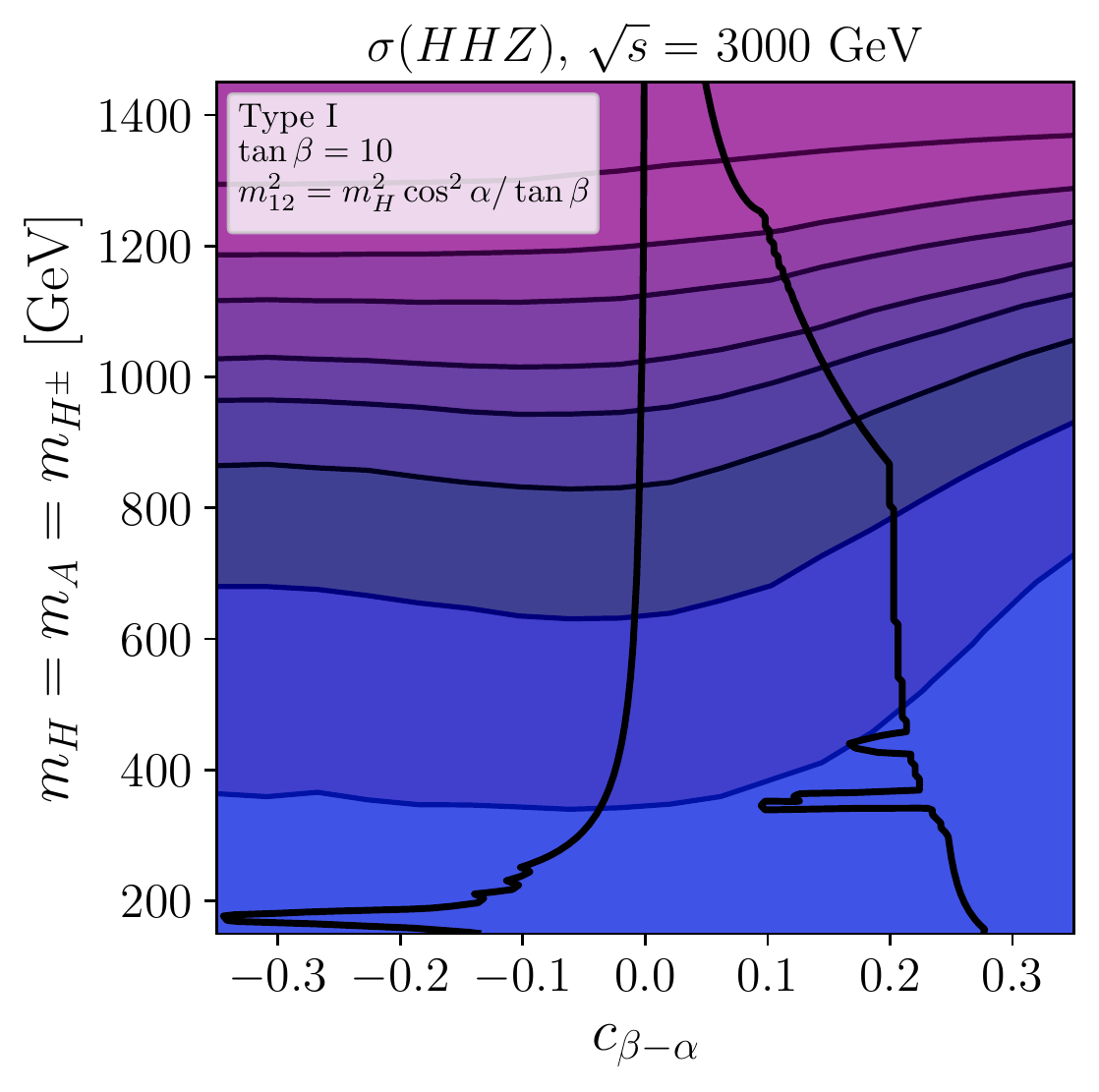}
  \includegraphics[width=0.48\textwidth]{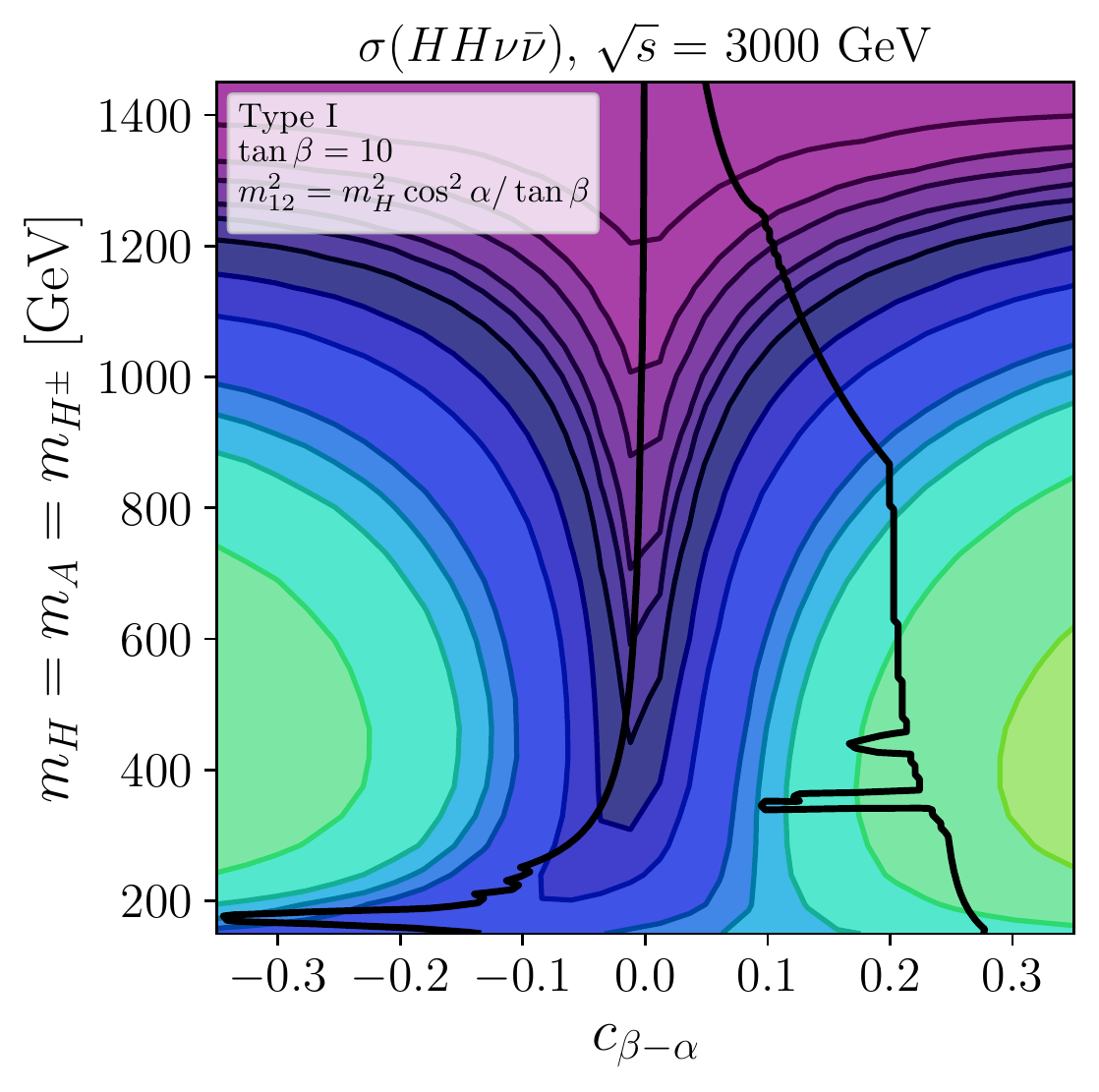}
\end{subfigure}	
\begin{subfigure}[b]{0.18\textwidth}
\includegraphics[height=0.42\textheight]{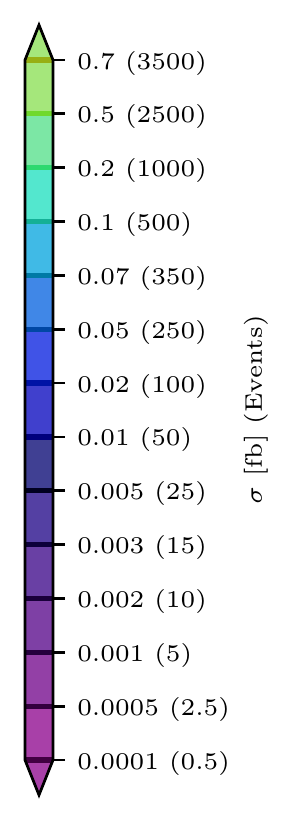}
\vspace{0.18\textheight}
\end{subfigure}
\end{center}
\caption{Cross sections for $e^+e^-\to HHZ$ (left) and
  $e^+e^-\to HH\nu\bar{\nu}$ (right) for $\sqrt{s}=3000\gev$
  for the benchmark planes 1-3 (top to bottom). The
  total allowed regions is given by the solid black line. The color code
  indicates the absolute cross section in fb.
}
\label{fig:xs_HH_3000-I}
\end{figure}

For $\sqrt{s} = 3000 \gev$ we show all benchmark planes~1-3 (top to
bottom) in \reffi{fig:xs_HH_3000-I}. At this high center-of-mass energy
the VBF dominated $HH\nu\bar\nu$ cross section (right) is larger
everywhere than the Higgs-strahlung mediated $HHZ$ cross section (left). The
latter one stays below $\sim 0.07\,\fb$, even for the smallest $\MH$ and
largest $\CBA$ values. Comparing $HH\nu\bar\nu$ and $hH\nu\bar\nu$
cross sections, 
the first process has generically smaller cross sections than the second one,   
basically due to the heavier final state.
The $HH\nu\bar\nu$ cross section can reach up to $\sim 0.15\,\fb$
in benchmark plane~2 for the largest allowed $\CBA$ values, and up to
$\sim 0.3\,\fb$ in benchmark plane~3 for $\MH \sim 300 \gev$ and again
the largest allowed $\CBA$ values. 
While $\laHHH$ stays below $\sim 1$ in the allowed parameter ranges,
$\lahHH$ can reach values larger than $\sim 10$. The size of the
$HH\nu\bar\nu$ cross section largely follows the behavior of $\lahHH$,
indicating a dominating contribution from the VBF type diagram mediated
by an $h$~boson. As will be discussed in \refse{sec:HH-sens} this opens
up the possibility of access to the $\lahHH$ coupling in this part of
the parameter space. On the other hand, the impact of $\laHHH$ remains
small, and thus this coupling seems to remain inaccessible.


\subsection{2HDM type II}
\label{sec:ee-II}

\begin{figure}[hbt!]
\begin{center}
\includegraphics[width=0.4\textwidth]{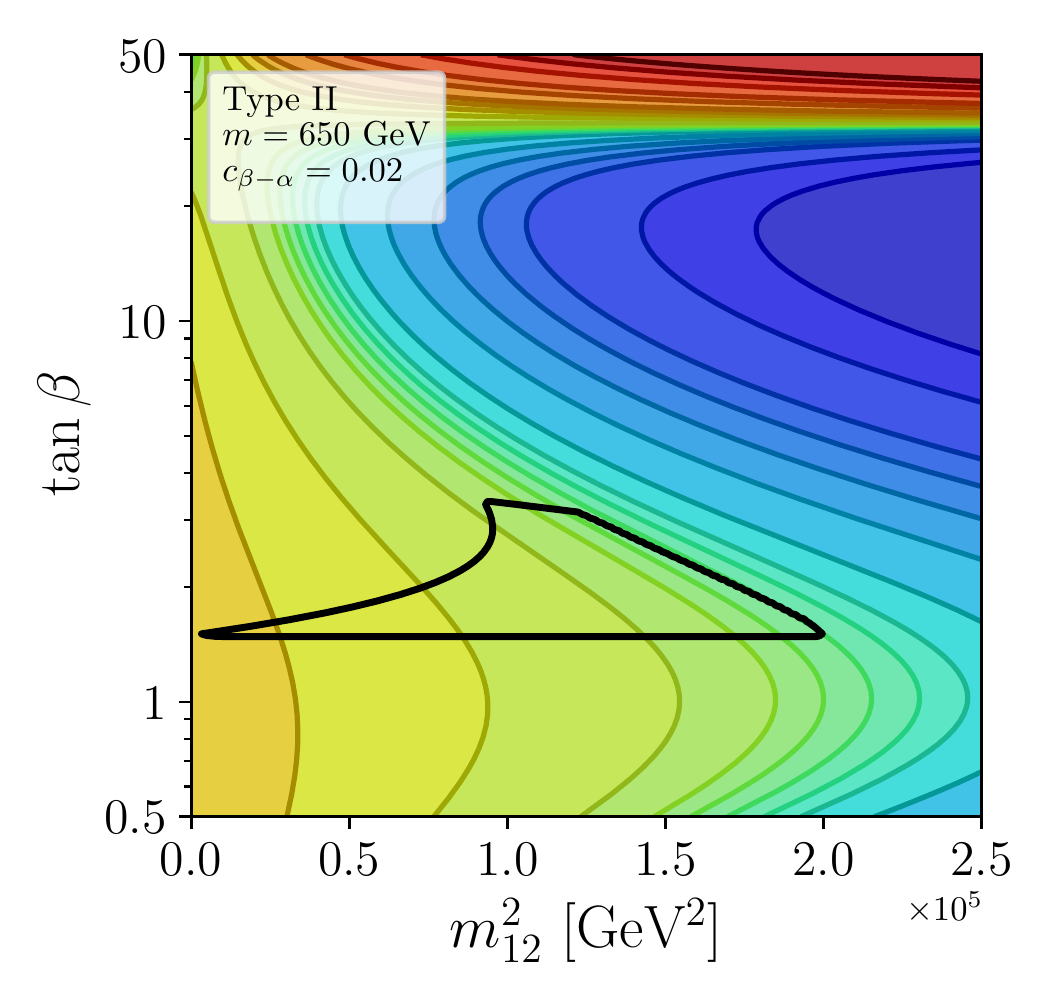}
\includegraphics[width=0.4\textwidth]{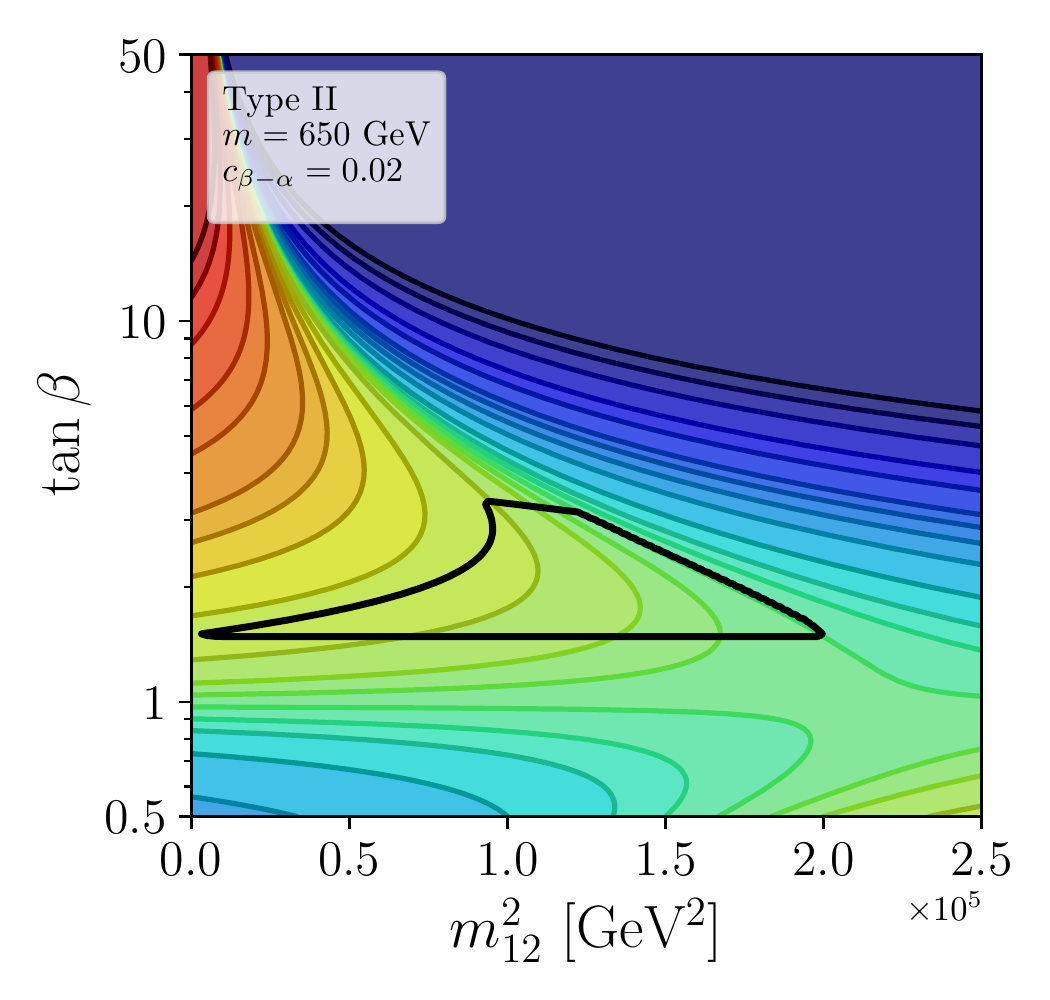}
\includegraphics[width=0.4\textwidth]{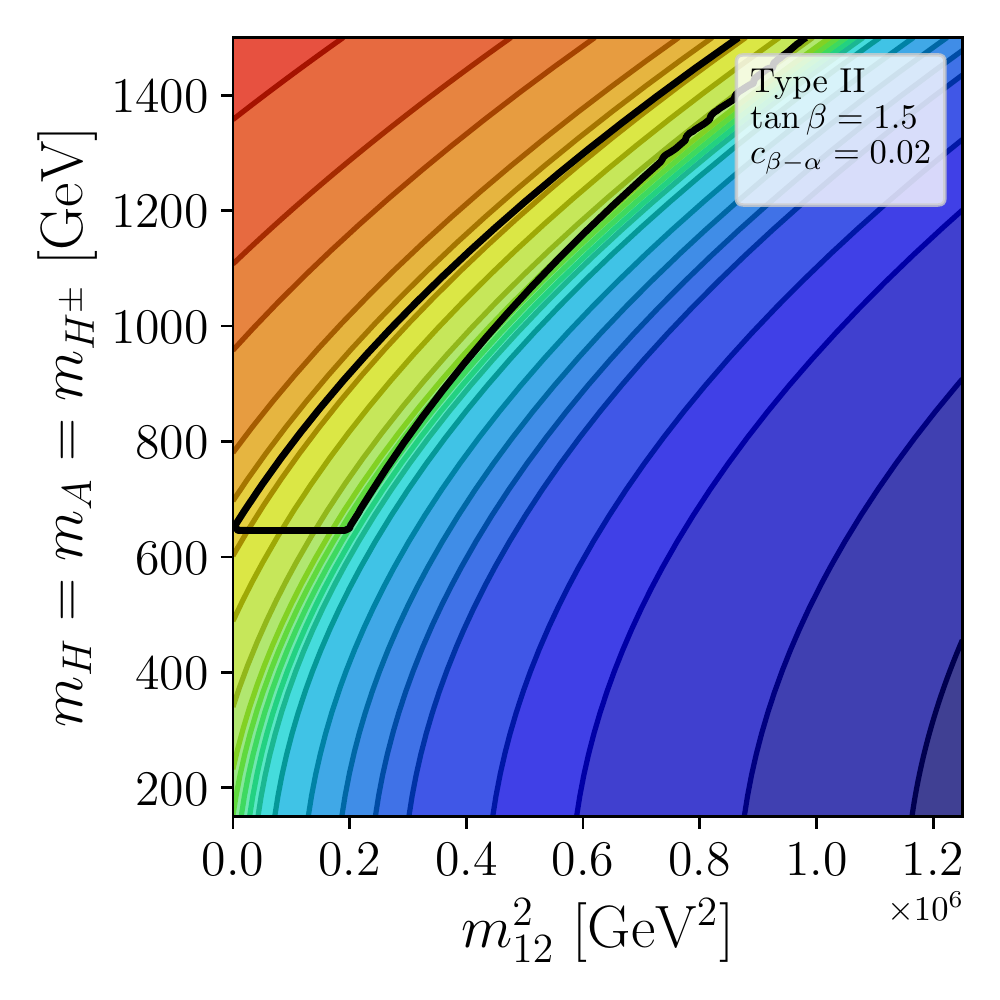}
\includegraphics[width=0.4\textwidth]{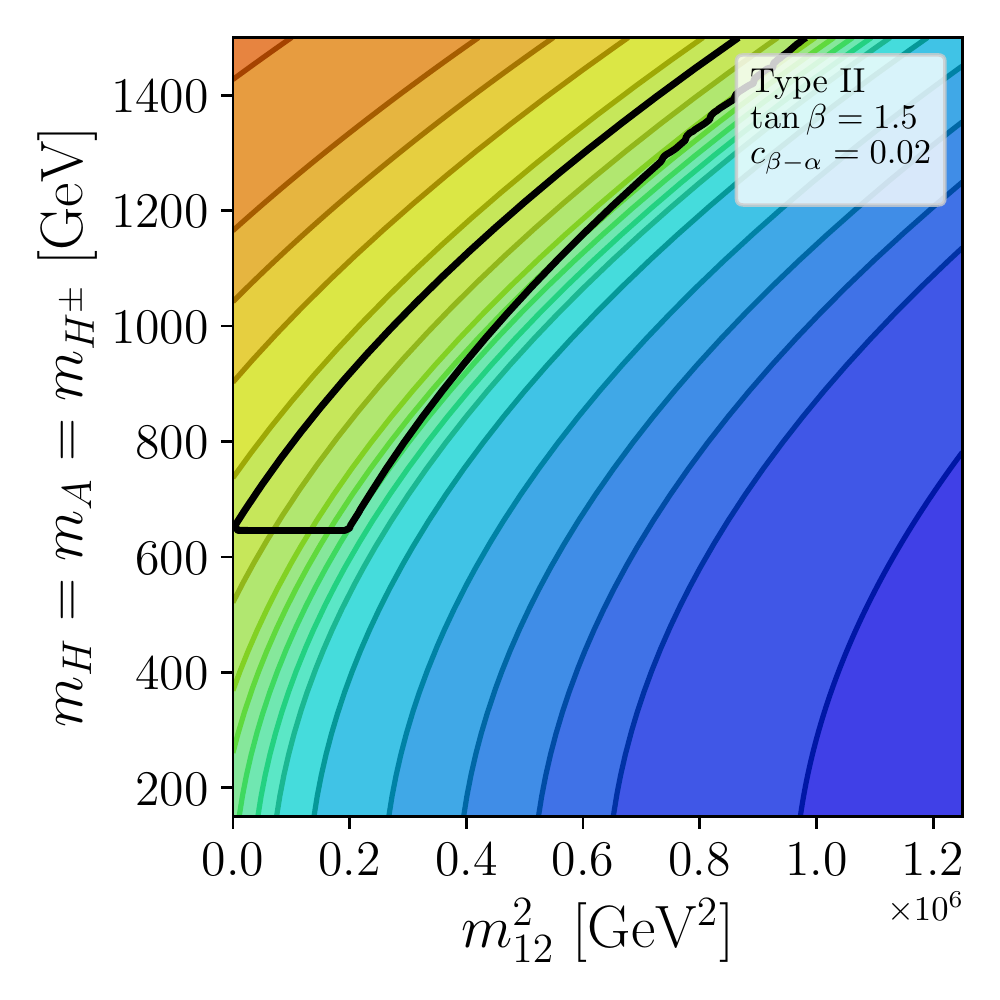}
\includegraphics[width=0.4\textwidth]{h1h2h2_colorbar}
\includegraphics[width=0.4\textwidth]{h2h2h2_colorbar}
\end{center}
\caption{Predictions for
  $\lahHH \simeq \lahAA$
  and $\laHHH \simeq \laHAA$
  in the 2HDM type~II benchmarks planes~4 (upper row) and 5~(lower row).
The regions inside the solid black lines are the parts of the
parameter space that are 
allowed taking all theoretical and experimental constraints (see text).
}
\label{fig:la-II}
\end{figure}

\begin{figure}[t!]
\begin{center}
\begin{subfigure}[b]{0.8\textwidth}
  \includegraphics[width=0.5\textwidth]{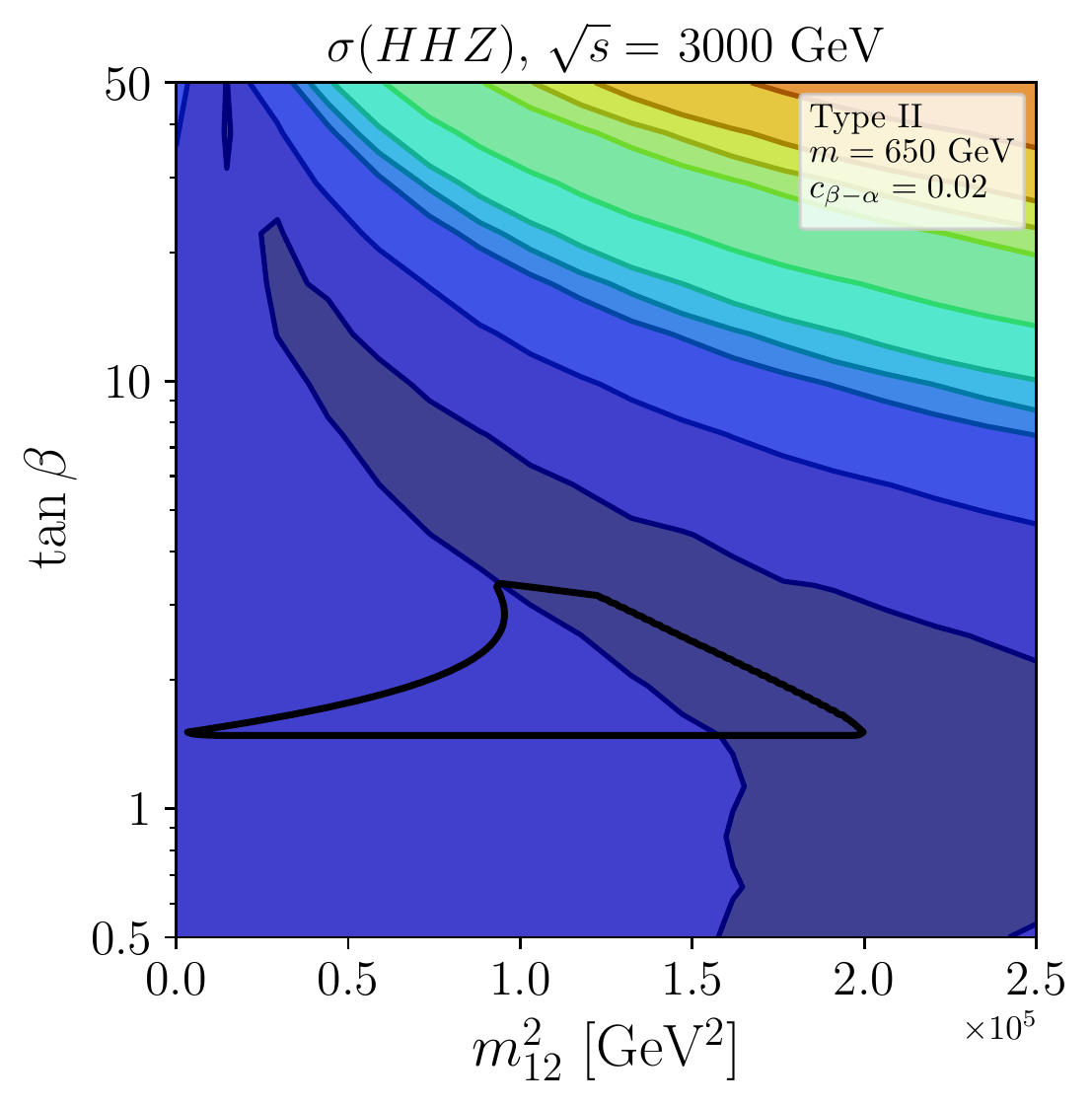}
  \includegraphics[width=0.5\textwidth]{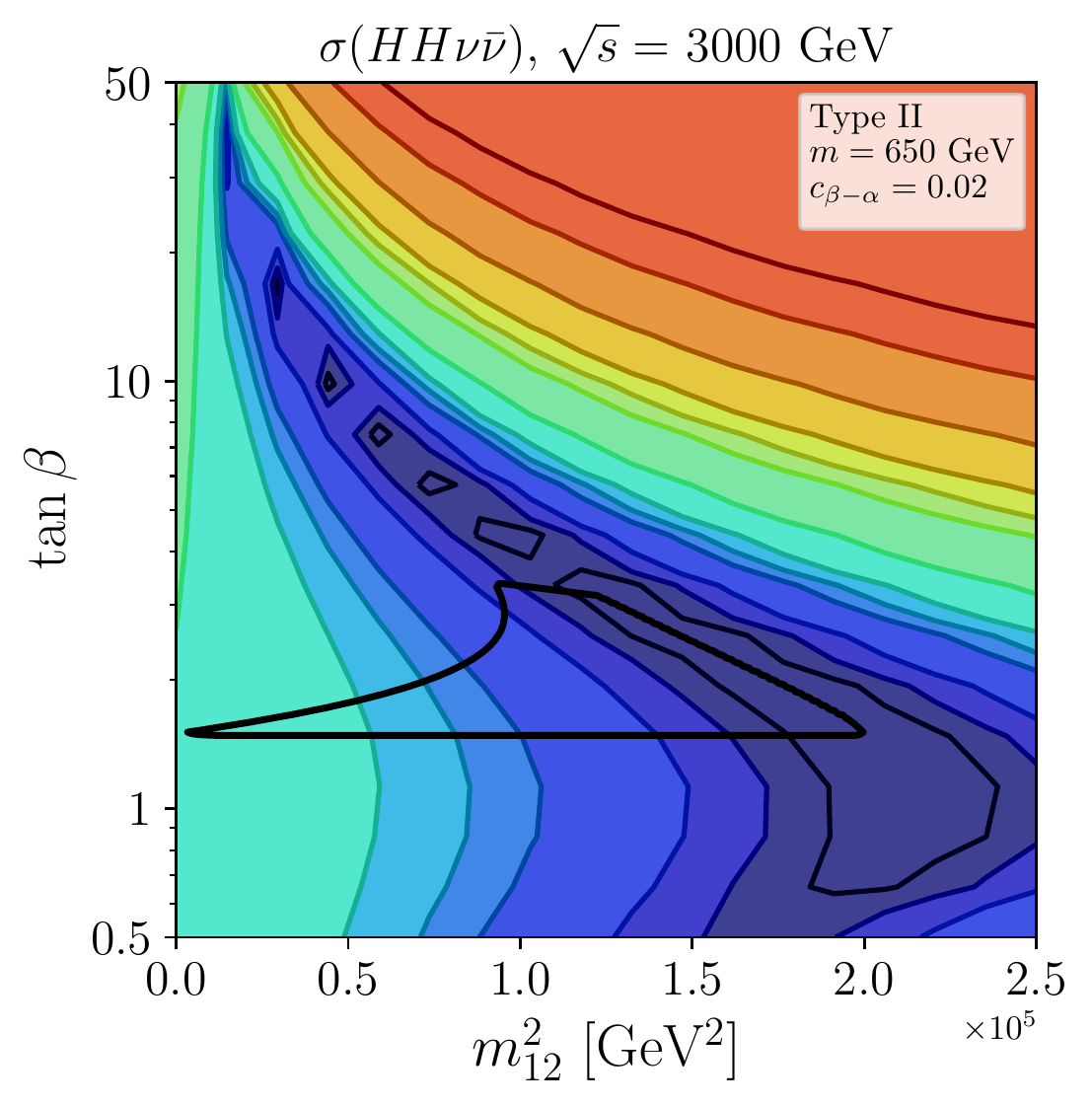}	
  \includegraphics[width=0.5\textwidth]{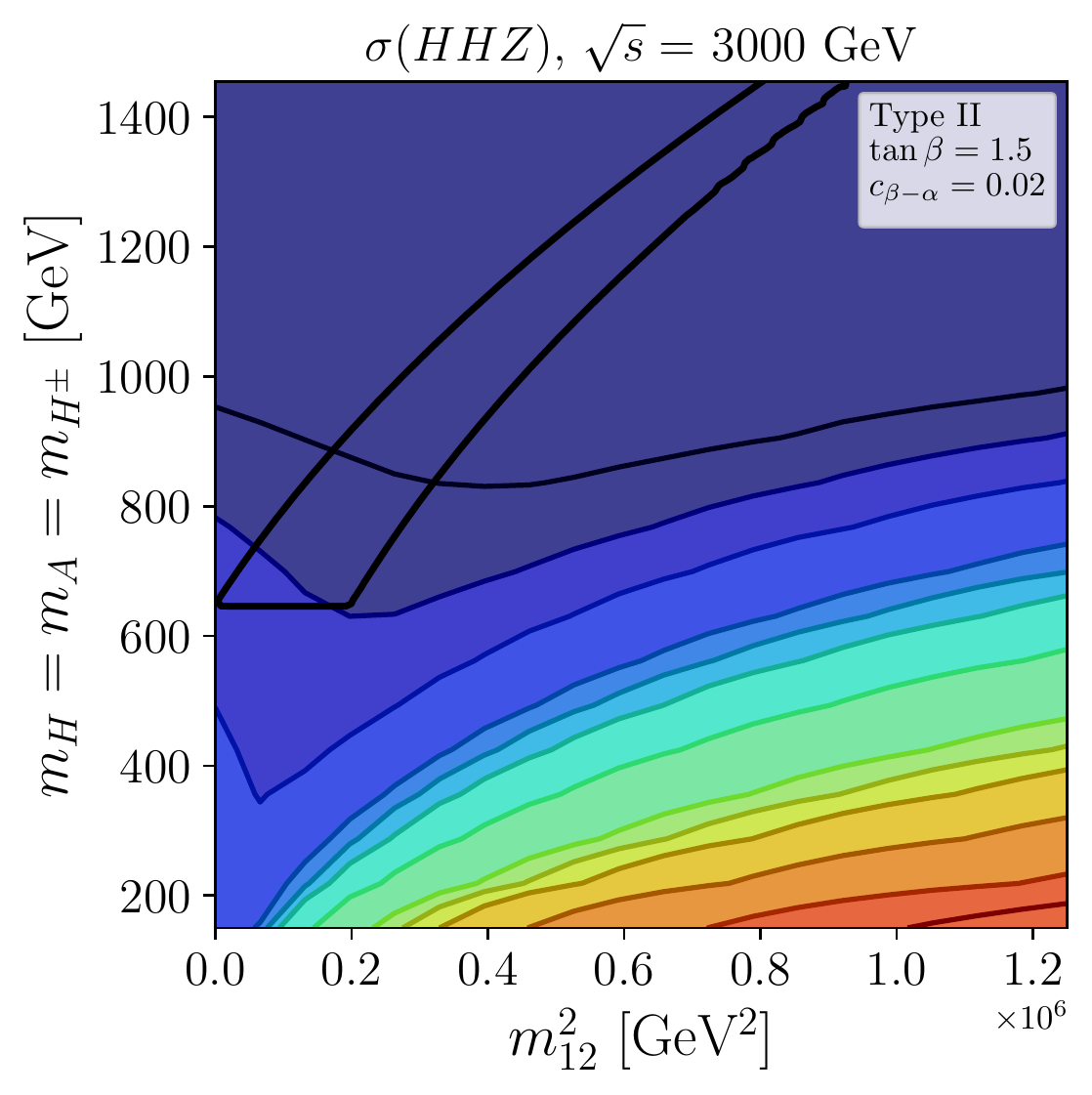}
  \includegraphics[width=0.5\textwidth]{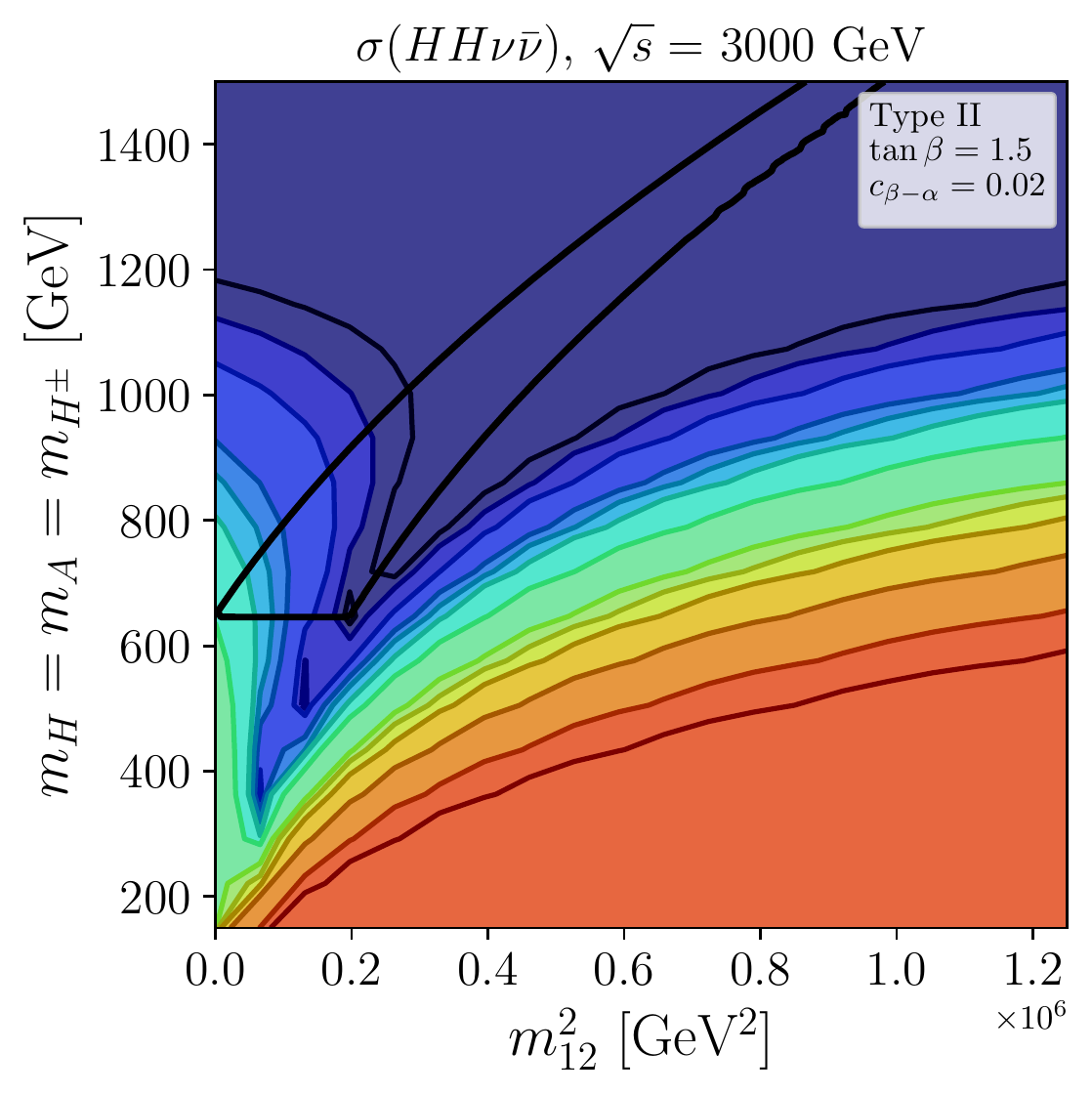}	
\end{subfigure}	
\begin{subfigure}[b]{0.18\textwidth}
\includegraphics[height=0.42\textheight]{colorbar_noSM_3000}
\vspace{0.10\textheight}
\end{subfigure}
\end{center}
\caption{Cross sections for $e^+e^-\to HHZ$ (left) and
  $e^+e^-\to HH\nu\bar{\nu}$ (right)
  at $\sqrt{s}=3000\gev$ for our benchmark planes 4, 5 (top, bottom). The
  total allowed regions is given by the area inside the solid black
  line. The color code 
indicates the absolute cross section in fb.}
\label{fig:xs_HH_3000-II}
\end{figure}

We now turn to the two benchmark planes defined for the 2HDM type~II,
see \refse{sec:planes} (planes~4 and~5).
Other planes were found to not exhibit any relevant variation. 
Furthermore, most of the plots also in these two planes
showing the results for the 
various production cross sections as given in \refeqs{eeZhh} and
(\ref{eenunuhh}) and various energies and integrated luminosities
according to \refta{tab:ee}, are also omitted, since the results do not
shows any relevant variation either.  
Within the allowed parameter space,  taking
into account all theoretical and experimental constraints as given in
\refse{sec:constraints},  we find numerical results for the cross
sections of $e^+e^- \to hhZ$ and $e^+e^- \to hh\nu\bar\nu$,
at all the studied energies given in \refta{tab:ee},
are in general very close to the respective SM predictions,  with a
deviation of at most $\sim 5\%$.
On the other hand,  the $hH$ production channels yield cross
sections below $0.01\,\fb$ with no relevant variation in the allowed
parameter space, for all center-of-mass energies.
The same applies to the heavy di-Higgs production (where again
the results for $HH$ are very similar to the results for $AA$
production) for $\sqrt{s} = 500, 1000, 1500 \gev$.  
Only for $e^+e^- \to HHZ$ and $e^+e^- \to HH\nu\bar\nu$ at the
highest energy, $\sqrt{s} = 3000 \gev$,  some relevant
production cross sections exceeding $\sim 0.1\,\fb$
within the allowed parameter space are obtained.
Thus, we choose this $\sqrt{s} = 3000 \gev$ center-of-mass
energy in our discussion below.

As we have said previously,  in the two channels for $HH$
production, $HHZ$ and $HH\nu \bar \nu$, the two involved triple
couplings are $\lahHH$ and $\laHHH$, and they  enter via the diagrams
with an intermediate off-shell $h$ and $H,$  respectively.  Thus, the
size of these two couplings will be the most relevant
issues for the $HH$ production rates.
We show the results for these two triple Higgs couplings (with 
$\lahHH \simeq \lahAA$ and $\laHHH \simeq \laHAA$) in \reffi{fig:la-II},
for the two chosen planes,  4 and 5. 
As in the previous subsection in each plane
we indicate with a solid black line the part of the parameter space
(inside the black line) 
that is allowed taking all theoretical and experimental constraints as
given in \refse{sec:constraints}. 
As can be seen in these plots,  within this allowed area,  the two
triple Higgs couplings roughly fulfill $\lahHH \sim 2\,\laHHH$.
Besides, the largest values of $\lahHH \sim 6-8$
are reached for smaller $\msq$ and larger $\MH = \MA = \MHp$ values. 

Finally, in \reffi{fig:xs_HH_3000-II} we present the predictions
for the $HHZ$ (left) and $HH\nu\bar\nu$ cross sections (right) in the
benchmark planes~4 and~5 for $\sqrt{s} = 3000 \gev$. The color code
indicates the total cross section in fb. While the $HHZ$ cross
section can reach values larger then $10\,\fb$ in the two benchmark
planes,  however,   within the allowed parameter space,  the cross
sections do not reach  
values larger than $0.015\fb$ for $\msq \lesssim 150000\gev^2$ in 
both benchmark planes.  For larger values of $\msq$ the cross
sections become even smaller. 

In the case of the $HH\nu\bar{\nu}$ channel, 
the cross sections can reach large values $\gsim 10\,\fb$ in the two benchmark
planes.  However,  focusing again on the allowed parameter region
smaller values,  slightly below $0.2\fb$,  are found.   These rates are
for the lower allowed values of $\msq$,  in both  
benchmark planes~4 and 5. The regions where the 
cross section becomes larger correspond to the regions where 
$\lahHH$ reaches its largest ``allowed'' values around $\sim6-8$.
This indicates that the total cross section receives a relevant
contribution from the diagrams containing $\lahHH$, in particular from
the VBF-like contributions, which dominate at this large center-of-mass energy. 
On the other hand, over large parts of the allowed parameter space the
cross section does not exceed $0.05\,\fb$, and no access to triple Higgs
couplings can be expected. Because of these small cross sections for the
2HDM type~II, even in the best case of $\sqrt{s} = 3000 \gev$, we
will study only one ``best-case'' scenario in the following section.



\section{Sensitivity to triple Higgs couplings}
\label{sec:sensitivity}

\subsection{Benchmark points}
\label{sec:bench}

After having explored the total cross sections  at the future $e^+e^-$
colliders for the several channels of two Higgs bosons production in
the 2HDM of type~I and~II,  we now turn to the potential
sensitivity to the involved triple Higgs couplings.  As was
discussed above,  each  channel involves different triple Higgs
couplings.
The processes with two light Higgs bosons $hh$ in the final state
involve $\lahhh$ (with $\kala=\lahhh/\laSM$) and $\lahhH$.  The
processes with  $hH$ involve $\lahhH$ and  
$\lahHH$.  Those with $HH$ involve $\lahHH$ and $\laHHH$, and those with
$AA$ involve  $\lahAA$ and $\laHAA$.   
Since the predictions for $HH$ are very similar to those of $AA$ 
we will focus here just on cases of $hh$, $hH$ and $HH$.   
As we anticipated before, the total cross section of these 
processes is not sufficient to infer the effects of the triple Higgs
couplings. In this work we propose to access to these couplings
through another observable: the differential cross section with
respect to the invariant mass of the final Higgs pair $h_ih_j$, 
which we will study in the following sections.

\begin{table}[h!]
\begin{center}
\begin{tabular}{|c|c|c|c|c|c|c|c|c|c|c|c|}
\hline 
Point & Type & $m$  & $\tan\beta$ & $c_{\beta-\alpha}$ & $\msq$  & $\kala$ & $\lahhH$ & $\lahHH$ & $\laHHH$ & $\Gamma_H$  & $\Gamma_A$ \tabularnewline
\hline \hline
BP1 & I & 300  & 10 & 0.25 & \refeq{eq:m12special} & 1.1 
& -0.2 & 2 & 0.3 & 0.84 & 0.18 \tabularnewline
\hline 
BP2 & I & 500  & 7.5 & 0.1 & 32000 & 0.8 & 0.3 & 2 & 0.6 & 0.88 & 0.71 \tabularnewline
\hline 
BP3 & I & 600  & 10 & 0.2 & \refeq{eq:m12special}  & 1.0 & -0.5 & 6 & 0.6 & 5.1 & 2.6 \tabularnewline
\hline 
BP4 & I & 1000  & 8.5 & 0.08 & \refeq{eq:m12special}  & 0.5 & 1.1 & 6 & -0.2 & 5.8 & 2.6\tabularnewline
\hline \hline
BP5 & II & 650  & 1.5 & 0.02 & 10000 & 1 & -0.1 & 7 & 3 & 10.1 & 14.1\tabularnewline
\hline
\end{tabular}
\caption{Benchmark points of the 2HDM selected for the study of the
  sensitivity to the triple Higgs couplings (masses and widths are given
  in GeV).} 
\label{tab:BP}
\end{center}
\end{table}

\begin{figure}[p!]
\begin{center}
	\includegraphics[width=0.48\textwidth]{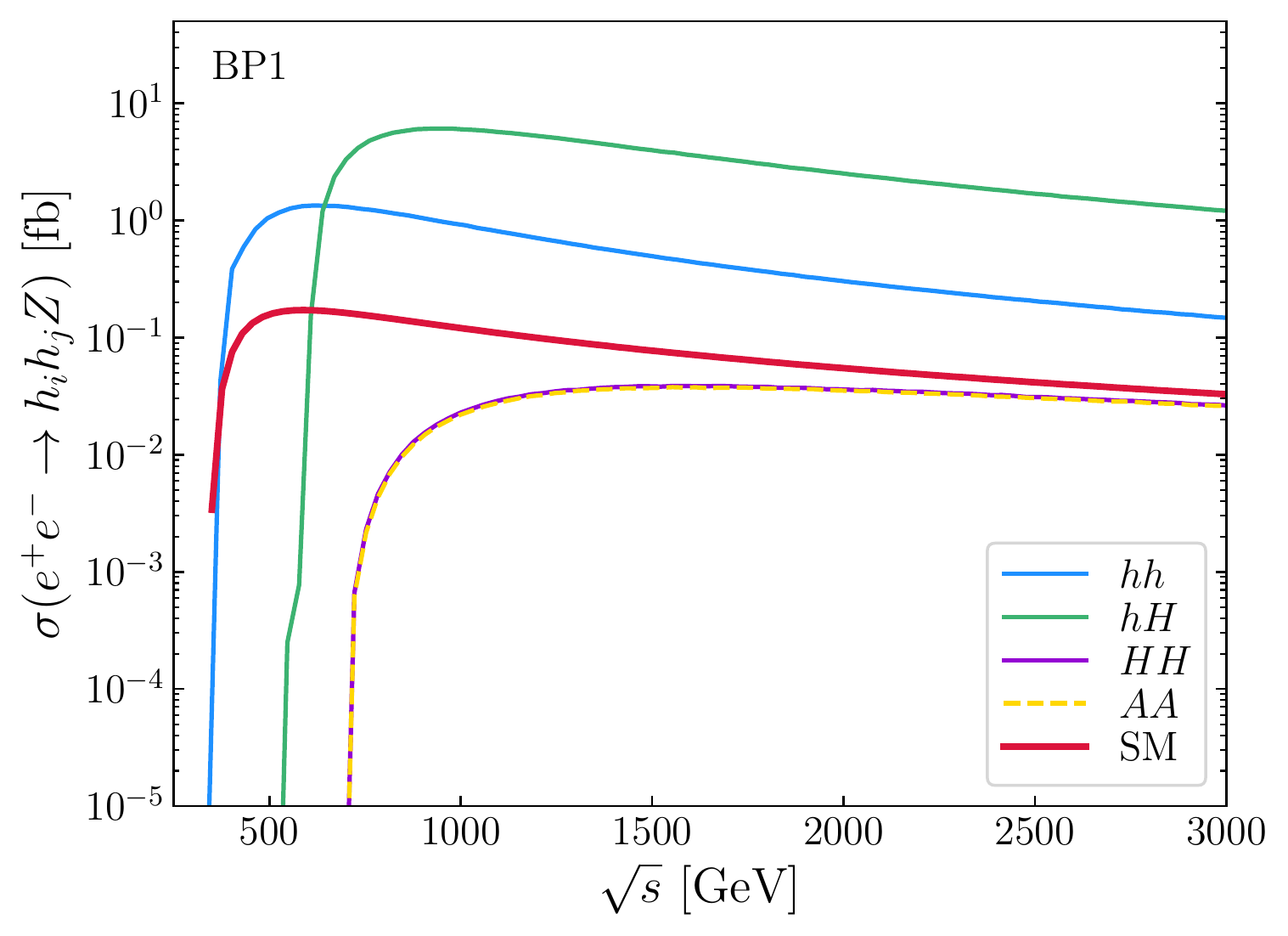}
	\includegraphics[width=0.48\textwidth]{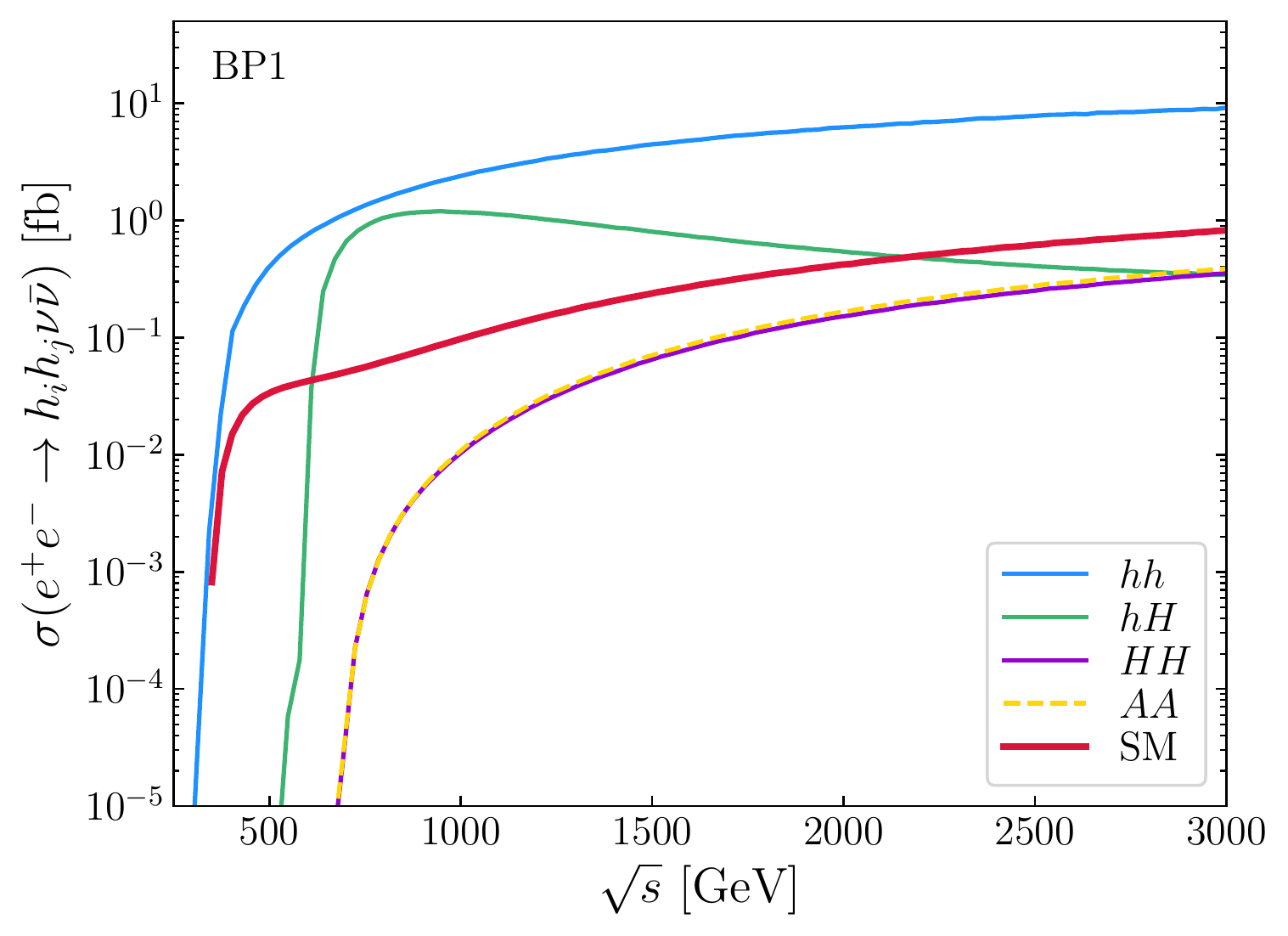}	
	
	\includegraphics[width=0.48\textwidth]{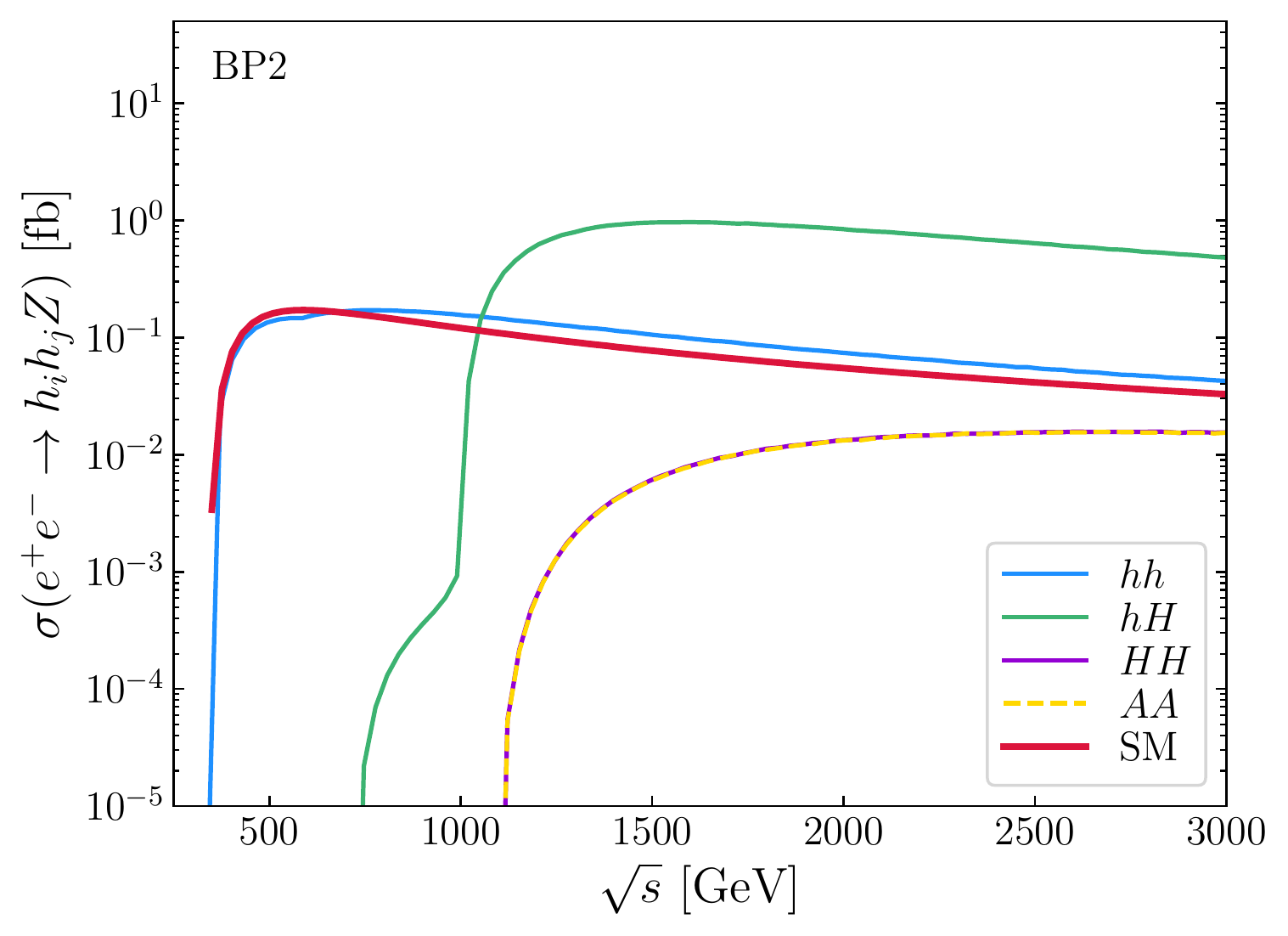}
	\includegraphics[width=0.48\textwidth]{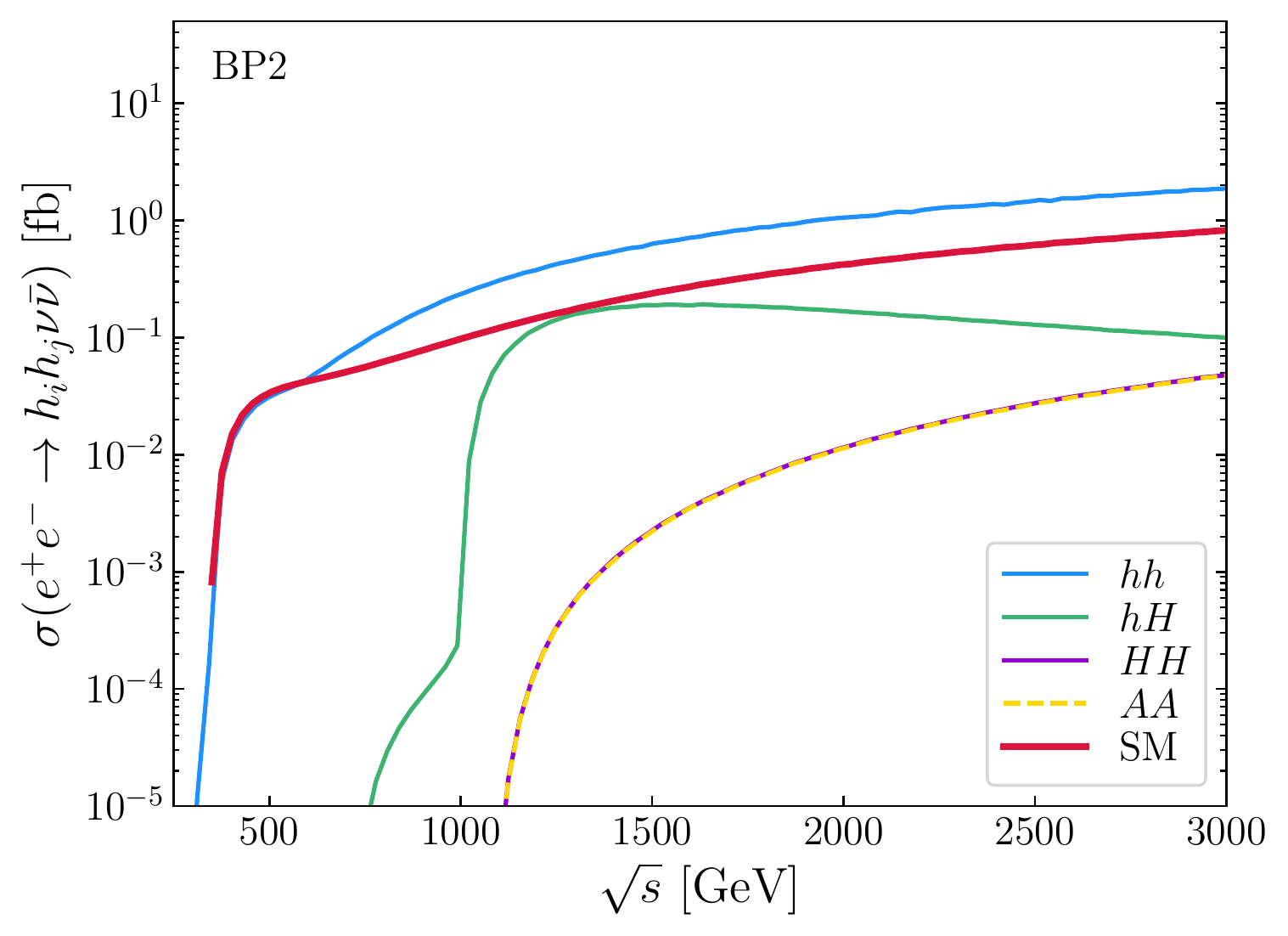}	
	
	\includegraphics[width=0.48\textwidth]{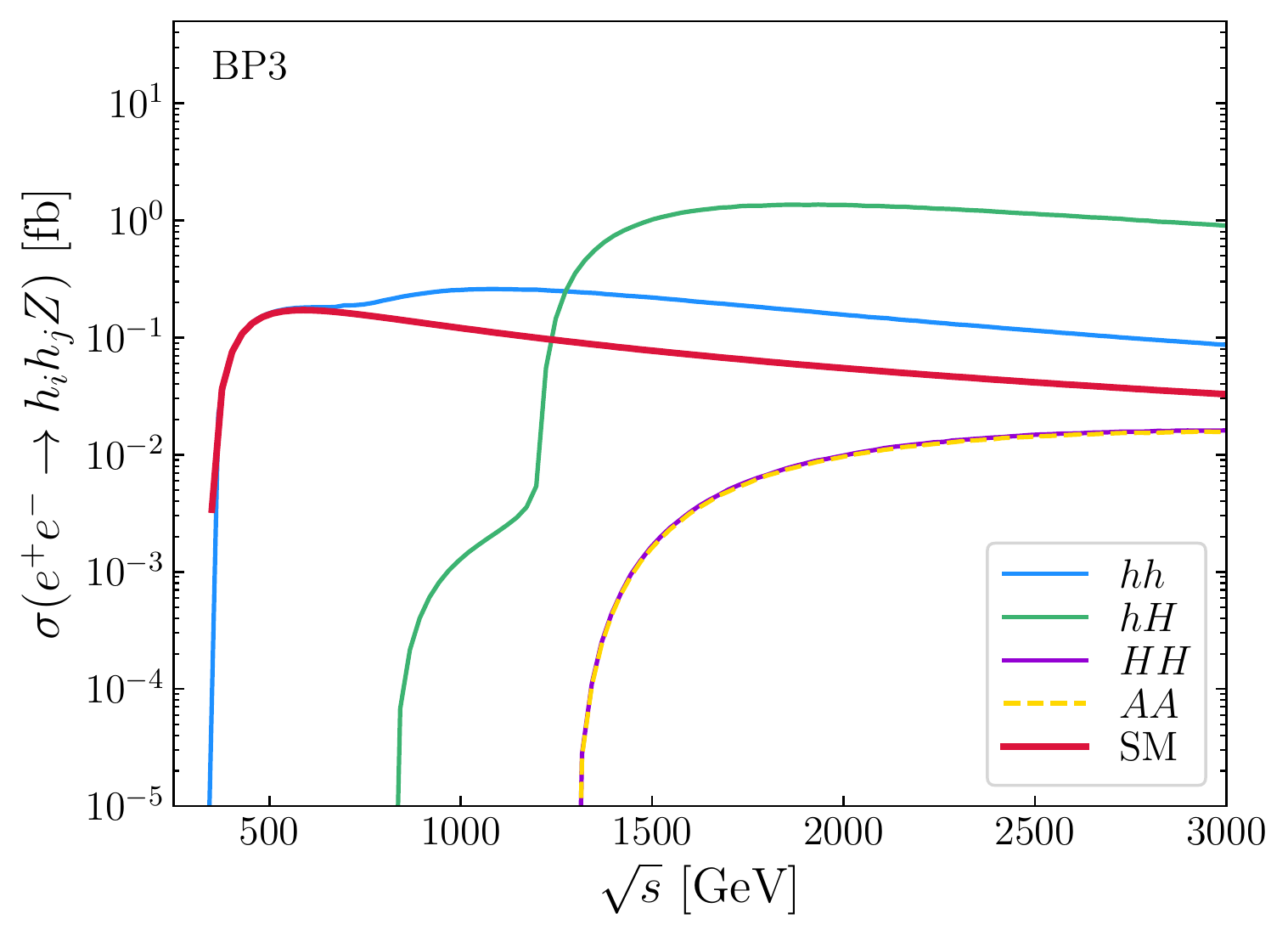}
	\includegraphics[width=0.48\textwidth]{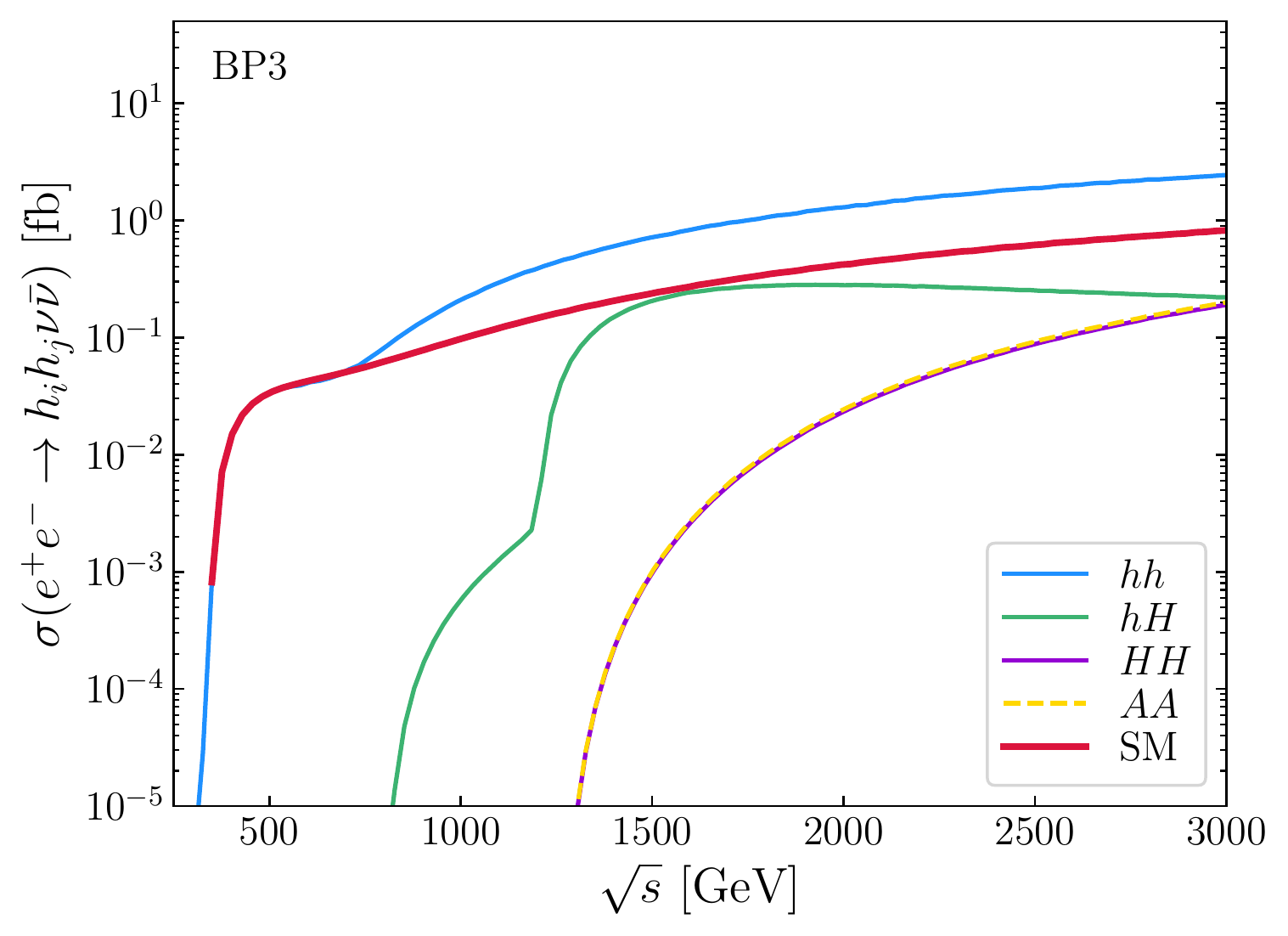}	
	
	\includegraphics[width=0.48\textwidth]{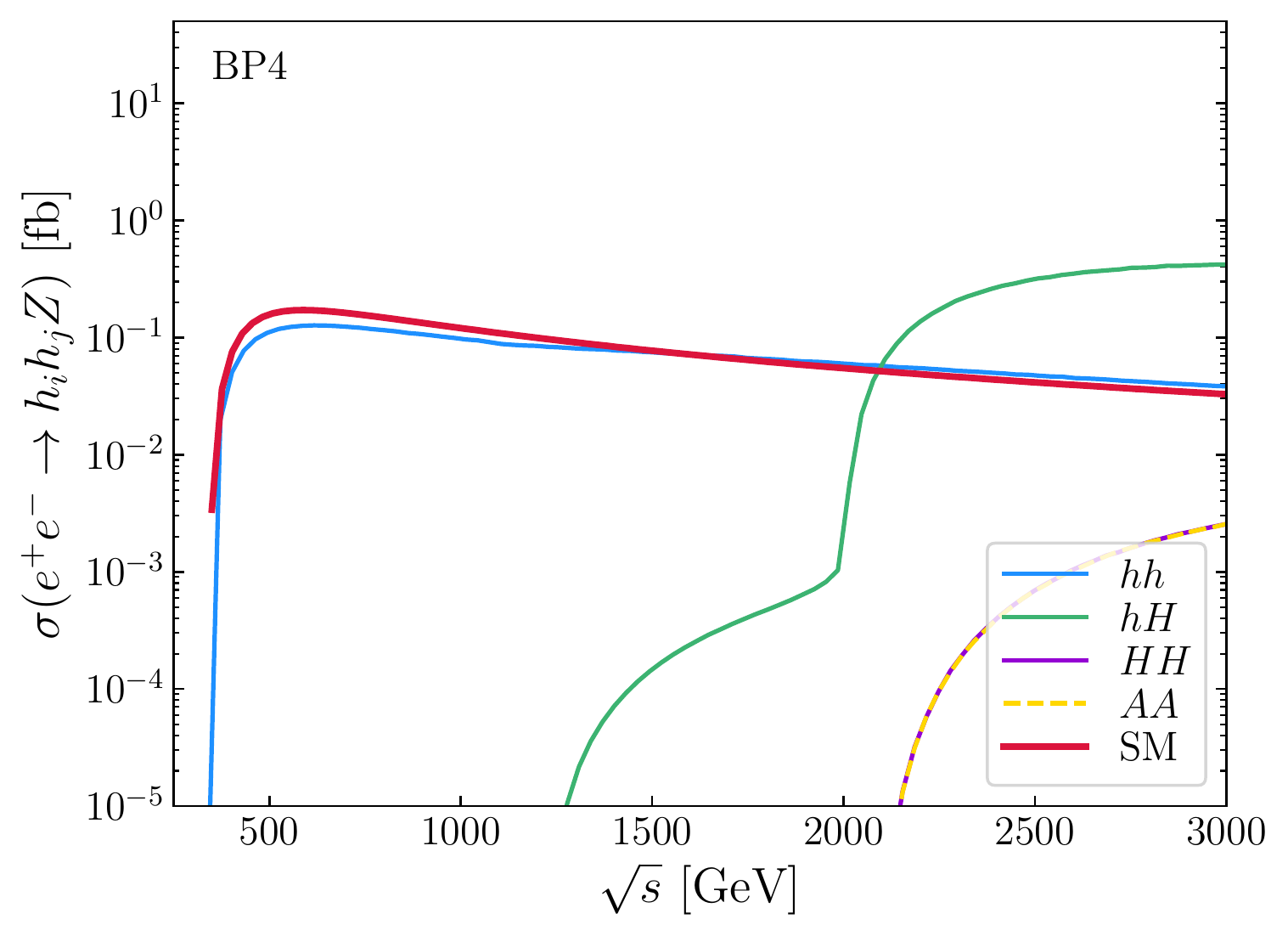}
	\includegraphics[width=0.48\textwidth]{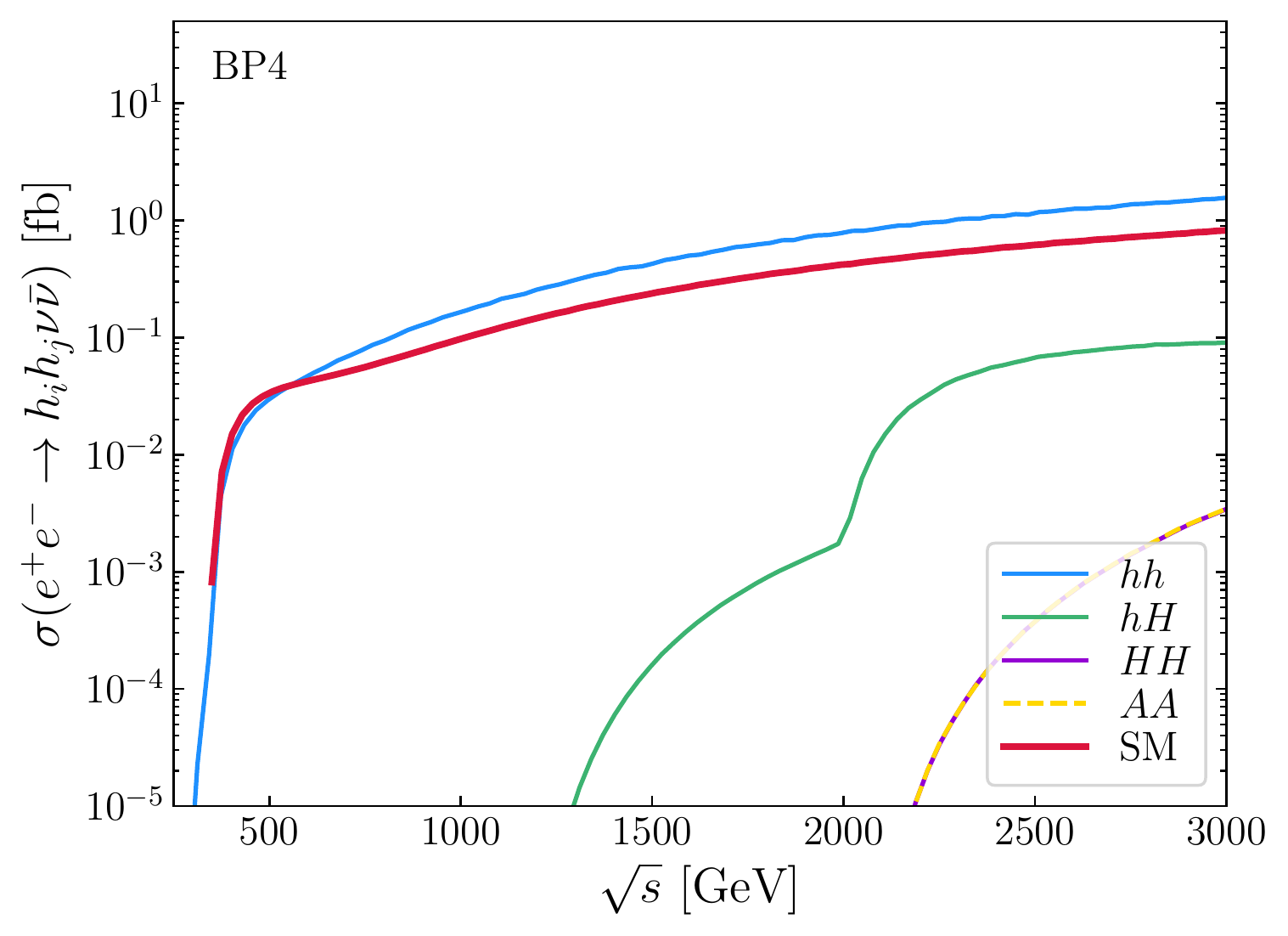}	
\end{center}
\caption{Cross sections as a function of the center-of-mass energy
  $\sqrt{s}$ for the processes $e^+e^-\to h_ih_jZ$ (left) and
  $e^+e^-\to h_ih_j\nu\bar{\nu}$ (right) for BP1, BP2, BP3 and BP4 (type I). } 
\label{fig:BP1-4-sqrt}
\end{figure}

\begin{figure}[htb!]
\begin{center}
	\includegraphics[width=0.48\textwidth]{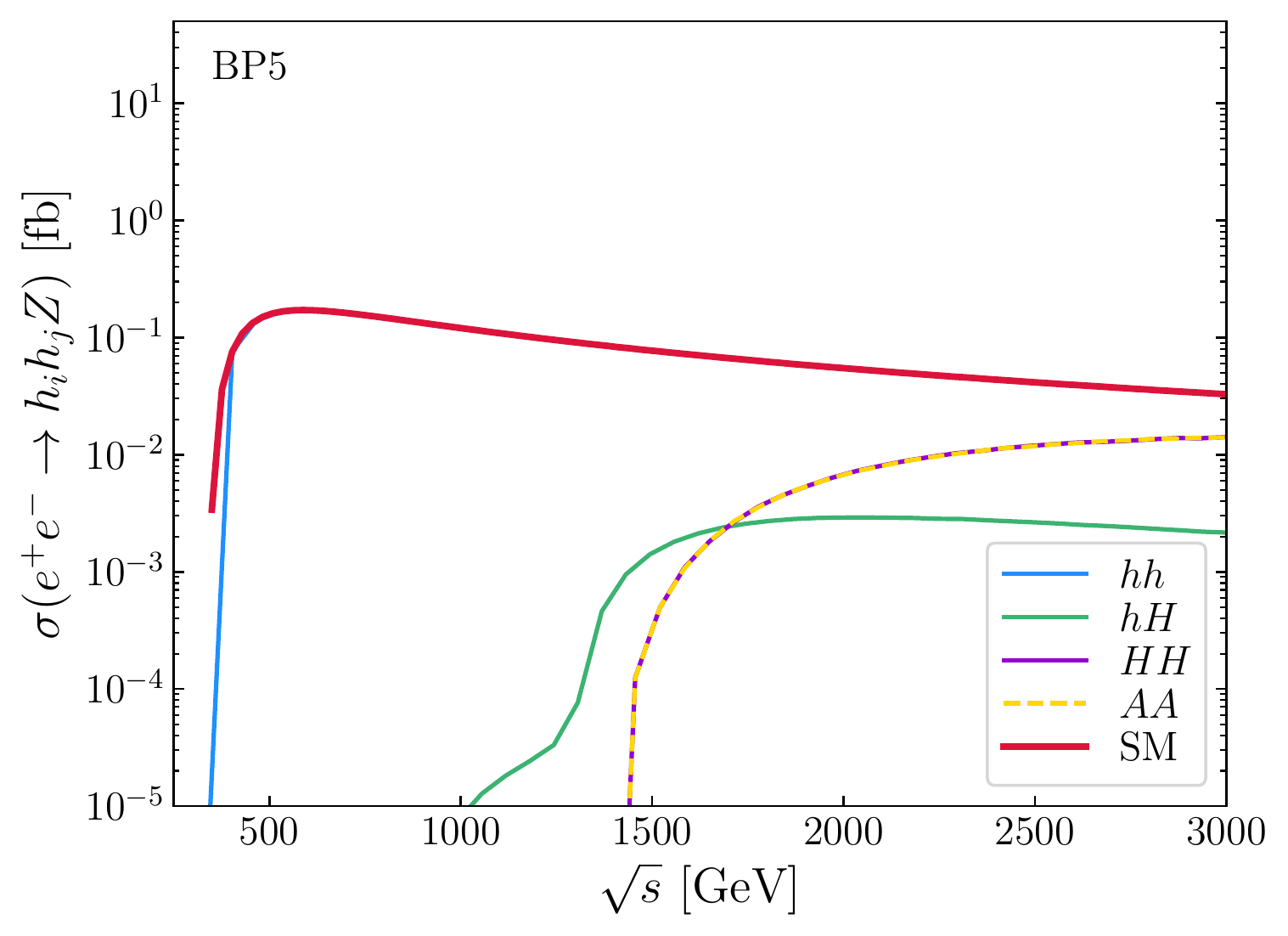}
	\includegraphics[width=0.48\textwidth]{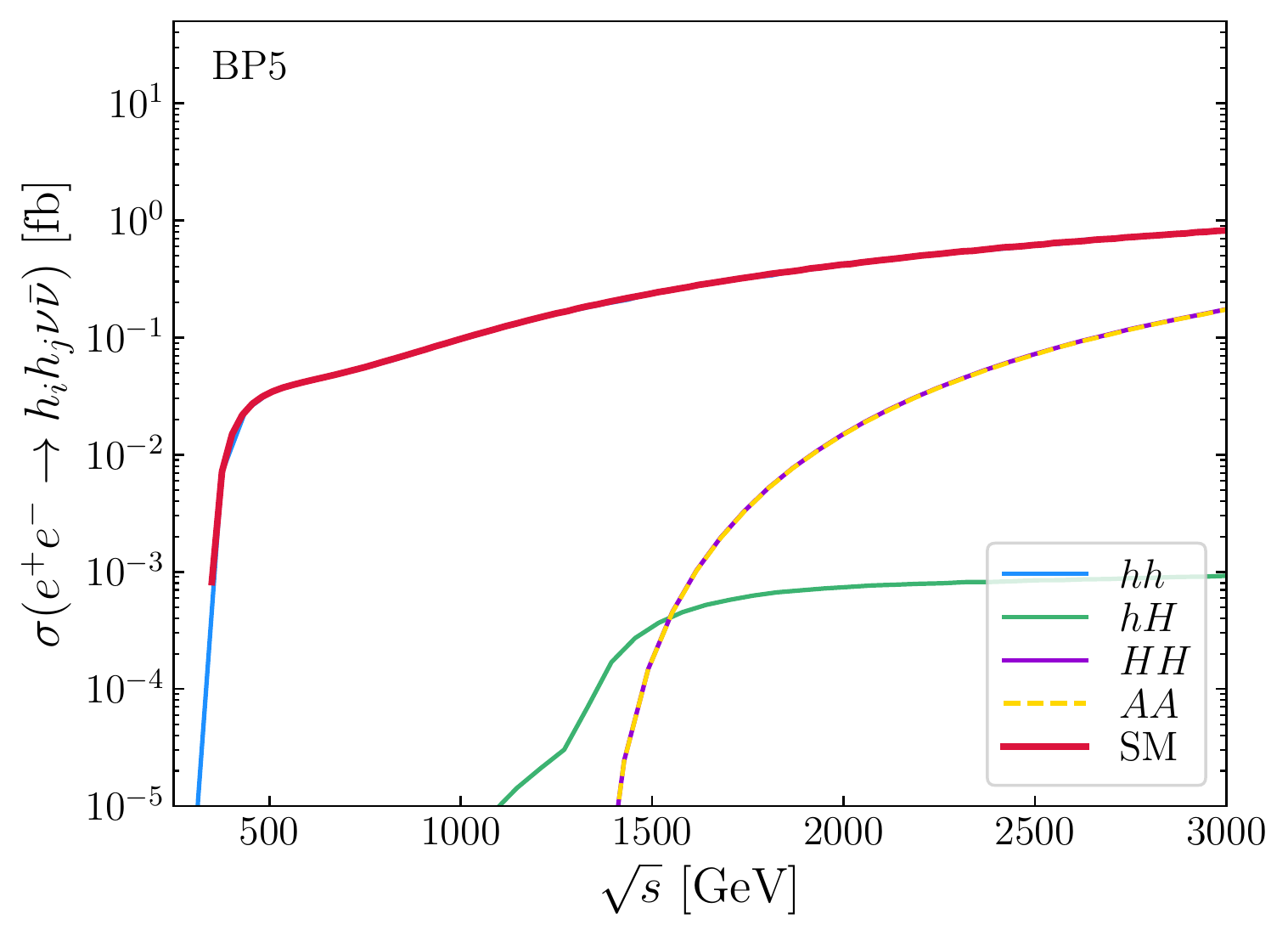}	
\end{center}
\caption{Cross sections as a function of the center-of-mass energy
  $\sqrt{s}$ for the processes $e^+e^-\to h_ih_jZ$ (left) and
  $e^+e^-\to h_ih_j\nu\bar{\nu}$ (right) for BP5 (type II). } 
\label{fig:BP5-sqrt}
\end{figure}

In order to  study the sensitivity to the various $\lambda_{h_ih_jh_k}$
we have chosen specific benchmark points (BPs) within the 2HDM,
which are in agreement with all
present data. These BPs have been selected with
simplified input Higgs mass parameters $m_H=m_A=m_{H^{\pm}}$ covering
typical values from $300 \gev$ to $1000 \gev$.
This allows to study
the different reach to the intermediate Higgs bosons and their
corresponding triple Higgs couplings at the various
future $e^+e^-$ colliders. The mass of the light Higgs in all these BPs
is fixed to $m_h=125 \gev$. 
The other parameters have been fixed to values that are
allowed by present theoretical and experimental constraints
such that  sizeable values of the triple
Higgs couplings are realized.
The five selected  
BPs and their input parameters are specified in \refta{tab:BP}.  The
first four,  BP1, BP2, BP3 and BP4 are for type~I and the fifth one, BP5,
is for type~II.  We have also included in this table the corresponding
output values of the relevant triple Higgs couplings and the relevant
heavy Higgs boson widths.  It should be noted that  
in all these points the largest triple Higgs coupling is 
$\lahHH$, so that one might expect the largest sensitivity to
this coupling. However, this is not the case, as will be discussed
in this section, 
since the other input parameters, especially the mass of the heavy
Higgs bosons and the value of $c_{\beta-\alpha}$,  also play a very
relevant role and will enter significantly in the conclusion to the
sensitivity to the triple Higgs couplings.   

To complete the description of these five BPs we show the
predictions of the total cross section for the various channels  as a
function of the $e^+e^-$ collider energy  in \reffi{fig:BP1-4-sqrt},
for BP1 through BP4,  and in  \reffi{fig:BP5-sqrt} for BP5.  As a
general remark,  one can see that a similar pattern of the cross section
behavior with energy is found here for these five selected BPs as
for the point shown in \reffi{fig:XSsqrts}, see the discussion in
\refse{sec:eehh}. Generically,  at high energies  
the rates for $h_ih_jZ$ decrease with energy and for $h_ih_j\nu \bar \nu$ 
increase with energy.  With the exception of $hH\nu \bar \nu$ that first
decreases with energy (as seen in our figures) up to around $3 \tev$,  and
then reaches a nearly flat (but slightly increasing)
behavior at very high energies above $3 \tev$ (not
shown in our plots). We have checked this flat behavior with
energy also at the WBF subprocess level,  i.e.,  the cross section
$\sigma(WW \to hH)$ tends to a constant value at very high $\sqrt{s}$.
This flat behavior also happens in the $\sigma(WW \to hh)$ case,
similarly to the SM case,  but it is reached at lower energies in the
$hh$ channel than in the $hH$ one.   

We find the following hierarchies in the size of the cross sections: 
{\it (i)} the  $hhZ$ channel provides the largest
$\sigma(h_ih_jZ)$ in the low 
energy region from threshold up to $\sim 600 \gev$ for BP1,
up to $\sim 1000 \gev$ for BP2, up to $\sim 1200 \gev$ for
BP3, up to $\sim 2000 \gev$ for BP4, and in the full 
energy range for BP5; {\it (ii)} $hHZ$ provides the largest
$\sigma(h_i h_jZ)$ 
at  energies above those mentioned values;
{\it (iii)} $hh \nu \bar \nu$ provides the largest  $\sigma(h_ih_j
\nu \bar \nu)$ for  all the studied BPs and at all energies.  However,
as we will discuss below,  just  looking for the largest cross
sections is not sufficient to reach 
the best sensitivity to the triple Higgs couplings.

Our strategy proposed here to explore the accessibility to the triple
Higgs-boson couplings is outlined as follows.  It is obvious
that for this analysis the relevant quantity is not that of the total
cross section for production of $h_ih_j$ pairs, but another quantity
where the effect  from the triple Higgs coupling is enhanced. There are
several of those quantities, but we choose here what we believe is the
most promising one: the
cross section  distribution with respect to the invariant mass of the
produced Higgs boson pair. The main idea with this distribution is 
to look for specific windows of this invariant mass where the 2HDM
rates will be the most sensitive ones to the triple Higgs coupling involved.
Obviously, the extreme case with maximum sensitivity will happen when
the invariant mass of the produced $h_ih_j$ pair, $m_ {h_ih_j}$ is close
to the intermediate Higgs boson mass, $m_{h_k}$,  since  in that case
the diagram carrying the $\lambda_{h_k h_i h_j}$ coupling dominates over
the others due to the resonant behavior of this diagram (see
\reffi{fig:diagrams}).  Here it should be noted that we are not
assuming the 
on-shell production of this intermediate Higgs boson $h_k$.  
As we have checked explicitly, it is not a good approximation to
estimate the total rates for $h_ih_j \,X$ production by using the naive
NWA, i.e., by calculating the $h_k \,X$  production rates times the
$\br(h_k \to h_i h_j)$. The reason for this bad approximation is
the fact that outside the window of the $m_{h_ih_j}$ invariant mass
that is close to $m_{h_k}$ the rest of diagrams (not carrying
$\lambda_{h_k h_i h_j}$) 
contribute very significantly, as will be shown in the following.
In the next subsections we will discuss separately the results
for the cases of $hh$ and $hH$,~$HH$.


\subsection{Sensitivity to triple Higgs couplings in
  \boldmath{$hh$} production}

In this section we analyze the sensitivity to triple Higgs
couplings in the $hh$ production channels.  We
present the results of the cross section distributions as a
function of the invariant mass of the $hh$ pair,  $m_{hh}$,  for all
the collider energies considered in this work.  First we comment on the
distributions for the points selected in 2HDM-type I and latter we
present the results for the point selected in 2HDM-type II.   

\begin{figure}[t!]
	\begin{center}
\includegraphics[width=0.48\textwidth]{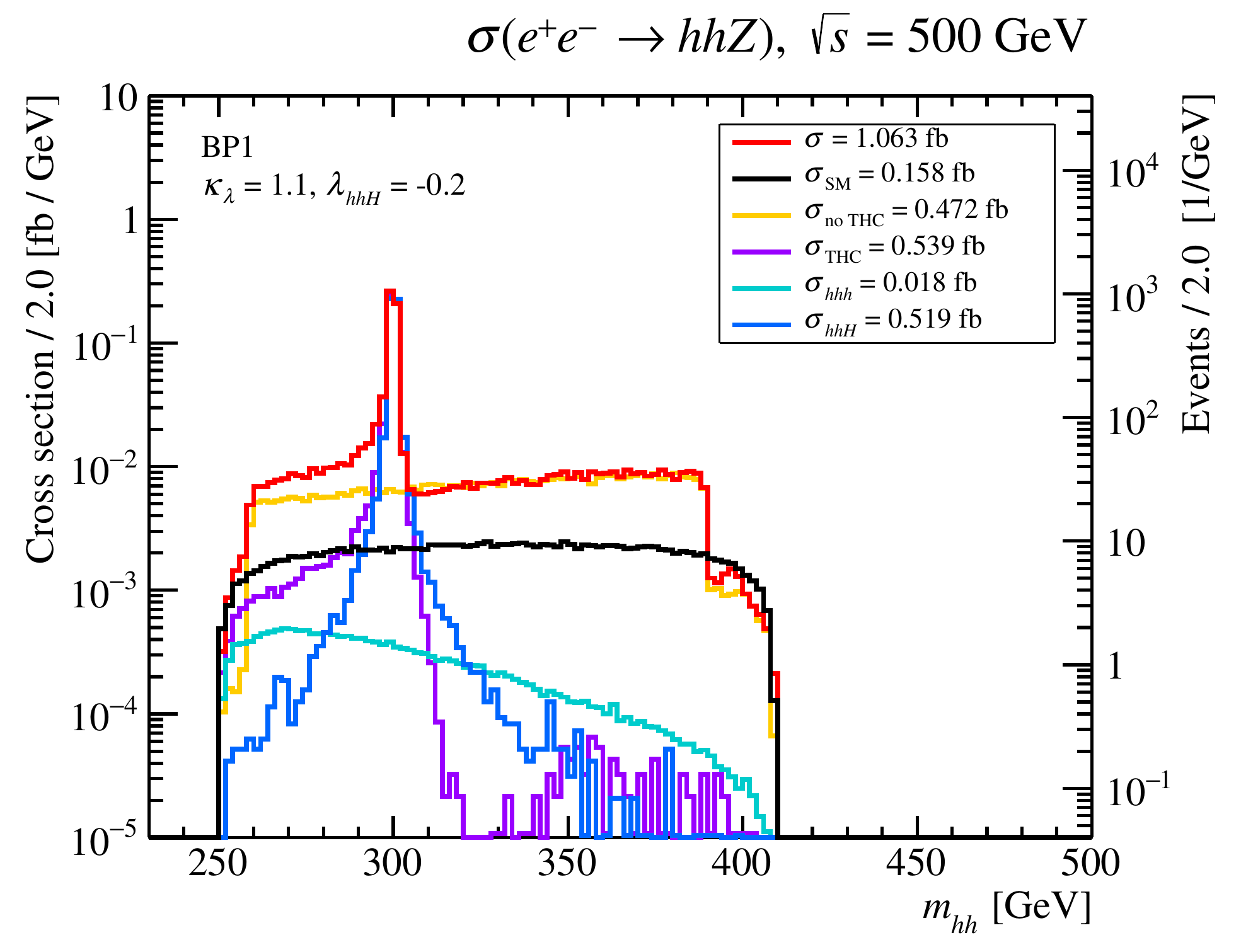}
\includegraphics[width=0.48\textwidth]{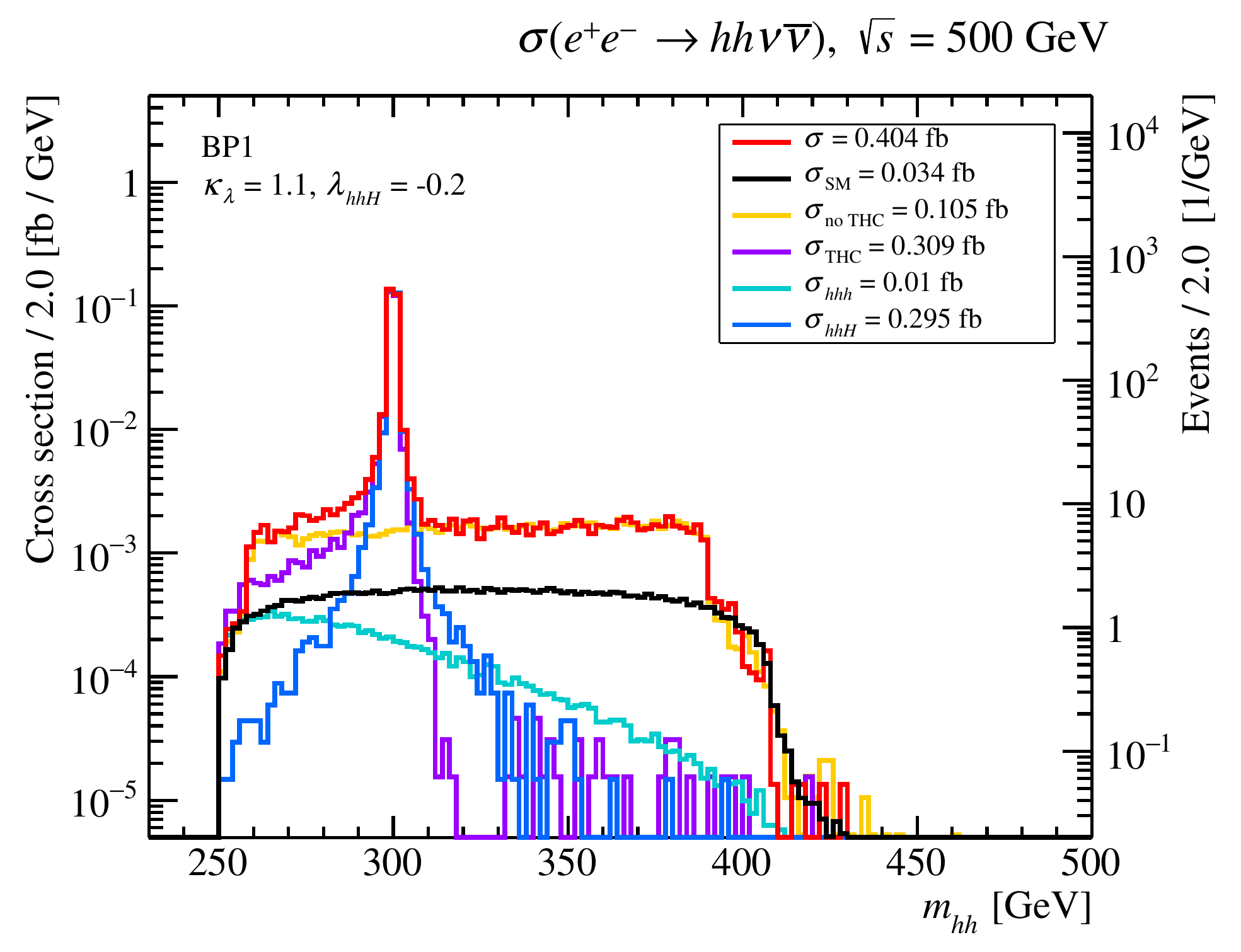}
	\end{center}
\caption{Distribution on the invariant mass of the final $hh$ pair in the 
process $e^+e^-\to hhZ$ (left) and $e^+e^-\to hh\nu\bar{\nu}$ (right) at
$\sqrt{s}=500\gev$ for BP1. } 
\label{fig:mhh500}
\end{figure}

\begin{figure}[th!]
	\begin{center}
\includegraphics[width=0.48\textwidth]{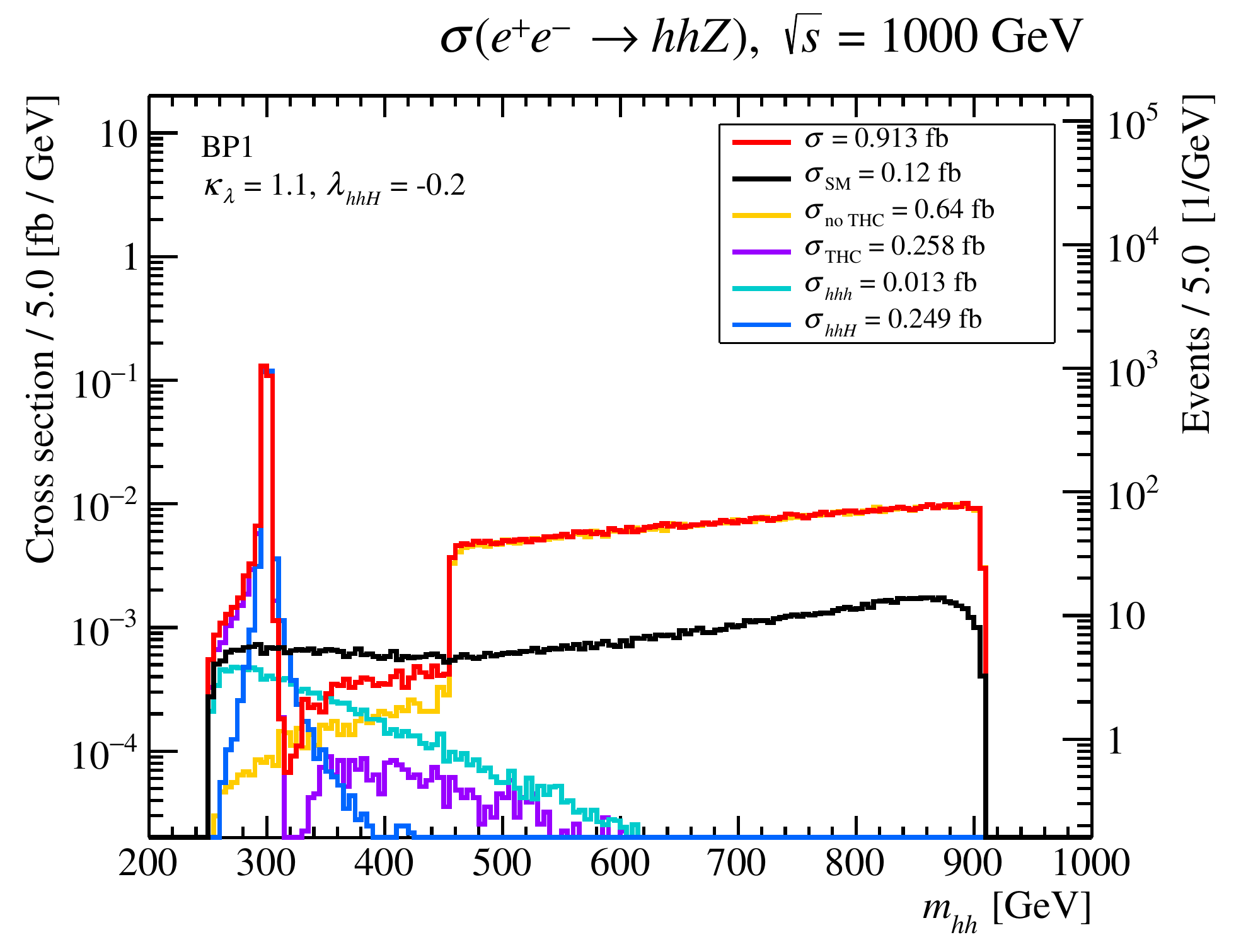}
\includegraphics[width=0.48\textwidth]{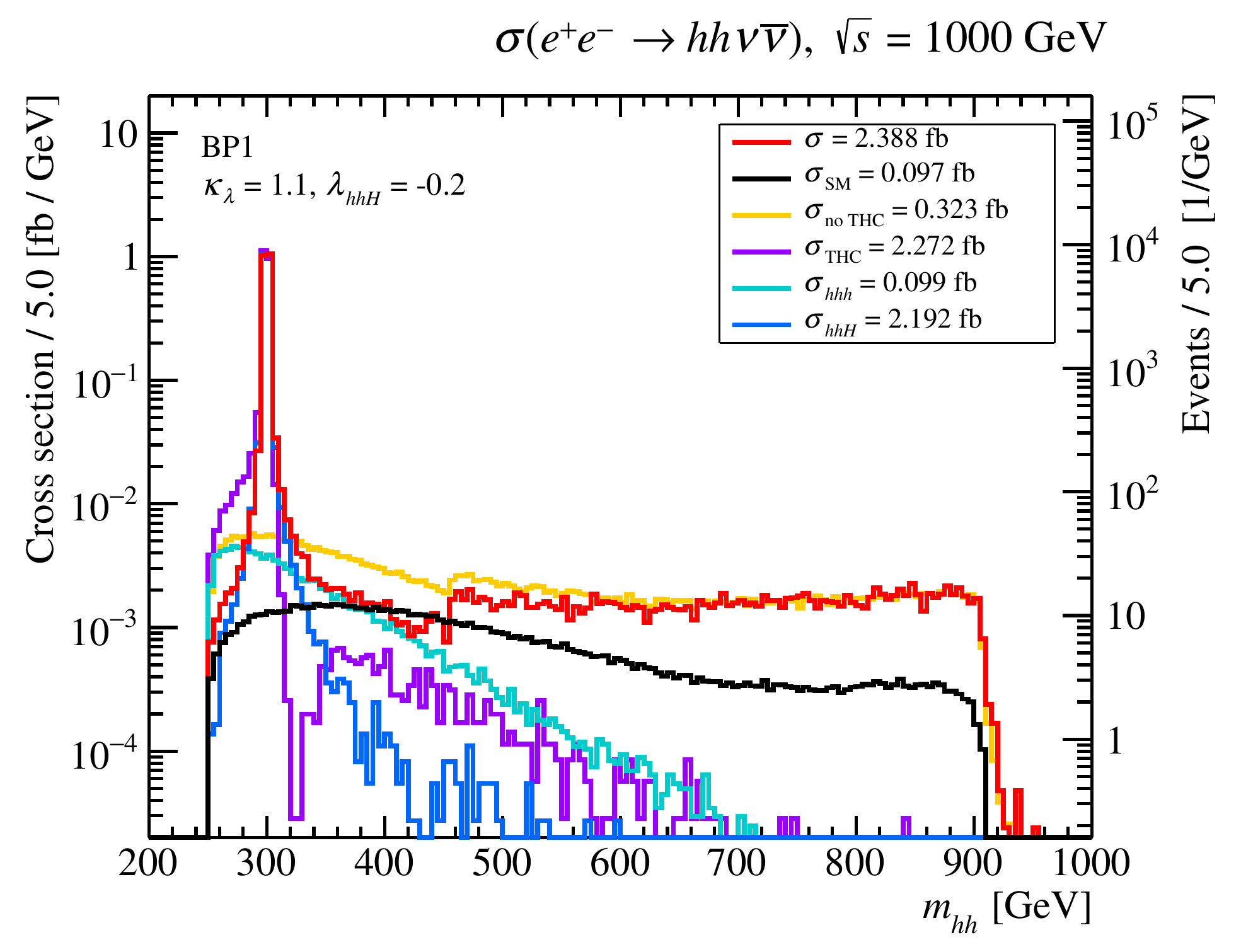}\\[2em]

\includegraphics[width=0.48\textwidth]{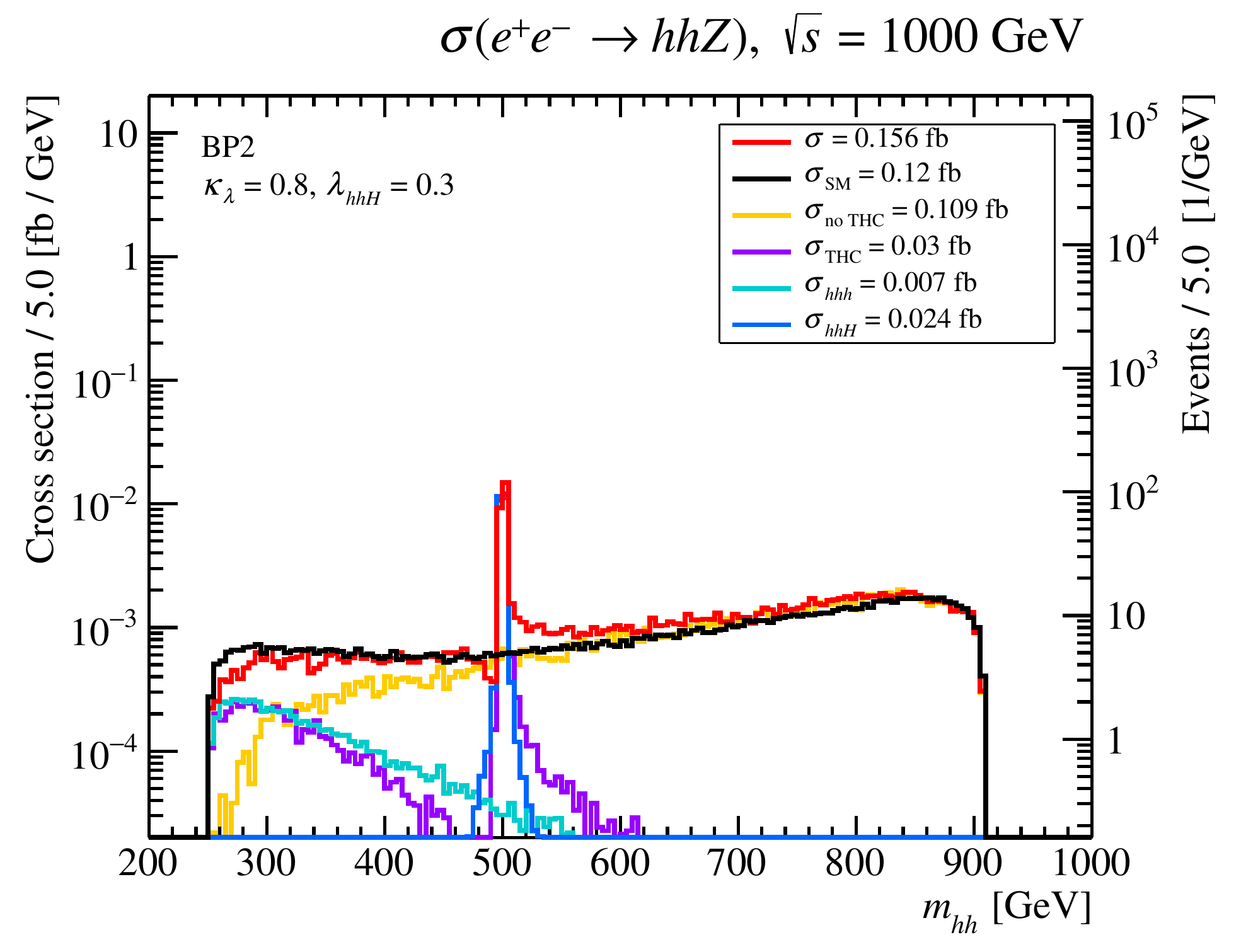}
\includegraphics[width=0.48\textwidth]{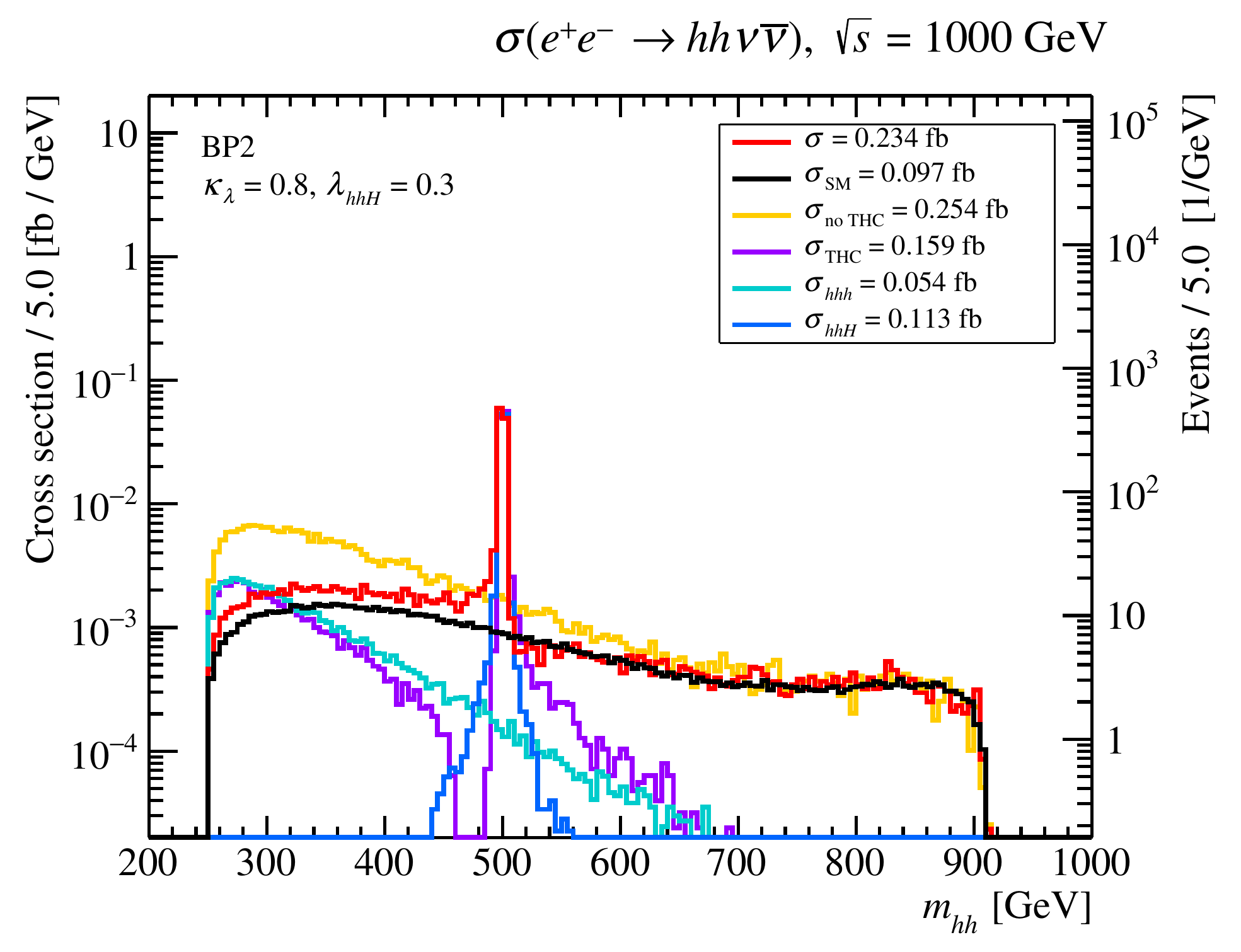}\\[2em]

\includegraphics[width=0.48\textwidth]{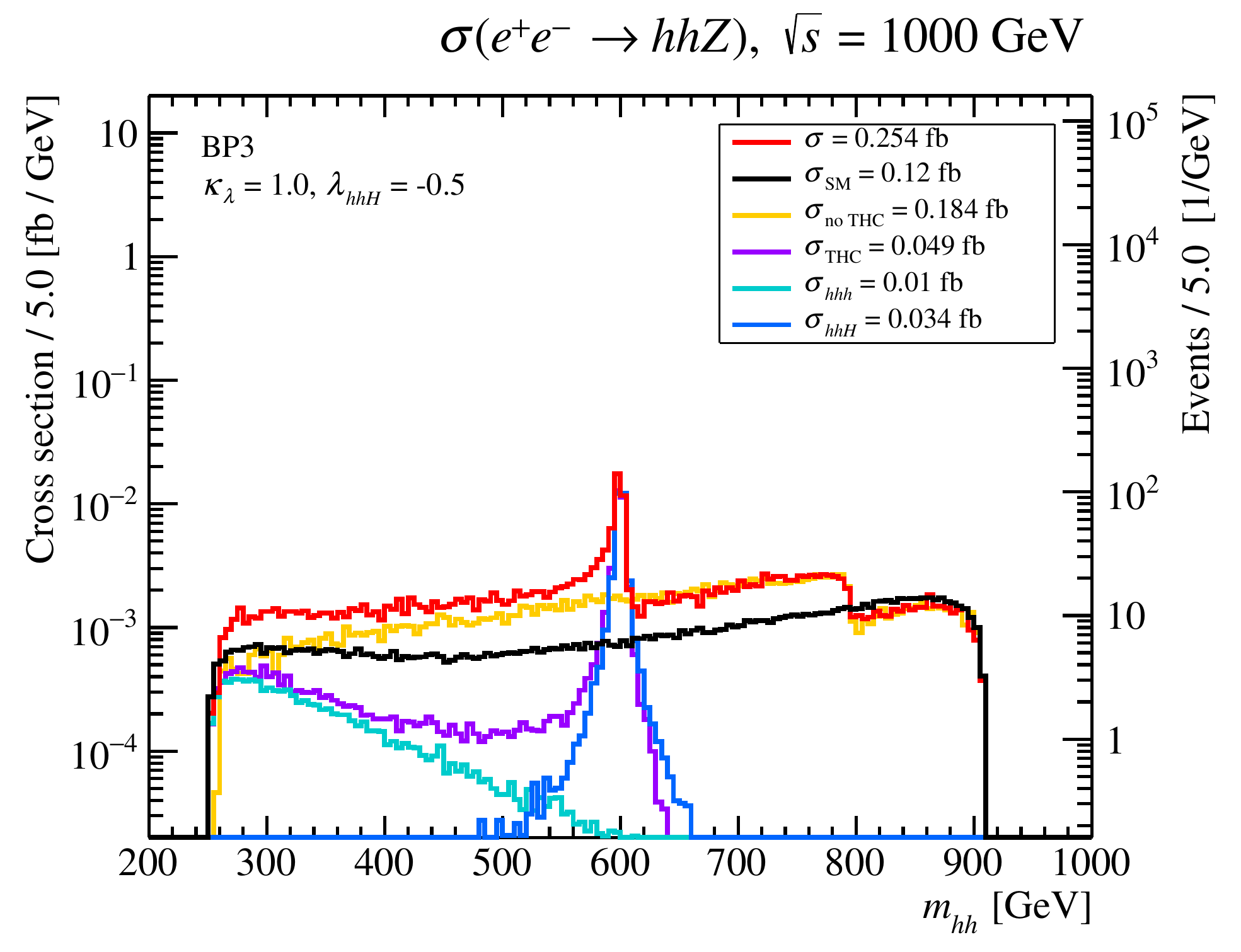}
\includegraphics[width=0.48\textwidth]{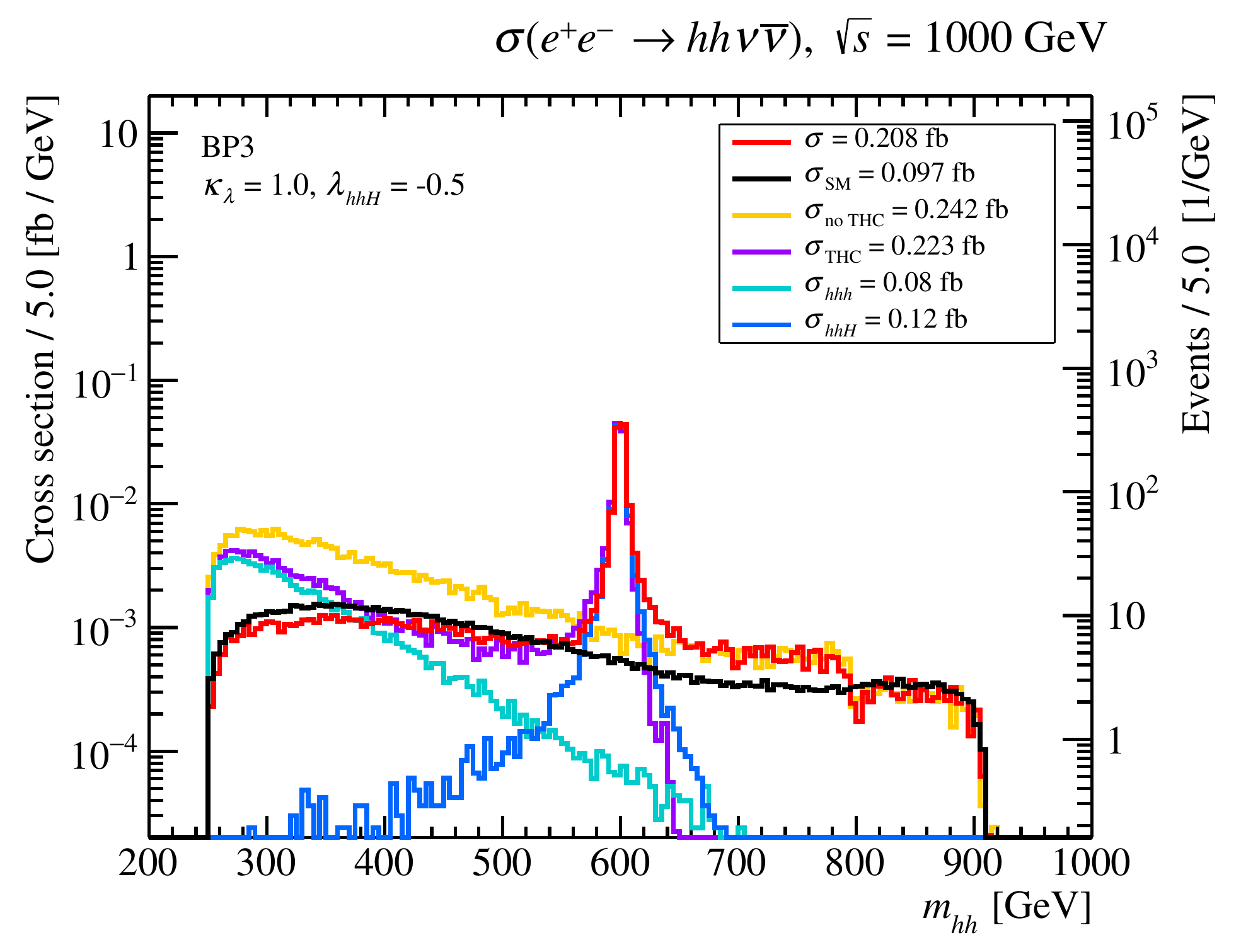}
	\end{center}
\caption{Distribution on the invariant mass of the final $hh$ pair in the 
process $e^+e^-\to hhZ$ (left) and $e^+e^-\to hh\nu\bar{\nu}$ (right) at
$\sqrt{s}=1000\gev$ for BP1, BP2 and BP3. } 
\label{fig:mhh1000}
\end{figure}

\begin{figure}[p!]
	\begin{center}
\includegraphics[width=0.44\textwidth]{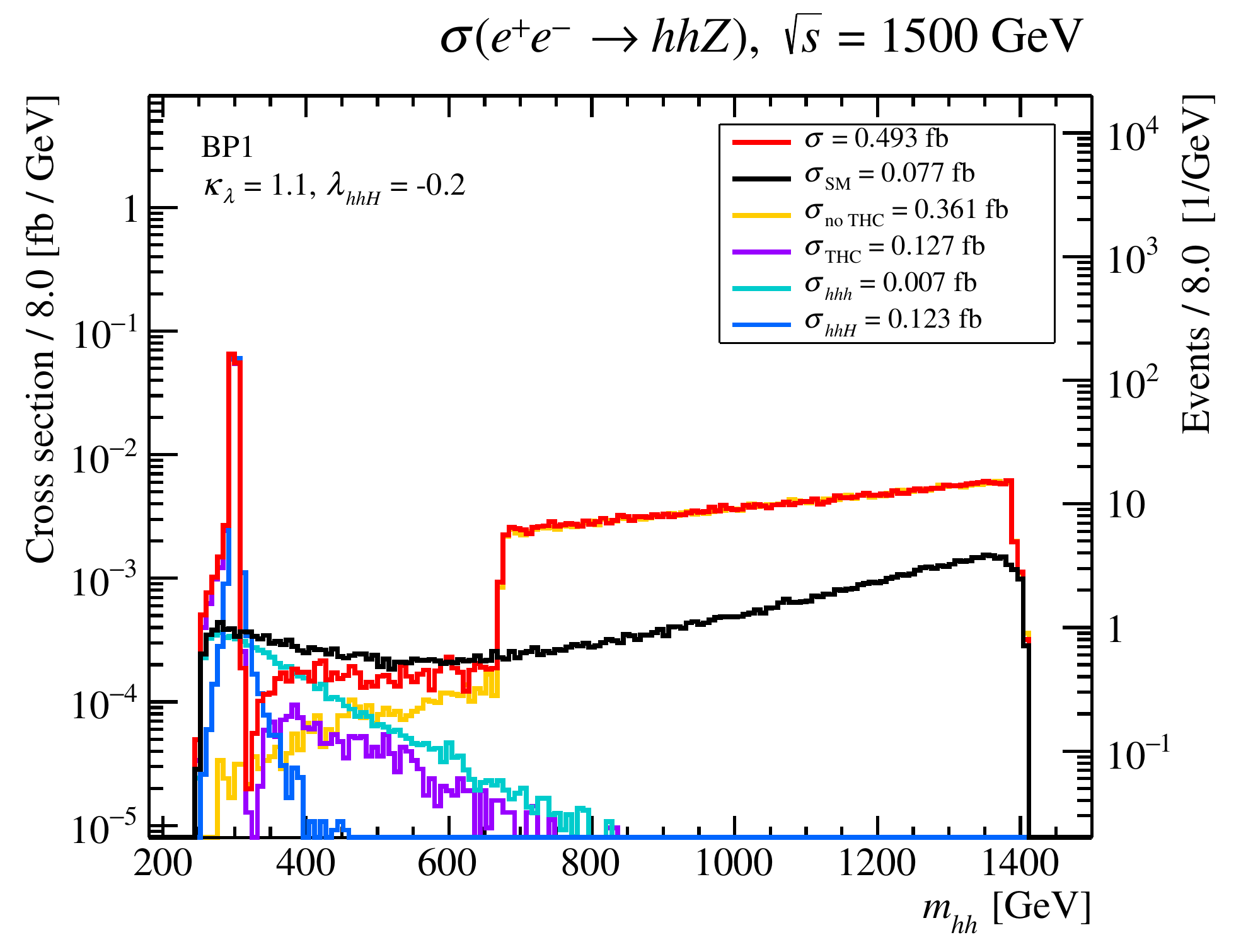}
\includegraphics[width=0.44\textwidth]{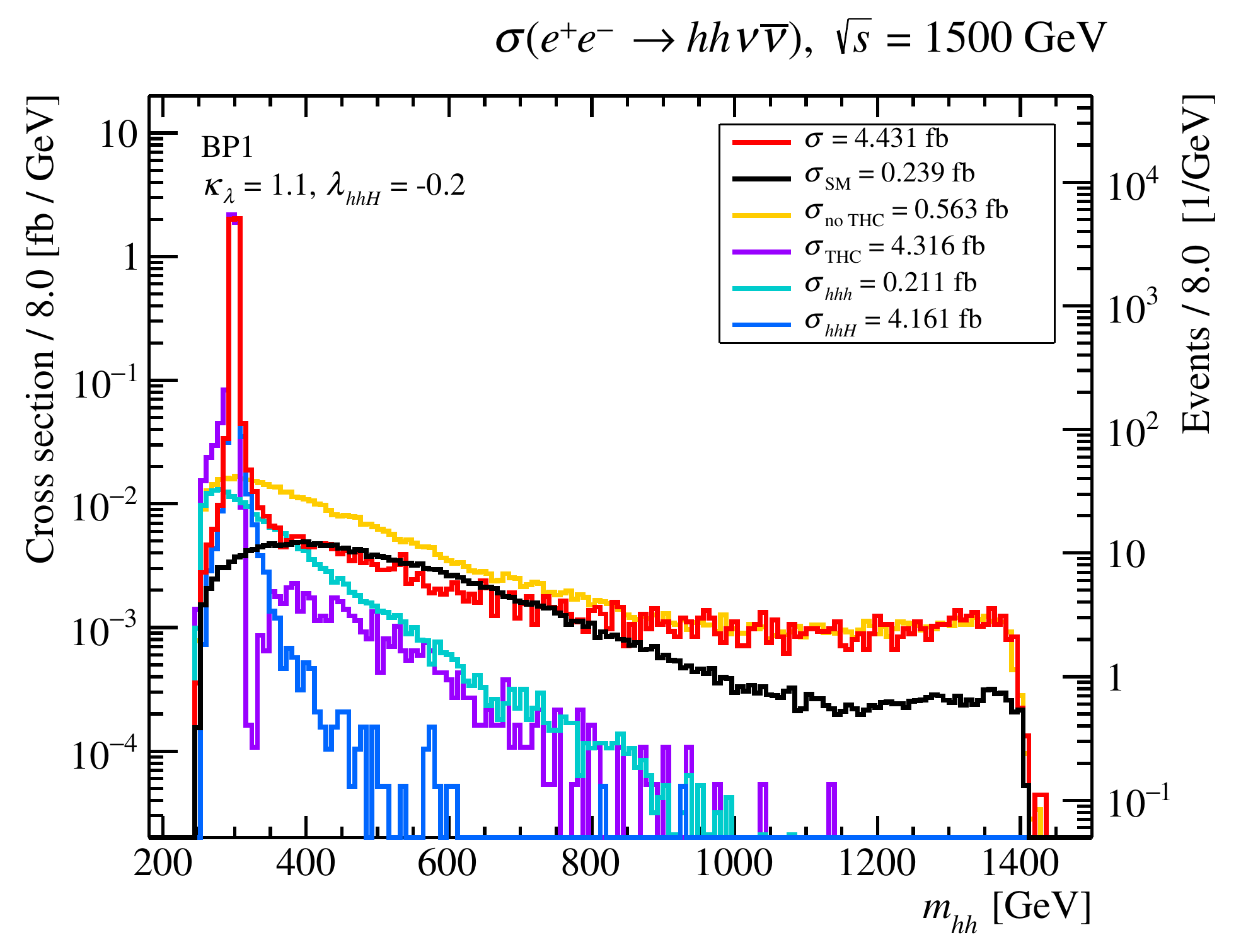}

\includegraphics[width=0.44\textwidth]{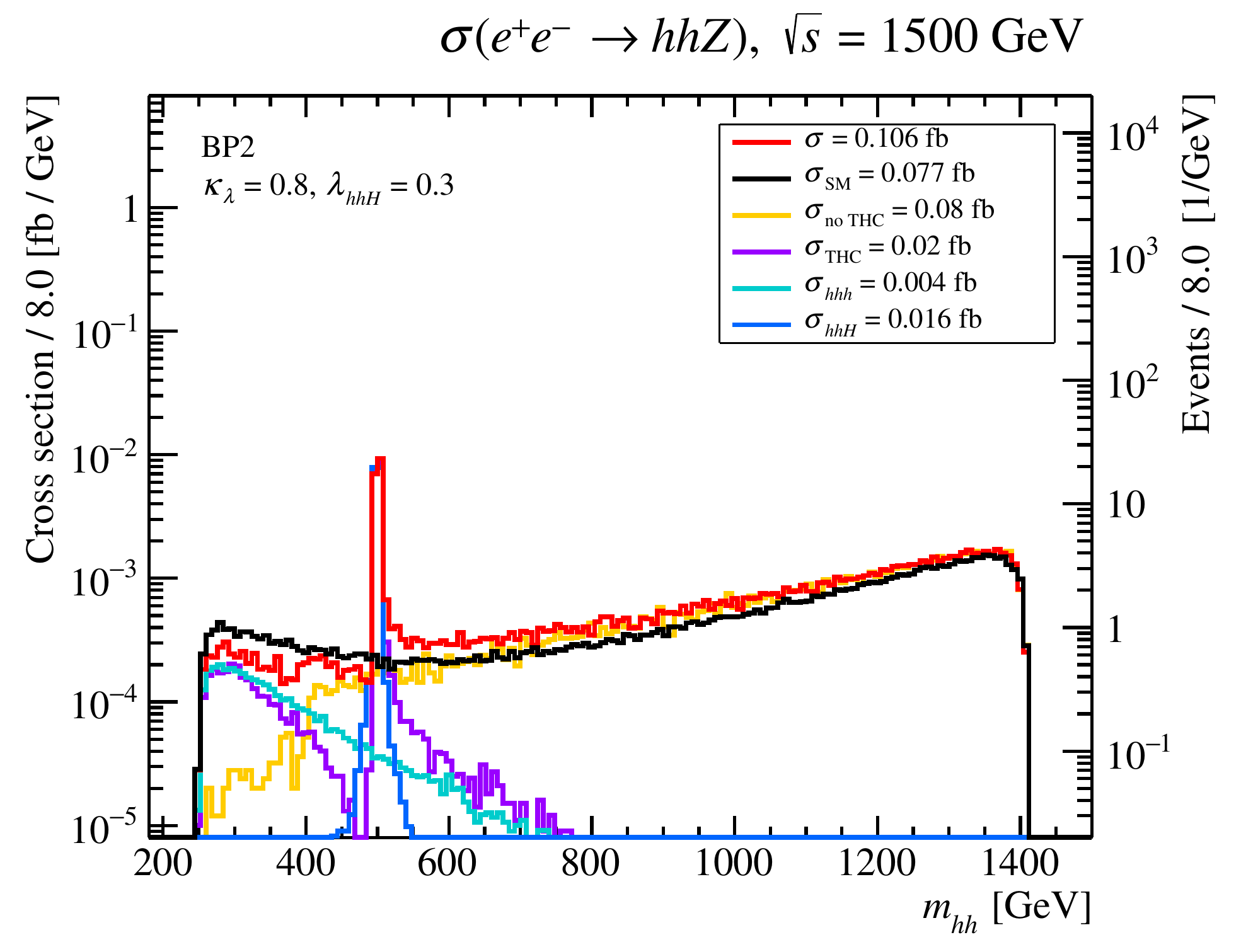}
\includegraphics[width=0.44\textwidth]{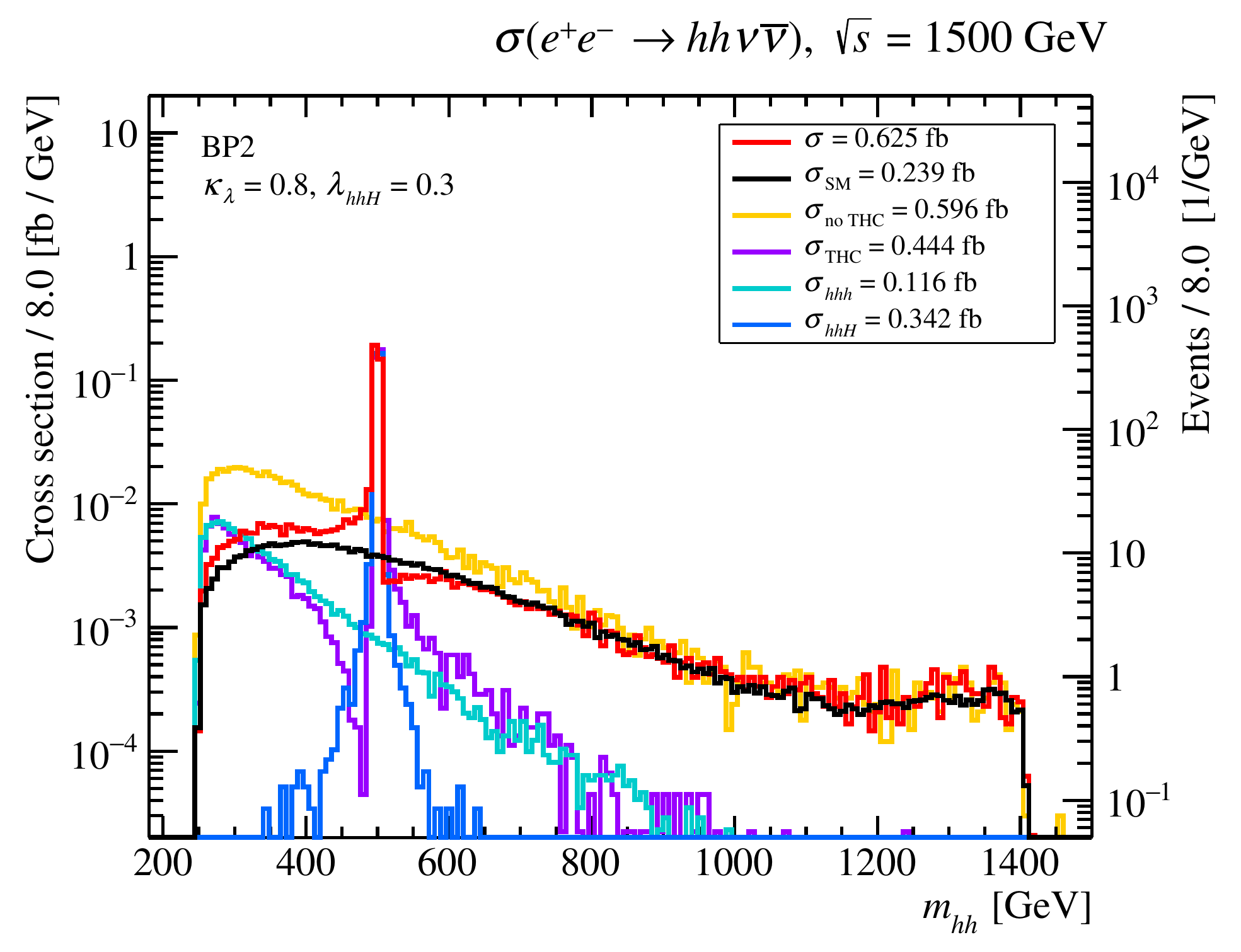}

\includegraphics[width=0.44\textwidth]{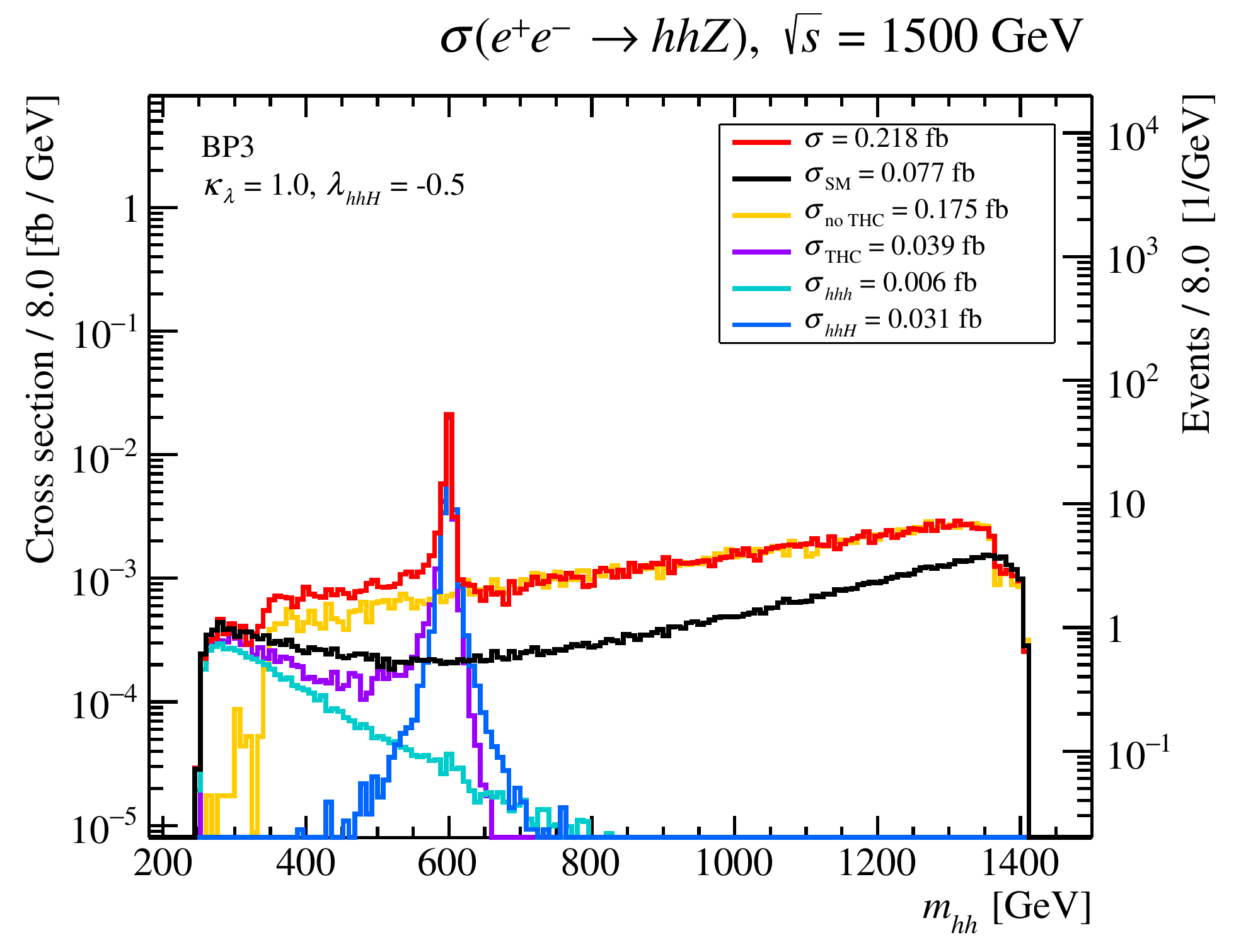}
\includegraphics[width=0.44\textwidth]{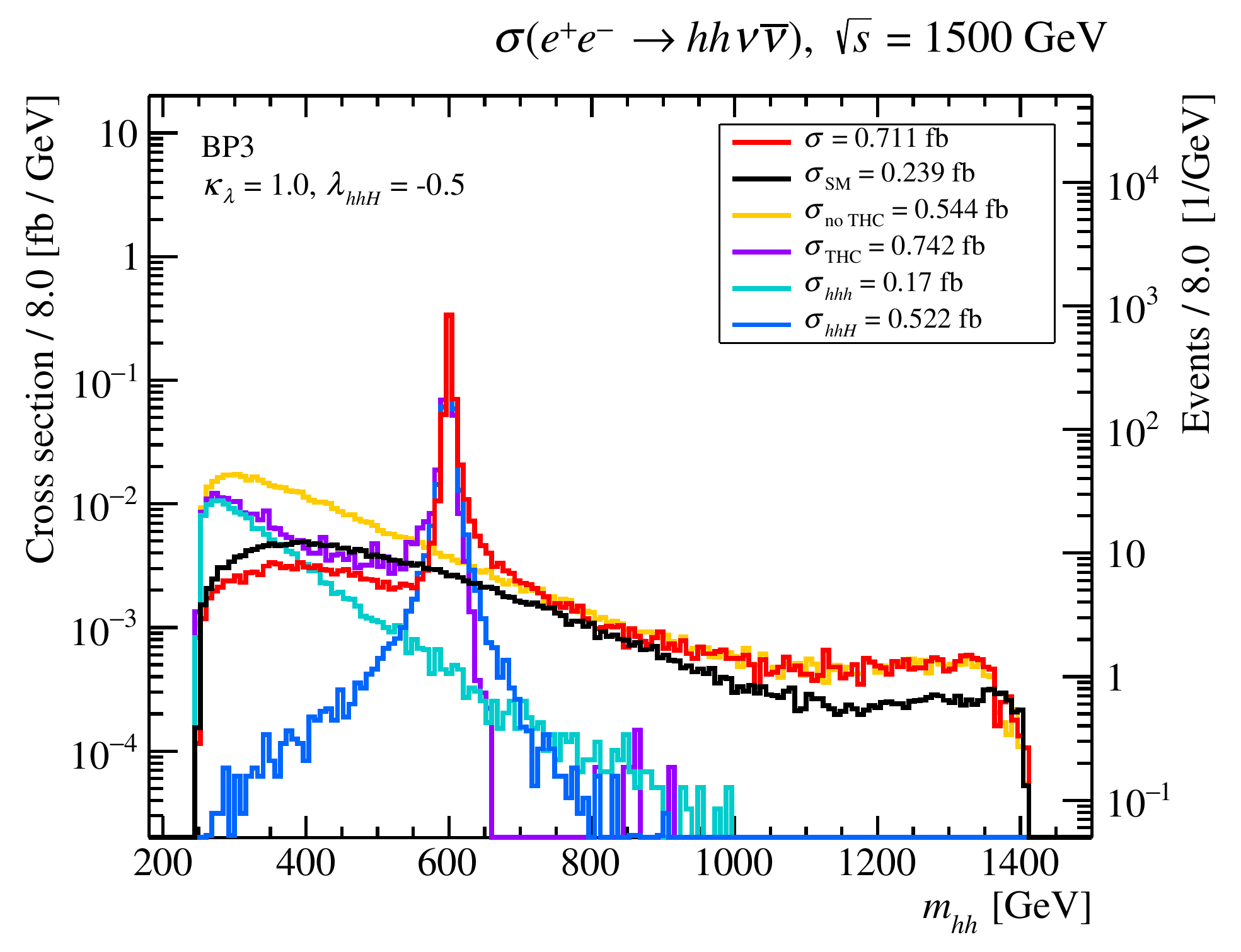}

\includegraphics[width=0.44\textwidth]{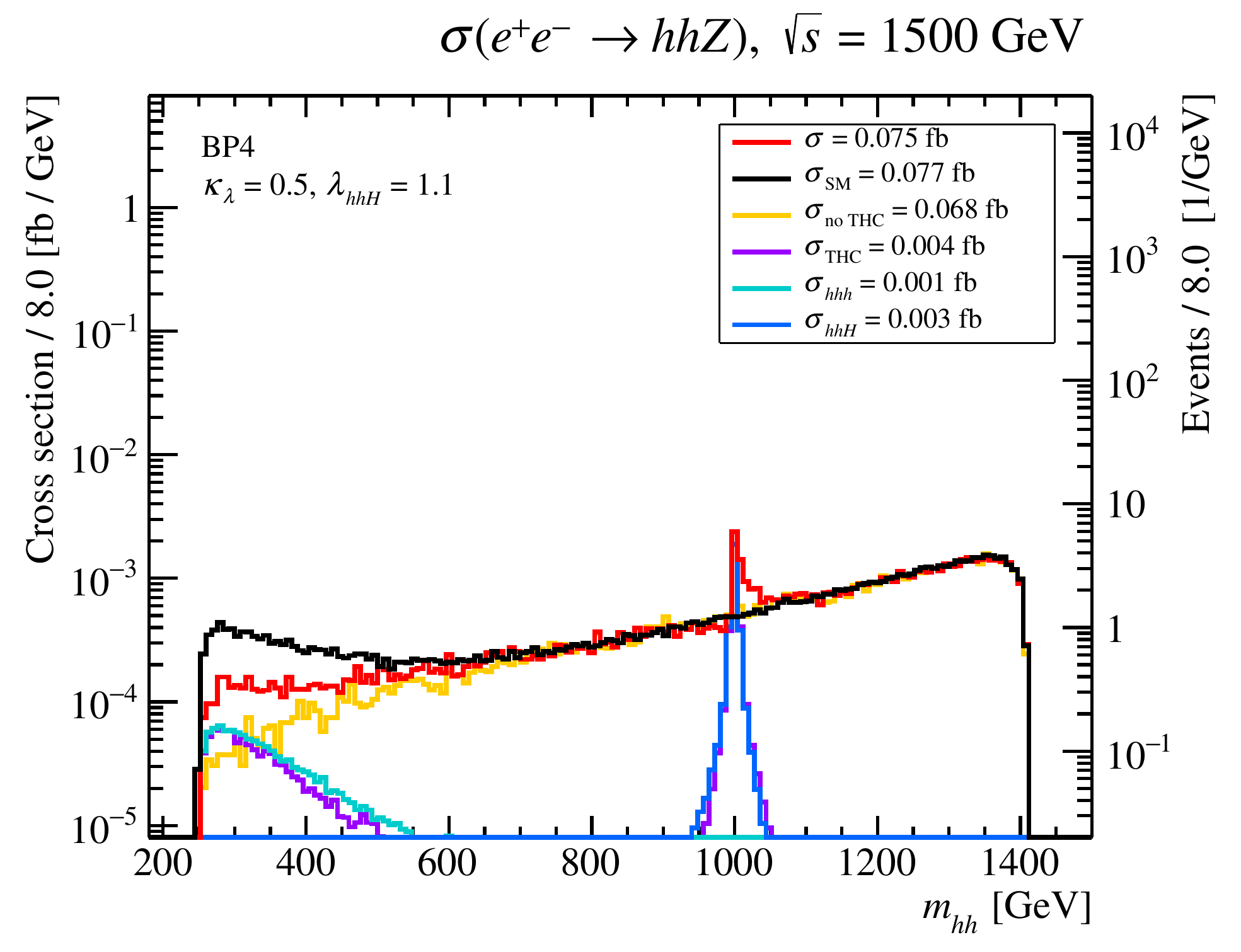}
\includegraphics[width=0.44\textwidth]{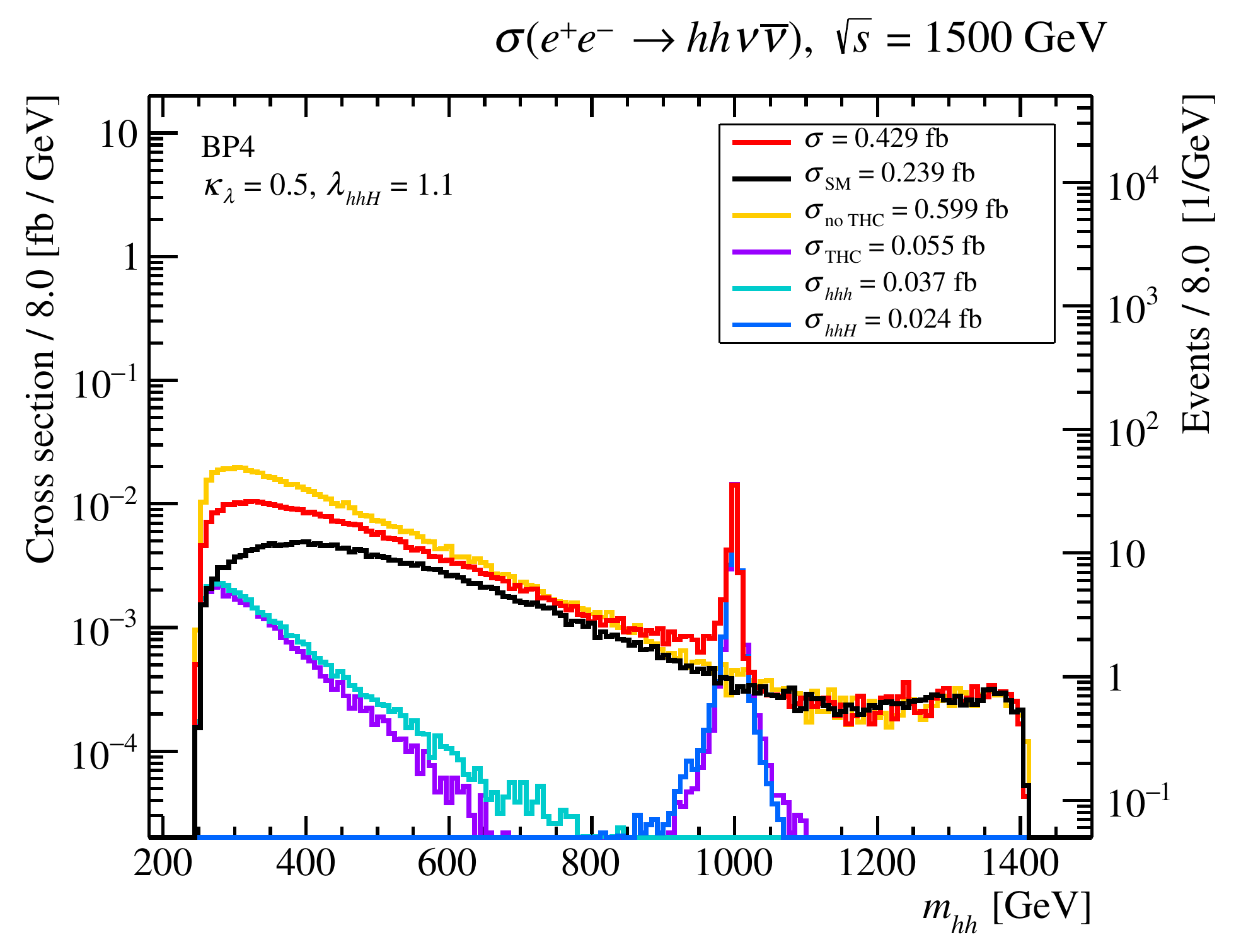}
	\end{center}
\caption{Distribution on the invariant mass of the final $hh$ pair in the 
process $e^+e^-\to hhZ$ (left) and $e^+e^-\to hh\nu\bar{\nu}$ (right) at
$\sqrt{s}=1500\gev$ for BP1, BP2, BP3 and BP4. } 
\label{fig:mhh1500}
\end{figure}

\begin{figure}[p!]
	\begin{center}
\includegraphics[width=0.44\textwidth]{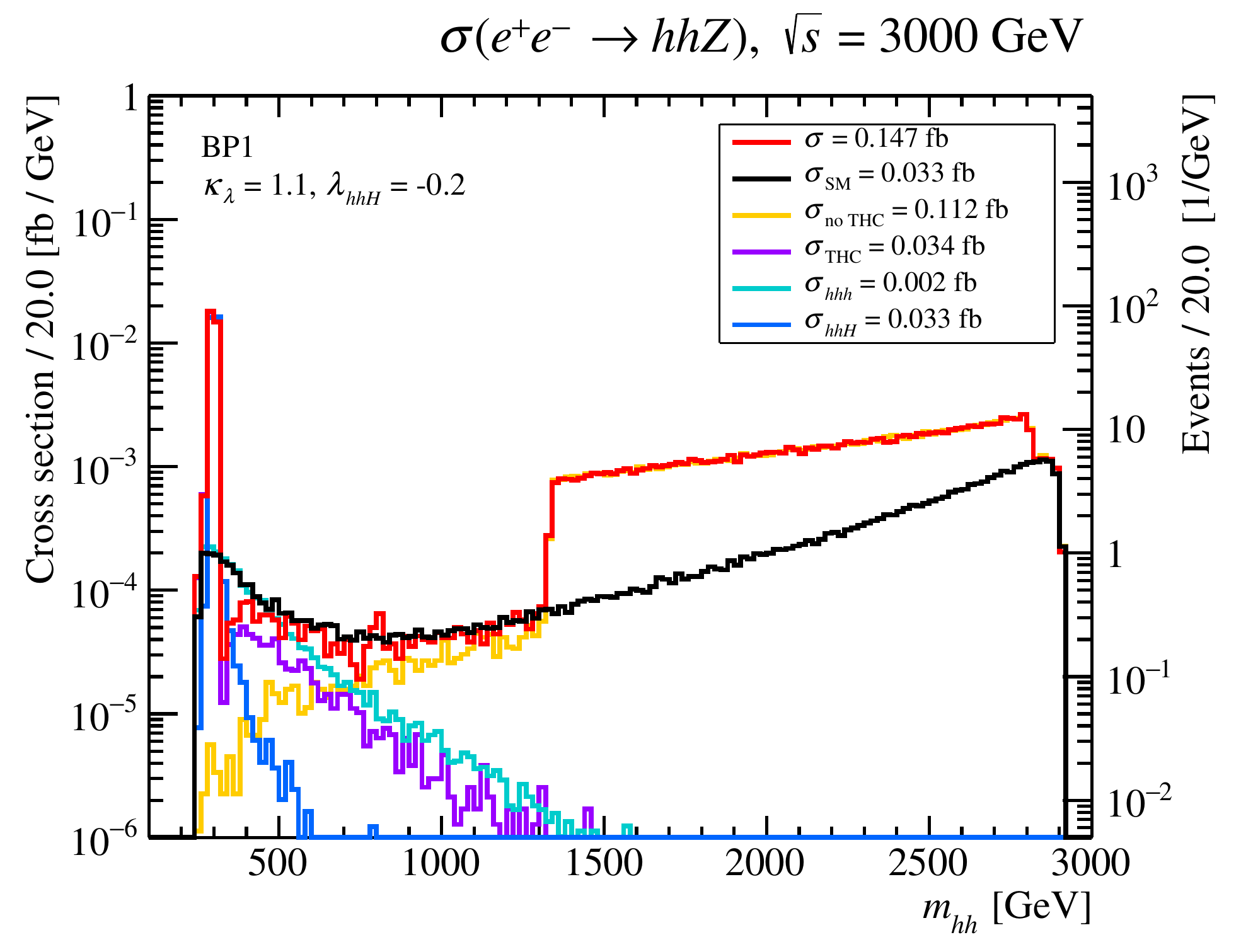}
\includegraphics[width=0.44\textwidth]{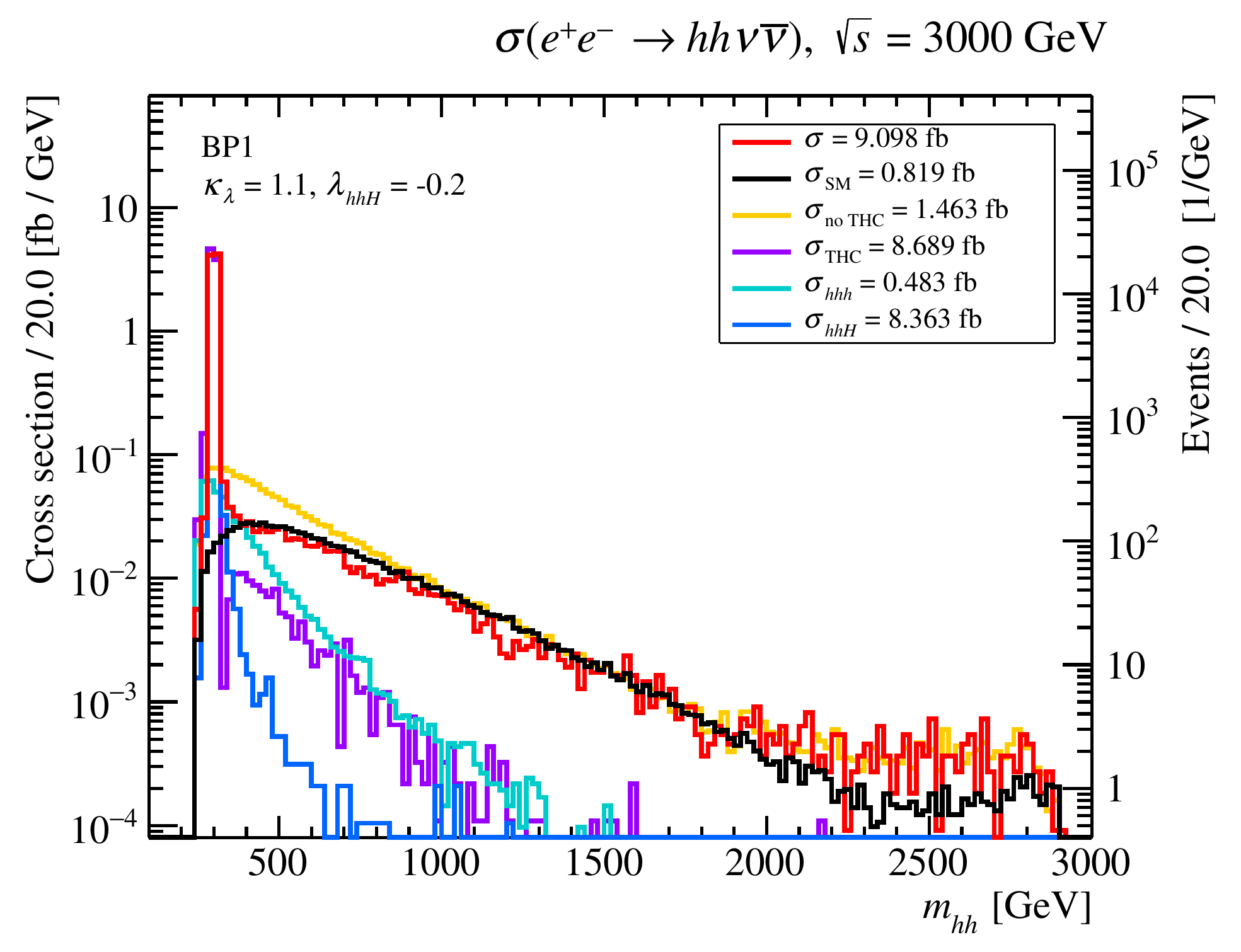}

\includegraphics[width=0.44\textwidth]{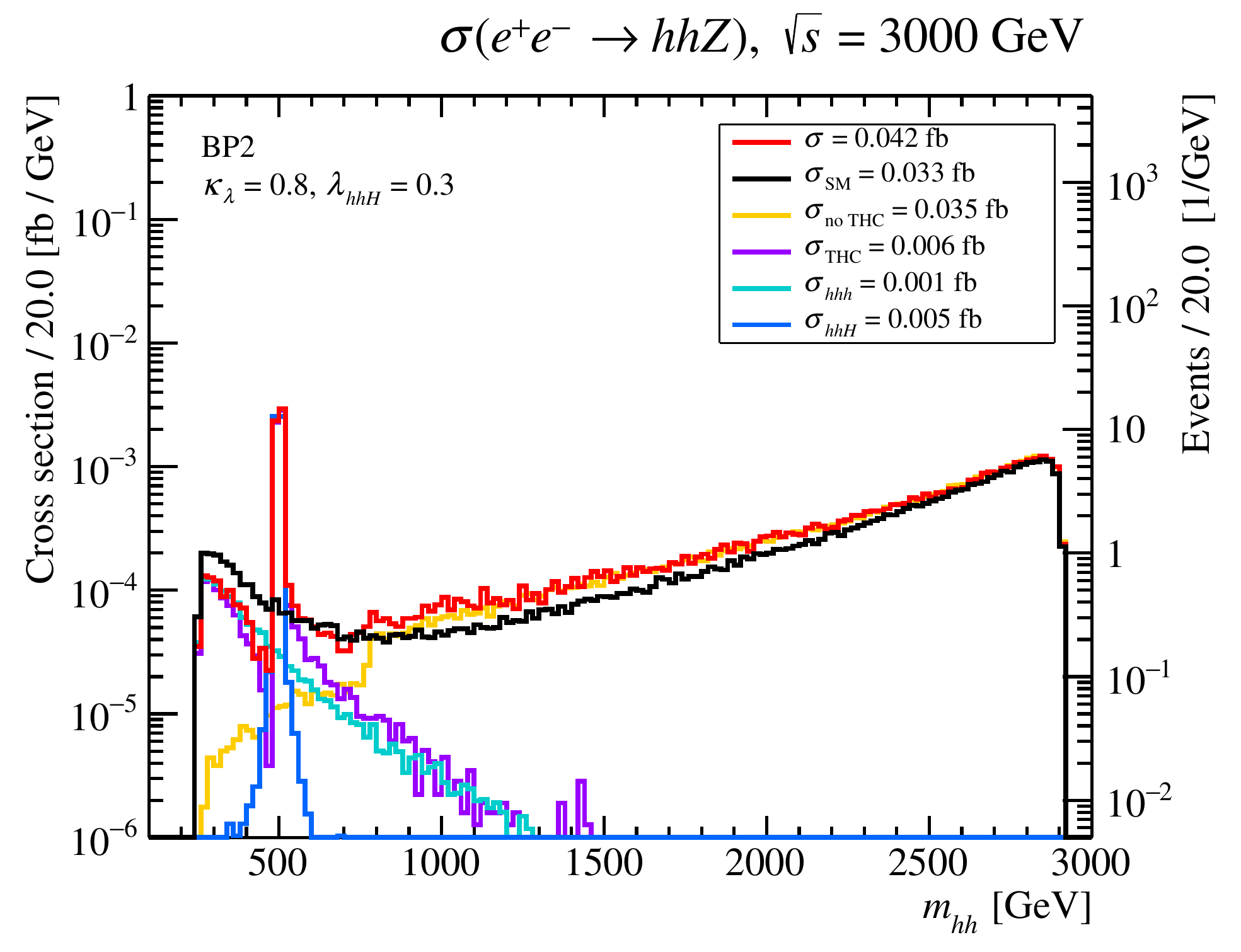}
\includegraphics[width=0.44\textwidth]{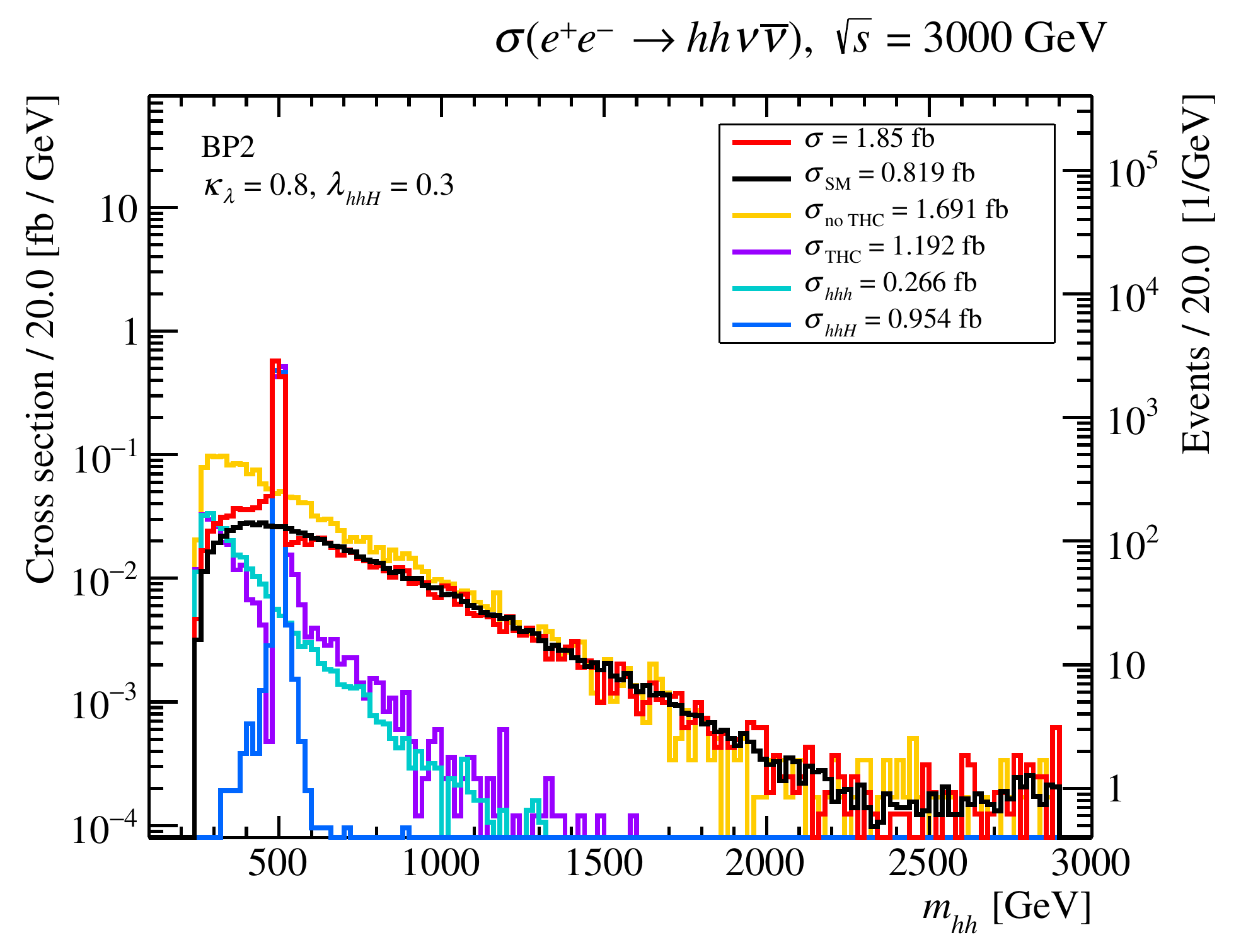}

\includegraphics[width=0.44\textwidth]{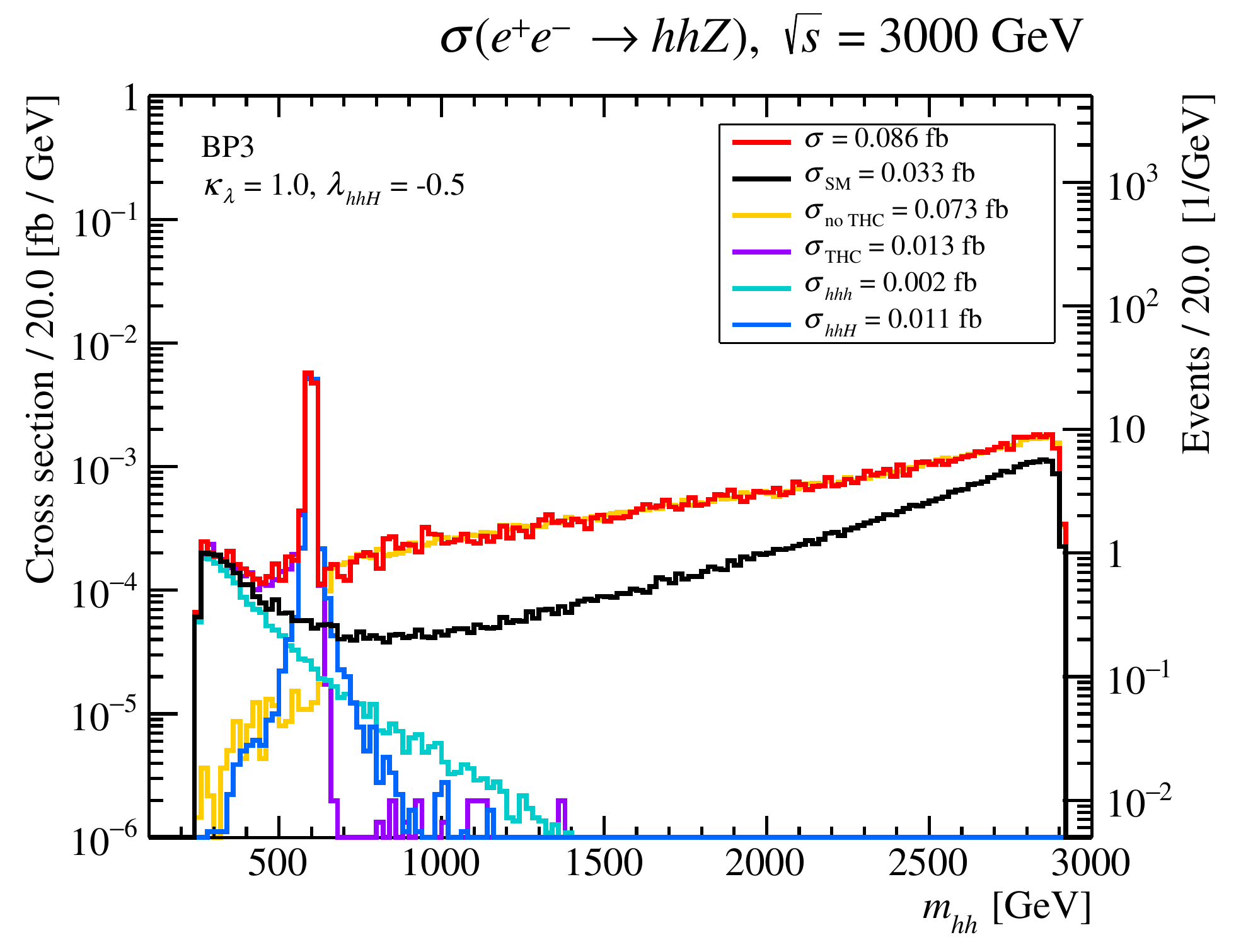}
\includegraphics[width=0.44\textwidth]{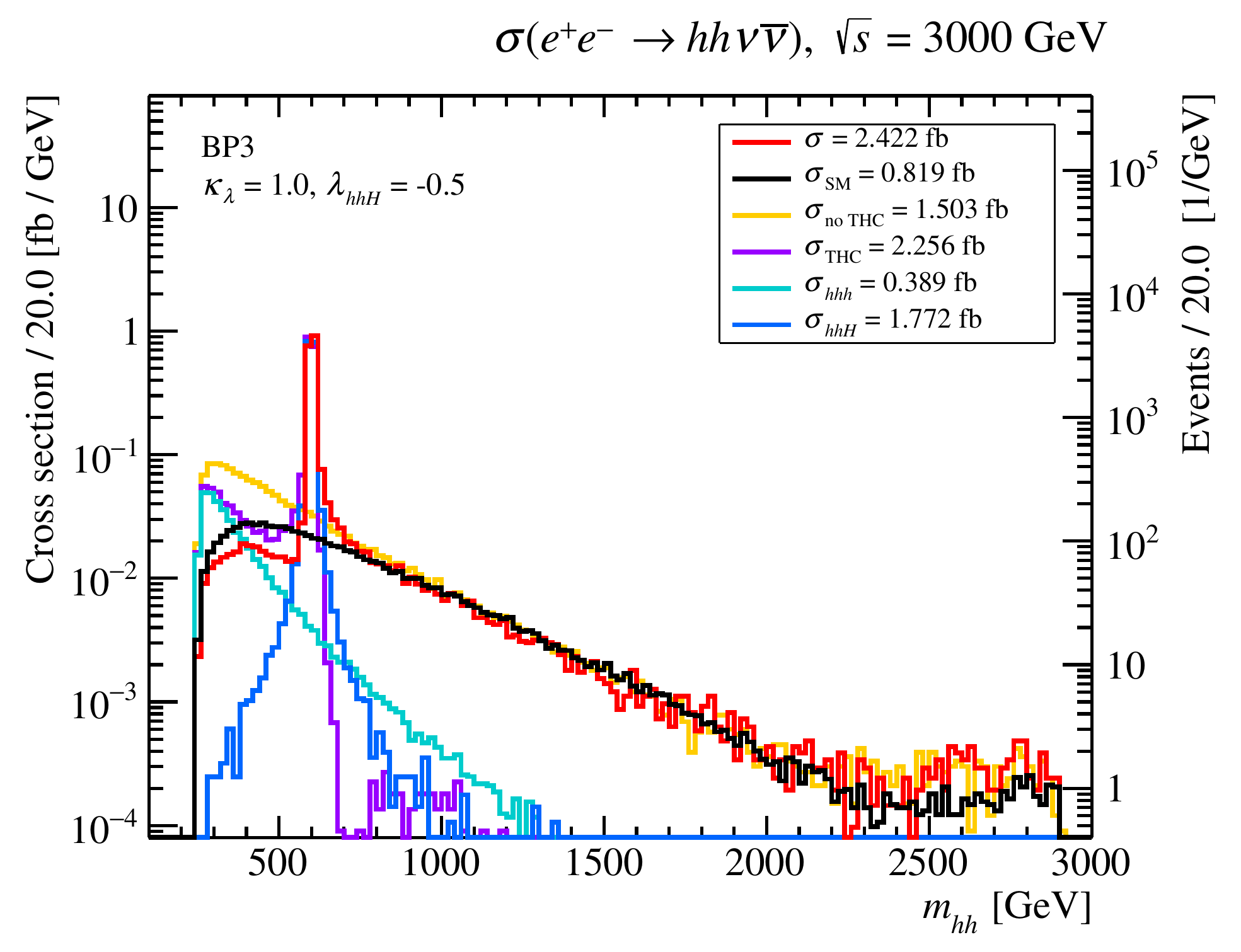}

\includegraphics[width=0.44\textwidth]{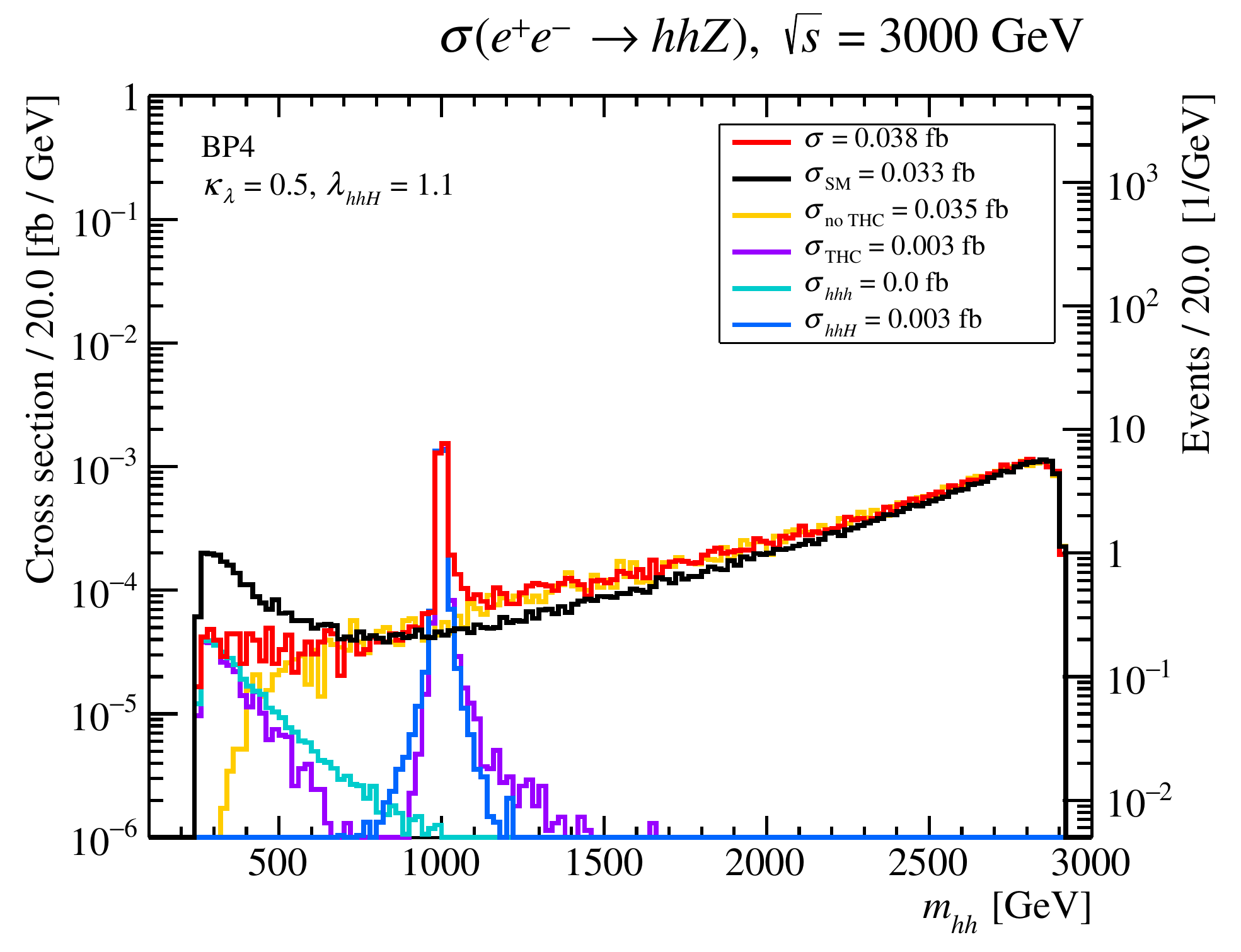}
\includegraphics[width=0.44\textwidth]{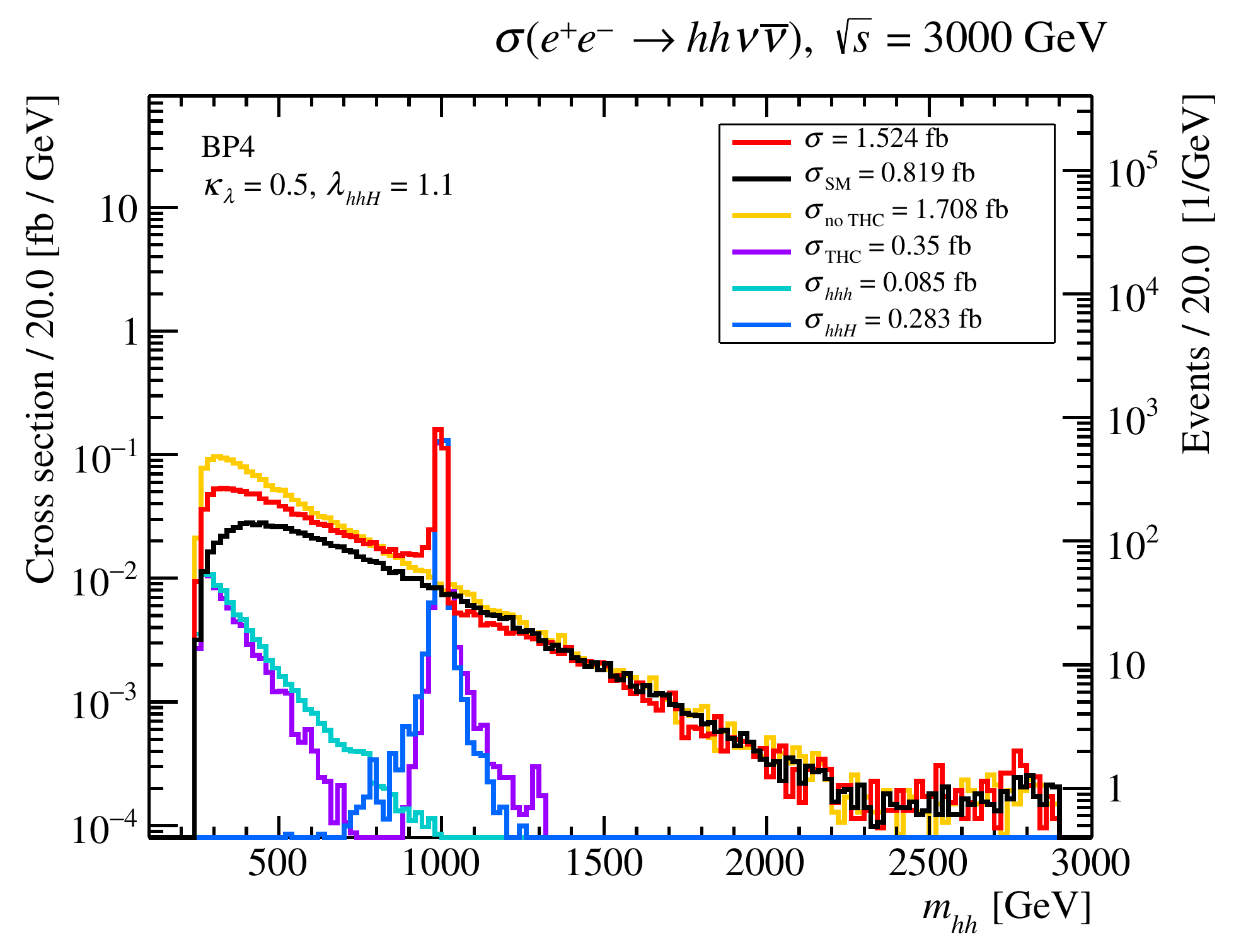}
	\end{center}
\caption{Distribution on the invariant mass of the final $hh$ pair in the 
process $e^+e^-\to hhZ$ (left) and $e^+e^-\to hh\nu\bar{\nu}$ (right) at
$\sqrt{s}=3000\gev$ for BP1, BP2, BP3 and BP4. } 
\label{fig:mhh3000}
\end{figure}

The set of plots in \reffi{fig:mhh500},  
 \reffi{fig:mhh1000},  \reffi{fig:mhh1500}, \reffi{fig:mhh3000},  are the 
results in the  2HDM-type I case,  for $\sqrt{s}=$ 500 GeV,  1000 GeV,
1500 GeV, and 3000 GeV, respectively. In the vertical axis on the right
of these plots,  the corresponding $hh$ event rates using the
luminosities of \refta{tab:ee} are also shown.  The plots in the left
of all these figures correspond to the $hhZ$ channel, and the plots on
the right correspond to the $hh \nu \bar \nu$ channel.  It should be
noted that for
each energy we have only included the predictions for the BPs where the
corresponding heavy Higgs boson mass $m_H$ is clearly reachable.
Concretely, BP1 with $m_H=300 \gev$ is reachable at the four energies;
BP2 with $m_H=500 \gev$ and BP3 with $m_H=600 \gev$  
are reachable at $1000 \gev, 1500 \gev$ and $3000 \gev$; and BP4 with
$m_H=1000 \gev$ is only reachable at $1500 \gev$ and $3000 \gev$.   For
completeness, we also include the corresponding predictions of the 2HDM
total cross section in all these cases,  both in the interior boxes of
these figures and in \refta{tab:RhhZ}  and  \refta{tab:Rhhnunubar} for
the $hhZ$ and $hh\nu \bar \nu$ cases,  respectively.  The predictions
for the SM case are included as well,  for comparison.  For the
forthcoming discussion, it is also important to analyze separately the
contributions from the various diagrams in these $m_{hh}$ distributions.
Thus,  in all these figures we display separately:  the
complete $d\sigma /d m_{hh}$ 2HDM rates (in red),  the contributions
from the diagrams containing the $\lahhh$ (in light blue), those
from the diagrams containing the $\lahhH$ (in dark blue),  the
sum of these two latter containing the two triple Higgs couplings (in
purple),  the sum of all the rest of diagrams not containing the triple
couplings (in yellow), and the complete $d\sigma /d m_{hh}$ SM rates (in
black). Notice that comparing the purple lines with the dark blue and light
blue lines one could also determine the effect of the interference between the 
$h$ and $H$ mediated diagrams.
One can also see from this comparison that the 
mentioned interference effect is negligible in the mass 
invariant $m_{hh}$ region close to the resonant peaks 
where the $H$ mediated diagrams clearly dominate.

We start with the discussion regarding the sensitivity to $\lahhh$ in
all these distributions. 
First,  we find that the maximum sensitivity to this triple coupling
appears in the low $m_{hh}$ region, 
slightly above the 250 GeV threshold of $hh$ production. This feature
also happens in the SM case, and is well known in the literature,
see, e.g., \citere{Roloff:2019crr}. The contribution from $\lahhh$
modifies the profile of the distribution in that region
close to threshold,  and the distortion is larger for larger values of
$\kala$.  In fact the contribution from the diagram containing this
$\lahhh$ produces an important interference effect which is
constructive  in the $hhZ$ channel, whereas it is destructive in the
$hh\nu \bar \nu$ case. This is seen in our plots by comparing the red 
lines (total) with the yellow lines (without triple Higgs couplings) and
with the light blue lines (with just $\lahhh$) in that region close to
the $hh$ threshold. The red lines are above the yellow lines in $hhZ$, but
they are below the yellow lines in $hh \nu \bar \nu$.  The size of this
interference effect from $\lahhh$  is larger for larger triple coupling
values (or equivalently larger $\kala$ values) and is clearly correlated
with the contribution from the diagram with an intermediate virtual $h$
(light blue line). This interference effect from the triple Higgs
coupling  also happens in the prediction of the cross section
distribution for the case of SM di-Higgs production,  as it is well
known in the literature (see, for instance,
\cite{Roloff:2019crr,Gonzalez-Lopez:2020lpd} and references therein).
Indeed,  it has been 
studied  as a very efficient strategy  at CLIC to determine the value of
$\kala$  within the context of the SM,  with a high precision of around
$\pm 0.1$ \cite{Roloff:2019crr}.  In the present 2HDM case, 
the figures show that the effect from $\lahhh$ is larger than
in the SM in the $hh\nu\bar \nu$ channel,  at the colliders with the
largest energies,  i.e.\ at  $\sqrt{s}=3 \tev$.  The biggest
contribution from $\lahhh$ and the largest size of the interference
effect  can be seen  in  \reffi{fig:mhh3000} in comparison to
\reffi{fig:mhh500}, \reffi{fig:mhh1000} and \reffi{fig:mhh1500}. 
Comparing the various BPs we see that the maximum effect is
produced in
the points with largest $\lahhh$ (hence, largest $\kala$), as expected.
In particular,  we have checked that the point BP3 with $\kala=1$
provides similar rates as the SM in this close to threshold $m_{hh}$
region (compare red and black lines).  Overall,  we find that the effect
from $\lahhh$ within the 2HDM is similar to the corresponding one in the
SM case,  and therefore the sensitivity to this triple Higgs coupling is
expected to be comparable to that concluded in the literature from the
studies of the sensitivity to $\kala$ in di-Higgs SM production at
$e^+e^-$ colliders. The only exception can occur when $m_H$ is not
much larger than $2\,m_{h}$ and the contribution from an intermediate
$H$~boson gives an important contribution in the relevant part of the
$m_{hh}$ distribution (as discussed below).

\smallskip
We next comment on the sensitivity to $\lahhH$ that appears,  as
previously mentioned,  via the diagrams with an intermediate $H$.  The
first clear signal from this intermediate heavy Higgs boson $H$
contribution  in all these figures is seen as a resonant peak emerging
in the invariant mass region with $m_{hh}$ close to $m_H$. The dark blue
line accounts for this $H$ mediated contribution which  displays the
resonant behavior and provides the dominant contribution to the total
result (red line) in the narrow region around the resonance .  The
second observation,  as already mentioned,  is that using the NWA to
compute the total cross section by
$\sigma(hhZ)=\sigma(HZ)\times \br(H \to hh)$ and
$\sigma(hh\nu\bar\nu)=\sigma(H\nu\bar\nu) \times \br(H \to hh)$   
does not provide an accurate prediction.  We have checked this failure of 
the NWA prediction,  using the corresponding cross section for single $H$ 
production and the $\br(H \to hh)$ for each BP. 
The main reason for this failure is that 
the remaining diagrams, other than the one mediated by the intermediate
heavy $\cp$-even Higgs boson, contribute very significantly to
the total cross section.  These remaining diagrams together explain the
non-resonant part of these  
distributions,  i.e.\ they explain the red lines outside the resonant
region (usually called 'the continuum').    In particular,  we have
checked explicitly that one of the most relevant contributions outside
the resonant peak is provided by the diagram where the $\cp$-odd
$A$ boson is the
intermediate  boson.  We have checked that the 'big step shape'
that is best seen to the right of the resonant peak in some of the
distributions is  due to the $A$ mediated contribution, which
indeed dominates the total production cross section. This can be
seen, for instance, in the distributions in the $hhZ$ channel for BP1
at $1000 \gev, 1500 \gev$ and $3000 \gev$, that display most clearly
these 'big steps'.  
 
The beginning and ending of these 'steps' due to the $A$ boson
mediated diagrams depend on the momentum and energy of the two
Higgs bosons in the final state
coming from this type of diagrams,  which can be written in short as
$e^+e^- \to h A^{(*)} \to hhZ$. In that case, we can differentiate one of
the final light Higgs boson, labeled as $h_1$, that comes from the
process $e^+e^-\to h_1A$ and the other Higgs boson, labeled as $h_2$,
that comes from the decay $A\to h_2Z$. On one side, the
momenta of $h_1$ and $A$ are completely determined by $\sqrt{s}$,
$m_A$ and $m_h$ and they are given by
\begin{equation}
p_{h_1}=p_A=\frac{1}{2\sqrt{s}}\sqrt{\left(s-m_A^2-m_h^2\right)^2-4m_A^2m_h^2}~.
\end{equation}
On the other side,  the momentum of $h_2$ depends on the energy and
momentum of its mother particle $A$, namely $p_A$ and $E_A$, the mass of
the final states $m_{h}$ and $m_Z$ and on the angle between the 
$h_2$ and the $A$, $\theta_{h_2A}$: 
 \begin{equation}
   p_{h_2}=\frac{(m_A^2+m_h^2-m_Z^2)p_A\cos\theta_{h_2A}
     \pm 2E_A\sqrt{m_A^2p_\mathrm{CM}^2-m_h^2p_A^2\sin^2\theta_{h_2A}}}
   {2(m_A^2+p_A^2\sin^2\theta_{h_2A})},
 \label{eq:ph2}
 \end{equation}
where $p_\mathrm{CM}$ is the final momentum of $h_2$ and $Z$ in the
center-of-mass frame of the~$A$.  If
$m_Ap_\mathrm{CM}/m_hp_A<1$, the maximum and the minimum of $p_{h_2}$
is reached for $\theta_{h_2A}=0$ for the positive and negative sign of the
\refeq{eq:ph2} respectively.  
Otherwise, if $m_Ap_\mathrm{CM}/m_hp_A>1$ the only physical solution is
the positive sign of \refeq{eq:ph2} and the maximum is reached for
$\theta_{h_2A}=0$, whereas the minimum is reached for
$\theta_{h_2A}=\pi$. Therefore, these extrema in $p_{h_2}$ define the
range in the invariant mass of the two final light Higgs bosons where
the effects from the $A$ boson manifest, and are given by 
 \begin{equation}
   (m_{hh}^\mathrm{max,min})^2=2m_h^2+2E_{h_1}E_{h_2}^\mathrm{max,min}
   -2p_{h_1}p_{h_2}^\mathrm{max,min} \cos\theta_{h_1h_2}~,
   \label{eq:steps}
 \end{equation}
where $E_{h_i}$ is the energy of $h_i$, with $i=1,2$, and the angle
$\theta_{h_1h_2}$ must satisfy $\theta_{h_1h_2}+\theta_{h_2A}=\pi$.  
For instance,  for BP1 at $1000 \gev$ we find the beginning
and ending of the step at $m_{hh}^\mathrm{min,max}= 456, 907 \gev$,
respectively, in agreement with what is seen in \reffi{fig:mhh1000}.

\begin{figure}[th!]
	\begin{center}
\includegraphics[width=0.48\textwidth]{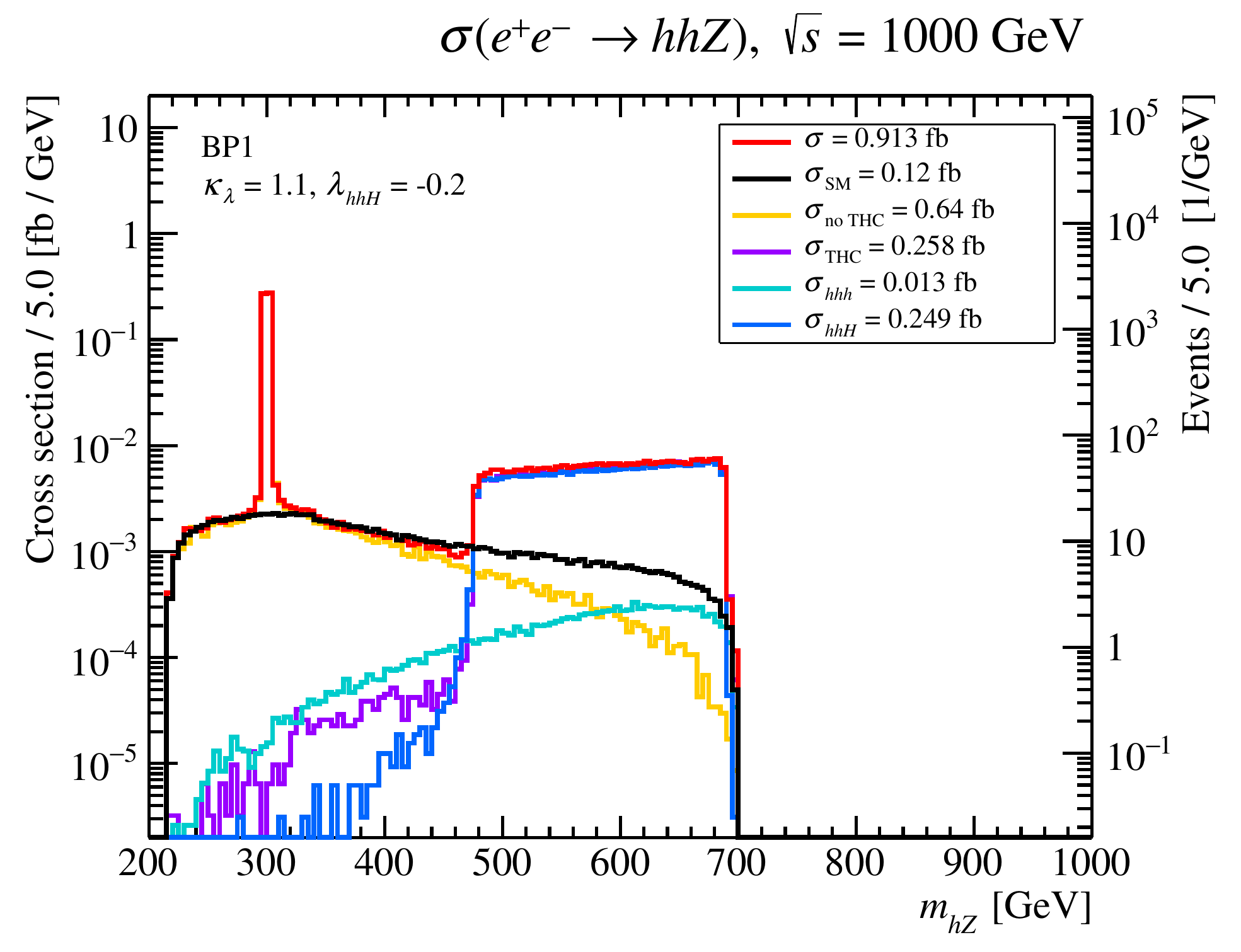}
\includegraphics[width=0.48\textwidth]{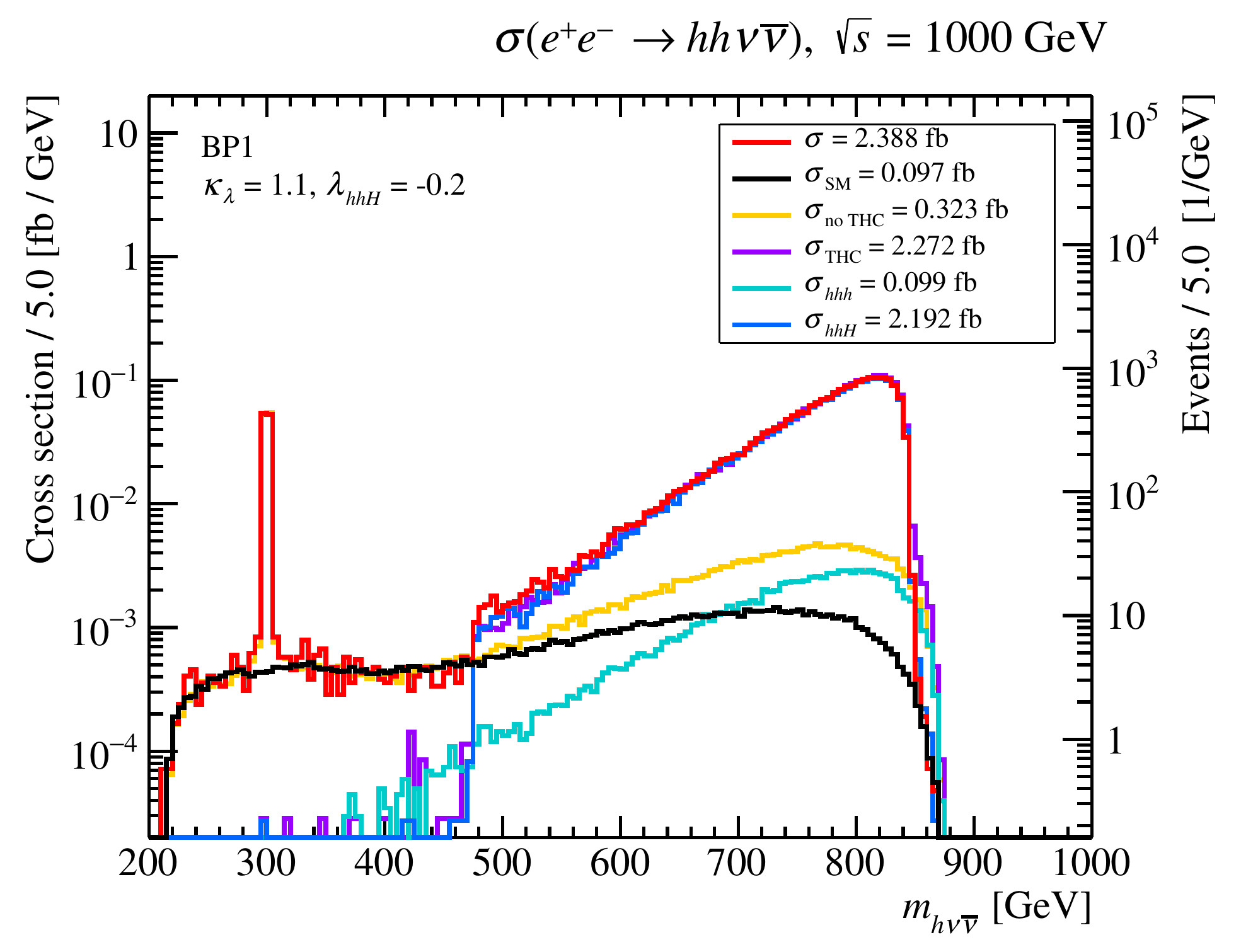}

	\end{center}
\caption{Distribution on the invariant mass of the final $hZ$ pair in the 
process $e^+e^-\to hhZ$ (left) and distribution on the invariant mass 
of the final $h\nu\bar{\nu}$ system in the process 
$e^+e^-\to hh\nu\bar{\nu}$ (right) at $\sqrt{s}=1000\gev$ for BP1. 
The Higgs boson $h$ in these two invariant mass variables $m_{hZ}$ and
$m_{h\nu\bar\nu}$ refer to the less energetic Higgs boson of the final pair.}
\label{fig:mhZ1000}
\end{figure}

On the other hand, this relevant $A$ mediated diagram  for the $hhZ$
channel provides the largest contribution when the  intermediate $A$ is
produced on-shell and then it decays to $hZ$.   
Indeed, we have found the corresponding emergent $A$ resonant peak which
can be seen clearly in the
distributions with the alternative invariant mass variable, $m_{hZ}$
(but not in the distributions with $m_{hh}$). Here we restrict 
ourselves to one example for the $m_{hZ}$ invariant mass distribution
displayed in \reffi{fig:mhZ1000} for BP1 at $\sqrt{s}=1000\gev$.
However, we do not investigate 
further the appearance of these peaks here because they are not
sensitive to the triple Higgs couplings which are the main objective
in our analysis. These $A$ resonant peaks obviously could provide
interesting information on the BSM physics induced by the $A$ bosons
within the 2HDM,  other than the triple Higgs couplings.

\begin{figure}[ht!]
	\begin{center}
\includegraphics[width=0.48\textwidth]{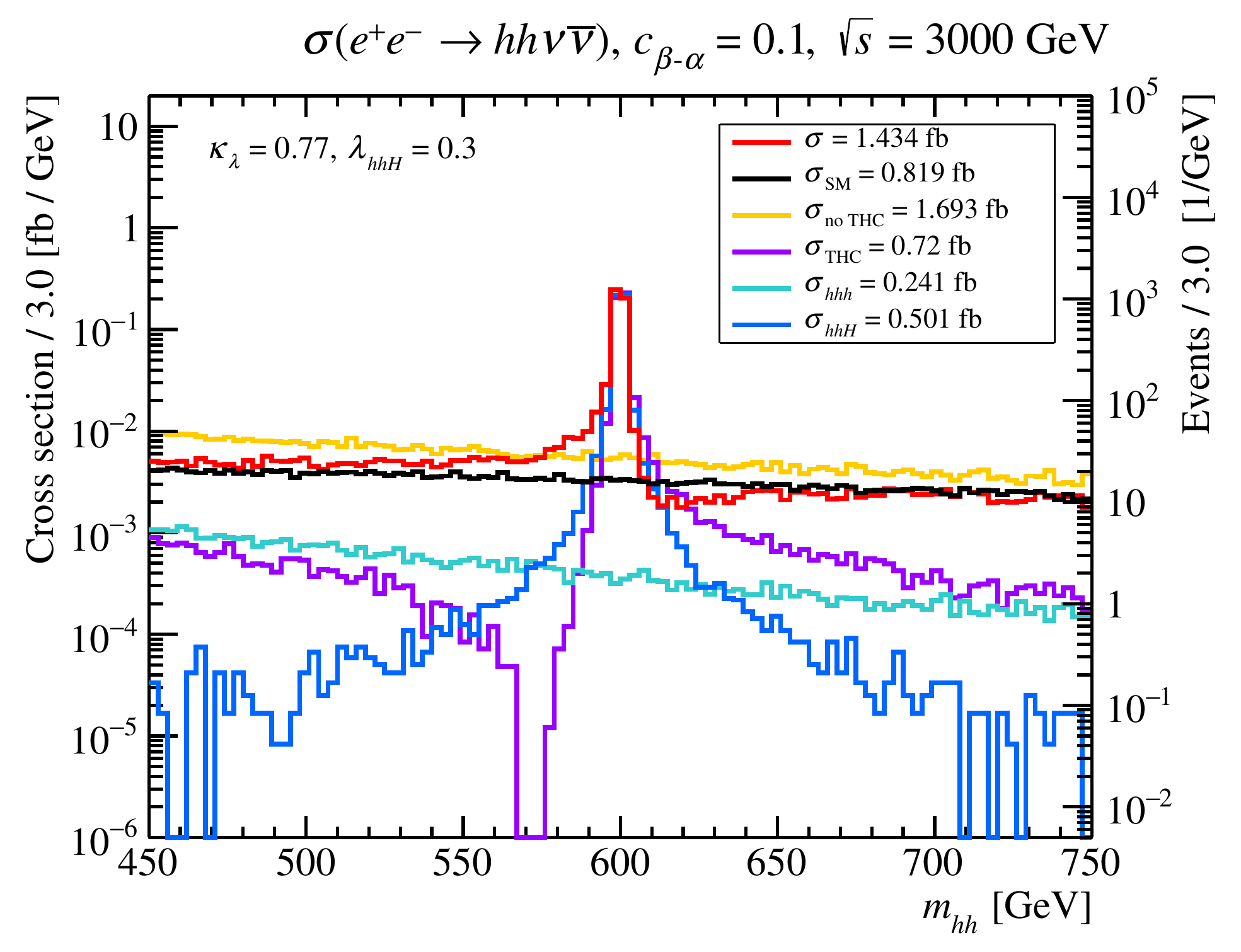}
\includegraphics[width=0.48\textwidth]{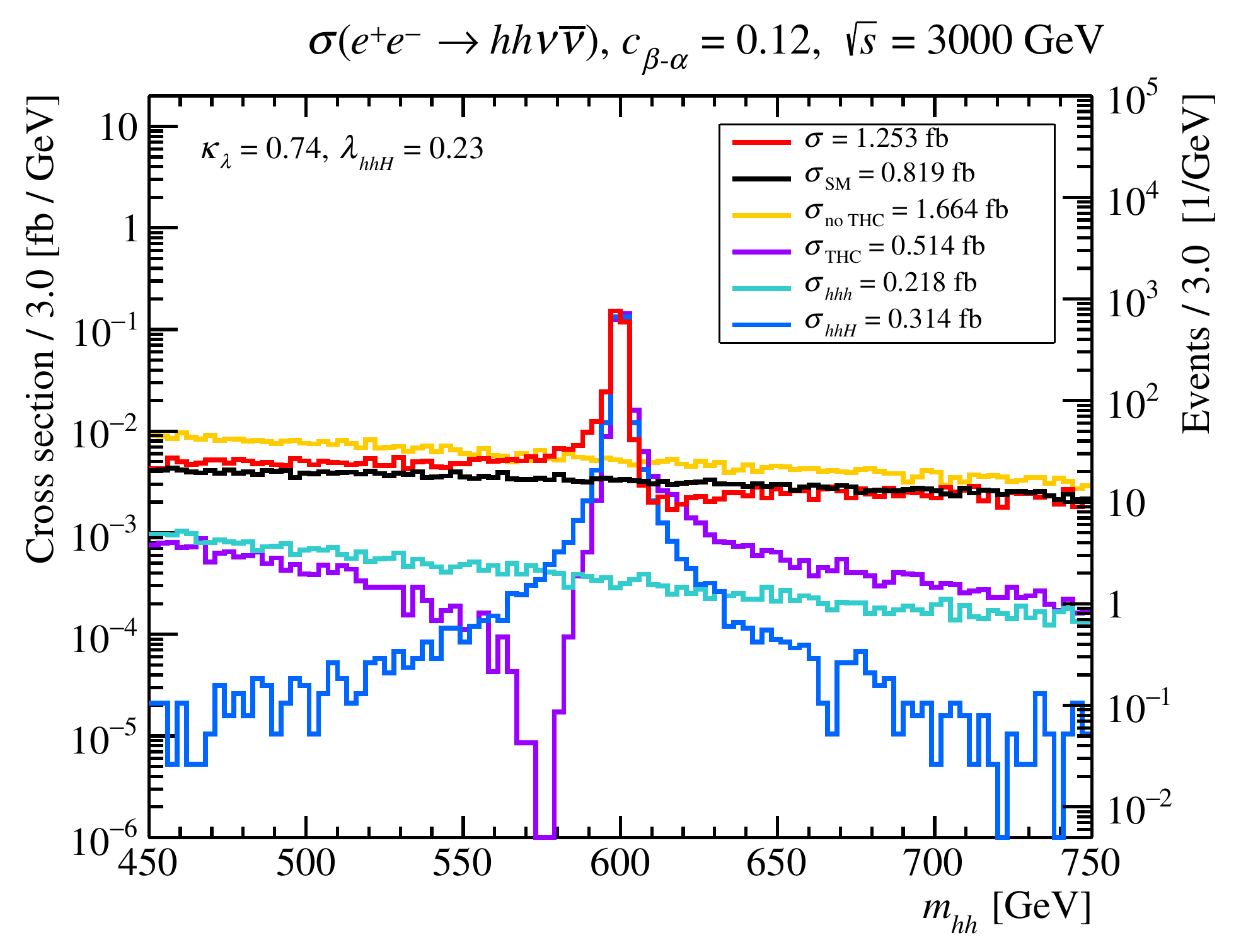}\\[2em]

\includegraphics[width=0.48\textwidth]{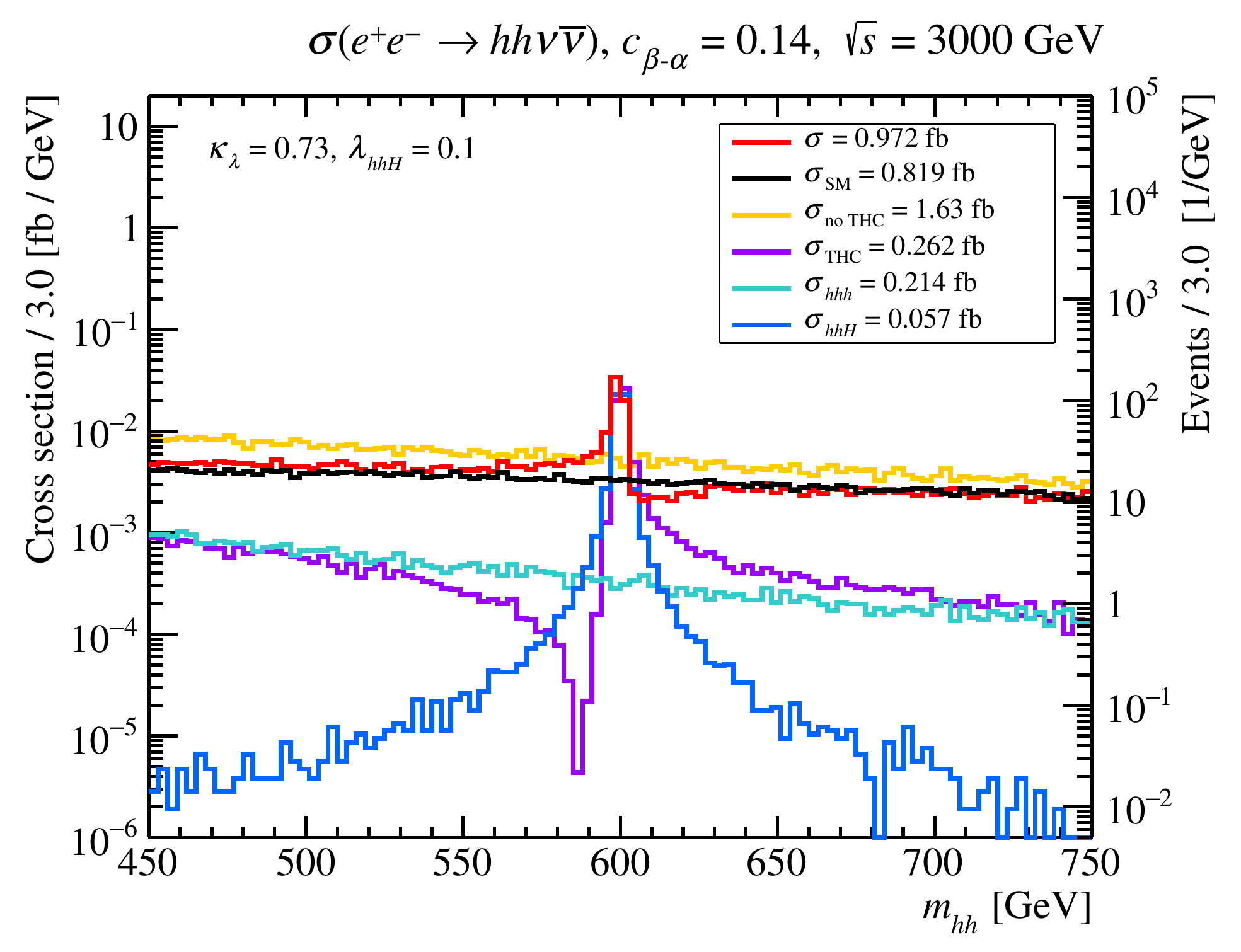}
\includegraphics[width=0.48\textwidth]{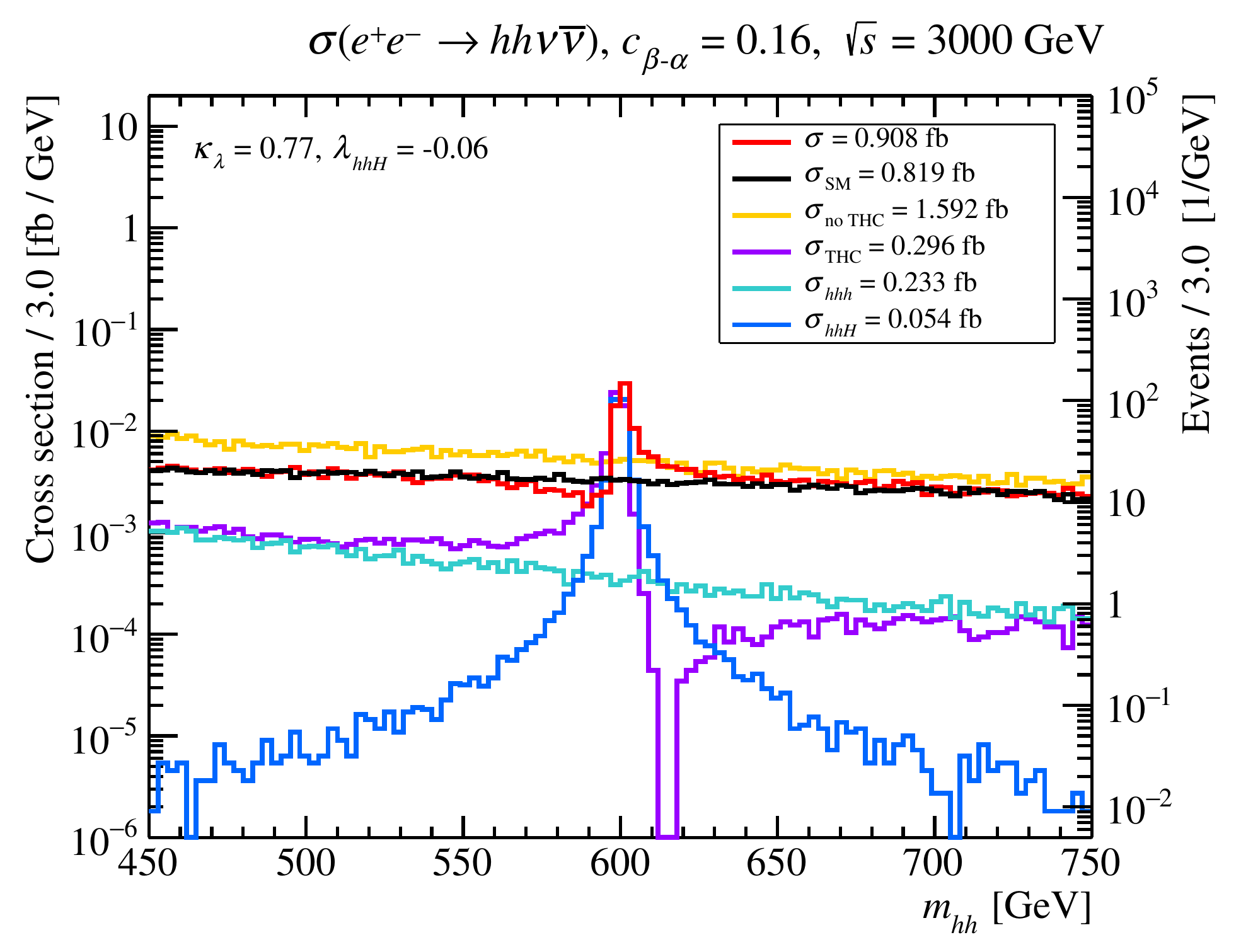}\\[2em]
		
\includegraphics[width=0.48\textwidth]{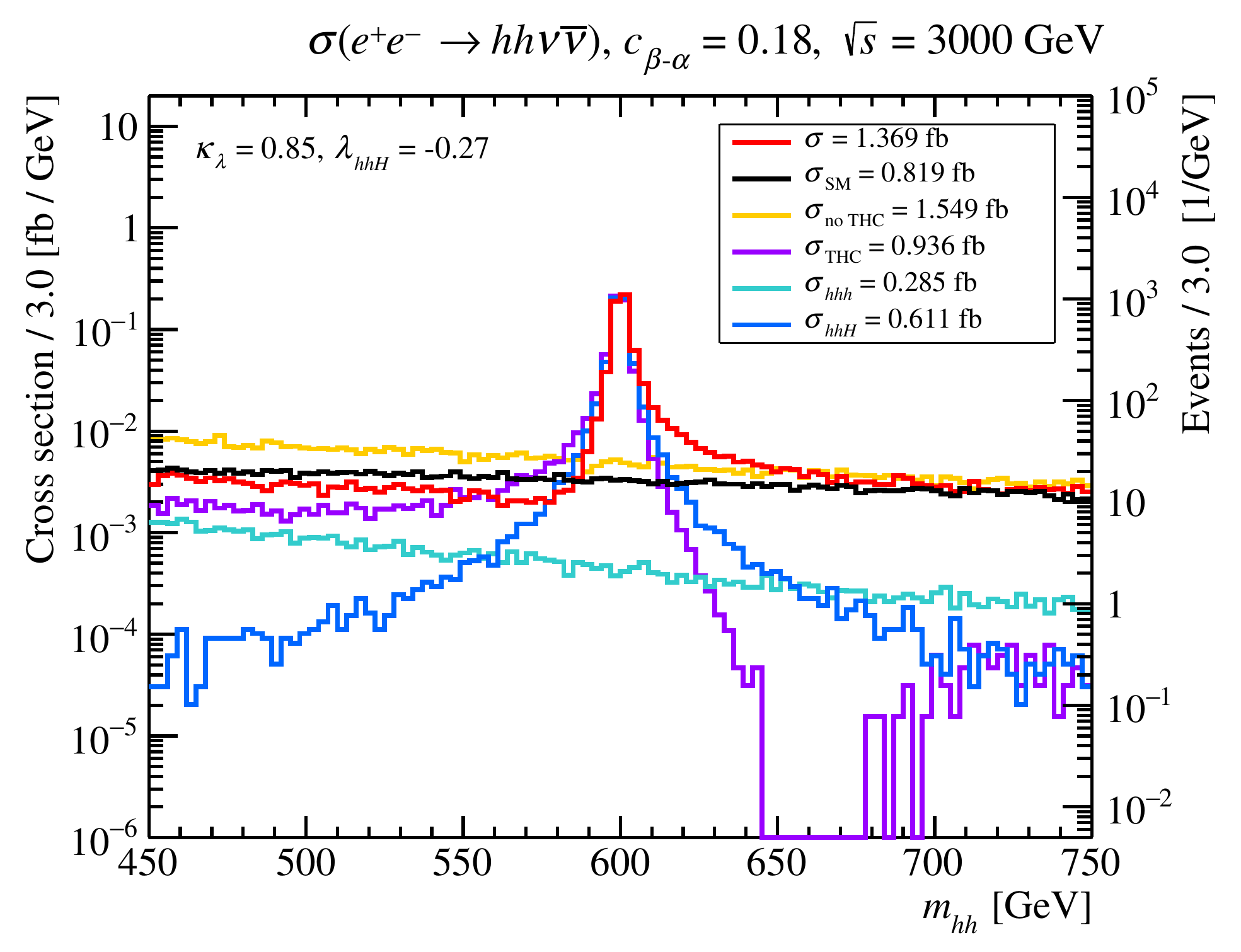}
\includegraphics[width=0.48\textwidth]{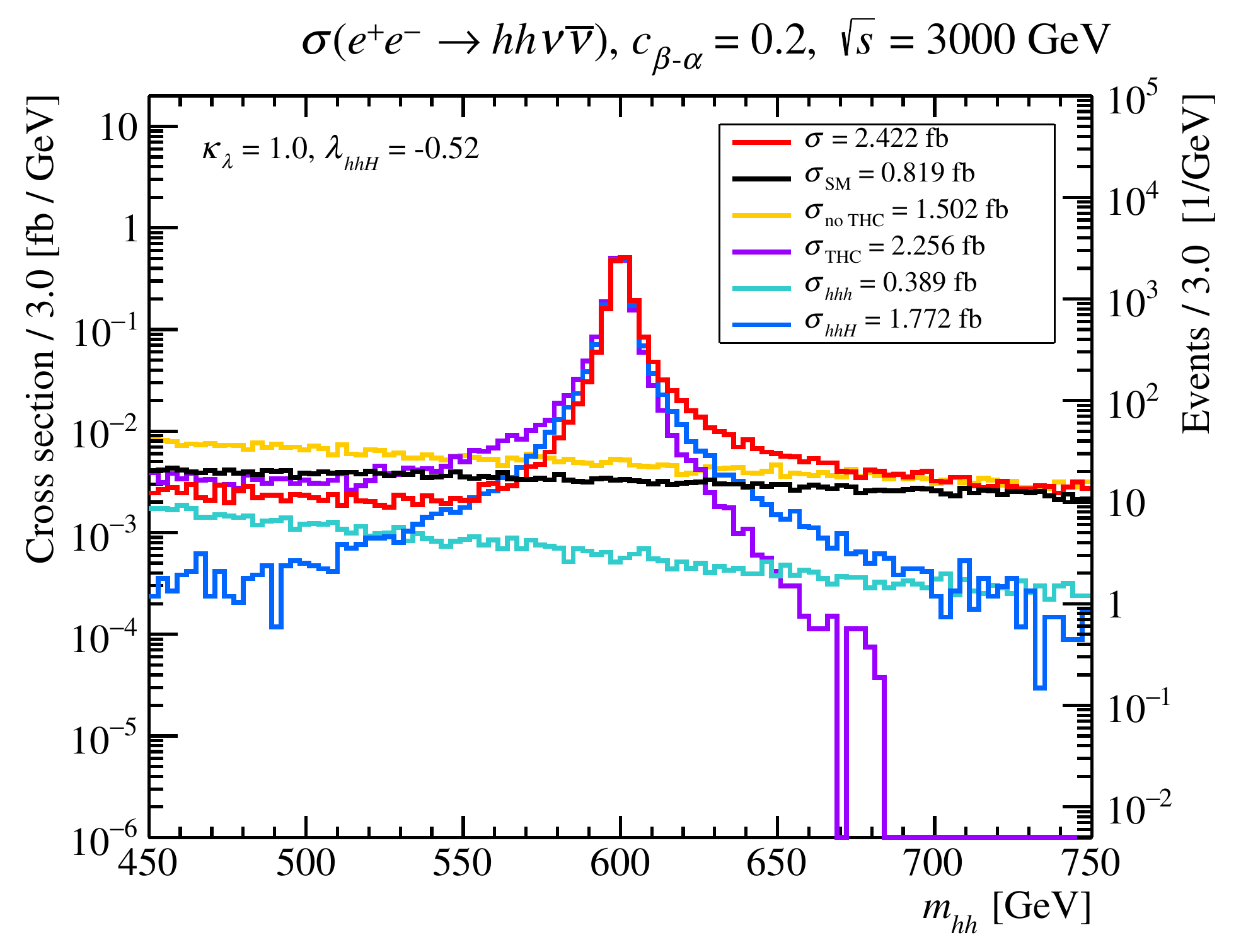}
	\end{center}
\caption{Evolution with $\CBA$ in benchmark point 3 of the invariant mass 
of the final $hh$ pair around the mass of the $H$ boson in the process
$e^+e^-\to hh\nu\bar{\nu}$ at $\sqrt{s}=3000\gev$.} 
\label{fig:BP3res}
\end{figure}

Overall,  the results of the various contributions to the differential
cross section given by the color lines in our plots in
\reffi{fig:mhh500},  \reffi{fig:mhh1000},  \reffi{fig:mhh1500} and
\reffi{fig:mhh3000} demonstrate explicitly that the dominant
contribution in the region of $m_{hh}$ close to the $m_H$ resonance  
stems clearly from the diagram carrying the $\lahhH$ (dark blue
line), and therefore the height of these resonant peaks in comparison
with the basis lying on the 
bottom of the resonance (given by the yellow line) provides the most
efficient access to this particular $\lahhH$.    
We furthermore remark that these figures display some sensitivity
to the sign of $\lahhH$, and not only to its absolute value.
There is an asymmetry in the peak of the $H$ boson
since the non-resonant diagrams are not negligible and
the relative sign of the resonant diagram changes when $m_{hh}=\MH$
(caused by a sign flip in the $H$-boson propagator).
This effect can be seen even more clearly in \reffi{fig:BP3res} (where we
show the ``family of BP3'' for various values of $\CBA$, which will
be introduced later at the end of this subsection): 
in the points where $\lahhH>0$ (corresponding to the points
with $\CBA=0.1,\ 0.12,\ 0.14$) the cross section at the left of the
$H$ peak is visibly larger than at the right of the peak. On the
other hand, in the cases with $\lahhH<0$ (corresponding to the points
with $\CBA=0.16,\ 0.18,\ 0.2$) the cross section at the left of the
peak is larger than at the right. The opposite effect due to the sign of
$\lahhH$ happens in the ``$H$ peak'' of the $hhZ$ channel, as can
be seen for example in the left panels of \reffis{fig:mhh500},
\ref{fig:mhh1000}, \ref{fig:mhh1500} and \ref{fig:mhh3000}.

To provide a more quantitative estimate of the sensitivity to
the leading triple coupling $\lahhH$ in this case of $hh$ production,
and in absence of a more realistic study including real backgrounds from
detector effects etc., we
propose here a more theoretical approach to define 'the significance of
the signal'.  'The signal' here refers to the specific events under the
resonant peaks which are the only ones really carrying the sensitivity
to  $\lahhH$.  Concretely,  we set this
interval as the one in between the crossings of the yellow and dark
blue lines. The events considered in this study are those
after the Higgs decays into $b \bar b$ pairs,  therefore producing
typically a final state signature with either four $b$-jets  and a $Z$
boson (presumably easily detectable at the clean $e^+e^-$ collider
environments) in the case of $hhZ$; or with four $b$-jets and
missing transverse energy, $\cancel{E}_{T}$,  
(from the undetected neutrinos) in the case of $hh \nu \bar \nu$.
The  number of events under the resonant peaks,  $N_{4bZ}^{R}$ for
$hhZ$, and  $N_{4b\cancel{E}_{T}}^{R}$ for $hh \nu \bar \nu$,  are then
extracted from these plots for the four BPs and the various considered
energies.   We also extract from these plots the event rates from the
non resonant contributions (the so-called continuum contributions),  in
the corresponding  $m_{hh}$ resonant region, named  $N_{4bZ}^{C}$ and
$N_{4b\cancel{E}_{T}}^{C}$,  for $hhZ$ and $hh\nu \bar \nu$
respectively,  which are defined by the yellow lines in these plots.
Finally,  we also extract  the  predictions of the corresponding 
SM event rates,  $N_{4bZ}^{\mathrm{SM}}$ and
$N_{4b\cancel{E}_{T}}^{\mathrm{SM}}$,  respectively,  for comparison.
All these event rates are obtained for the corresponding luminosities in
\refta{tab:ee} and include the BR factors of the two Higgs decays,
i.e.\  $(0.58)^2$.
 
More realistically, 
this $m_{hh}$ invariant mass should rather be considered as the
invariant mass of the 4-$b$-jets,  
$m_{4b}$ out of which each $bb$ pair reconstructs each of the two Higgs
bosons. This could be done,  presumably,  by  the proper cuts in the
$m_{2b}$ invariant mass window close to $m_h$.  But as stated above, we
simplify our analysis and choose to work with the theoretical variable
$m_{hh}$.  However,  
since the more realistic experimental detection of these signals will
require the tagging of the $b$-jets and the tagging of the missing
transverse energy in the $hh\nu \bar \nu $ case,  or the $Z$  boson in
the $hhZ$ case,  we have re-evaluated all the above rates, taking into
account  the detection acceptance,  ${\cal A}$,  and the $b$-jet tagging
efficiency,  $\epsilon_b$.  Concretely,  we evaluate the acceptance as
follows: 
\begin{equation}
{\cal A}=\frac{N_{\rm with \,\,cuts}}{N_{\rm without \,\,cuts}},
\end{equation}
where $N_{\rm without \,\,cuts}$ and $N_{\rm with \,\,cuts}$ are the
total event rates predictions without and with cuts applied,
respectively. 
For this study we again use {\tt MadGraph} and apply the following cuts to the
$b$-jets (meaning cuts applied to the $b$-quarks since we are not
dealing with jets),  missing transverse energy (for the neutrino
channel) and $Z$ transverse momentum (for the $Z$ channel),  which are
similar to those given  in~\cite{Abramowicz:2016zbo,Contino:2013gna}: 
 \begin{equation}
\label{cuts}
p_T^b>20\enspace \mathrm{GeV} ;\,\,\,\,
\vert\eta^b\vert<2 ;\,\,\,\,
p_T^Z>20\enspace \mathrm{GeV} ;\,\,\,\,
\Delta R_{bb}>0.4 ;\,\,\,\,
E\!\!\!\!/_T>20\enspace \mathrm{GeV},
\end{equation} 
where,  $p_T^b$,  $\eta^b$ are the transverse momentum and pseudo
rapidity of each of the four $b$-jets,  $p_T^Z$ the transverse
momentum of the $Z$, and
$\Delta R_{bb}\equiv \sqrt{(\Delta \eta_{bb})^2+(\Delta \phi_{bb})^2}$,
with $\Delta\eta_{bb}$ and $\Delta\phi_{bb}$,  are
the separations in pseudorapidity and azimuthal angle of the two
$b$-jets,  respectively,  which are identified coming from one light
Higgs decay.
With these cuts we get the values of the acceptances for the various
BPs and for the four studied energies that are displayed  in
\refta{tab:RhhZ} for the $hhZ$ channel and in \refta{tab:Rhhnunubar}
for the $hh\nu \bar \nu$ channel.  As one can see,  the acceptance
varies with energy, as expected,  and we find the best values of around
0.70 for the lowest energies.  At the highest energy of $3000 \gev$ the
acceptance for $hhZ$ gets reduced to around $0.1$ and for $hh\nu \bar\nu$
to around $0.5$.  These acceptances are close to the SM ones
(slightly better, indeed) which we have also included in these tables,
for comparison.   

For the numerical estimates of the more realistic event rates that
are involved in our study of the resonant peaks for the various BPs,
$\bar N$,   we then apply the reduction factors, given by $ {\cal A} $
and $\epsilon_b$,  to the previous event rates, $N$,  as follows: 
\begin{equation}
\bar N =N \times {\cal A} \times (\epsilon_b)^4,
\end{equation}
where we set the value of  the $b$-jet tagging efficiency in our
numerical evaluations to $\epsilon _b= 0.8$  (see, e.g., 
\citeres{Asner:2013psa,Abramowicz:2016zbo}).  
To evaluate the size of the $\lahhH$ effect we compute the following
ratio~$R$:  
\begin{equation}
R=\frac{{\bar N}^R-{\bar N}^C}{\sqrt{{\bar N}^C}}~,
\end{equation}
which is our theoretical estimator of the 'sensitivity' to $\lahhH$.  

We  collect all the results for the corresponding
${\bar N}^R$,  ${\bar N}^C$ and $R$ in \refta{tab:RhhZ} and
\refta{tab:Rhhnunubar} for $hhZ$ and $hh\nu \bar \nu$,  respectively.   
As one can see in these tables,  the 
number of signal events are quite significant in most cases. For the
$hhZ$ channel,  ${\bar N}_{4bZ}^{R}$ decreases with the collider energy
and leads the largest $R$ value for the BP1 at $1000 \gev$.  For BP2 and
BP3 the rates are clearly lower and the largest $R$ values found are
again at $1000 \gev$.  For BP4 the rates are too low to be detected and we
find no sensitivity to $\lahhH$ in this $hhZ$ channel. For the  
$hh\nu \bar \nu$ channel the signal rates are clearly larger (except for
BP1 at $500 \gev$) and,  in contrast to  $hhZ$,  increase with the energy
collider.  We find high sensitivity to $\lahhH$ in all the studied
cases, except for BP4 at $1500 \gev$, where the rates are too low.  The
highest sensitivities, corresponding to the highest $R$ values
above 100, are found for BP1 for the three center-of-mass energies
$1000, 1500$ and $3000 \gev$.  BP3 at $3000 \gev$ also shows a
large $R$ value of 100.  BP2 and BP4 reach their maximum $R$ values, 48
and 34 respectively,  also  at $3000 \gev$.

\begin{table}[t!]
\begin{center}
\begin{tabular}{|c|c|c|c|c|c|}
\hline 
$hhZ$ & $\sqrt{s}$ {[}GeV{]} & $\sigma_{\mathrm{2HDM}}$ / $\sigma_{\mathrm{SM}}$ {[}fb{]} & $\bar{N}_{4bZ}^{R}$ / $\bar{N}_{4bZ}^{C}$ / $\bar{N}_{4bZ}^{\mathrm{SM}}$ & $\mathcal{A}_{\mathrm{2HDM}}/$ $\mathcal{A}_{\mathrm{SM}}$ & $R_{4bZ}$ \tabularnewline
\hline 
\hline 
\multirow{4}{*}{BP1} & 500 & 1.063 / 0.158 & 193 / 10 / 3 & 0.70 / 0.68 & 58\tabularnewline
\cline{2-6} \cline{3-6} \cline{4-6} \cline{5-6} \cline{6-6} 
 & 1000 & 0.913 / 0.120 & 206 / 1 / 4 & 0.70 / 0.71 & 205\tabularnewline
\cline{2-6} \cline{3-6} \cline{4-6} \cline{5-6} \cline{6-6} 
 & 1500 & 0.493 / 0.077 & 22 / $<1$ / 1 & 0.51 / 0.62 & -\tabularnewline
\cline{2-6} \cline{3-6} \cline{4-6} \cline{5-6} \cline{6-6} 
 & 3000 & 0.147 / 0.033 & 1 / $<1$ / $<1$  & 0.05 / 0.05 & -\tabularnewline
\hline 
\hline 
\multirow{3}{*}{BP2} & 1000 & 0.156 / 0.120 & 20 / 1 / 1 & 0.73 / 0.71 & 19\tabularnewline
\cline{2-6} \cline{3-6} \cline{4-6} \cline{5-6} \cline{6-6} 
 & 1500 & 0.106 / 0.077 & 4 / $<1$ / $<1$  & 0.65 / 0.62 & -\tabularnewline
\cline{2-6} \cline{3-6} \cline{4-6} \cline{5-6} \cline{6-6} 
 & 3000 & 0.042 / 0.033 & $<1$ / $<1$ / $<1$  & 0.07 / 0.05 & -\tabularnewline
\hline 
\hline 
\multirow{3}{*}{BP3} & 1000 & 0.254 / 0.120 & 29 / 5 / 2 & 0.71 / 0.71 & 11\tabularnewline
\cline{2-6} \cline{3-6} \cline{4-6} \cline{5-6} \cline{6-6} 
 & 1500 & 0.218 / 0.077 & 8 / 1 / $<1$  & 0.70 / 0.62 & 7\tabularnewline
\cline{2-6} \cline{3-6} \cline{4-6} \cline{5-6} \cline{6-6} 
 & 3000 & 0.086 / 0.033 & 1 / $<1$ / $<1$  & 0.08 / 0.05 & -\tabularnewline
\hline 
\hline 
\multirow{2}{*}{BP4} & 1500 & 0.075 / 0.077 & 1 / $<1$ / $<1$  & 0.64 / 0.62 & -\tabularnewline
\cline{2-6} \cline{3-6} \cline{4-6} \cline{5-6} \cline{6-6} 
 & 3000 & 0.038 / 0.033 & $<1$ / $<1$ / $<1$  & 0.07 / 0.05 & -\tabularnewline
\hline 
\end{tabular}
\caption{$R_{4bZ}$ for the BPs at all relevant $\sqrt{s}$.
    Also shown are total cross sections, $\sig$, event numbers, $\bar N$,
    and acceptances, ${\cal A}$, in the 2HDM and the SM, as defined in
    the text. ``-'' indicates values that cannot be evaluated due to a too
small number of events.}
\label{tab:RhhZ}
\par\end{center}
\vspace{1em}
\end{table}

\begin{table}[ht!]
\vspace{-1em}
\begin{center}
\begin{tabular}{|c|c|c|c|c|c|}
\hline 
$hh\nu\bar{\nu}$ & $\sqrt{s}$ {[}GeV{]} & $\sigma_{\mathrm{2HDM}}$ / $\sigma_{\mathrm{SM}}$ {[}fb{]} & $\bar{N}_{4b\cancel{E}_{T}}^{R}$ / $\bar{N}_{4b\cancel{E}_{T}}^{C}$
/ $\bar{N}_{4b\cancel{E}_{T}}^{\mathrm{SM}}$ & $\mathcal{A}_{\mathrm{2HDM}}/$ $\mathcal{A}_{\mathrm{SM}}$ & $R_{4b\cancel{E}_{T}}$\tabularnewline
\hline 
\hline 
\multirow{4}{*}{BP1} & 500 & 0.404 / 0.034 & 119 / 4 / 1 & 0.70 / 0.68 & 58\tabularnewline
\cline{2-6} \cline{3-6} \cline{4-6} \cline{5-6} \cline{6-6} 
 & 1000 & 2.391 / 0.097 & 1510 / 24 / 0 & 0.65 / 0.55 & 303\tabularnewline
\cline{2-6} \cline{3-6} \cline{4-6} \cline{5-6} \cline{6-6} 
 & 1500 & 4.423 / 0.239 & 794 / 13 / 2 & 0.58 / 0.41 & 217\tabularnewline
\cline{2-6} \cline{3-6} \cline{4-6} \cline{5-6} \cline{6-6} 
 & 3000 & 9.098 / 0.819 & 2425 / 46 / 6 & 0.44 / 0.25 & 351\tabularnewline
\hline 
\hline 
\multirow{3}{*}{BP2} & 1000 & 0.234 / 0.097 & 79 / 3 / 1 & 0.65 / 0.55 & 44\tabularnewline
\cline{2-6} \cline{3-6} \cline{4-6} \cline{5-6} \cline{6-6} 
 & 1500 & 0.625 / 0.239 & 70 / 3 / 1 & 0.56 / 0.41 & 39\tabularnewline
\cline{2-6} \cline{3-6} \cline{4-6} \cline{5-6} \cline{6-6} 
 & 3000 & 1.850 / 0.819 & 282 / 28 / 9 & 0.41 / 0.25 & 48\tabularnewline
\hline 
\hline 
\multirow{3}{*}{BP3} & 1000 & 0.208 / 0.097 & 85 / 5 / 3 & 0.66 / 0.55 & 36\tabularnewline
\cline{2-6} \cline{3-6} \cline{4-6} \cline{5-6} \cline{6-6} 
 & 1500 & 0.709 / 0.239 & 111 / 5 / 3 & 0.61 / 0.41 & 47\tabularnewline
\cline{2-6} \cline{3-6} \cline{4-6} \cline{5-6} \cline{6-6} 
 & 3000 & 2.422 / 0.819 & 577 / 30 / 11 & 0.47 / 0.25 & 100\tabularnewline
\hline 
\hline 
\multirow{2}{*}{BP4} & 1500 & 0.428 / 0.239 & 4 / $<1$ / $<1$  & 0.50 / 0.41 & -\tabularnewline
\cline{2-6} \cline{3-6} \cline{4-6} \cline{5-6} \cline{6-6} 
 & 3000 & 1.523 / 0.819 & 72 / 4 / 3 & 0.38 / 0.25 & 34\tabularnewline
\hline 
\end{tabular}
\caption{$R_{4b\cancel{E}_{T}}$ for the BPs at all relevant
    $\sqrt{s}$. 
    Also shown are total cross sections, $\sig$, event numbers, $\bar N$,
    and acceptances, ${\cal A}$, in the 2HDM and the SM, as defined in
    the text. ``-'' indicates values that cannot be evaluated due to a too
small number of events.}
\label{tab:Rhhnunubar}
\par\end{center}
\end{table}

All in all,  we conclude that  the process $e^+e^- \to hh \nu \bar \nu$
seems very promising to give access to the triple $\lahhH$ coupling  at
all the studied energies,  since very high values of our estimator $R$
are found in all the studied 2HDM points (except BP4 at 1500 GeV).  The
highest sensitivities,  indicated by the highest values of $R$ in this
$hh\nu \bar \nu$ channel, are found at $3000 \gev$.  The channel $hhZ$ can
also give access to this triple  $\lahhH$ coupling,  but at the lower
energy colliders  and for the 2HDM points with a relatively light
$H$~boson. The  highest sensitivities in this case are reached for BP1 at
$500 \gev$ and $1000 \gev$.

\begin{figure}[ht!]
	\begin{center}
\includegraphics[width=0.48\textwidth]{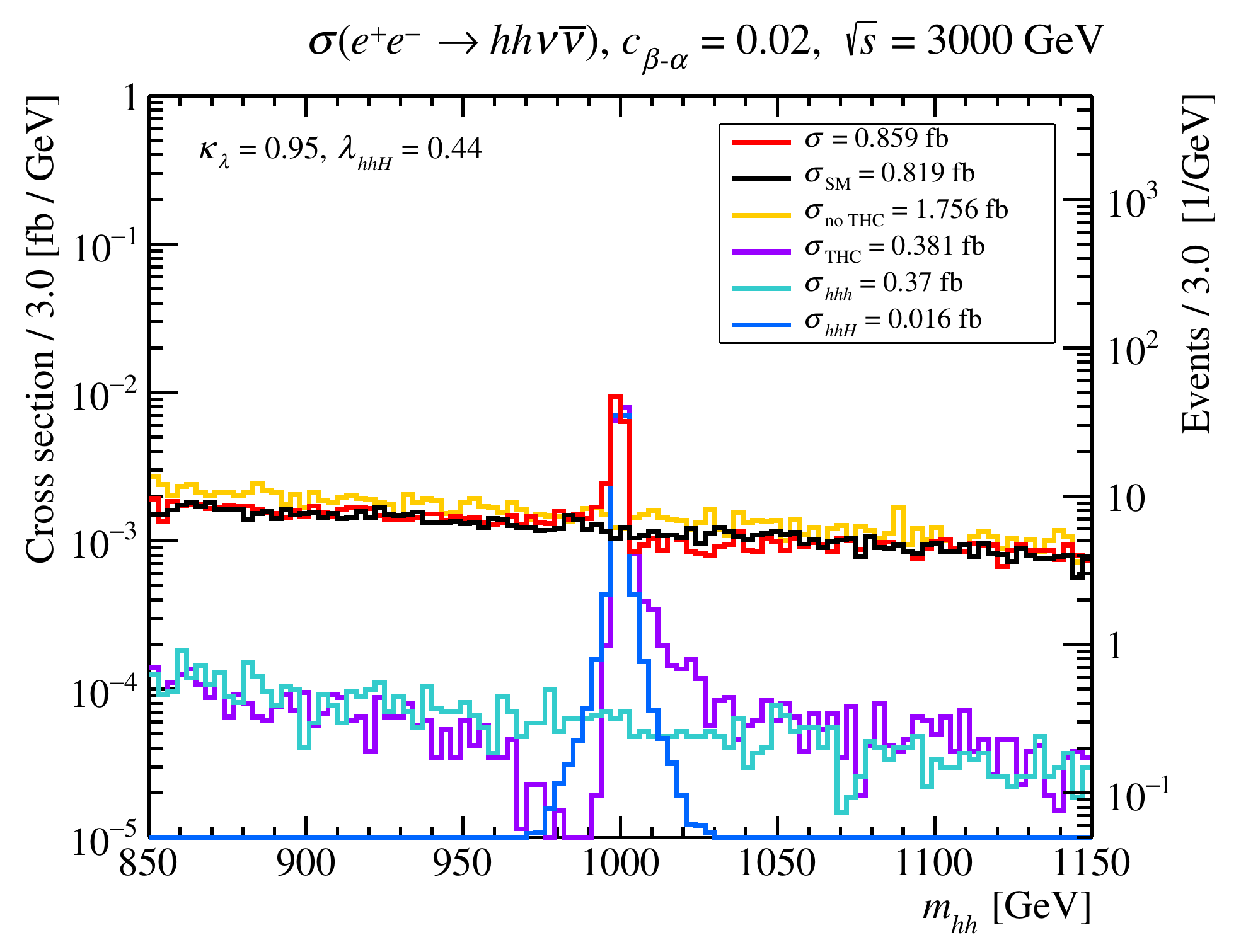}
\includegraphics[width=0.48\textwidth]{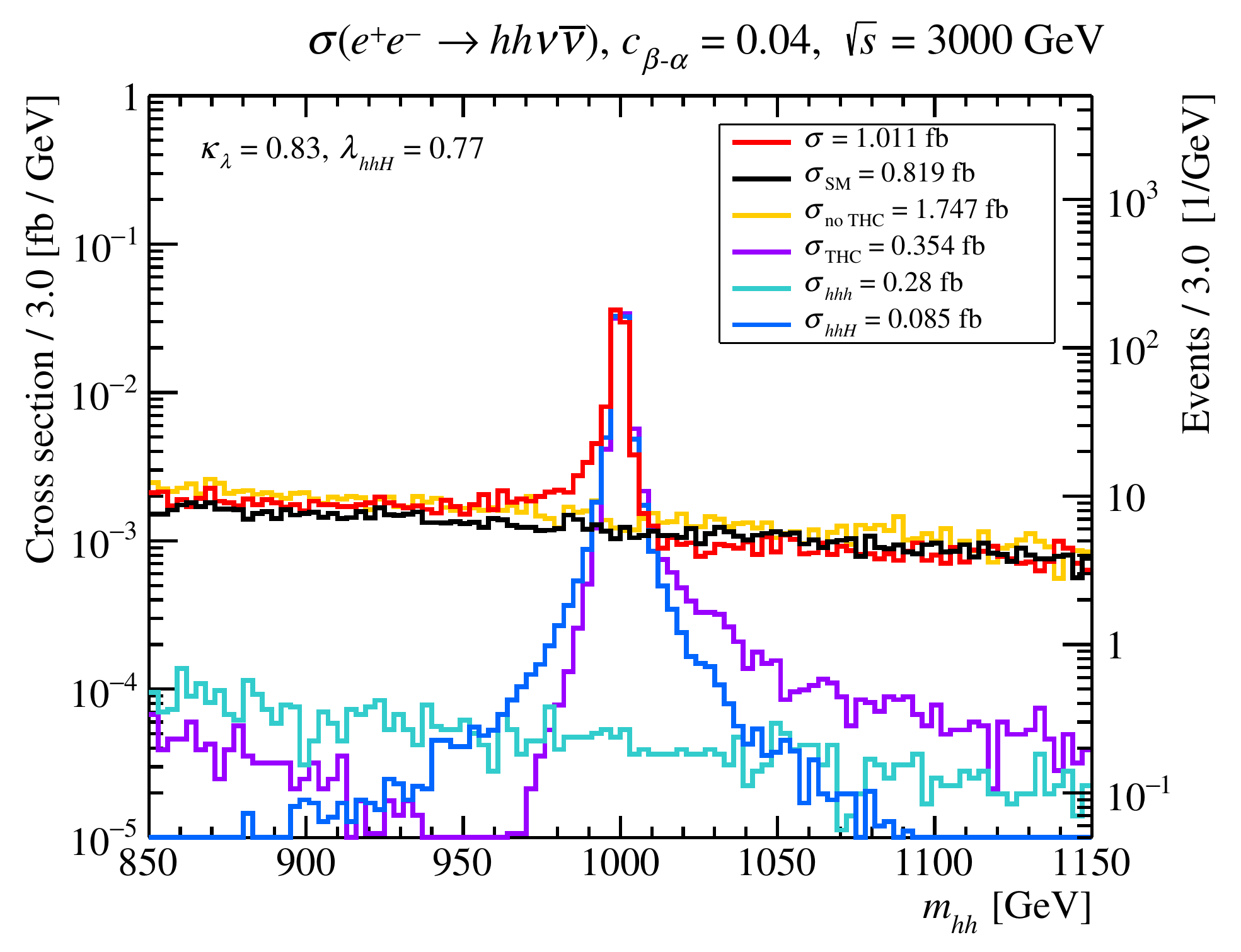}\\[2em]

\includegraphics[width=0.48\textwidth]{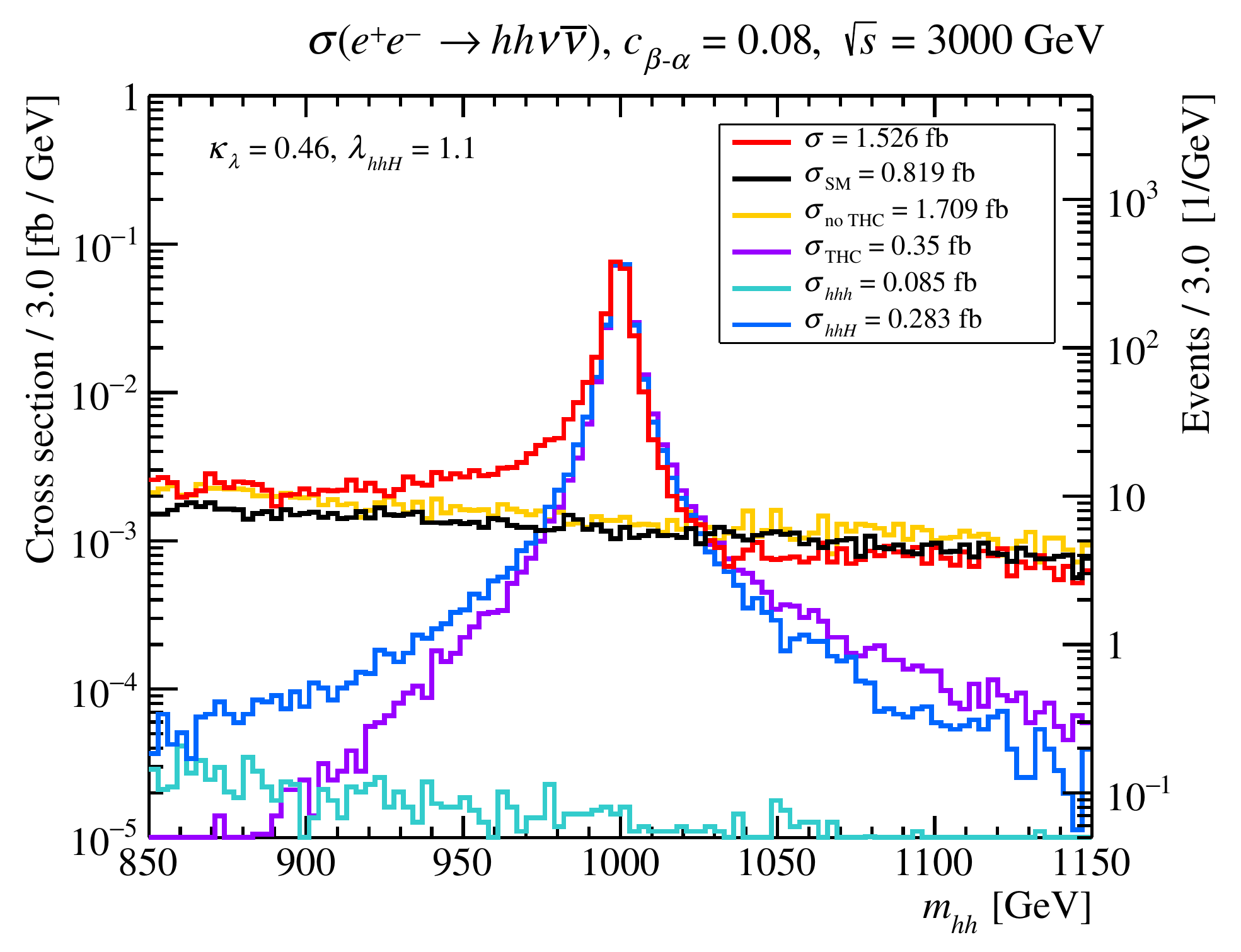}
\includegraphics[width=0.48\textwidth]{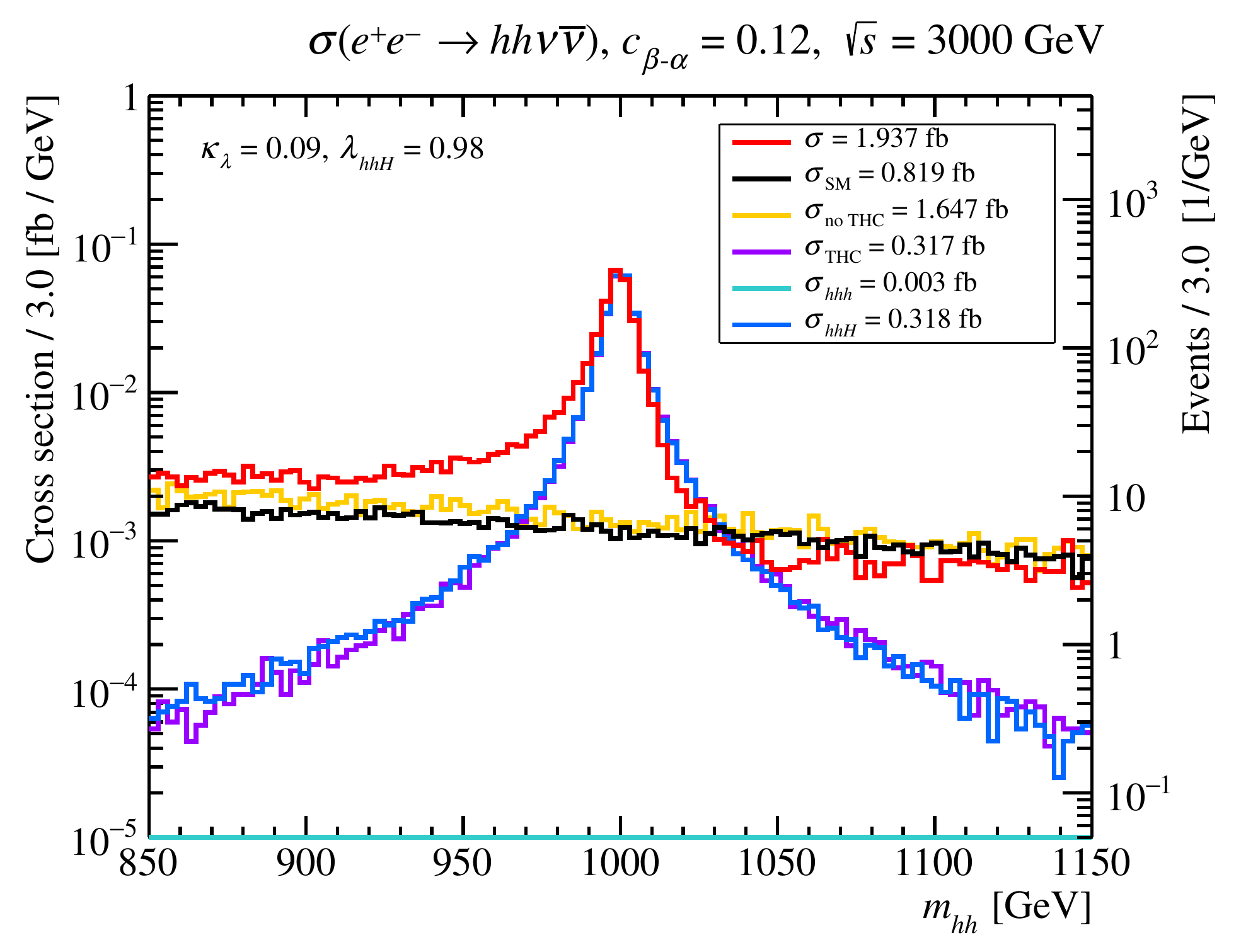}\\[2em]
		
\includegraphics[width=0.48\textwidth]{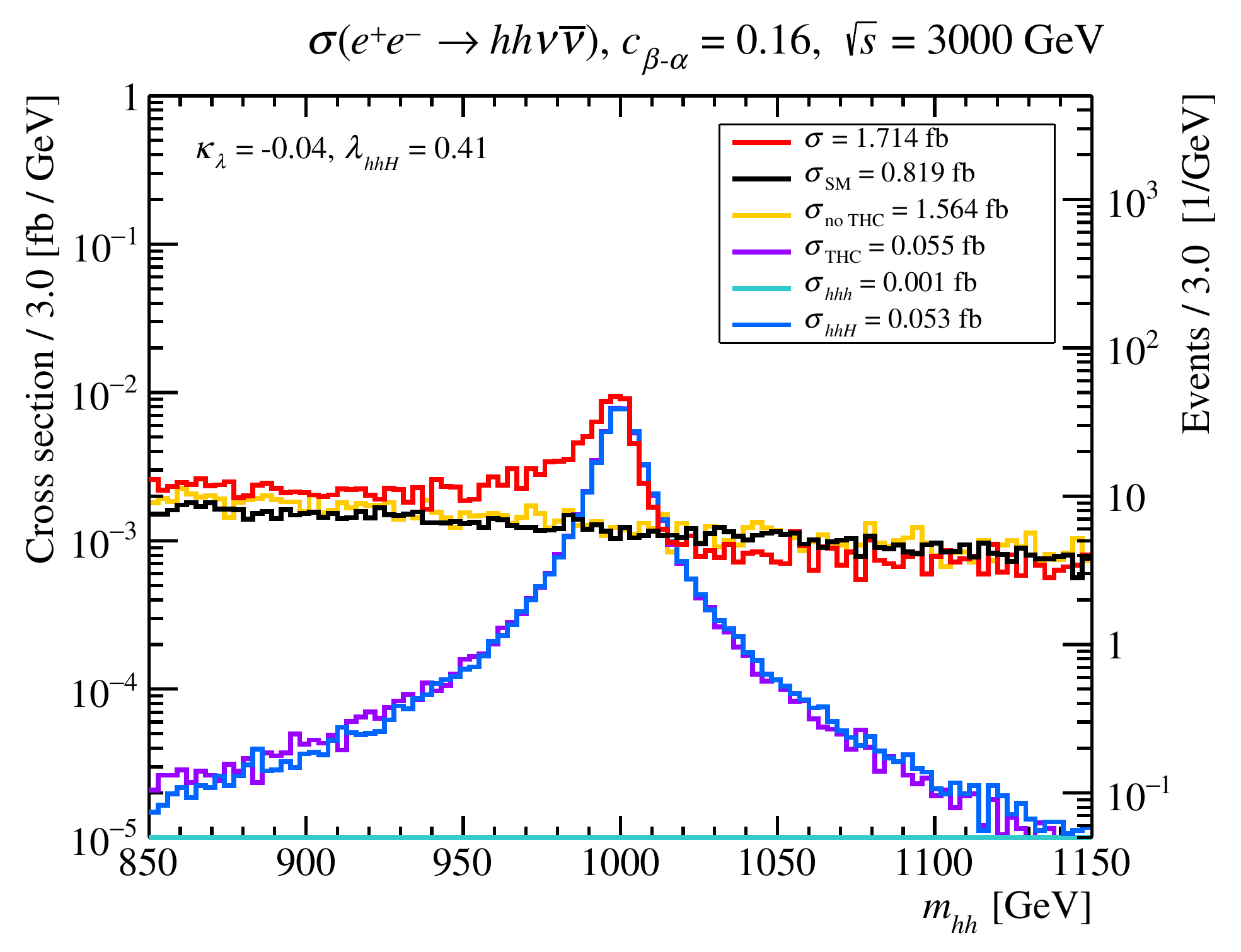}
\includegraphics[width=0.48\textwidth]{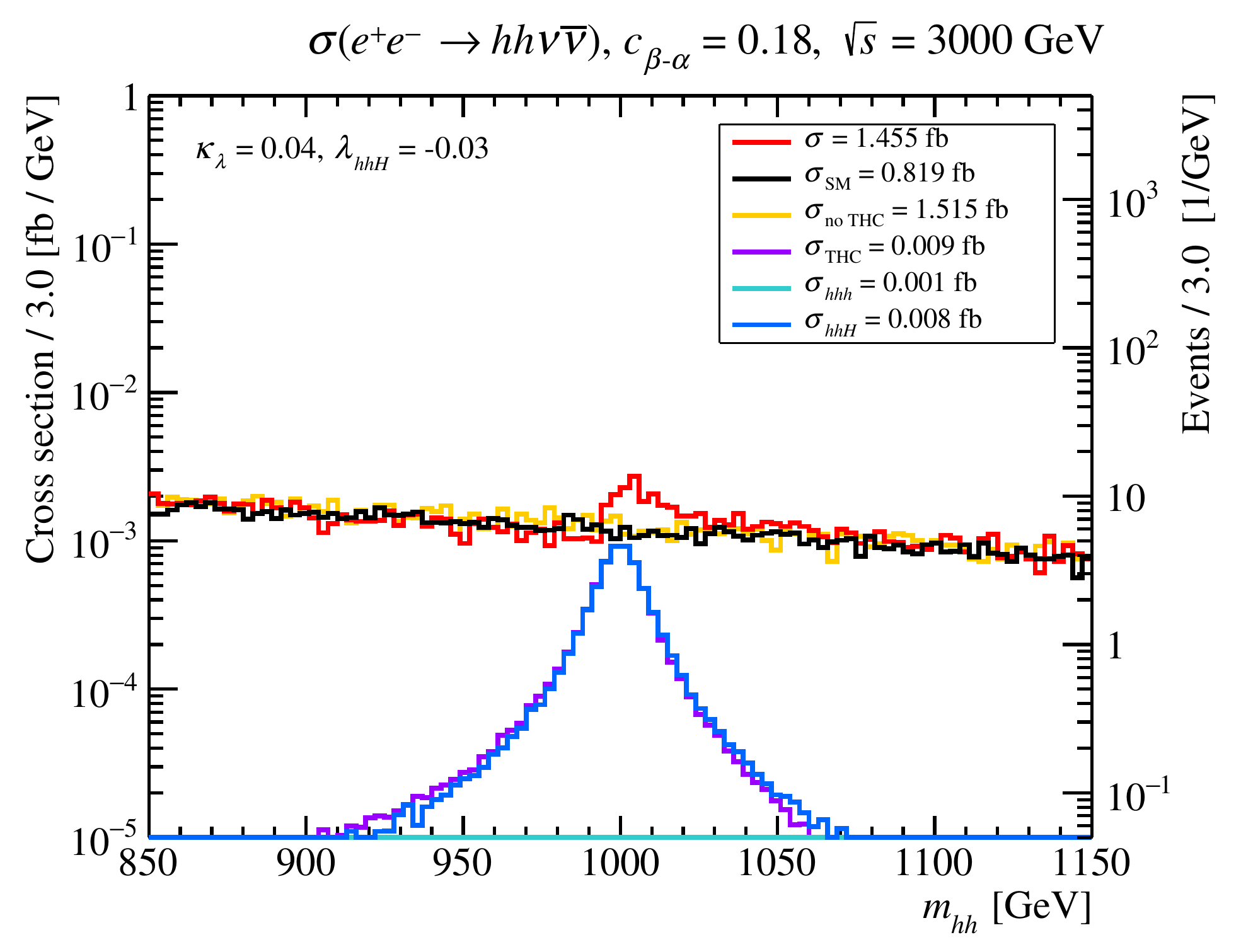}
	\end{center}
\caption{Evolution with $\CBA$ in benchmark point 4 of the invariant mass 
of the final $hh$ pair around the mass of the $H$ boson in the process
$e^+e^-\to hh\nu\bar{\nu}$ at $\sqrt{s}=3000\gev$.} 
\label{fig:BP4res}
\end{figure}

\begin{table}[t!]
\vspace{-1em}
\begin{center}
\begin{tabular}{|c|c|c|c|c|c|}
\hline 
$c_{\beta-\alpha}$ & $\lahhH$ & $\Gamma_{H}$ {[}GeV{]} & $\sigma_{\mathrm{2HDM}}$ {[}fb{]} & $\bar{N}_{4b\cancel{E}_{T}}^{R}$ / $\bar{N}_{4b\cancel{E}_{T}}^{C}$
/ $\bar{N}_{4b\cancel{E}_{T}}^{\mathrm{SM}}$ & $R_{4b\cancel{E}_{T}}$\tabularnewline
\hline\hline
0.1 & 0.30 & 1.24 & 1.434 & 167 / 7 / 2 & 60\tabularnewline
\hline 
0.12 & 0.23 & 1.51 & 1.253 & 97 / 7 / 2 & 34\tabularnewline
\hline 
0.14 & 0.10 & 1.88 & 0.972 & 17 / 3 / 1 & 8\tabularnewline
\hline 
0.16 & -0.06 & 2.48 & 0.908 & 15 / 3 / 1 & 7\tabularnewline
\hline 
0.18 & -0.27 & 3.47 & 1.369 & 195 / 13 / 5 & 50\tabularnewline
\hline 
0.2 (BP3) & -0.52 & 5.08 & 2.422 & 577 / 30 / 11 & 100\tabularnewline
\hline 
\end{tabular}
\caption{$R_{4b\cancel{E}_{T}}$ for BP3 for different $\CBA$ at
    $\sqrt{s} = 3000 \gev$. 
    Also shown are $\lahhH$, the total widths $\Gamma_H$, total cross
    sections, $\sig$, and event numbers, $\bar N$, as defined in
    the text. ``-'' indicates values that cannot be evaluated due to a too
small number of events.}
\label{tab:Rhhnunubar-BP3primes}
\par\end{center}
\end{table}

\begin{table}[t!]
\begin{center}
\begin{tabular}{|c|c|c|c|c|c|}
\hline 
$c_{\beta-\alpha}$ & $\lahhH$ & $\Gamma_{H}$ {[}GeV{]} & $\sigma_{\mathrm{2HDM}}$ {[}fb{]} & $\bar{N}_{4b\cancel{E}_{T}}^{R}$ / $\bar{N}_{4b\cancel{E}_{T}}^{C}$
/ $\bar{N}_{4b\cancel{E}_{T}}^{\mathrm{SM}}$ & $R_{4b\cancel{E}_{T}}$\tabularnewline
\hline\hline
0.02 & 0.44 & 1.05 & 0.855 & 4 / 1 / 0 & 3\tabularnewline
\hline 
0.04 & 0.77 & 2.38 & 1.009 & 21 / 2 / 1 & 13\tabularnewline
\hline 
0.08 (BP4) & 1.10 & 5.80 & 1.526 & 72 / 4 / 3 & 34\tabularnewline
\hline 
0.12 & 0.98 & 8.77 & 1.935 & 82 / 7 / 4 & 28\tabularnewline
\hline 
0.16 & 0.41 & 12.2 & 1.706 & 12 / 3 / 2 & 5\tabularnewline
\hline 
0.18  & -0.03 & 15.2 & 1.450 & - & -\tabularnewline
\hline 
\end{tabular}
\caption{$R_{4b\cancel{E}_{T}}$ for BP4 for different $\CBA$ at
    $\sqrt{s} = 3000 \gev$. 
    Also shown are $\lahhH$, the total widths $\Gamma_H$, total cross
    sections, $\sig$, and event numbers, $\bar N$, as defined in
    the text. ``-'' indicates values that cannot be evaluated due to a too
small number of events.}
\label{tab:Rhhnunubar-BP4primes}
\par\end{center}
\end{table}

After having explored the sensitivity to $\lahhH$ in all the four
BPs of the 2HDM type~I, we will now analyze the relevance of the
choice of $\CBA$ for the found sensitivities. We
have then repeated 
our previous  study  by means of the $R$ variable,  but  now varying
$\CBA$ for each BP,  whereas the other input parameters are kept at
their original value. In this part of the analysis we
focus now on the channel and energy with the highest
sensitivity, namely $hh \nu \bar \nu$  at $3000 \gev$, and present two
examples of BP related points: the case of 'the family of BP3'  in
\refta{tab:Rhhnunubar-BP3primes}, and the case of 'the family of
BP4' in \refta{tab:Rhhnunubar-BP4primes}. 
We have also included in these tables the corresponding predictions of
the triple coupling 
$\lahhH$ for those 'family points'.  The corresponding
predictions for the cross section distributions with the invariant mass
$m_{hh}$ are displayed in \reffi{fig:BP3res} and \reffi{fig:BP4res} for
'the family of BP3' and the family of BP4', respectively.  In these
figures, we see clearly how both the resonant peak and 'the continuum'
evolve with $\CBA$ in each of the two studied cases. 
To conclude on the significances of these new resonant signals and their
sensitivity to $\lahhH$, we again compare the rates under the resonant
peak (red lines) with the rates under the continuum (yellow lines), in
the $m_{hh}$ interval defined by the two crossings of the yellow and
dark blue lines. Then we compute the corresponding
$R_{4b\cancel{E}_{T}}$ value using the same acceptances and $b$-tagging
efficiencies as before.  The corresponding results are shown in
\refta{tab:Rhhnunubar-BP3primes} and \refta{tab:Rhhnunubar-BP4primes}
for 'the family of BP3' and 'the family of BP4', respectively.
One can see that the sensitivity, given by $R_{4b\cancel{E}_{T}}$,
directly correlates with the value of $|\lahhH|$. Since the benchmark
points were defined to reach large values of $|\lahhH|$ (within each
family), we find that 
the highest sensitivity found is precisely for the parent BP3 and BP4.
Similar results and conclusions
are found for the other 'family points' (not shown for brevity).



\subsection{Sensitivity to triple Higgs couplings in
\boldmath{$hH$} and \boldmath{$HH$}\\ production} 
\label{sec:HH-sens}

In this section  we study the sensitivity to the triple Higgs couplings
in $hH$ and $HH$ production.  We have selected in these two cases the
highest collider energy of $3 \tev$, where the effects of the triple
couplings have been found to be the largest ones.  

\begin{figure}[ht!]
\vspace{-1em}
  \begin{center}
\includegraphics[width=0.41\textwidth]{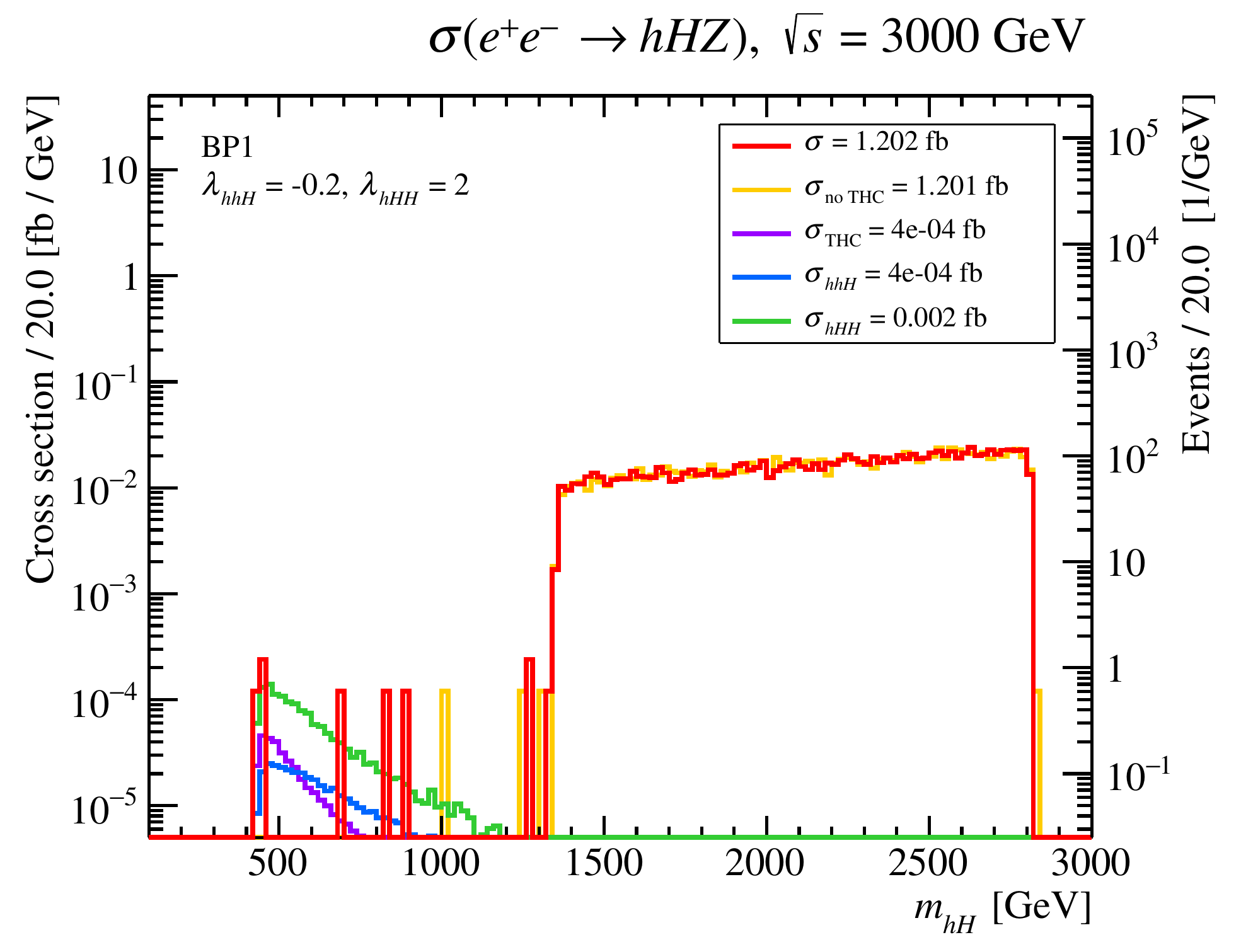}
\includegraphics[width=0.41\textwidth]{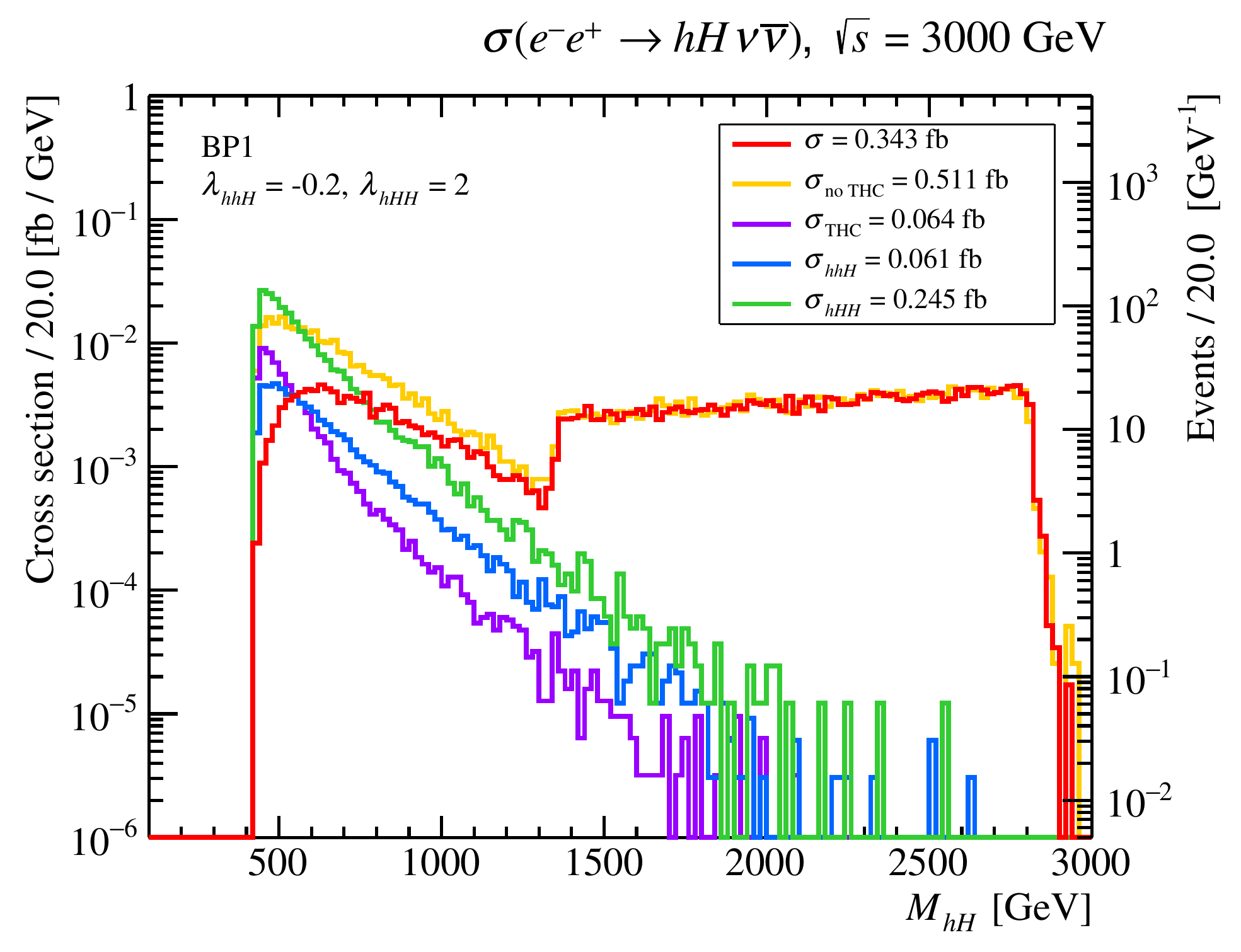}

\includegraphics[width=0.41\textwidth]{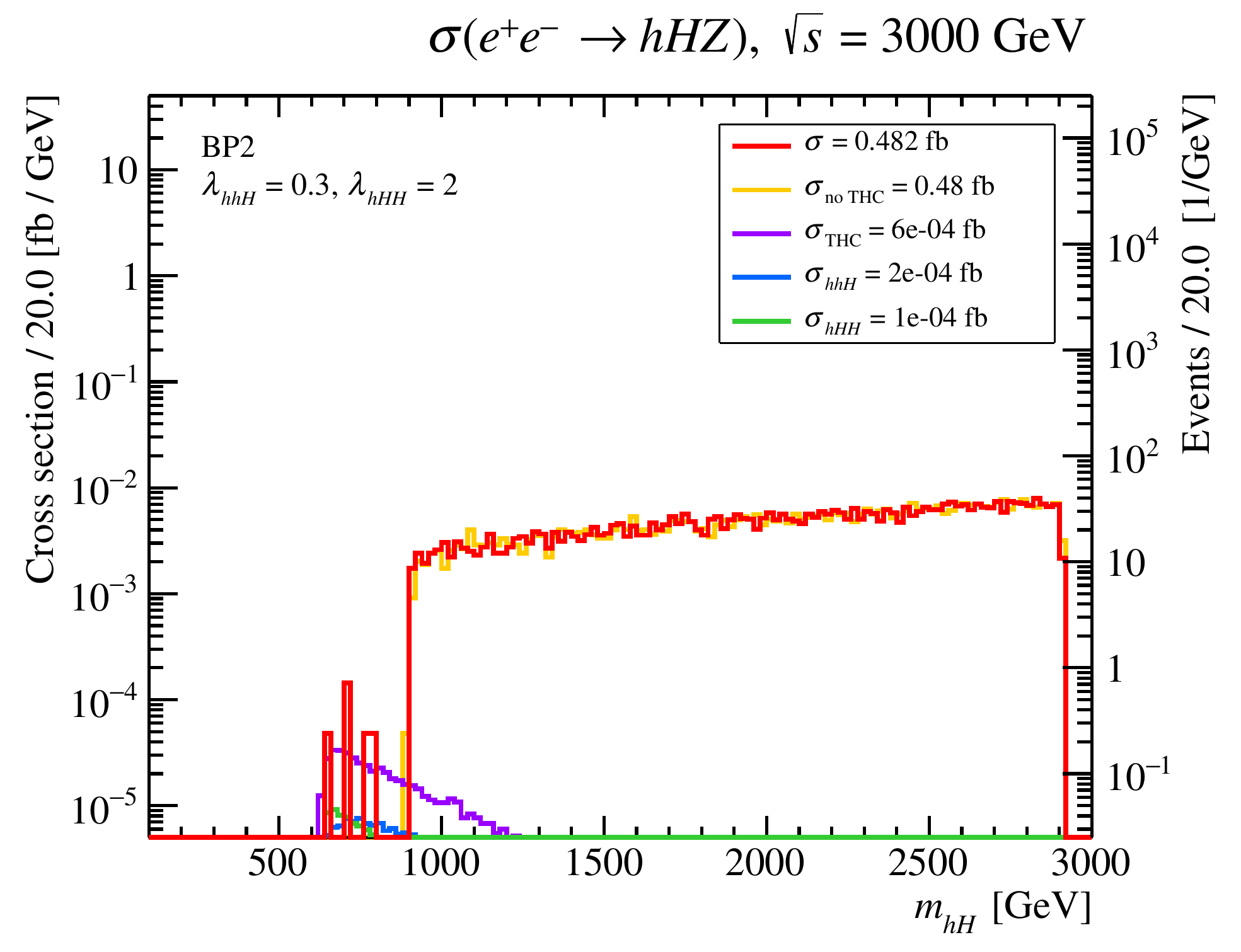}
\includegraphics[width=0.41\textwidth]{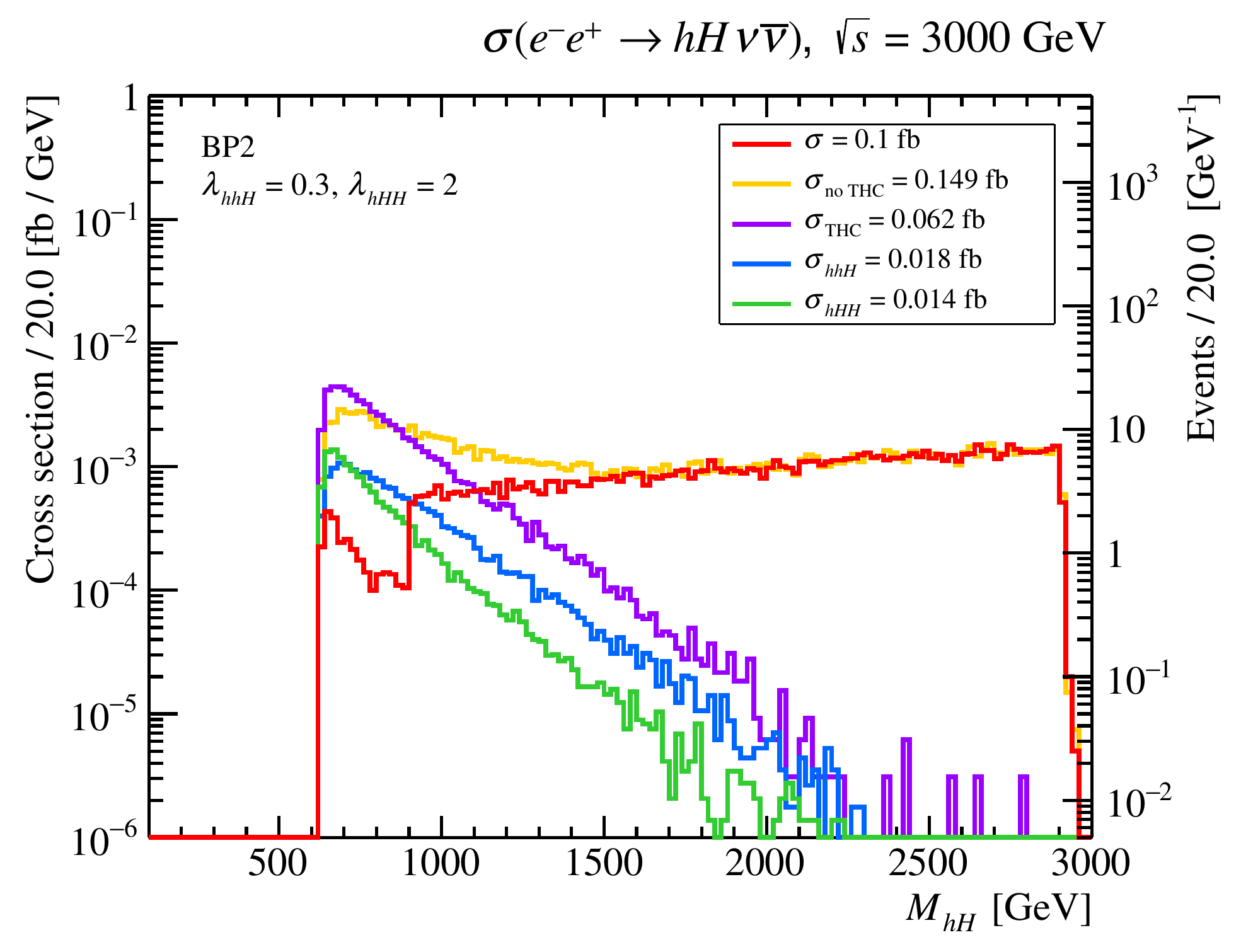}

\includegraphics[width=0.41\textwidth]{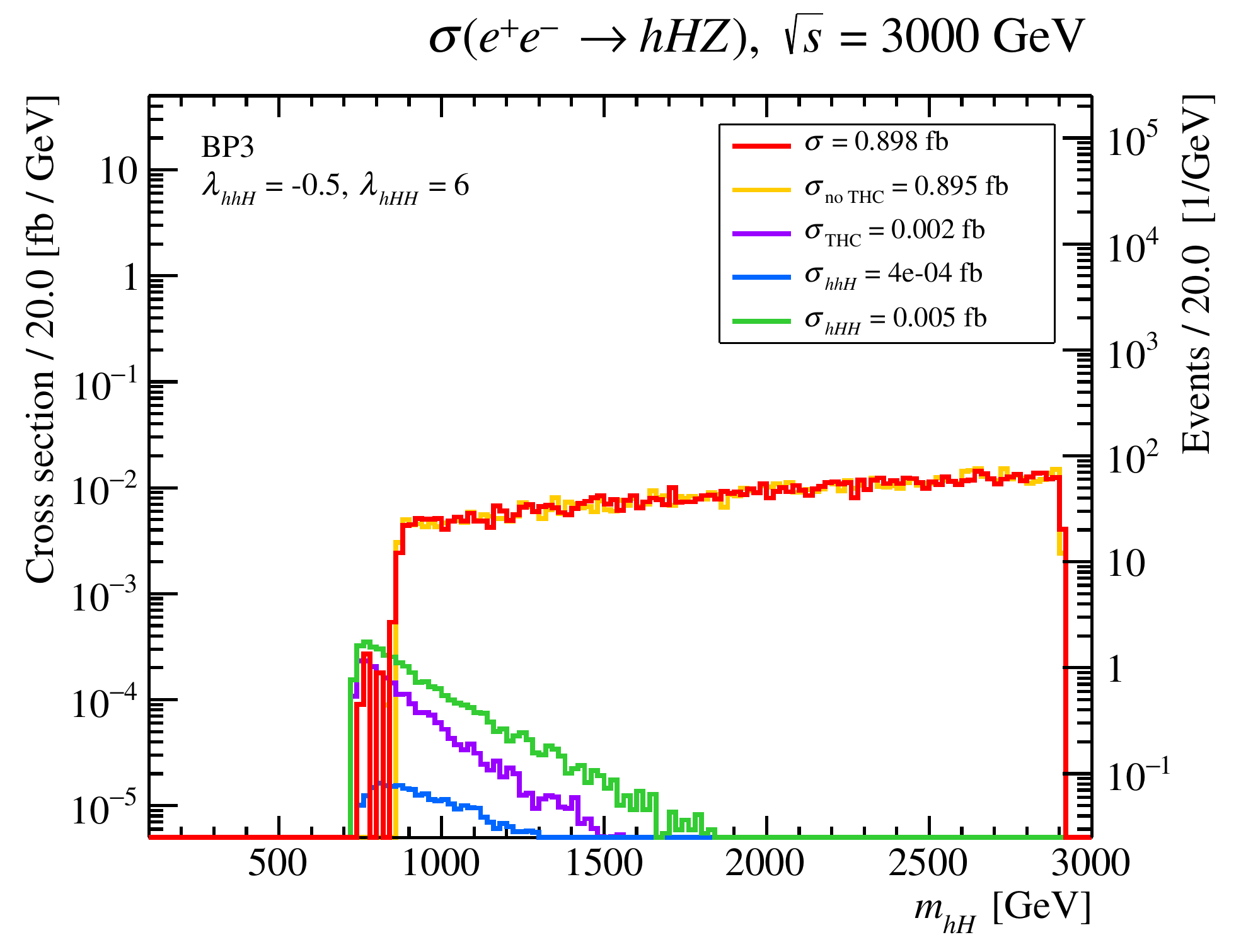}
\includegraphics[width=0.41\textwidth]{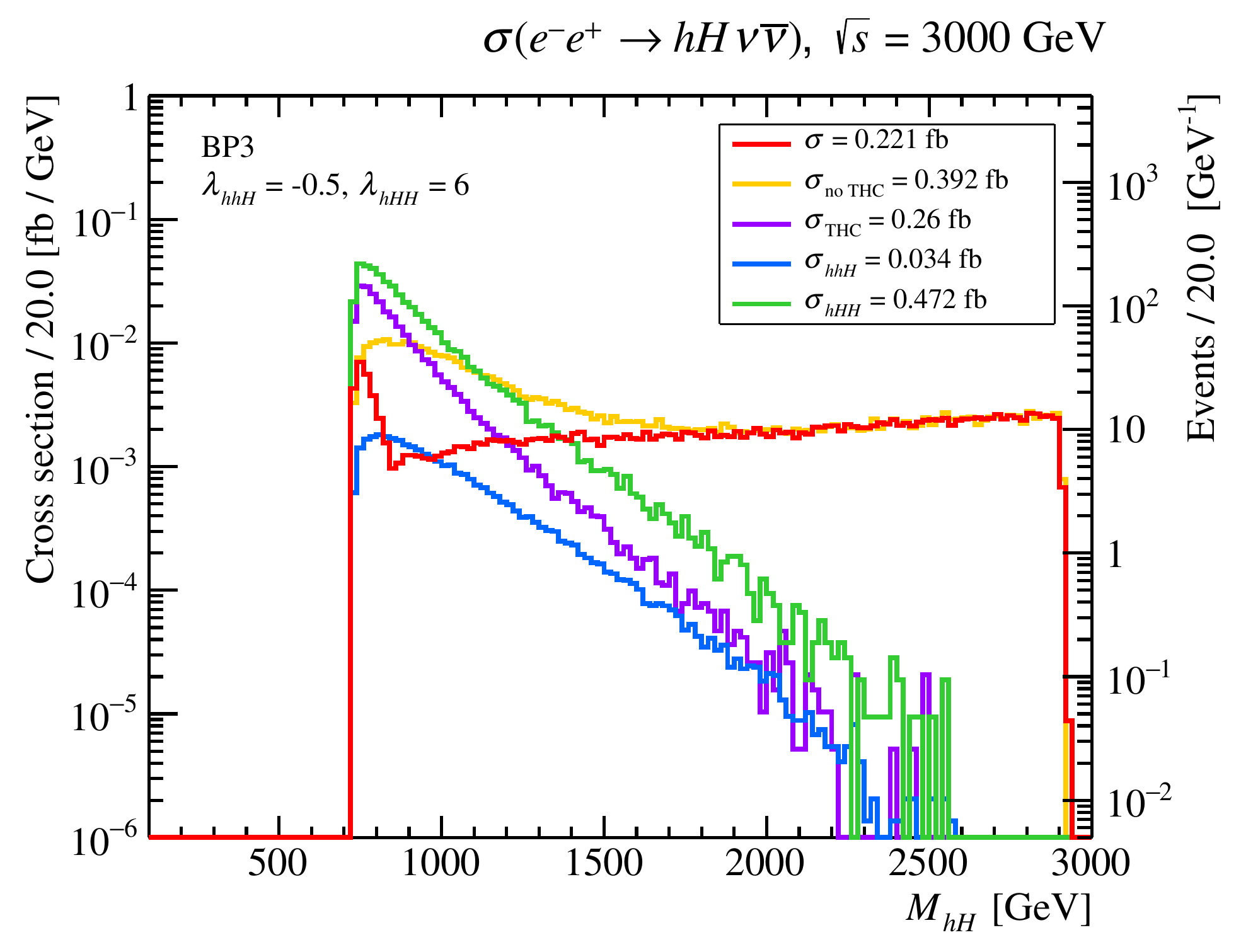}

\includegraphics[width=0.41\textwidth]{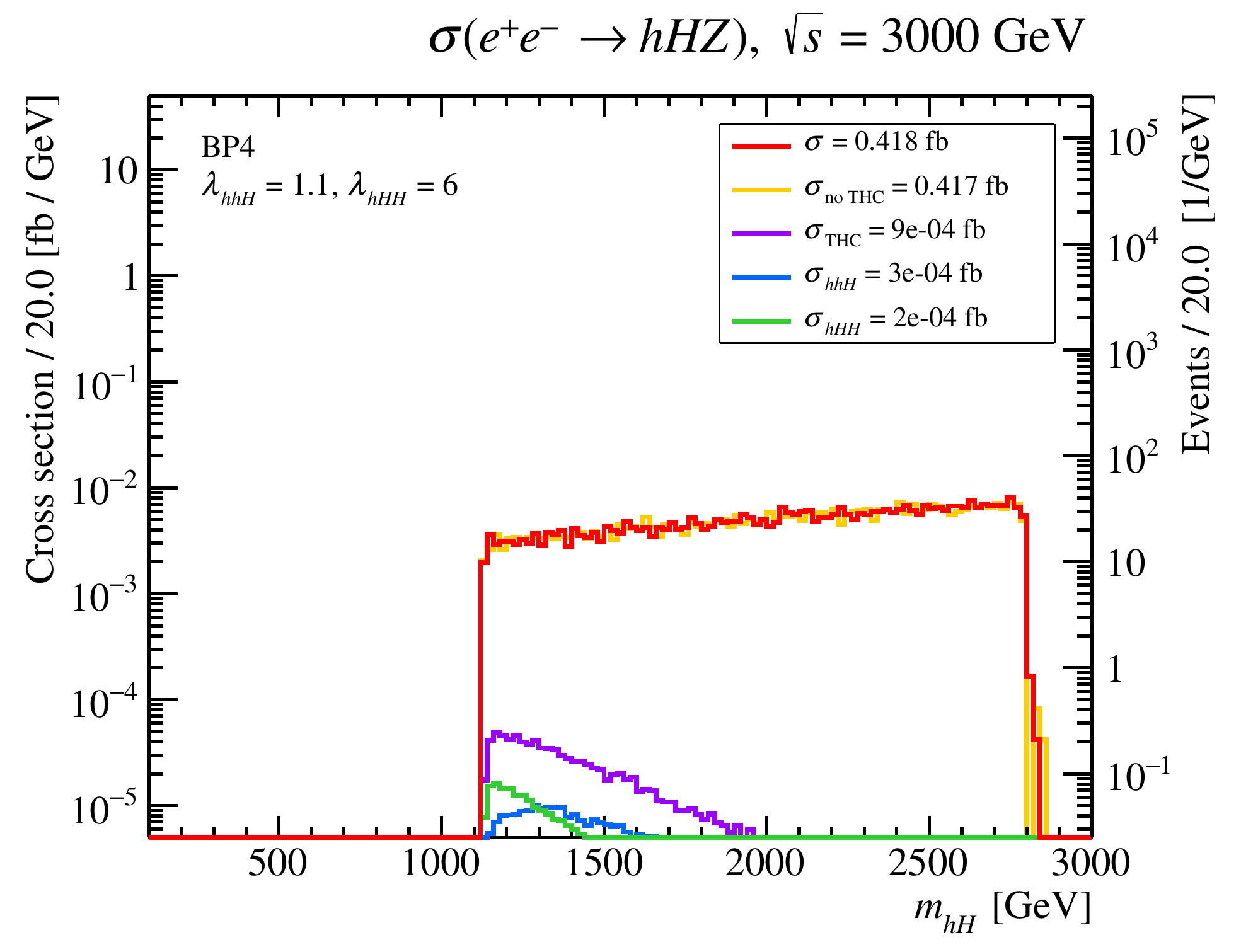}
\includegraphics[width=0.41\textwidth]{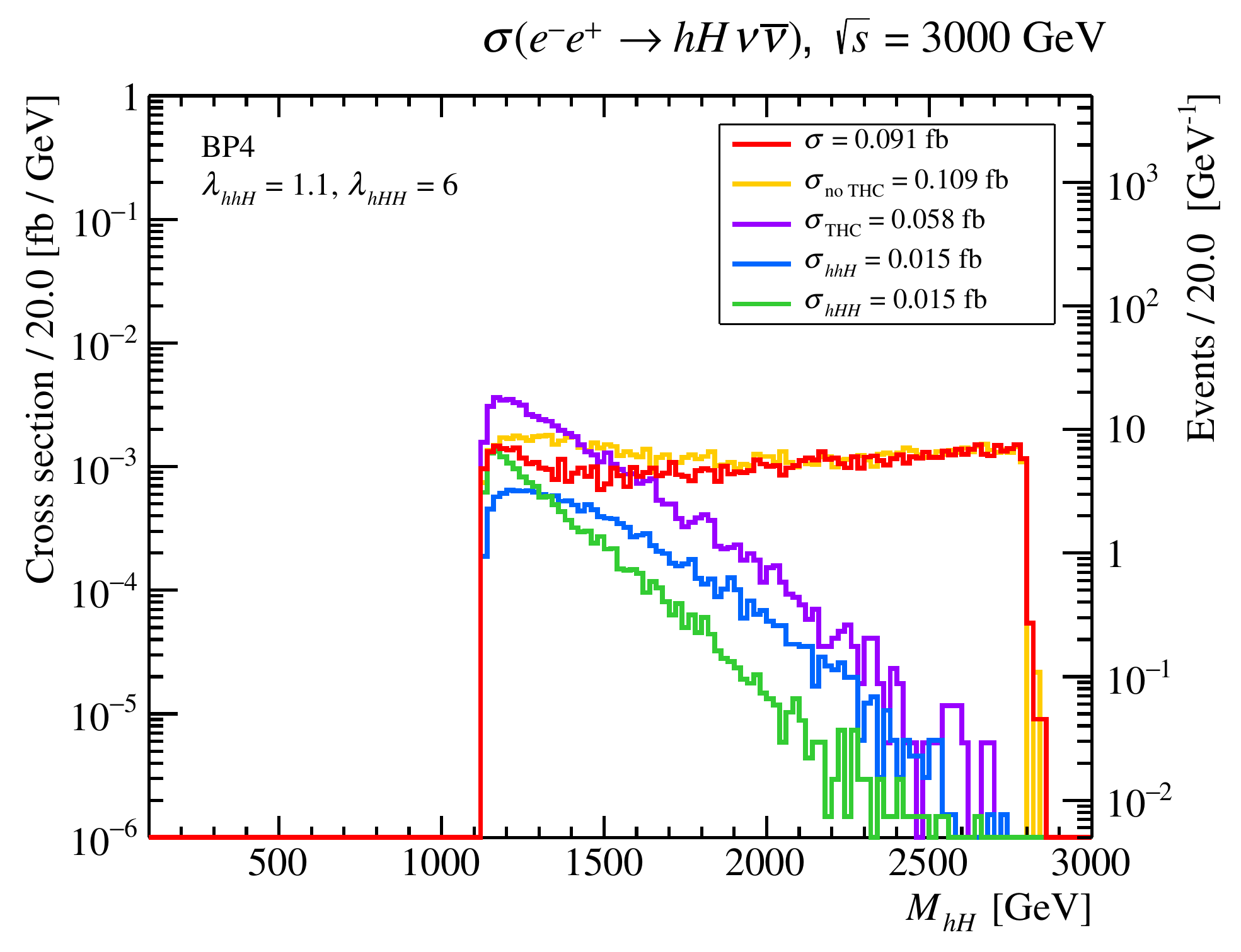}
	\end{center}
\caption{Distribution of the invariant mass of the final $hH$ pair in the 
process $e^+e^-\to hHZ$ (left) and $e^+e^-\to hH\nu\bar{\nu}$ (right) at
$\sqrt{s}=3000\gev$ for BP1, BP2, BP3 and BP4. } 
\vspace{-1em}
\label{fig:disthH}
\end{figure}

We start with the $hH$ case.   In this channel the two couplings
involved are $\lahhH$ and $\lahHH$, and their effects appear via their
contributions in the diagrams that are mediated by either an
intermediate $h$ or $H$,  respectively.  However, these
intermediate Higgs bosons are always off-shell. Therefore, they do not
produce resonant peaks in the relevant invariant mass distribution,
which in this case refers to the invariant mass of the final $hH$ pair,
$m_{hH}$.  The results of the distributions $d\sigma/dm_{hH}$ and the
corresponding event rates for both channels $hHZ$ and $hH\nu \bar \nu$
are shown in \reffi{fig:disthH} in the left and right column,
respectively.  The color code in 
\reffi{fig:disthH} is as follows: red lines are the complete 
$d\sigma/dm_{hH}$ taking into account all diagrams,  dark blue lines are
the contributions from the diagrams mediated by$h$,  green lines are
those from the diagrams mediated by$H$, purple lines are the
contributions from the sum of the two latter mediated by $h$ and $H$,
and yellow lines are the contributions from the rest of diagrams other
than the $h$ and $H$ mediated ones.  
The first conclusion from these plots is that the $hHZ$ channel does not
provide sensitivity to any of the two involved triple Higgs couplings.
This channel is dominated by the diagram $e^+e^-  \to Z^* \to H\,A \to H\,hZ$, 
as we commented in \refse{sec:xshH-I}. In fact, the cross section
distributions shows a ``step" shape similar to the one analyzed in the case
of $hh$ production and the beginning and ending
of this contribution can be reproduced if $\MH$ is introduced accordingly in
\refeq{eq:steps}.
  The second conclusion from
\reffi{fig:disthH}  is that the $hH\nu \bar \nu$ channel can provide
sensitivity to the triple couplings.  The relevant effect from
these two couplings,  $\lahhH$ (dark blue lines) and of $\lahHH$ (green
lines) appear in the close to threshold region, i.e.\   with $m_{hH}$
slightly above $m_h+m_H$.   The two contributions together provide the
largest contributions in this region,  leading to the purple lines in
these plots,  and interfere negatively with the remaining contributions
(yellow lines), producing a decrease in the total rates (red lines
lying  below the yellow lines).  Comparing the contributions of the two
triple couplings separately,  we see that in all the BPs the largest
contribution is from the green line,  indicating that the largest
sensitivity found in this channel is from to the triple coupling
$\lahHH$. This feature is in clear correlation with the fact that for
these chosen points this particular $\lahHH$ coupling gets large values,
varying from 2 in BP1 and BP2 to 6 in BP3 and BP4.
The relative interference of the $h$ and $H$ contribution depends
on the relative sign of $\lahhH$ and $\lahHH$. A positive (negative)
interference is found for $\lahhH \cdot \lahHH$ positive (negative).

\begin{figure}[ht!]
\vspace{-1em}
  \begin{center}
\includegraphics[width=0.41\textwidth]{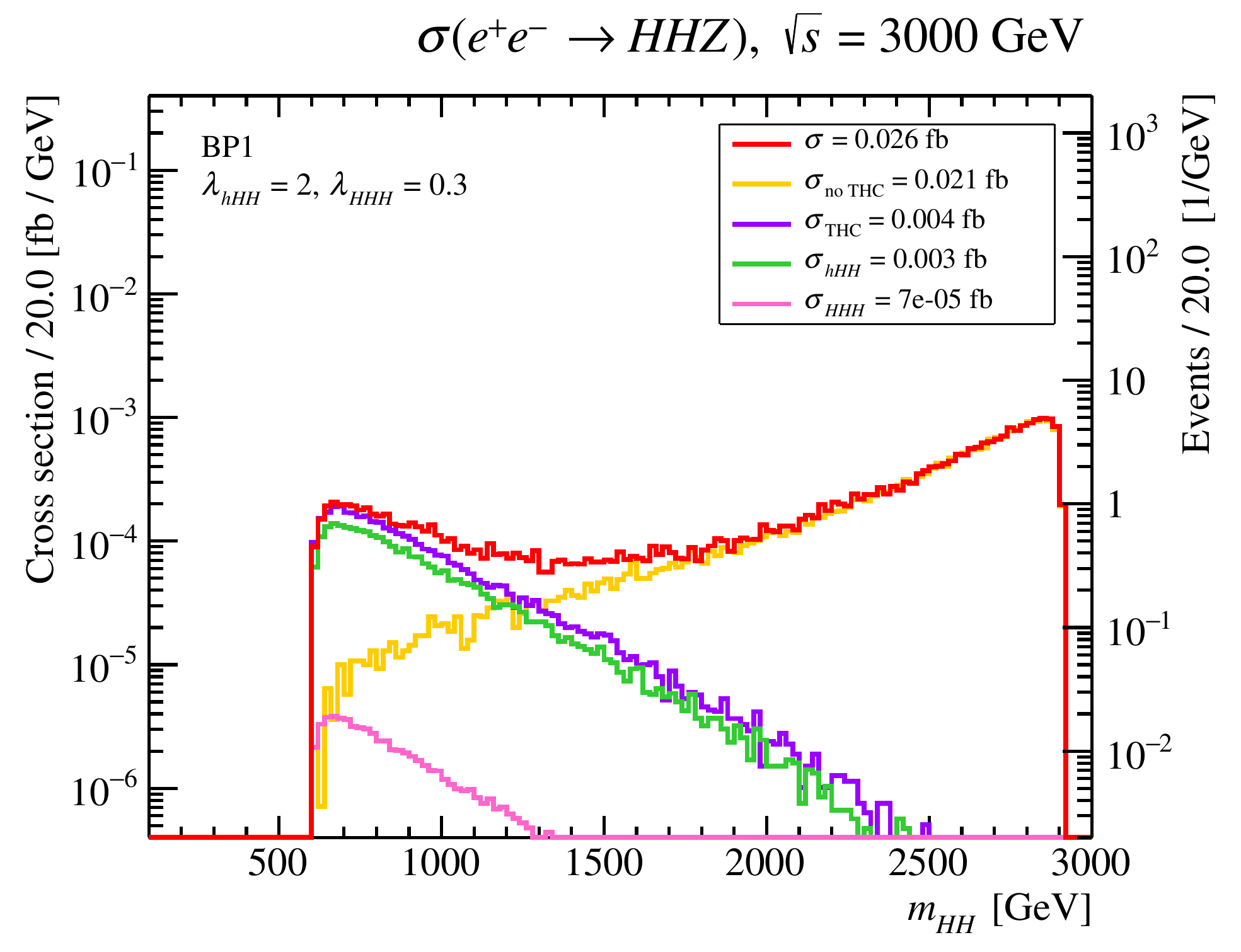}
\includegraphics[width=0.41\textwidth]{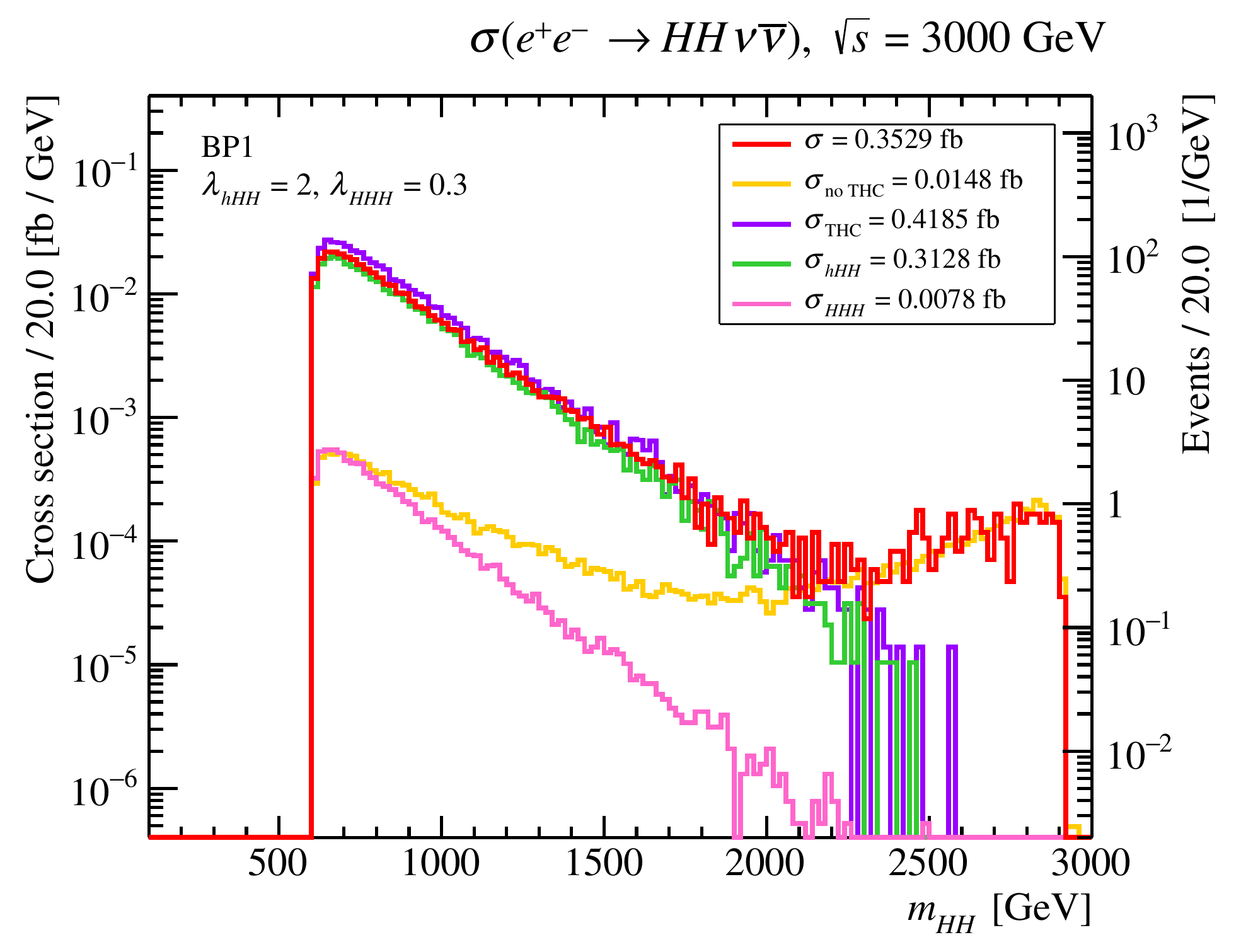}

\includegraphics[width=0.41\textwidth]{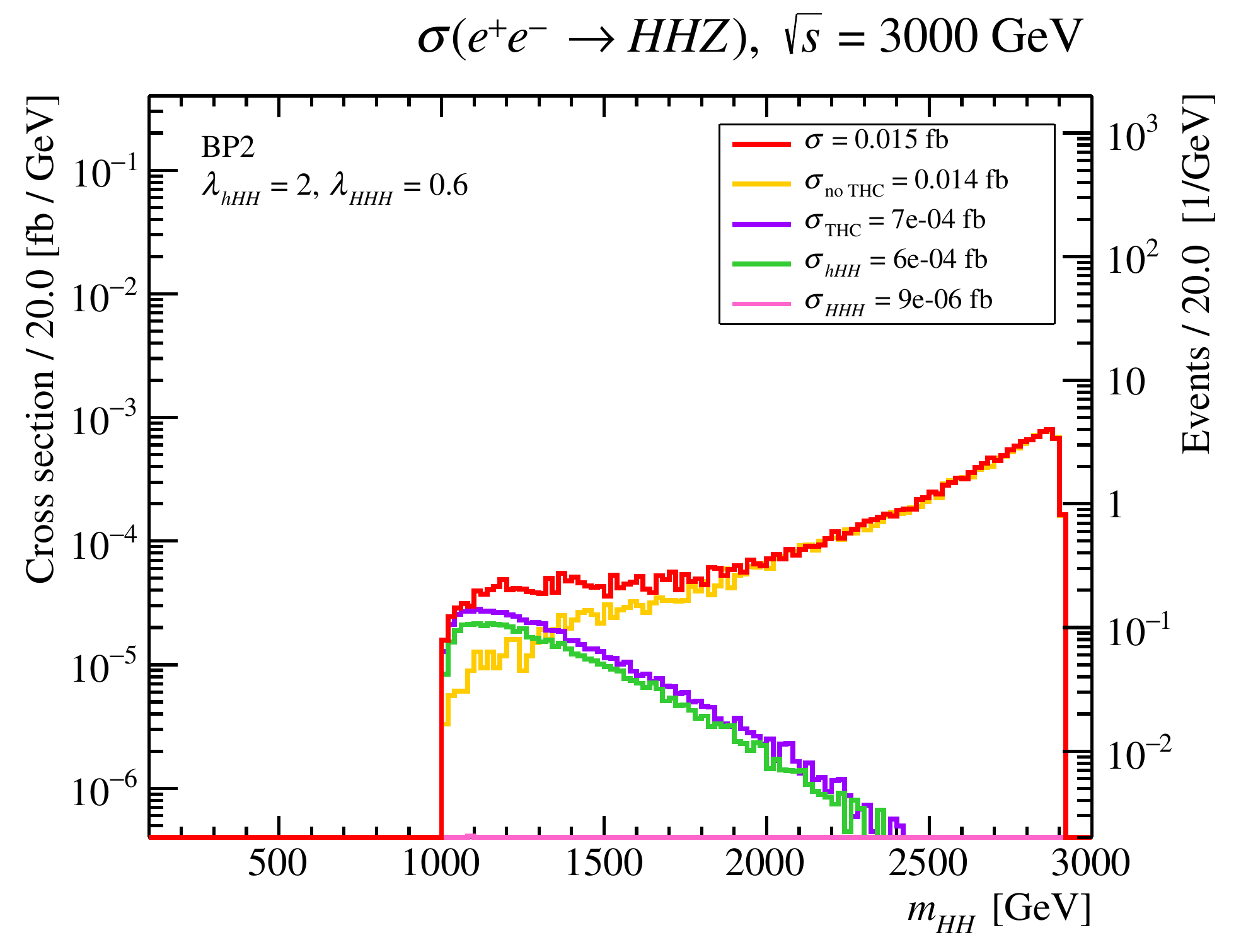}
\includegraphics[width=0.41\textwidth]{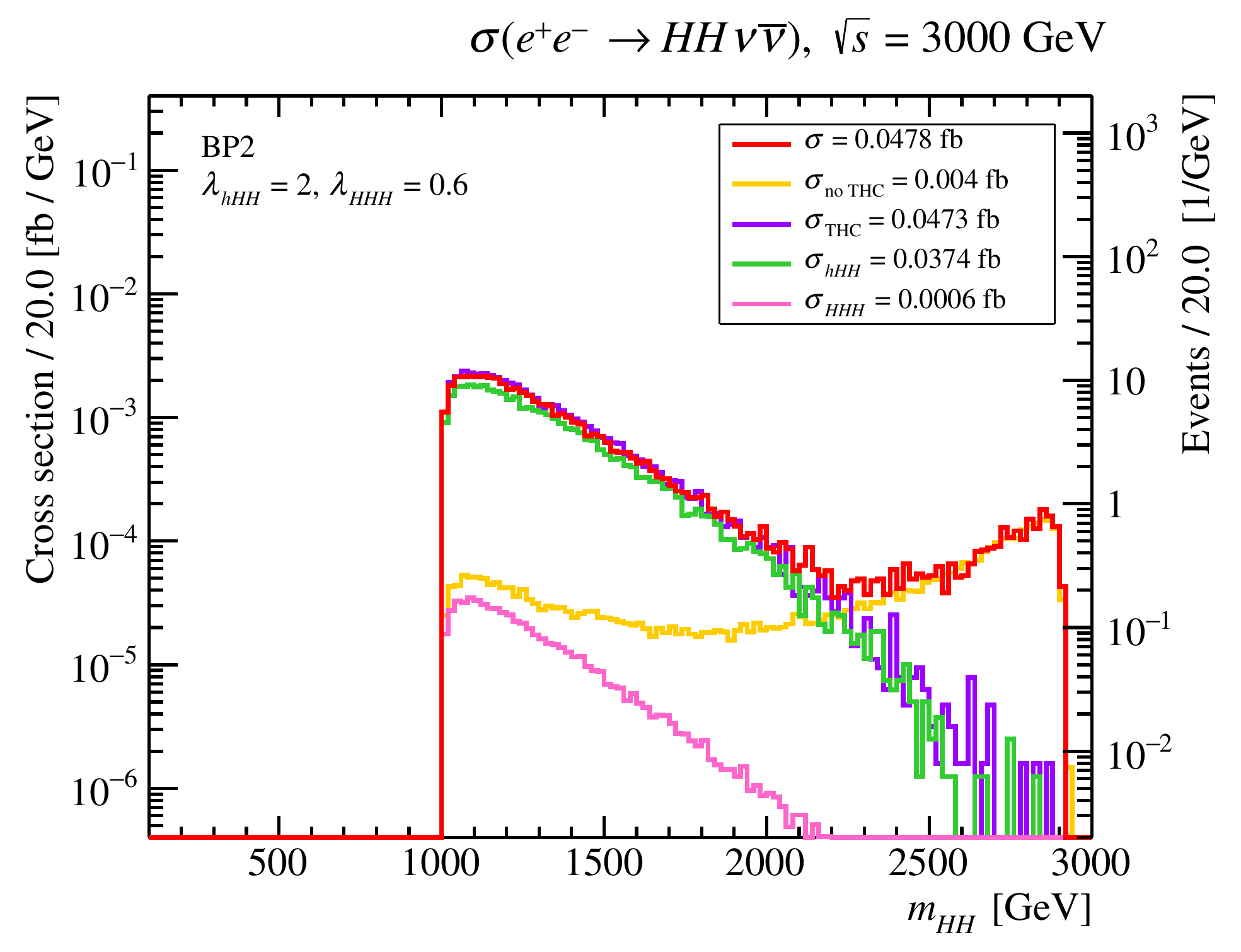}

\includegraphics[width=0.41\textwidth]{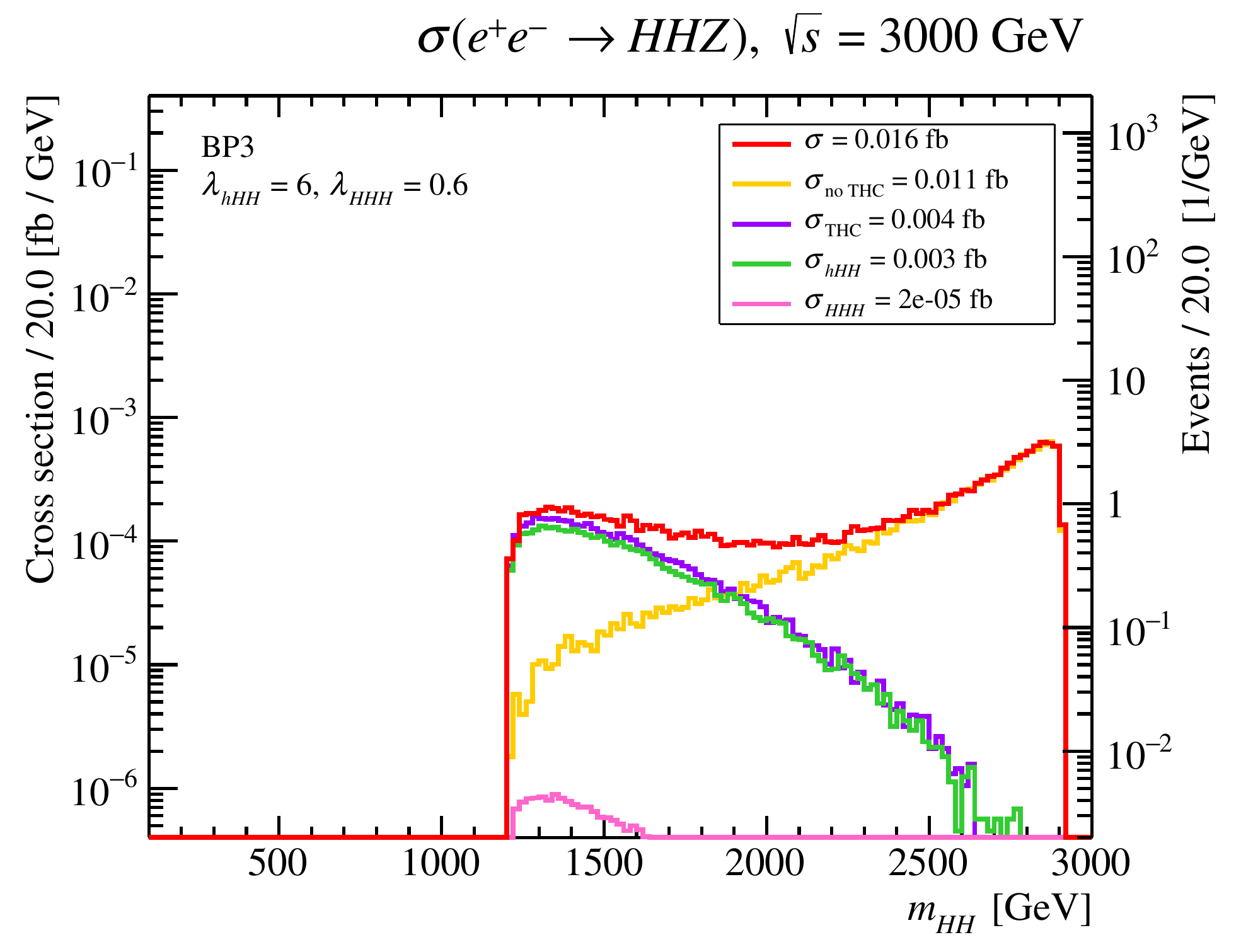}
\includegraphics[width=0.41\textwidth]{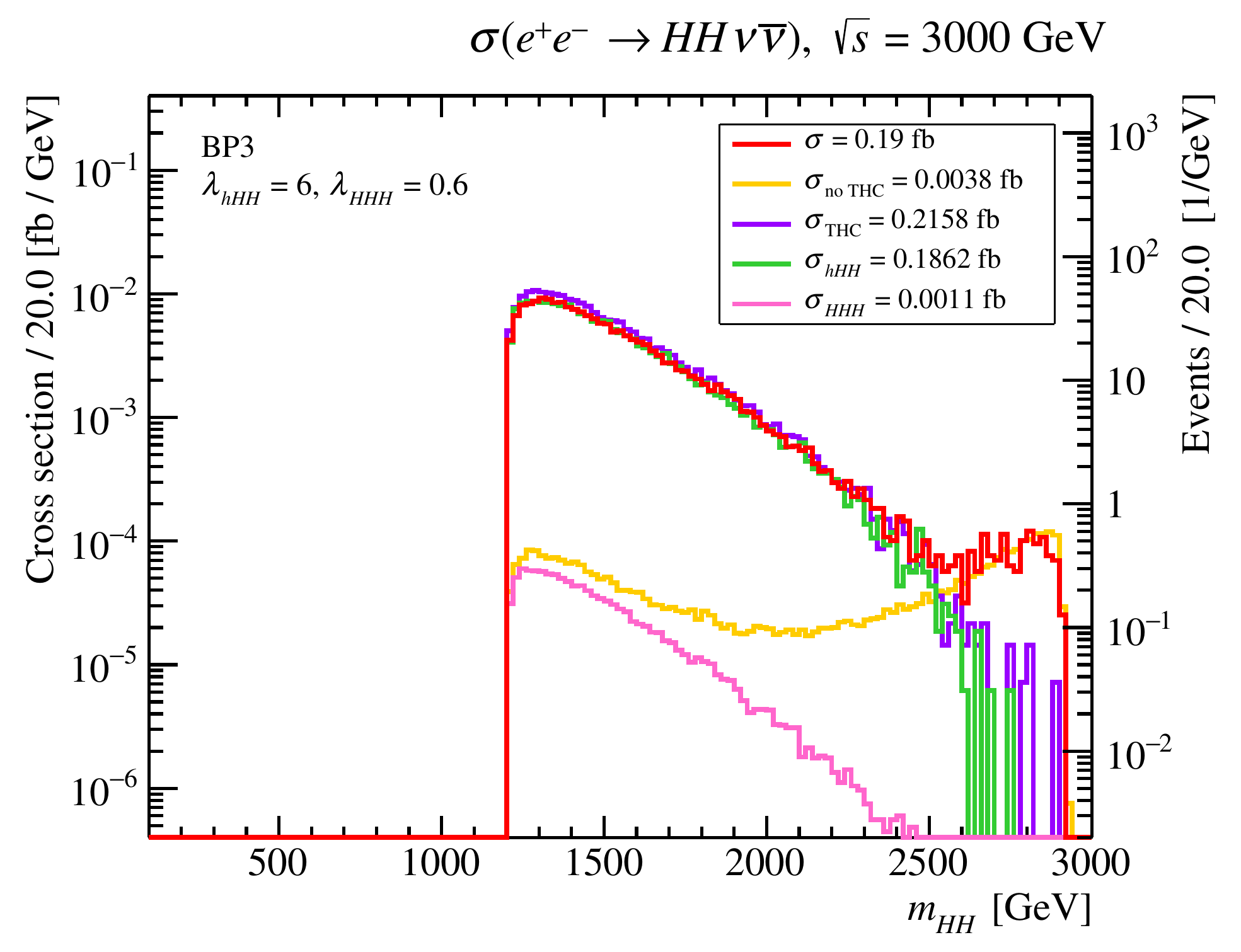}

\includegraphics[width=0.41\textwidth]{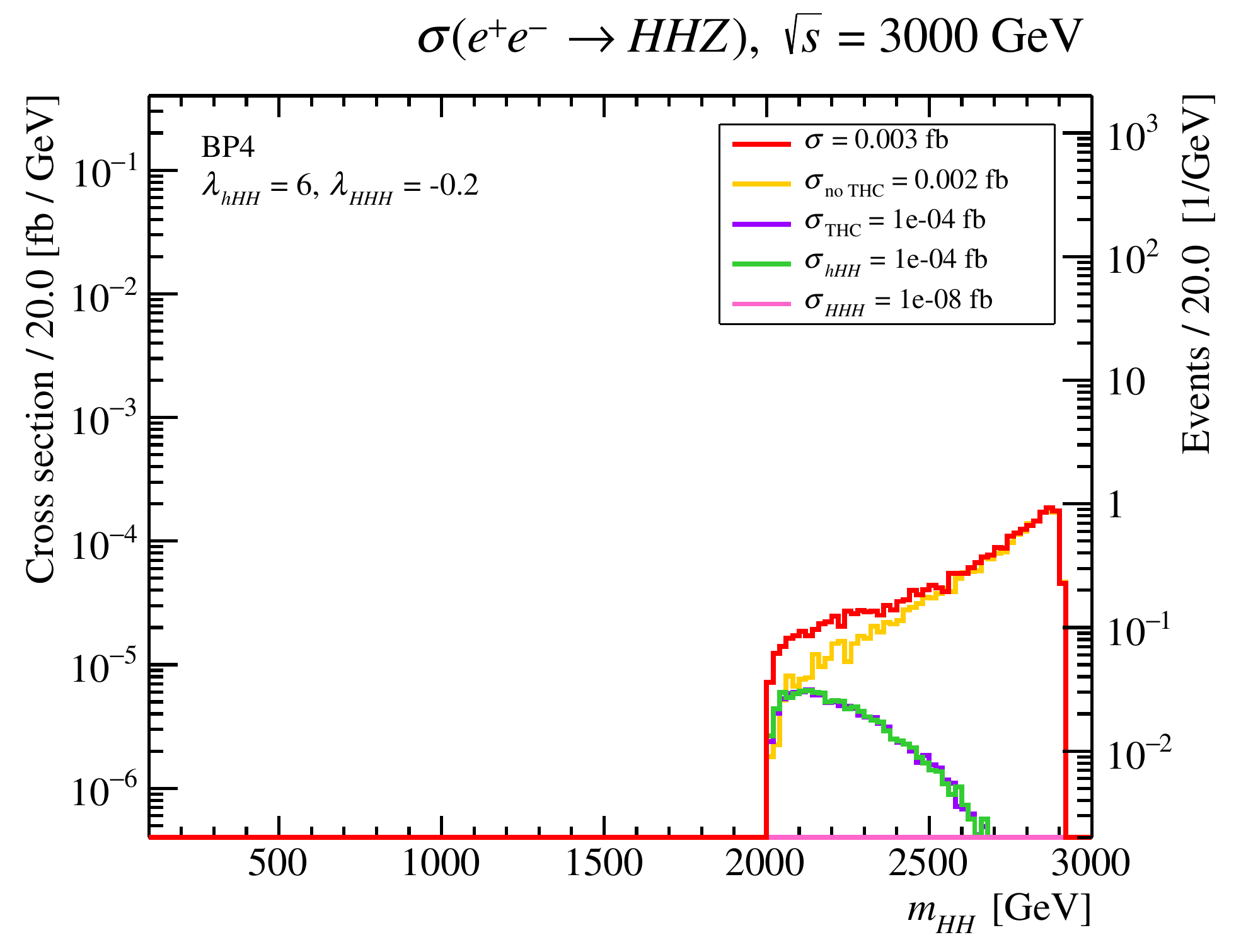}
\includegraphics[width=0.41\textwidth]{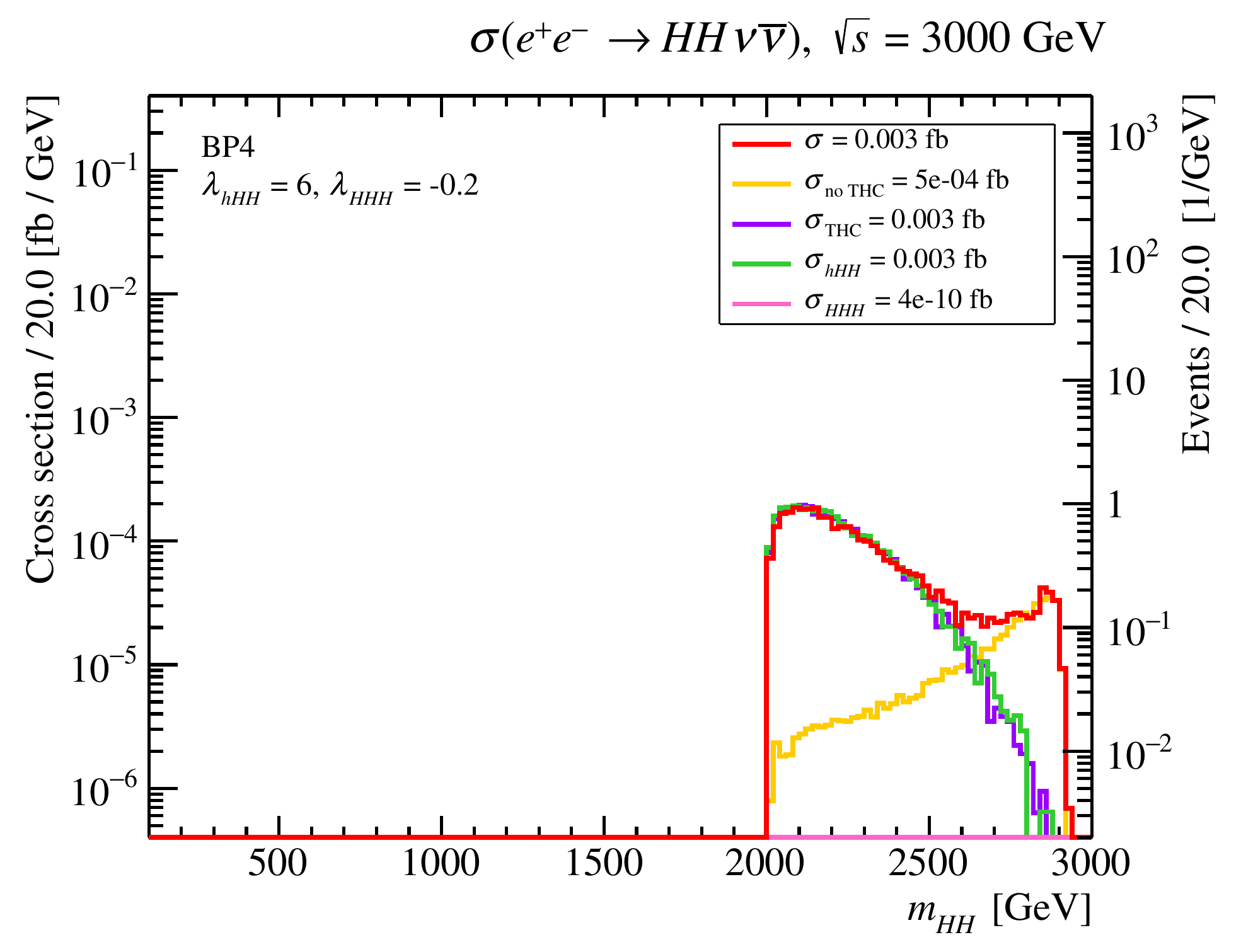}
	\end{center}
\caption{Distribution on the invariant mass of the final $HH$ pair in the 
process $e^+e^-\to HHZ$ (left) and $e^+e^-\to HH\nu\bar{\nu}$ (right) at
$\sqrt{s}=3000\gev$ for BP1, BP2, BP3 and BP4. } 
\label{fig:distHH}
\vspace{-1em}
\end{figure}

Next we comment on the sensitivity to the triple couplings in $HH$
production.  In this case  the two triple couplings involved are
$\lahHH$ and $\laHHH$, and their effects appear via their contributions
in the diagrams that are mediated by either an intermediate boson $h$ or
$H$,  respectively.  
As before,  these intermediate Higgs bosons are off-shell.
Therefore, they 
do not produce resonant peaks in the relevant invariant mass
distribution,  which in this case refers to the invariant mass of the
final $HH$ pair, $m_{HH}$.  The results of the distributions
$d\sigma/dm_{HH}$ and the  corresponding $HH$ event rates for both
channels $HHZ$ and $HH\nu \bar \nu$ are shown in
\reffi{fig:distHH},  in the left and right column, respectively. The
color code in these plots is as follows: red lines are the complete 
$d\sigma/dm_{HH}$ taking into account all diagrams,  green lines are the
contributions from the diagrams mediated by $h$,  pink lines are those
from the diagrams mediated by $H$, purple lines are the contributions
from the sum of the two mediated by $h$ and $H$,  and yellow lines
are the contributions from the rest of diagrams other than the $h$ and
$H$ mediated ones.  The most relevant effects from these two couplings,
$\lahHH$ (green lines) and of $\laHHH$ (pink lines) appear as
before in the 
region close the threshold of $HH$ production, which in this case means
$m_{HH}$ slightly above $2m_H$.  The first important conclusion from
these plots is that both channels, $HHZ$ and $HH \nu \bar \nu$,  notice
the effects of these triple Higgs couplings.  Comparing  the two
involved couplings, we see that the largest effect is clearly from  
$\lahHH$ which is the largest among all the studied triple Higgs
couplings in all the four BPs. Its values range from 2 to 6,  as can
be seen in \refta{tab:BP}.   Accordingly,  the green lines in
\reffi{fig:distHH} provide a much larger contribution than the pink
lines in the two channels, $HHZ$ and $HH \nu \bar \nu$ for all the BPs.
Comparing the event rates at  $3 \tev$ (see the right vertical
axis), both the total rates and the ones in the mentioned close to
threshold region, one can see that they are sizeable  (except for BP4) in
$HH \nu \bar \nu$.  On the other hand, they remain relatively small
for $HHZ$ production. Consequently, the
sensitivity to $\lahHH$ at this large energy appears mainly in the
channel with neutrinos whereas there is no sensitivity to $\laHHH$.
Comparing the red,  the yellow  and the purple
lines in these $HH\nu\bar\nu$ plots, we see that the purple ones
dominate over the 
yellow lines,  i.e.\  the contributions from the rest of diagrams
other than the ones mediated by $h $ or $H$,  except for the 
largest values of $m_{HH}$.  Since the
purple lines are slightly above the red lines in  $HH \nu \bar \nu$, we
conclude that the interference from the triple Higgs coupling is
destructive in this channel,  but with a small effect due to the
large numerical difference between the sets of contributing diagrams.
In contrast, in the $HHZ$ channel, the
purple line is below the red line, therefore, the interference is
constructive.  As in the previous $hH$ case,  we do not discuss further
the sensitivity to the triple Higgs couplings. 

To estimate the effects of the triple Higgs couplings on the cross
section distributions for the $hH$ and $HH$ channels 
more quantitatively,  as we
have done for $\lahhH$ in the $hh$ case,  is interesting
but involved.
This would require a more refined analysis in the close
to threshold $m_{hH}$ and $m_{HH}$ regions respectively, including the study of the more realistic
final states considering the Higgs bosons decays and the corresponding
backgrounds, which is beyond the scope of this work.  Thus, we do not go
further in these $hH$, and $HH$ channels.


\subsection{Sensitivity to triple Higgs couplings in the 2HDM type II }

\begin{figure}[t!]
	\begin{center}
\includegraphics[width=0.44\textwidth]{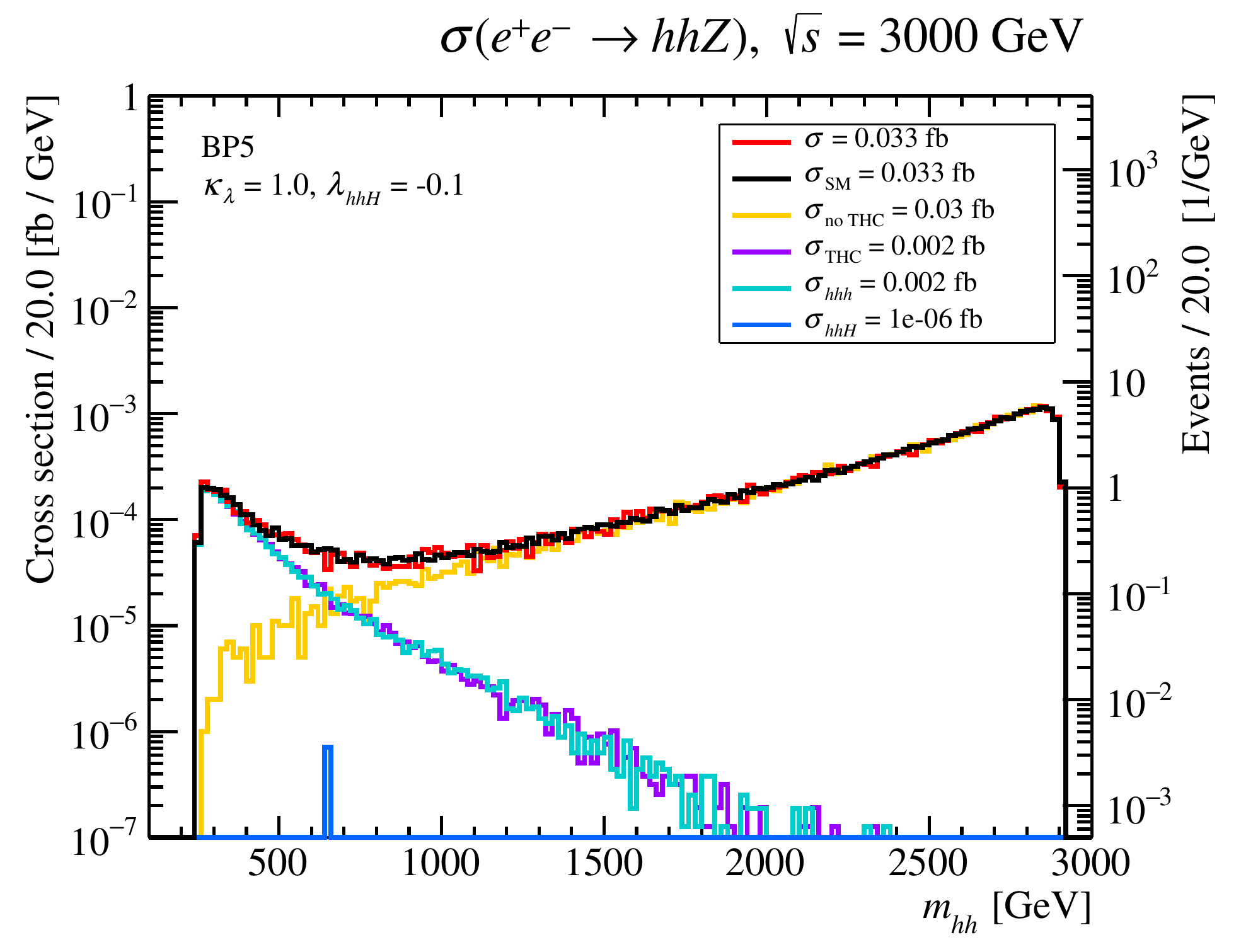}
\includegraphics[width=0.44\textwidth]{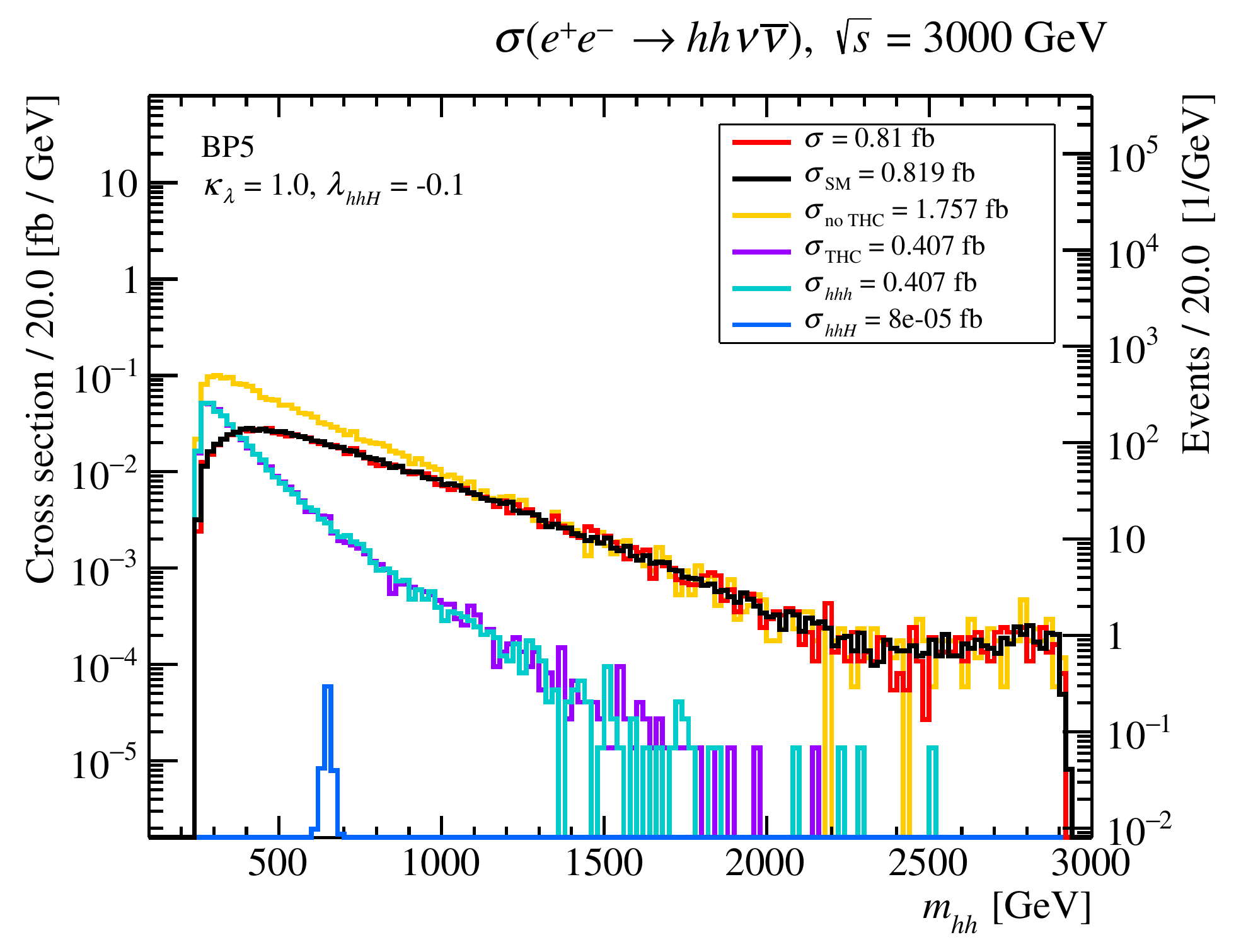}\\[3.3em]
\includegraphics[width=0.44\textwidth]{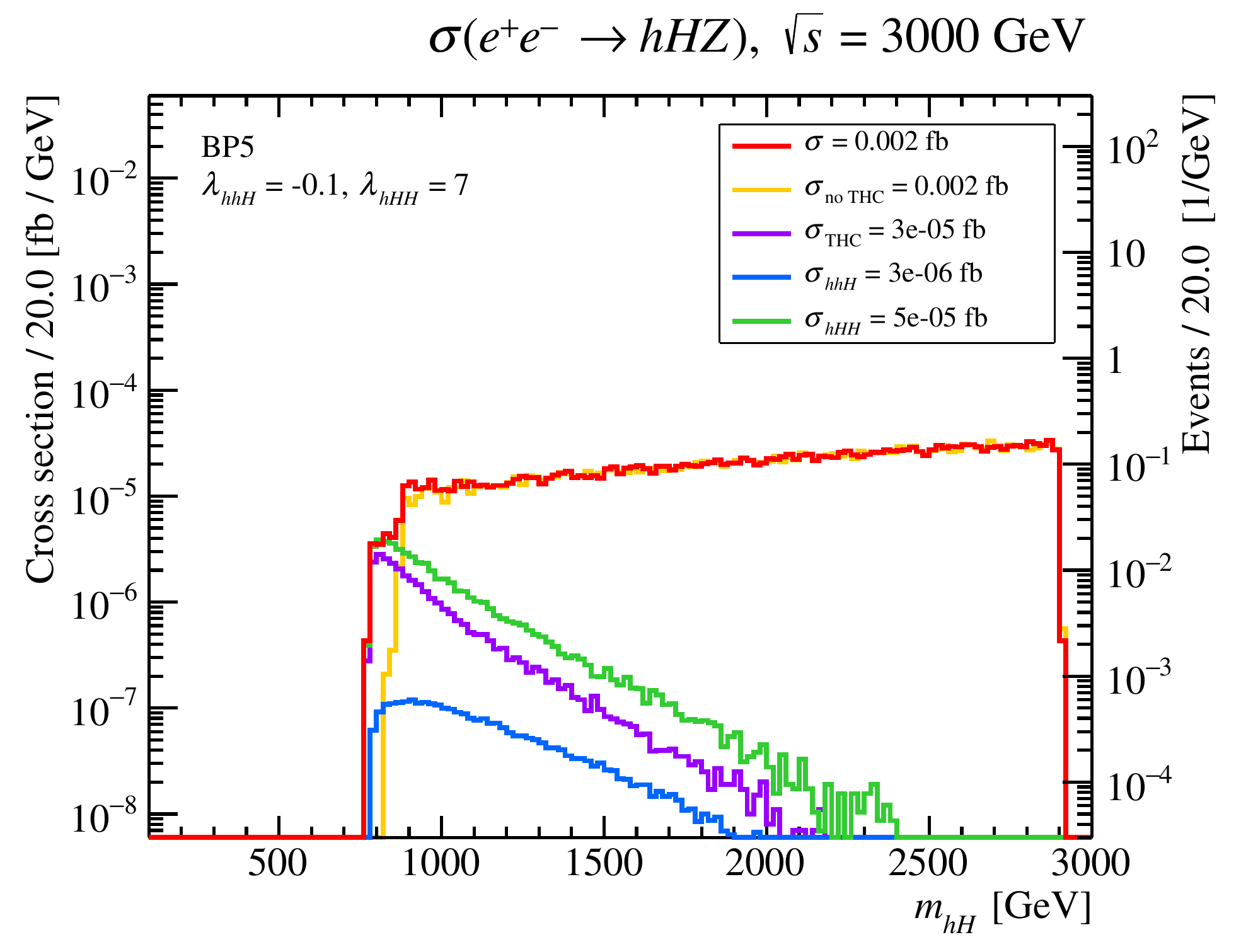}
\includegraphics[width=0.44\textwidth]{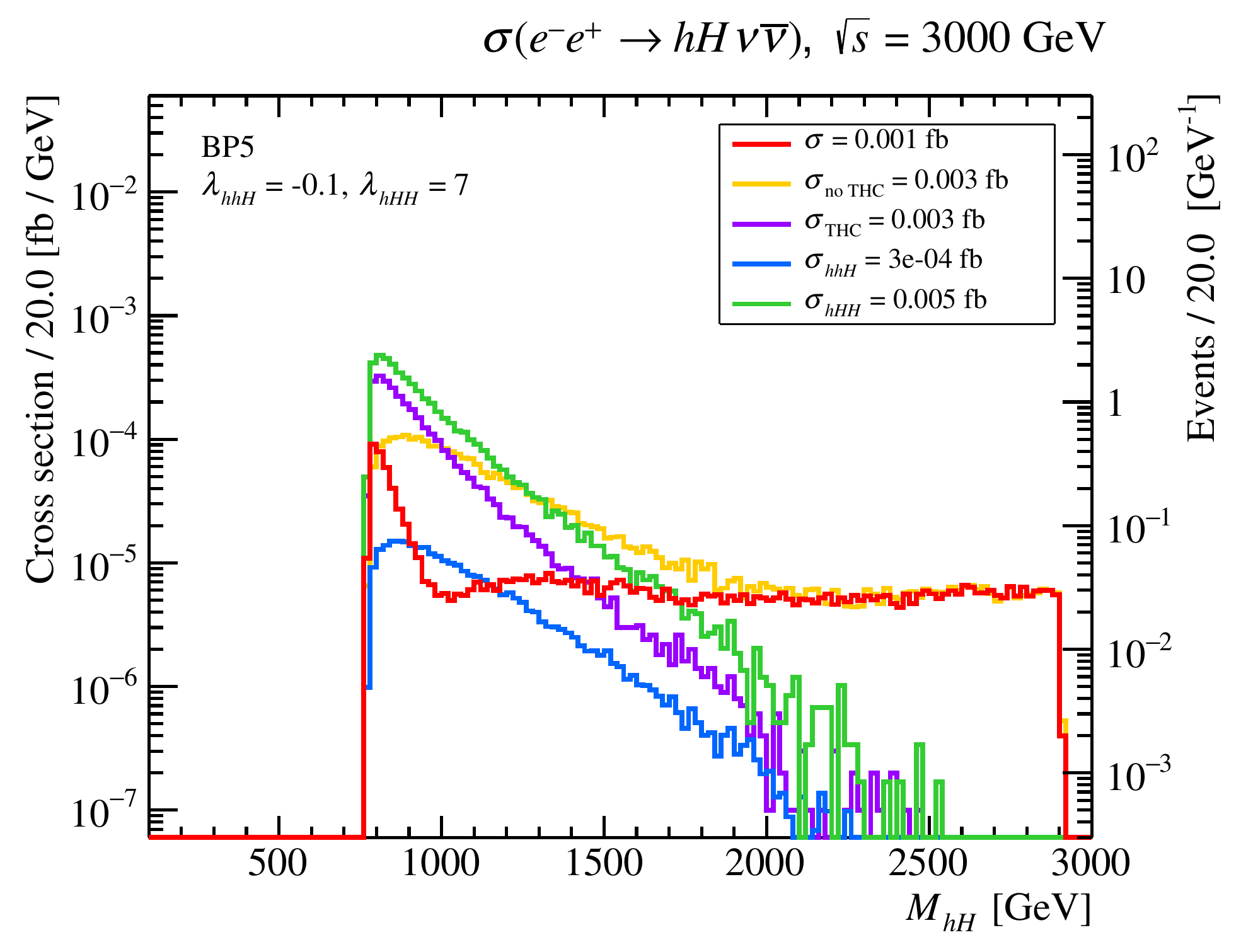}\\[3.3em]
\includegraphics[width=0.44\textwidth]{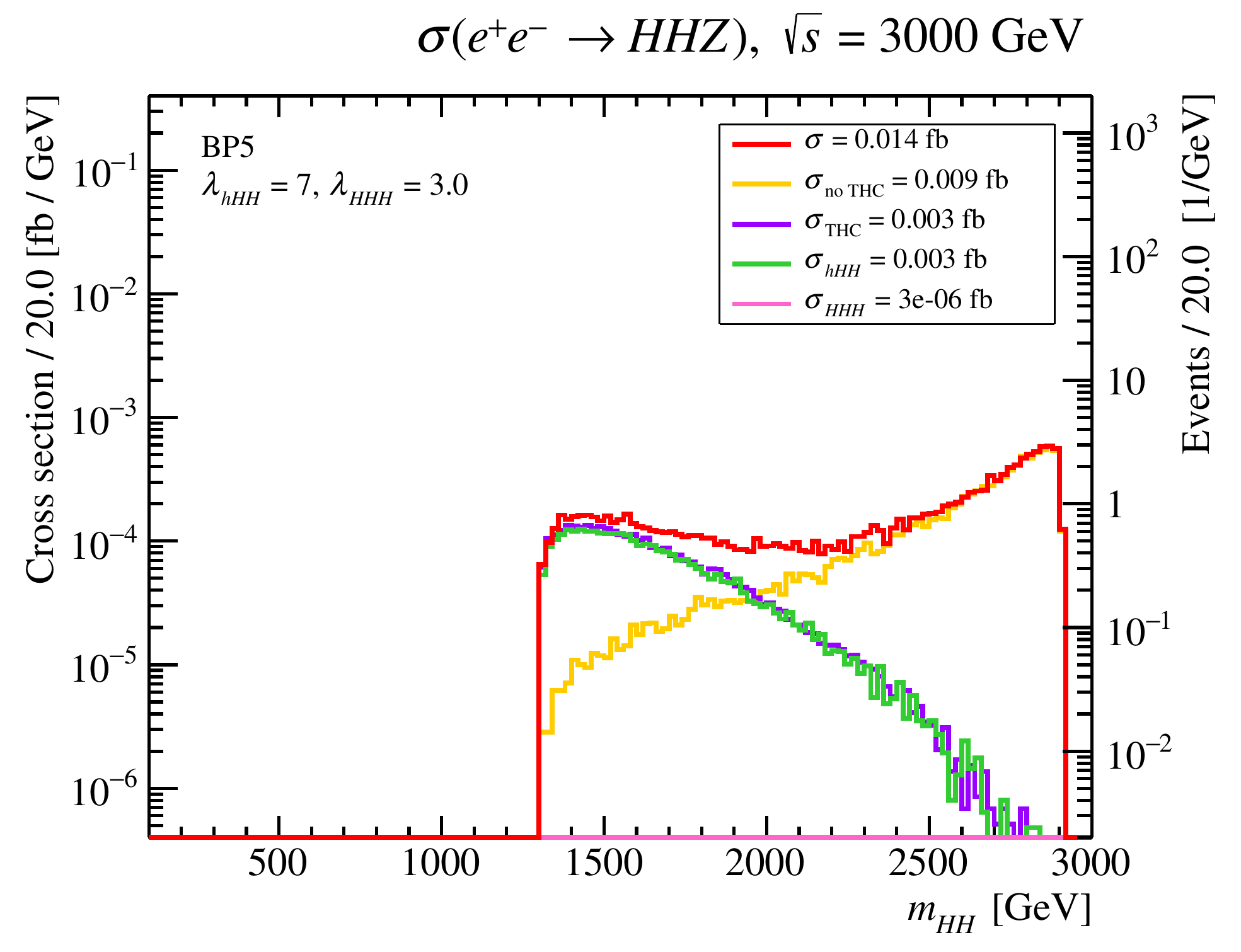}
\includegraphics[width=0.44\textwidth]{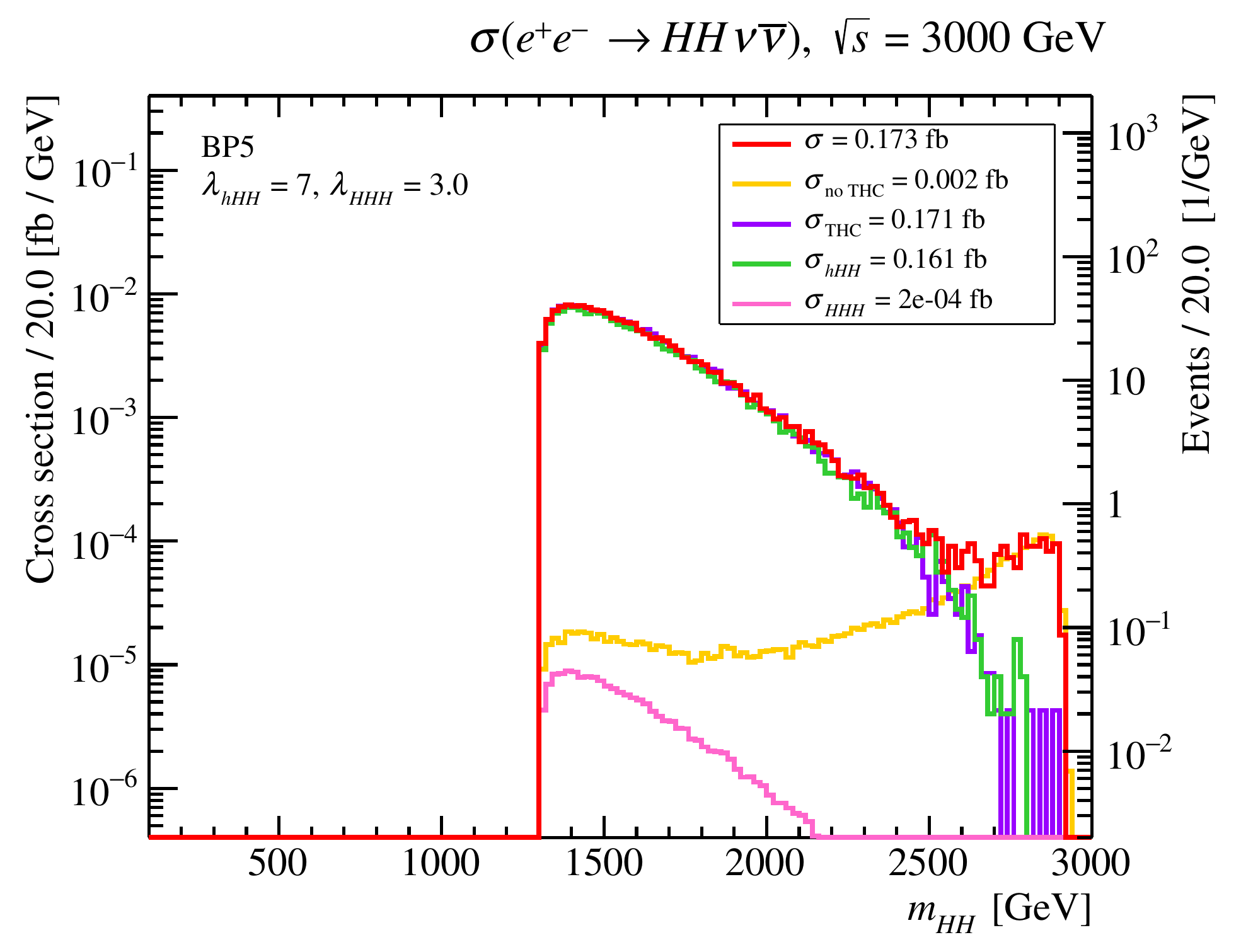}
	\end{center}
\caption{Distribution on the invariant mass of the final $h_ih_j$ pair in the 
process $e^+e^-\to h_ih_jZ$ (left column) and $e^+e^-\to
h_ih_j\nu\bar{\nu}$ (right column) at $\sqrt{s}=3000\gev$ with
$h_ih_j=hh$ (upper row), $h_ih_j=hH$ (middle row) and $h_ih_j=HH$ (lower
row) for BP5 (type II). } 
\label{fig:distBP5}
\end{figure}

In this section we focus in the 2HDM type~II and present the results of
the relevant cross section distributions for $h_ih_j$ production in the
corresponding invariant mass $m_{h_ih_j}$ for the benchmark point BP5
(see input values in \refta{tab:BP}).
As before we focus on the highest envisaged $e^+e^-$ collider energy
of $3 \tev$.  The predictions for all the channels are summarized in
\reffi{fig:distBP5}.  These include:  $hhZ$ (top left),  $hHZ$ (middle
left),  $HHZ$ (bottom left),   
$hh \nu \bar \nu$ (top right),  $hH \nu \bar \nu$ (middle right),  and
$HH \nu \bar \nu$ (bottom right). Both,  the differential cross sections
(vertical left axes) and the $h_ih_j$ event rates (vertical right
axes) are displayed.  The color code in this figure follows the
pattern as in \reffis{fig:disthH} and \ref{fig:distHH}:
The red lines show the complete $d\sig/dm_{h_ih_j}$,
light blue lines are the contribution from diagrams involving $\lahhh$
(two upper plots), dark blue indicates the contribution involving
$\lahhH$ (upper and middle row), green lines show the $\lahHH$
contributions (middle and lower row), pink shows the contribution from
$\laHHH$ (lower row), purple lines are the sum of all diagrams
involving triple Higgs couplings, and yellow lines show the contribution
from the other diagrams, i.e.\ not involving triple Higgs couplings.
Black lines are the corresponding predictions in the SM.

In our analysis of BP5 in the 2HDM type~II we start
with  the $h_ih_jZ$ channels.  We see some (potential)
sensitivity to $\lahhh$ in
the close to $hh$ threshold region for $hhZ$ production, see the light
blue line that clearly  dominates in this region.  However, the event
rates are too low and therefore unable to provide sensitivity to this
coupling.  The case of $hHZ$ is clearly not sensitive to the triple
Higgs couplings at this high energy.  In the case of $HHZ$
production there is
an appreciable effect from the green line,  involving $\lahHH$ that
gives the dominant contribution in the close to threshold region.
However the predicted
rates are yet too small and do not provide sensitivity enough to this
coupling.
We now turn to the $h_ih_j\nu\bar\nu$ channels shown in the right
column of \reffi{fig:distBP5}, where, as expected, overall larger
production cross sections are obtained.
We find relevant effects in the $hh \nu \bar \nu$ channel from
$\lahhh$, as can be seen in the light blue line in the region close
to the $m_{hh} = 2 m_h$ threshold. These contributions interfere
negatively with the diagrams not involving triple Higgs couplings
(yellow lines), yielding a sizable number of events.
However, the prediction
for the complete differential cross section  (red line) is very close to
the SM one (black line),  reflecting the fact that $\kala \simeq 1$  in BP5.
Consequently,  we do not expect deviations in BP5
from the sensitivity expected to $\kala$ within the SM.
In the case of $hH \nu \bar \nu$,  although there is an apparent effect
from $\lahHH$, and to a lesser extent from $\lahhH$,
they do not seem statistically significant, because the rates are also
very low in this case.  Finally,  the 
channel $HH \nu \bar \nu$ seems more promising  at this high
energy. The green line,  involving $\lahHH$  clearly dominates over the
rest and provide practically  the complete cross section (red line).
Here one should note that for BP5 $\lahHH$ is the largest one among
all triple Higgs couplings with a value of $\sim 7$.
The event rates are yet significant, indicating that the effect from this
large  $\lahHH$ could be measured in this channel.


\section{Conclusions}
\label{sec:conclusions}

An important task at future colliders is the investigation of the
Higgs-boson sector. Here the measurement of the triple Higgs
coupling(s) plays a special role. 
Based on previous analyses~\cite{Arco:2020ucn}, within the framework of
Two Higgs Doublet Models (2HDM) type~I and~II we have analyzed several
benchmark planes that are over large parts in agreement with all
theoretical and experimental constraints.
These comprise electroweak precision data, tree-level  unitarity,
stability of the vacuum, bounds from direct searches for BSM Higgs
bosons, the rate measurements of the $125 \gev$ Higgs boson at the LHC,
as well as flavor constraints.
While three benchmark planes
are based on the 2HDM type~I, for type~II we defined two benchmark
planes. The relevant parameters are summarized in \refse{sec:planes}.
As one important characteristic, we have set all heavy Higgs-boson
masses equal, $m_H = m_A = m_{H^\pm}$ to simplify the analysis. 
For these planes we investigated the di-Higgs production cross sections
w.r.t.\ triple Higgs couplings
at future high-energy $e^+e^-$ colliders, such as ILC or CLIC.
We considered two different channels
$e^+e^- \to h_i h_j Z$ and $e^+e^- \to h_i h_j \nu \bar\nu$ with 
$h_i h_j=hh,hH,HH,AA$.
The first one is similar to the ``Higgs-strahlung'' channel of single
Higgs production. The second one has an important contribution from the
vector-boson fusion mediated subprocess, $W^+W^- \to h_ih_j$. It also
receives a  
contribution from the $Z$ mediated subprocess, $e^+e^- \to Zh_ih_j$, with 
$Z \to \nu \bar\nu$, which is usually smaller than the contribution from
$WW$ fusion at the high energy colliders. The triple Higgs couplings
relevant for our analysis are $\lahhh$ (with $\kala = \lahhh/\laSM$),
$\lahhH$, $\lahHH$ and $\laHHH$. 
Due to the fact that $HH$ production is very similar to $AA$
production we investigated only the former. Conclusions for $\lahAA
(\laHAA)$ are very similar as for $\lahHH (\lahAA)$.

In the first part of our study we have evaluated the $h_ih_jZ$ and
$h_ih_j\nu\bar\nu$ total  production cross sections for
$\sqrt{s} = 500, 1000, 1500$ and $3000 \gev$, i.e.\ the (relevant)
energy stages foreseen for the ILC and CLIC.  We have
discussed the various production cross sections as a function of
the 2HDM parameters in the selected benchmark planes.

As expected from the known SM cross sections,  our study of the
2HDM cross sections as a function of the center-of-mass energy shows
that, in general,  the 
$h_ih_jZ$ cross sections decrease with increasing center-of-mass energy,
while $h_ih_j\nu\bar\nu$ cross sections increase,  despite different
types of diagrams can contribute for $i \neq j$.
We furthermore related the pattern of enhanced or
suppressed cross sections to the sizes of the relevant triple Higgs
couplings. 
As anticipated, the 2HDM $hh$ production cross section
approaches the SM value in the alignment limit, $\CBA \to 0$.  On the
other hand, large enhancements of $hh$ production w.r.t.\ the 
SM cross sections are found at some 2HDM parameter configurations. 
Our analysis indicates that this enhancement is due to the
additional contributions from the extended 2HDM Higgs sector,
including the effects from the new diagrams with intermediate heavy
Higgs bosons. The size of the enhancement depends importantly on
the choice of the 2HDM parameters, most prominently on $\CBA$ away
from the alignment limit. These intermediate heavy Higgs bosons are $H$ and
$A$ for the $hhZ$ channel, and $H$, $H^+$ and $A$ for the
$hh\nu \bar \nu$ channel.  We have found that the enhancement mainly originates
from the diagrams where the intermediate heavy Higgs bosons, $H$ and $A$,
can be produced on-shell (i.e.\ in the case of $hhZ$ for
$\sqrt{s} \gsim \MH + \MZ$ or $\gsim  \MA + \Mh$, respectively),
but values of $\kala \neq 1$ can also play a role.  Consequently, the
largest contributions to the enhancement w.r.t.\ the relevant invariant
mass distribution occur in the resonant region, 
i.e.\ for $m_{hh} \sim \MH$ or $m_{hZ} \sim \MA$, respectively. 
The effect of the $ H^\pm$ boson,  on the contrary,
does not produce such a resonant behavior in the
$e^+e^- \to hh \nu\bar\nu$ process,
and it does not involve any triple Higgs coupling.
The effect of the on-shell $A$ boson is resonant in the relevant mass
invariant distribution in 
both $hhZ$ and $hh \nu \bar \nu$ channels but it is not sensitive to
the triple Higgs couplings either. Out of all these intermediate
heavy Higgs boson contributions, the most relevant one in both $hhZ$
and $hh \nu \bar \nu$ channels is the one mediated by an on-shell
$H$, where the relevant mass invariant distribution shows the $H$
resonance at $m_{hh} \sim \MH$  and displays the unique sensitivity to
the triple Higgs coupling $\lahhH$.

Concerning $hH$ production, these channels do also receive relevant
contributions from diagrams with an intermediate $A$~boson:
$e^+e^-  \to Z^* \to H\,A^{(*)} \to H\,hZ$ and
$e^+e^- \to Z^* \to H\,A^{(*)} \to H\,hZ^{(*)} \to H\,h\nu\bar\nu$,  and
we find sizeable cross sections at all the studied collider settings.
But these $A$ mediated contributions are not sensitive either to the
triple Higgs couplings.  Again we find sensitivity only via the $H$
mediated contributions that in this case give access to probe $\lahHH$.
Regarding $HH$ production,  the cross sections at the planned
colliders are found to be small, $\lsim 1 \fb$,  in the allowed 
parameters space, and therefore the sensitivity to the involved
couplings, $\lahHH$ and $\laHHH$, is correspondingly smaller. 

In the second part of our analysis we investigated in detail the
mechanisms leading to 
the enhancements of the production cross sections identified in the
first part,  and explored how to reach the best sensitivity to the
involved triple Higgs couplings via the study of the cross sections
distributions with respect to the invariant mass of the corresponding
final Higgs bosons pair. 
We have concentrated on five benchmark points, which are contained
in our previously defined benchmark planes (with the exception of
BP2, which is inspired by benchmark plane~2, but has a slightly smaller
value of $m_H = m_A = m_{H^\pm}$ and a correspondingly slightly adjusted
value of $\msq$). Four are defined in 
the 2HDM type~I (BP1, BP2, BP3 and BP4) while one is defined in
type~II (BP5). 
For all production channels we found, depending on the center-of-mass
energy,  that the contributions from an intermediate $H$ Higgs boson
play a crucial role,  offering 
interesting access to (BSM) triple Higgs couplings involving this heavy
Higgs boson. Specially, when this $H$ is produced on-shell and
appears as a
resonant peak in the corresponding invariant mass distributions of the
final Higgs bosons pair. Furthermore, 
we outlined which process at which center-of-mass energy would be best
suited to probe the corresponding triple Higgs-boson couplings.
In general the 2HDM type~I allows for larger BSM triple Higgs couplings
and thus better prospects for their measurements than the 2HDM type~II.

Focusing on the 2HDM type~I, 
$hh$ production approximately offers access to
$\lahhh$ similar to the 
corresponding SM determination. On the other hand, in the region of
$m_{hh} \sim m_H$, a large (resonant) enhancement of the cross
section is found, 
giving clear access to $\lahhH$.  In order to quantify the
sensitivity to these triple Higgs couplings we chose to analyze the
event rates for the final states containing  four $b$-jets,  i.e.\  we
assumed the final $h$ bosons decays into $b\bar b$ pairs.  We then
evaluated the number of 'signal' events (corresponding to the
resonant diagrams involving $\lahhH$) and compared with the 'continuum'
events (corresponding to the other diagrams, which do not resonate),
convoluted the relevant acceptances and 
efficiencies.  Only benchmark points with low $m_H$ have a significant
dependence on $\lahhH$ in the $hhZ$ production channel. On the other
hand, the $hh\nu\bar\nu$ channel offers more significant dependence on
$\lahhH$, for all considered values of $m_H$,
in particular at the higher center-of-mass energies.  Therefore,
the $hh\nu\bar\nu$ channel shows significant sensitivity to $\lahhH$,
at CLIC. These 
findings were further substantiated by varying $\CBA$ for two of the
benchmark points (within the region allowed by all constraints). The
significance for $\lahhH$ in the $hh\nu\bar\nu$ channel scaled directly
with $|\lahhH|$.

Regarding $hH$ and $HH$ cross sections, they were evaluated for the highest
foreseen energy at CLIC, $\sqrt{s} = 3000 \gev$, as it is expected that
this center-of-mass energy will yield the best sensitivities to the BSM
triple Higgs couplings.  Some access to $\lahhH$ and $\lahHH$ may be
possible in the region close to threshold of
$m_{hH} \gsim m_h + m_H$ in the $hH\nu\bar\nu$ 
production channel.  Here $\lahHH$ was found to be more relevant than $\lahhH$. 
For $HH$ production also the $HHZ$ channel shows a relevant dependence
on $\lahHH$, but the number of expected events for the anticipated CLIC
luminosity remains too small for a measurement. $HH\nu\bar\nu$, on the
other hand, may give additional access to $\lahHH$, in particular in the
region close to threshold of $m_{HH} \gsim 2\,m_H$.

\smallskip
Overall we find that higher energy $e^+e^-$ colliders can offer access
to $\lahhh$ in the 2HDM similar to the SM case.  Depending on the BSM
Higgs-boson masses that lead to resonant BSM Higgs contributions in the
mass invariant distributions of di-Higgs production also a significant
dependence  on BSM triple Higgs couplings is found.
The best sensitivity found among all the triple Higgs couplings it to 
$\lahhH$  via $hh\nu \bar \nu$ production and reaches the highest values
at the highest energy colliders.  Considering all triple couplings,  ILC
and CLIC may both shed light on the Higgs potential of models with
extended Higgs-boson sectors.

  
\subsection*{Acknowledgements}

\begingroup 
The present work has received financial support from the `Spanish Agencia 
Estatal de Investigaci\'on'' (AEI) and the EU
``Fondo Europeo de Desarrollo Regional'' (FEDER) through the project
FPA2016-78022-P and from the grant IFT
Centro de Excelencia Severo Ochoa SEV-2016-0597.
F.A.\ and M.J.H.\  also aknowledge finantial support from the Spanish
``Agencia Estatal de Investigaci\'on'' (AEI) and the EU ``Fondo Europeo de
Desarrollo Regional'' (FEDER) 
through the project PID2019-108892RB-I00/AEI/10.13039/501100011033
and from the European Union's Horizon 2020 research and innovation
programme under the Marie Sklodowska-Curie grant agreement No 674896 and
No 860881-HIDDeN.
The work of S.H.\ was also supported in part by the
MEINCOP Spain under contract PID2019-110058GB-C21.
The work of F.A.\ was also supported by the Spanish Ministry of Science
and Innovation via an FPU grant with code FPU18/06634. 
\endgroup





\end{document}